\documentclass[a4paper,11pt]{article}
 \usepackage{jheppub}
 \usepackage[T1]{fontenc}
 \usepackage{lmodern}

\usepackage{lineno}
\usepackage{graphicx}
\usepackage{hyperref}
\usepackage{amssymb}
\usepackage{multirow}
\usepackage{soul}
\usepackage[low-sup]{subdepth}
\usepackage{amssymb}
\usepackage[dvipsnames]{xcolor}
\usepackage{exscale}
\usepackage{amsmath}
\usepackage{latexsym}
\usepackage{amsfonts}
\usepackage{float}
\usepackage{braket}

\usepackage{verbatim}
\usepackage{indentfirst}
\usepackage{tikz}

\usepackage{booktabs}
\usepackage{mathrsfs}
\usepackage{epsfig}
\usepackage{dcolumn}
\usepackage{bm}
\usepackage{color}
\usepackage{extarrows} 
\usepackage{subcaption}
\usepackage{changepage}
\usepackage{bigints}

\allowdisplaybreaks

\newcommand{\lsim}{{\;\raise0.3ex\hbox{$<$\kern-0.75em\raise-1.1ex\hbox{$\sim$}}\;}}
\newcommand{\gsim}{{\;\raise0.3ex\hbox{$>$\kern-0.75em\raise-1.1ex\hbox{$\sim$}}\;}}
\newcommand{\bea}{\begin{equation}}
\newcommand{\eea}{\end{equation}}
\newcommand{\beq}{\begin{equation}}
\newcommand{\eeq}{\end{equation}}
\newcommand{\ba}{\begin{array}}
\newcommand{\ea}{\end{array}}
\newcommand{\beqa}{\begin{equation}}
\newcommand{\eeqa}{\end{equation}}
\newcommand{\beqs}{\begin{subequations}}
\newcommand{\eeqs}{\end{subequations}}

\def\baa{\begin{array}}
\def\eaa{\end{array}}

\mathchardef\minus="002D
\def\nn{\nonumber}

\def\dis{\displaystyle}

\renewcommand{\rm}{\mathrm}
\def\MP{M_{\rm{Pl}}}

\def\geqq{\geqslant}
\def\({\left(}
\def\){\right)}
\def\[{\left[}
\def\]{\right]}

\def\pd{\partial}
\def\d{\td }

\def\to{\rightarrow}
\def\ito{\!\rightarrow\!}

\newcommand{\LG}{\mathcal{L}}

\newcommand{\td}{\text{d}}

\newcommand{\hc}{\text{ H.c. }}

\def\vs{\vspace*{1mm}}

\def\hp{\hspace*{0.3mm}}

\def\hsm{\hspace*{-0.3mm}}
\def\hsmx{\hspace*{-0.5mm}}

\def\hs{\hspace*{-0.05cm}}
\def\fNL{f_{\rm{NL}}}
\def\Hinf{H_\rm{inf}^{}}
\def\fNLCC{f_{\rm{NL}}^{\rm{CC}}}
\def\ii  {\rm{i}}

\def\tnu{\tilde{\nu}}
\def\tlambda{\tilde{\lambda}}
\def\dalpha{{\dot{\alpha}}}
\def\dbeta{{\dot{\beta}}}
\def\dgamma{{\dot{\gamma}}}

\def\sech{\text{sech}}
\def\csch{\text{csch}}

\title{\Large\hspace*{-3mm}Cosmological~Collider~Signatures~from~Right-Handed~Neutrino~Loop\hspace*{-6mm}}

\author[a]{Jingtao You,}
\author[a]{~Linghao Song,}
\author[b]{~Chengcheng Han,}
\author[a,c]{~Hong-Jian He,}
\author[d]{~Xingang Chen,}
\author[e]{~Zhong-Zhi Xianyu}

\affiliation[a]{Tsung-Dao Lee Institute \& School of Physics and Astronomy, \\
Key Laboratory for Particle Astrophysics and Cosmology,\\
Shanghai Key Laboratory for Particle Physics and Cosmology,\\ 
Shanghai Jiao Tong University, Shanghai, China}
\affiliation[b]{Department of Physics, Sun Yat-Sen University, Guangzhou 510275, China;\\
Asia Pacific Center for Theoretical Physics, Pohang 37673, Korea}
\affiliation[c]{Department of Physics, Tsinghua University, Beijing, China;\\
Center for High Energy Physics, Peking University, Beijing, China}
\affiliation[d]{Institute for Theory and Computation, 
Harvard-Smithsonian Center for Astrophysics,\\ 
60 Garden Street, Cambridge, MA 02138, USA}
\affiliation[e]{Department of Physics, Tsinghua University, Beijing 100084, China;\\
Peng Huanwu Center for Fundamental Theory, Hefei, Anhui 230026, China}

\emailAdd{119760616yjt@sjtu.edu.cn}
\emailAdd{lh.song@sjtu.edu.cn}
\emailAdd{hanchch@mail.sysu.edu.cn}
\emailAdd{hjhe@sjtu.edu.cn}
\emailAdd{xingang.chen@cfa.harvard.edu}
\emailAdd{zxianyu@tsinghua.edu.cn}

\abstract{\\[1mm]
We study cosmological collider (CC) signatures generated by right-handed neutrino loops in the setup of inflation combined with the neutrino seesaw mechanism.\ We formulate the inflaton interaction with the right-handed neutrinos through a unique dimension-5 operator respecting shift symmetry, which induces an effective chemical potential in the slow-roll background, leading to helicity-dependent production of right-handed neutrinos and enhancing the CC signatures.\ The right-handed neutrino is described by a two-component Weyl spinor with a Majorana mass term.\ Using the Schwinger-Keldysh formalism, we derive a set of seed integrals for fermion propagators of the right-handed Majorana neutrino.\ With these, we compute the factorized nonlocal contributions to the three-point inflaton correlator generated by the triangle loop of right-handed neutrinos.\ We show that the chemical potential can substantially soften the heavy-mass Boltzmann suppression and 
amplify the oscillatory non-Gaussianity signatures associated with the dominant helicity mode.\ We systematically analyze the spinor structure, the factorization of fermion triangle loop, the loop-momentum integrals and the time seed integrals, which give a controlled computation of the nonlocal contributions to the CC signals of the right-handed neutrinos.
\\[2mm]
[{\hp}arXiv:2605.21419{\hp}]
}

\makeatletter
\def\@fpheader{\relax}
\makeatother

\date{\today}

\begin{document} 
\maketitle
\flushbottom
\setcounter{page}{2}

\section{\hspace*{-2.5mm}Introduction}
\label{sec:1}
\label{introduction}

Recent studies have highlighted the inflationary universe and its quantum fluctuations \cite{Starobinsky:1980te,Guth:1980zm,Linde:1981mu,Linde:1983gd,Mukhanov:1981xt} as an important arena for probing high-energy particle interactions.
Unlike terrestrial experiments that are limited by achievable center-of-mass energies, inflation is sensitive to particles with masses comparable to the Hubble scale $H$.\ If such heavy states are present and interact with the inflaton, they can leave characteristic imprints on primordial curvature perturbations and the corresponding correlation functions.\ In particular, higher-point correlators of primordial fluctuations may exhibit non-analytic momentum-dependence in special kinematic limits, thereby encoding direct information about the masses, spins, and couplings of the particles present during inflation.\ This observation underlies the program of \emph{cosmological collider physics}\,\cite{Chen:2009we,Chen:2009zp,Baumann:2011nk,Noumi:2012vr,Arkani-Hamed2015,Chen:2012ge,Pi:2012gf,Gong:2013sma,Chen:2015lza,Baumann160703735,Chen:2016uwp,Kumar:2017ecc,Kumar:2019ebj,Alexander:2019vtb}, in which the primordial non-Gaussianity\,\cite{Chen:2010xka, Wang:2013zva, Meerburg:2019qqi, Achucarro:2022qrl} are used as a spectroscopic probe of ultraviolet physics in the early universe.\ Current CMB measurements have already placed constraints 
on the non-Gaussianity parameter at the level of 
$f_{\rm{NL}}\!\lesssim\! {O}(10)$, depending on the shape of the non-Gaussian signal.\
Future observations of large-scale structure may push this sensitivity down to 
${O}(1)\!-\!{O}(0.1)$, and potentially to ${O}(0.01)$ with 21{\hp}cm tomography \cite{Munoz:2015eqa,Meerburg:2016zdz,Meerburg:2019qqi,Achucarro:2022qrl}.

\vs 

A central prediction of cosmological collider physics is that massive fields can generate oscillatory or non-analytic power-law signals in momentum-space correlators of primordial fluctuations. 
Schematically, in the squeezed limit $k_L^{}\!\ll\! k_S^{}\hp$, one expects the 3pt correlator of the primordial curvature perturbation $\zeta$ to take the form \cite{Chen:2009zp,Noumi:2012vr,Arkani-Hamed2015}, 
\begin{equation}
\langle\zeta^3\rangle\simeq B_\zeta(k_L,k_S)
\sim
\mathcal{A}(k_L,k_S)
\!\hsm\(\!\!\frac{~k_L^{}\,}{\,k_S^{}\,}\hsm\!\)^{\!\!\alpha +\ii\hp\omega}
\!+\mathrm{c.c.}\,,
\end{equation}
where the oscillation frequency or power-law exponent set by $\omega$ is determined by the heavy-particle spectrum.\ 
Such non-analytic dependence on the momentum ratio $k_L/k_S$ cannot be generated 
by local effective interactions of light fields alone, and thus constitutes a distinctive signature 
of particle production and propagation during inflation.\ 
For this reason, these features are often referred to as the \emph{cosmological collider signals} or 
\emph{clock signals}.\ 
Hence, the detection of such cosmological collider signals would open up a new window to probe physics 
at energy scales far beyond those accessible in laboratory experiments.

\vs 

Realizing the goal of quantitatively probing physics at the inflation scale requires theoretical control over   
the relevant inflationary correlators.\  Similar to flat-space scattering amplitudes, inflationary correlators 
can be computed perturbatively with a Feynman-diagram expansion.\ 
In recent years, much progress has been made in analytic treatments of scalar-mediated cosmological collider signals.\ 
These developments mainly rely on a detailed understanding of scalar integrals associated with bulk scalar propagators.\ 
The tree-level processes were first tackled with a differential-equation-based 
approach\,\cite{Arkani-Hamed:2018kmz,Baumann:2019oyu,Baumann:2020dch,Baumann:2022jpr}, 
and the full tree-level solutions have been found 
later\,\cite{Xianyu:2023ytd,Fan:2024iek,Liu:2024str,Xianyu:2025lbk}.\ 
At the loop level, some new techniques have been developed for analytic integration, 
with a focus on the oscillatory signals, including partial or complete  Mellin-Barnes representations\,\cite{Sleight:2019hfp,Xianyu220501692,Qin:2022fbv,Qin:2024gtr}, 
spectral decomposition\,\cite{Xianyu:2022jwk,Zhang:2025nzd}, dispersion relations\,\cite{Liu:2024xyi,Werth:2024mjg}, 
and the factorization theorems and cutting rules for nonlocal signals\,\cite{Tong:2021wai,Xianyu230413295,Qin:2023nhv,Colipi-Marchant:2025oin,Das:2025qsh,Das:2026vfv}.\ 
These developments have improved both the conceptual understanding and the computational methods  
of scalar cosmological collider observables.

\vs 
 
In contrast, the studies on computing fermion-loop contributions remain very limited.\ 
This is because fermion loops involve nontrivial spinor structures, helicity dependence, and 
a more intricate analytic structure than scalar loops.\ 
These complexities make the quantitative treatment of fermion loops highly challenging.\ 
Some previous studies investigated heavy-fermion loop contributions to the primordial 
non-Gaussianity\,\cite{Chen:2018xck,Hook:2019zxa,Hook:2019vcn,Lu190707390,Aoki:2026olh}, but these 
calculations often rely on strong assumptions that oversimplify the complexity of fermion-loop integrals.\ 
In particular, existing estimates often adopt a saddle-point approximation 
for the hard internal propagator and infer the loop contribution 
from its characteristic non-analytic soft-momentum scaling, 
rather than evaluating the full loop integral explicitly.\  
While such approximations might capture the exponential parametric dependence, 
the associated power-law behaviors are still poorly understood.\ 
Hence, a more precise treatment of the fermion-loop integrals is necessary 
for reliable signal predictions.\ 
In this work, we perform a computation of the cosmological collider signal 
from the three-point inflationary correlator induced by fermion loops.\ 
With a systematic treatment of the spinor and helicity structures and by combining the recent 
techniques of scalar-loop integrals, we derive a controlled formula including both 
the exponential dependence and associated power-law dependence on the mass and chemical potential.\

\vs 

In addition to the technical difficulty, there is also an important question regarding the observability.\ 
For a heavy state with mass $m\!\gtrsim\!H$, the particle production
in the de\,Sitter space is typically suppressed exponentially, and this suppression 
is inherited by the nonlocal signal 
through an approximate Boltzmann factor, 
\begin{equation}
\exp(-\pi m/H) \hp. 
\end{equation}
As a result, even when heavy particles generate distinctive oscillatory signatures, the overall amplitude 
could be too small to be observable if their production rate is strongly suppressed.\ 
A particularly interesting possibility for fermions is that this suppression can be substantially reduced  
in the presence of a \emph{chemical potential} \cite{Chen:2018xck,Hook:2019zxa,Hook:2019vcn,Wang:2019gbi,Wang:2020ioa,Wang:2020uic,Bodas:2020yho,Sou:2021juh,Tong:2022cdz,Qin:2022fbv,Bodas:2024hih,Bodas:2025vpb,Aoki:2026olh}.\ 
In an inflationary background, derivative couplings between the inflaton and heavy fields can induce 
an effective chemical potential through the time-dependent inflaton background.\ For fermions, such a 
coupling may schematically take the following form,
\begin{equation}
\frac{1}{\,\Lambda\,}\,\partial_\mu\phi\, J^\mu ,
\end{equation}
where $J^\mu$ is a chiral fermionic current and 
$\Lambda$ denotes the cutoff scale of the effective inflaton-fermion interaction.\ 
Once the inflaton develops a homogeneous slow-roll background, $\dot{\phi}_0\!\neq\! 0\hp$, 
this interaction generates an effective chemical potential,
\begin{equation}
\lambda \sim \frac{\,\dot{\phi}_0\,}{\,\Lambda\,}.
\end{equation}
Physically speaking, the chemical potential induces helicity-dependent fermion production: one helicity state 
can be exponentially enhanced, while the other is suppressed.\ 
For sufficiently large $\lambda$, production of the dominant helicity mode can be significantly amplified, weakening the usual Boltzmann suppression and thereby enhancing the observational prospects for fermionic cosmological collider signals.

\vs

To demonstrate how this possibility is explicitly realized, 
it is important to identify a well-motivated new physics candidate for heavy fermions from particle physics.\ 
In this work, we take the heavy right-handed neutrinos with the canonical seesaw 
mechanism\,\cite{Minkowski:1977sc,GellMann:1980vs,Yanagida:1979as,Glashow:1979nm}  
as a compelling new physics benchmark for this approach, where the seesaw mechanism naturally explains 
the tiny masses of the light neutrinos as observed by the neutrino oscillation experiments, 
and the right-handed neutrinos provide the last missing piece in the structure of 
the Standard Model (SM) of particle physics.\ 
We note\,\cite{You:2024hit,Han:2024qbw} that the scale of the canonical neutrino seesaw 
is naturally of $O(10^{13-14})$GeV, lying around the upper range of the inflation energy scale.\ 
When the heavy right-handed neutrinos couple to the inflaton through a unique derivative operator of 
dimension-5 \cite{Chen:2018xck,Wang:2019gbi,You:2024hit,Han:2024qbw}, 
the rolling inflaton background can induce a chemical potential 
for the right-handed neutrinos and enhance their production during inflation.\ 
In this way, the right-handed neutrinos (with the canonical seesaw mechanism) serve as 
a unique heavy fermion candidate and can be tested by studying the cosmological collider physics.

\vs 

This setup also provides an important complement to the commonly studied bosonic cosmological collider signals.\ The heavy fermion signals originate from the  propagation of heavy fermions in the loop and manifest themselves through non-analytic structures in appropriate kinematic limits of the cosmological correlator.\ The oscillation frequency probes the heavy-fermion mass spectrum, whereas the overall oscillation amplitude depends on the strength of the inflaton coupling and the size of the induced chemical potential.\ Hence, the loop-induced fermionic cosmological collider signals can encode the information about both the mass-spectrum of heavy fermions and their underlying interaction dynamics.

\begin{figure}[t]
\centering
\includegraphics[width=0.9\textwidth]{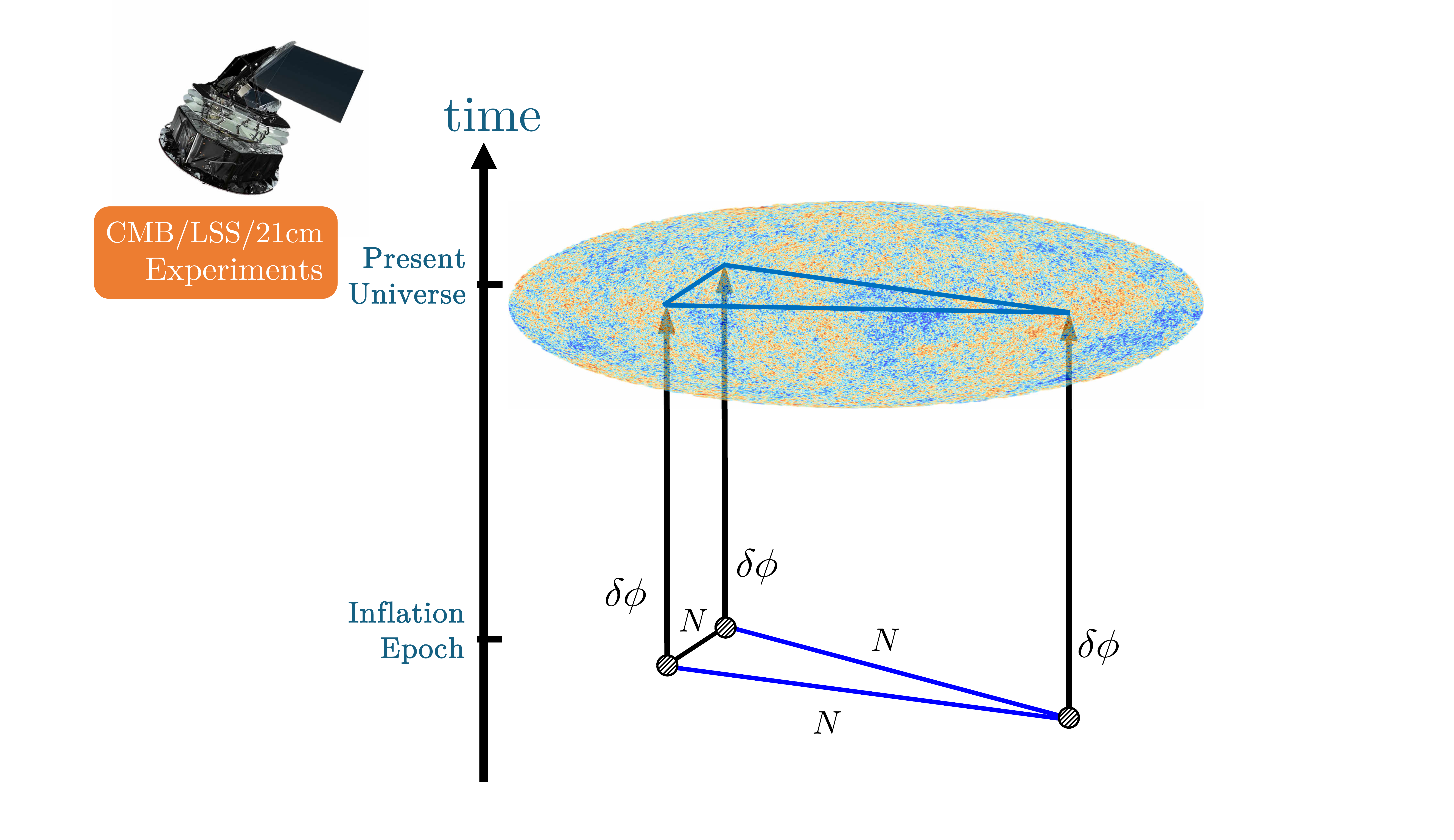}
\vspace*{-7mm}
\caption{\small Illustration of the primordial non-Gaussianity observable from the three-point inflaton correlator 
as generated by the heavy-neutrino triangle loop during inflation epoch and as (to be) measured 
in the present universe.}
\label{illustration}
\end{figure}

With these motivations, in this work we study the cosmological collider signatures generated 
by heavy right-handed neutrino loops through the inflaton-right-handed-neutrino interaction, 
where the right-handed neutrinos play the key role to realize the canonical neutrino seesaw mechanism.\ 
We will analyze how the right-handed neutrino loops modify the primordial correlation functions through their interaction with the inflaton.\ A key ingredient of our analysis is the chemical potential induced by the slow-roll inflaton background.\ We demonstrate that this chemical potential can exponentially enhance the production of one of the fermion helicity states and significantly amplify the resultant oscillatory signals in the primordial non-Gaussianity.\ 
Our analysis uses the two-component Weyl spinors and the Schwinger-Keldysh formalism, together with a factorized treatment of the loop diagram and a set of seed integrals for the fermion propagators.\ This provides a general framework for heavy-fermion loop signals in cosmological collider physics and demonstrates that the heavy right-handed neutrinos 
(whose masses set the seesaw scale) may leave observable imprints in the primordial non-Gaussianity, as illustrated in Fig.\,\ref{illustration}.\

\vs

In general, the full clock signal can receive both nonlocal and local contributions.\  
The nonlocal contribution originates from the long-distance propagation of gravitationally 
produced massive particles between separated bulk vertices and gives a characteristic 
non-analytic dependence on the external momenta in the soft limit.\ The local contribution 
arises from the short-distance part of the bulk dynamics.\ Although it may also contain 
oscillatory clock dependence, its coefficient and phase are controlled 
by short-distance dynamics and local interactions.
In the three-point squeezed limit, these two contributions can provide the same oscillatory 
momentum dependence and thus may interfere at the level of the total amplitude.\ 
For the computational part of this work, we compare with the estimate of \cite{Chen:2018xck}
and perform an improved computation of the nonlocal contribution from the fermion loops, 
including the computation of each factorized hard subdiagram and bubble-loop subdiagram.\ 
We estimate the nonlocal contribution in the late-time approximation for the bubble loop, 
which corresponds to the leading pole contribution in terms of the Mellin-Barnes representation.\ 
These improvements show the importance of the precise control over 
power-law parametric dependence in the final results, 
which can lead to order-of-magnitude changes from the previous estimates for signals 
with very large oscillation frequencies.\ 
The amplitude obtained in this work provides the best currently available estimate of the CC signal 
from the fermion loop contributions, and we will use it to discuss the observability of the CC signal.\ 
Nonetheless, we caution that it is possible that including contributions from other poles and the local contribution 
may change the order of magnitude of the result.\
On the other hand, the local contribution is technically much more difficult to compute at present, 
and we leave the remaining challenges for future work.\

The rest of this paper is organized as follows.\ 
In Section\,\ref{sec:2}, we introduce the inflationary setup with right-handed neutrinos, derive the effective chemical potential induced by the derivative coupling of the inflaton and right-handed neutrinos, and present the fermion propagators and the associated seed integrals in the inflationary background.\ In Section\,\ref{sec:3}, we compute the three-point inflaton correlator generated by the right-handed neutrino loops and extract the non-analytic cosmological collider signals.\ In Section\,\ref{sec:4}, we analyze the dependence of the cosmological collider signals on the heavy fermion mass, chemical potential, and coupling parameters.\ With these, we present estimates for the 
three-point non-Gaussian amplitude $\fNL$.\ Finally, we summarize our conclusions in Section\,\ref{Conclusion}.\ 
In Appendices\,\ref{app:A}-\ref{app:C}, we provide the technical derivations as needed for the analyses in the main text.\
In Appendix\,\ref{app:D}, we give a brief review to describe the main concepts 
and ingredients of the cosmological collider physics,  
which makes the current presentation accessible to a wider readership.

\vspace*{3mm}
\noindent 
\textbf{Notation and Conventions:}
\\[2mm]
We consider that the background spacetime is the inflationary patch of de\,Sitter spacetime, which provides the 
standard approximation to slow-roll inflation.\ We use the coordinates $(\tau,\mathbf{x})$, where 
$\tau\!\in\!(-\infty,0)$ is the conformal time and $\mathbf{x}$ denotes the comoving spatial coordinates.\ 
The metric tensor is given by the distance-square,
\begin{equation}
\d s^2 = a^2(\tau)(\hsm -\d\tau^2 \!+\hsm \d\mathbf{x}^2)
=\frac{\,-\td\tau^2 \!+\hsm \d\mathbf{x}^2\,}{(H\tau)^2}\,,
\end{equation}
where $a(\tau) \!=\! -1/(H\tau)$ is the scale factor in de\,Sitter spacetime, and the Hubble parameter $H$ (also denoted as $H_\rm{inf}$) during inflation is taken to be constant.\
For any functions $f(\tau)$ and $\phi(t)$, we denote the time-derivatives 
$f'(\tau)\!=\hsm \d f(\tau)/\d\tau\hp$
and $\dot{\phi}(t)\!=\hsm \d \phi(t)/\d t\hp$.\  

\vs 

Three-dimensional vectors will be denoted in boldface, for example, a spatial momentum is written as $\mathbf{k}$ 
and its magnitude is defined as $k\!=\!|\mathbf{k}|$.\ 
For clarity, some definitions of the momentum variables and momentum ratios 
(used in the following presentation) are given as follows: 
\begin{equation}
\label{momentum_{(d)}ef}
\begin{aligned}
k_s \hsm\equiv\hsm \lvert \mathbf{k}_1 \!+\! \mathbf{k}_2 \rvert\hp, ~~~
k_t \hsm\equiv\hsm \lvert \mathbf{k}_1 \!+\! \mathbf{k}_4 \rvert\hp, ~~~
k_{ij} \hsm\equiv\hsm k_i \!+\! k_j\hp, ~~~
r_1 \hsm\equiv\hsm k_s / k_{12}\hp, ~~~
r_2 \hsm\equiv\hsm k_s / k_{34}\hp,
\end{aligned}
\end{equation}
where the subscripts $i,j=1,2,3,\cdots$.\ 

\vs

The generalized hypergeometric function ${}_pF_q$ is defined as follows:
\begin{align}
{}_pF_q
\left[\!
\begin{array}{c}
a_1,\ldots,a_p\\
b_1,\ldots,b_q
\end{array}
\,\middle|\, z
\right]
=
\sum_{n=0}^{\infty}\!
\frac{\,(a_1)_n\cdots(a_p)_n\,}{(b_1)_n\cdots(b_q)_n}
\frac{\,z^n\,}{n!}\,,
\end{align}
where $(a)_n \!=\! \Gamma(a\!+\!n)/\Gamma(a)$ is the Pochhammer symbol.\ 
The corresponding regularized hypergeometric function ${}_p\tilde{F}_q$ 
is defined in the following formula:
\begin{align}
{}_p\tilde{F}_q
\left[
\begin{array}{c}
a_1,\ldots,a_p\\
b_1,\ldots,b_q
\end{array}
\,\middle|\, z
\right]
=
\frac{1}{\,\Gamma[b_1,\ldots,b_q]\,}
\,{}_pF_q
\left[
\begin{array}{c}
a_1,\ldots,a_p\\
b_1,\ldots,b_q
\end{array}
\,\middle|\, z
\right]\!.
\end{align}

\vspace*{2mm}
\section{\hspace*{-2.5mm}Setup of Inflation with Neutrino Seesaw}
\label{sec:2}

In the canonical neutrino seesaw mechanism, for a natural Yukawa coupling $y_{\nu}=O(1)$, 
the seesaw scale is typically around $10^{14}\,$GeV, which is a brand-new scale 
beyond the SM and can be naturally embedded in grand unification.\ 
Interestingly, this natural high-scale seesaw is also compatible with the upper range of the inflation scale\,\cite{BICEP2:2019upn,Planck9,Planck10}, characterized by the Hubble parameter $H$ during inflation.\ 
This observation motivates us to consider the appealing possibility that the inflaton field $\phi$ couples directly to the right-handed neutrino $N$.\ Such a coupling could provide an intriguing opportunity to probe the natural seesaw mechanism through inflationary observables in the early universe. 

In this section, we explore a natural interaction between the inflaton and right-handed neutrinos in a general slow-roll inflationary setup. We show that, in the presence of the inflaton background, this interaction alters the neutrino dynamics by generating an effective chemical potential. We also provide the fermion propagators in the Schwinger–Keldysh (SK) formalism, together with the associated seed integrals that will be employed in the computation of inflationary correlators in the following sections.

\subsection{\hspace*{-2.5mm}Inflaton Interaction with Right-Handed Neutrino}
\label{sec:2.1}
\label{Interaction between the Inflaton and the Right-handed Neutrino}

In many inflation models, the inflaton potential $V(\phi)$ must be sufficiently flat to satisfy the slow-roll conditions\,\cite{Linde:1981mu,Linde:1983gd} and agree with observations\,\cite{BICEP2:2019upn,Planck9,Planck10,BICEP3}.\ An approximate shift symmetry is often introduced to protect this flatness.\ As a result, the direct coupling between the inflaton and the right-handed neutrino, $y_\phi \phi (NN\!+\!\!\hc\!\!)$, is expected to be suppressed\footnote{The contraction of two Weyl spinors $\(N N\)$ is defined by $\(N N \equiv N^\alpha N_\alpha\)$, in accordance with the convention adopted in Ref.\,\cite{Dreiner:2008tw}.}.\ If instead one considers a derivative coupling between the inflaton and the right-handed neutrino, the leading interaction naturally arises as a dimension-5 operator.\ 
The relevant Lagrangian in the Friedmann-Lemaître-Robertson-Walker (FLRW) spacetime
(or the de\,Sitter spacetime)\footnote{We give a brief discussion about the Lagrangian of the Weyl spinors in curved spacetime 
in the Appendix\,\ref{details: helicity and eom}.} is given by
\begin{equation}
\begin{aligned}
\label{Total_Lagrangian_weyl}
\LG & = \sqrt{-g\,}
\bigg[\hsm\!-\!\frac{1}{2}\hp\partial_{\mu}\phi\hp\partial^{\mu}\phi\hp
-\!V(\phi)+\frac{\ii}{a(\tau)}  N^\dagger\bar{\sigma}^\mu\text{D}_\mu N + 
\frac{\ii}{a(\tau)} \nu_\text{L}^\dagger\bar{\sigma}^\mu\text{D}_\mu \nu_\text{L}^{}
\\
& \hspace*{13.5mm}
-\!\frac{1}{\,a(\tau)\Lambda\,} \pd_\mu \phi N^\dagger \bar{\sigma}^\mu N+\!\(\!\hsm -\frac{1}{2}M N\!N
\!- \frac{y_\nu h}{\sqrt{2}}\, \nu_\text{L} N +\!\!\hc\!\!\!\hsm\)\!\!\bigg]\hsm .
\end{aligned}
\end{equation}
In the above equation, $ V(\phi) $ denotes the inflaton potential, which is assumed to be sufficiently flat such that the inflaton can be treated as effectively massless during inflation. The fields $ \nu_\text{L} $ and $ N $ correspond to the left-handed and right-handed neutrinos, respectively, both of which are described using two-component Weyl spinors. The operator $\mathrm{D}_\mu $ represents the covariant derivative acting on spinors in de Sitter spacetime.\  
Finally, $ h $ denotes the remaining physical component of the Higgs field in the unitary gauge.\ 
The dimension-5 coupling can be expressed as follows:
\begin{equation}
\label{dimension5_operator}
    \Delta\mathcal{L}
    = \sqrt{-g\,}\,\frac{-1}{a(\tau)\Lambda} \,\partial_\mu \phi \, N^\dagger \bar{\sigma}^\mu N
    = a^3(\tau)\,\frac{-1}{\Lambda}\,\partial_\mu \phi \, N^\dagger \bar{\sigma}^\mu N \,,
\end{equation}
where we have used $\sqrt{-g\,} = a^4(\tau)$ for de Sitter space in conformal coordinates.\
Here, $\partial_\mu \phi$ denote derivatives with respect to conformal coordinates.

In the previous work\,\cite{Han:2024qbw,You:2024hit}, we studied the prediction of local-type non-Gaussianity $\fNL^\text{local}$ generated through Higgs-modulated reheating in the neutrino seesaw framework.\ 
In the present work, we instead investigate the impact of the dimension-5 coupling on the quantum fluctuations of the inflaton.\ 
This effect is encoded in the cosmological collider signals induced by the derivative coupling between the inflaton and the right-handed neutrino during inflation. Before proceeding to the calculation, we first introduce the theoretical framework for fermions in an inflationary background.
 
\subsection{\hspace*{-2.5mm}Fermions in the Slow-Roll Inflationary Background}
\label{sec:2.2}
\label{Fermions in an Inflationary Slow-roll Background}

In Eq.\eqref{Total_Lagrangian_weyl},  we introduce a minimal model of the canonical seesaw combined with the inflaton.\ Within this scenario, the right-handed neutrino, with a Majorana mass term $M$, is coupled to the inflaton through a dimension-5 operator.\ In this section, we study the dynamics of the neutrino field during inflation.\ We will illustrate that the right-handed neutrino would experience an effective chemical potential sourced from the inflaton background due to the dimension-5 coupling. 

During inflation, the inflaton field $\phi(t,\mathbf{x})$ is decomposed into a homogeneous background part $\phi_0(t)$ and a quantum fluctuation $\delta\phi(t,\mathbf{x})$,
\begin{equation}
\phi(t,\mathbf{x})=\phi_0(t)+\delta\phi(t,\mathbf{x})\,.
\end{equation}
Depending on the type of inflation models, the background inflaton field could be much larger or smaller than the Planck scale.
In a slow-roll background, the rolling rate of $\phi_0^{}(t)$, $\dot{\phi}_0\equiv\partial_t {\phi}_0$ with respect to the physical time $t$, satisfies the slow-roll condition and is also constrained by observations through the dimensionless power spectrum $P_{\zeta}^{}$ 
of the comoving curvature perturbation $\zeta$ as follows:
\begin{equation}
P_{\zeta}=\frac{H^{4}}{\,4 \pi^{2} \dot{\phi}_0^{2}\,}
=\frac{H^{2}}{\,8\pi^2\epsilon M_{\rm{Pl}}^2\,}\simeq 2\hsm\times\! 10^{-9}\,,
\end{equation}
where the slow-roll parameter $\epsilon$ is given by 
\begin{equation}
\epsilon\equiv-\frac{\dot{H}}{\,H^2\,}
=\frac{\dot{\phi}_0^2}{\,2H^2\MP^2\,} \,.
\end{equation}
Using the observed amplitude of the scalar power spectrum, one finds that the rolling rate of the inflaton background is approximately given by 
\begin{equation}
\label{eq:dot(phi)-value}
\dot{\phi}_0\simeq (60H)^2\,.
\end{equation}

In this case, the dimension-5 coupling to the inflaton induces a nontrivial modification of the quadratic part of the neutrino Lagrangian. In particular,
\begin{align}
\label{lagrangian_neutrino}
\Delta\mathcal{L}
&= \sqrt{-g\,}\,\frac{-1}{\,a(\tau)\Lambda\,} \,\partial_\mu \phi \, N^\dagger \bar{\sigma}^\mu N 
\nn\\
&\supset \sqrt{-g\,}\,\frac{-1}{\,a(\tau)\Lambda\,} \,\partial_0 \phi_0 \, N^\dagger \bar{\sigma}^0 N 
\nn\\
&= -a^4(\tau)\,\frac{\,\dot{\phi}_0\,}{\Lambda}\, N^\dagger \bar{\sigma}^0 N \,,
\end{align} 
where $\partial_0 \phi_0 = \partial_\tau \phi_0(\tau)$ denotes the derivative of the inflaton background $\phi_0(\tau)$ with respect to conformal time $\tau$, while $\dot{\phi}_0$ denotes the derivative with respect to physical time $t$ (referred to as the rolling rate in the above discussion). These are related by $\partial_0 \phi_0 = a\,\dot{\phi}_0$. We may then write the quadratic terms for the right-handed neutrino in Eq.\eqref{Total_Lagrangian_weyl} as
\begin{equation} 
\label{Lagrangian_weyl_free}
\Delta\LG
=
a^3 N^\dagger \ii D\!\!\!\!/\, N
\!-\!
a^4 \lambda N^\dagger \bar{\sigma}^0 N
\!-\!
\frac{1}{2} a^4 M_{\!R} (N N \!+\!\! \hc\!\!)\,,
\end{equation}
where $M_{\!R}$ is the heavy neutrino mass, and $\lambda\!\equiv\!\dot{\phi}_0/\Lambda$ arises from the dimension-5 coupling between the inflaton and the right-handed neutrino and plays the role of an effective chemical potential for the right-handed neutrino.\ It is convenient to redefine the field as $N(x)\!=\!a^{-3/2}\tilde{N}(x)\hp$, so that the covariant derivative acting on $N$ is reduced to an ordinary partial derivative acting on $\tilde{N}$. In terms of $\tilde{N}$, the above Lagrangian takes the following form: 
\begin{equation}
\label{free_lagrangian}
\Delta\LG
=
\tilde{N}^\dagger i\partial\!\!\!/\tilde{N}
-
a\lambda \tilde{N}^\dagger \bar{\sigma}^0 \tilde{N}
-
\frac{1}{2} a M_{\!R} (\tilde{N} \tilde{N} \!+\!\! \hc\!\!)\,.
\end{equation}
The equation of motion is obtained by varying the Lagrangian 
with respect to $\tilde{N}^\dagger$, yielding
\begin{equation}
i \bar{\sigma}^\mu \partial_\mu \tilde{N}
=
a\lambda \bar{\sigma}^0 \tilde{N}
+
a M_{\!R} \tilde{N}^\dagger\,.
\end{equation}

Then we decompose the spinor $\tilde{N}$ into eigenmodes of the three-momentum,
\begin{equation}
\label{mode_expansion}
\begin{aligned}
\tilde{N}_\alpha(\tau, \mathbf{x}) &= 
\int\!\! \frac{\td^3 \mathbf{k}}{(2\pi)^3} 
\sum_{s=\pm}\! \left[ \xi_{ s,\alpha} (\tau, \mathbf{k}) b_s (\mathbf{k}) e^{\ii \mathbf{k} \cdot \mathbf{x}} \!+\hsm \chi_{ s,\alpha} (\tau, \mathbf{k}) b_s^{\dagger} (\mathbf{k}) e^{-\ii \mathbf{k} \cdot \mathbf{x}} \right]\!,
\\
\tilde{N}^\dagger_\dalpha(\tau, \mathbf{x}) &= 
\int\!\! \frac{\td^3 \mathbf{k}}{(2\pi)^3} 
\sum_{s=\pm}\! \left[ \xi^\dagger_{ s,\dalpha} (\tau, \mathbf{k}) b_s^{\dagger} (\mathbf{k}) e^{-\ii \mathbf{k} \cdot \mathbf{x}} \!+\hsm  \chi^\dagger_{s,\dalpha} (\tau, \mathbf{k}) b_s (\mathbf{k}) e^{\ii \mathbf{k} \cdot \mathbf{x}} \right]\!,
\end{aligned}
\end{equation}
where $s\!=\!\pm 1$ denotes the sign of the spinor helicity, and $b_s$ and $b_s^\dagger$ are the annihilation and creation operators obeying the anticommutation relation,
\begin{equation}
\label{anticommutation relation}
\{b_s(\mathbf{k}), b_{s'}^\dagger(\mathbf{k'})\} = (2\pi)^3 \delta_{ss'} 
\delta^3(\mathbf{k} \!-\! \mathbf{k'}) \, .
\end{equation}
The mode functions $\xi_{s,\alpha}(\tau,\mathbf{k})$ and $\chi_{s,\alpha}(\tau,\mathbf{k})$ are associated with definite helicity. We decompose them in terms of the unit helicity eigenspinors $h_s$ as
\begin{equation}
\xi_{s,\alpha}(\tau, \mathbf{k}) = u_s(\tau,k) \hp h_{s,\alpha}(\mathbf{k})\hp , 
\qquad
\chi^{\dagger\dot\alpha}_{s}(\tau, \mathbf{k}) = v_s(\tau,k) \hp s \hp h^{\dagger\dot\alpha}_{-s}(\mathbf{k}) \hp .
\end{equation}
The helicity basis eigenspinors $h_s$ satisfy
\begin{equation}
\label{property_helicity}
\vec{\sigma} \hsm\cdot\!\hat{\,\bf k} \, h_s(\mathbf{k}) = s \, h_s(\mathbf{k}), 
\qquad
h_s^\dagger(\mathbf{k}) h_{s'}(\mathbf{k}) = \delta_{ss'}, 
\qquad
\sum_{s=\pm1}\!\! h_s(\mathbf{k}) h_s^\dagger(\mathbf{k}) = 1 .
\end{equation}
For the spatial momentum along the direction
\begin{equation}
\hat{\,\mathbf{k}}=(\sin\theta\cos\varphi,\hp 
\sin\theta\sin\varphi,\hp \cos\theta) \hp ,
\end{equation}
the helicity eigenspinors can be chosen as follows:
\begin{equation}
\label{helicity_value}
h_+^{}(\theta,\varphi)=
\!\begin{pmatrix}\!
e^{-\ii \frac{\varphi}{2}}\!\cos\!\frac{\theta}{2} \\[1mm]
\!
e^{\ii \frac{\varphi}{2}}\sin\!\frac{\theta}{2}
\end{pmatrix}\!,
\quad~~
h_-^{}(\theta,\varphi)=
\!\begin{pmatrix}
\!-e^{-\ii \frac{\varphi}{2}}\!\sin\!\frac{\theta}{2} \\
e^{\ii \frac{\varphi}{2}}\!\cos\!\frac{\theta}{2}
\end{pmatrix}\!.
\end{equation}
More details about these helicity eigenspinors
and the derivation of their relations are given in the Appendix\,\ref{details: helicity and eom}.

Then, from the equations of motion we derive  the equations for the $\tau$-dependent 
spinor coefficients $u_s$ and $v_s$ 
(with $s\hsm =\hsm\pm$):
\beqs 
\label{eq:uv-1st}
\begin{align}
\ii\hp u_\pm' \pm k u_\pm 
&= a \lambda u_\pm \!+\hsm a M_{\!R} v_\pm \,, 
\\
\ii\hp v_\pm' \mp k v_\pm 
&= - a \lambda v_\pm \!+\hsm a M_{\!R} u_\pm \,,
\end{align}
\eeqs
where $u'\!=\!\td u/\td\tau$ and 
$v'\!=\!\td v/\td\tau$.\ 
To solve the above mixed differential equations,
we recast them in the following second-order differential equations on $u_\pm^{}$ and $v_\pm^{}$ respectively,
\beqs 
\begin{align}
\label{Eq:2.18}
u_\pm'' \!-\hsm a H u_\pm' 
\!+\!\Big[\hsm (\pm k \!-\hsm a\lambda)^2 
\!+\hsm a^2 M_{\!R}^2 \!\pm\hsm \ii\hp a H k \Big] u_\pm &= 0 \,,
\\
v_\pm'' \!-\hsm a H v_\pm' 
\!+\! \Big[\hsm (\mp k \!+\hsm a\lambda)^2 
\!+\hsm a^2 M_{\!R}^2 \!\mp\hsm \ii\hp a H k \Big] v_\pm &= 0 \,. 
\end{align}
\eeqs  
Imposing the Bunch-Davies initial condition 
(for $\tau\!\to\!-\infty$)
together with the canonical normalization, 
the spinor coefficients $u_s$ and $v_s$ are solved
as follows:
\begin{equation}
\label{modefunction}
\begin{aligned}
u_+( \tau,k) &=  \frac{e^{+\pi \tlambda/2}}{\sqrt{-2k\tau\,}\,} \tilde{M}_{\!R}\hp
W_{\!-\frac{1}{2} - \ii\tlambda,\ii \tnu}
(\ii\hp 2k\tau)\,, 
\\
u_-( \tau,k)
&= \ii\frac{e^{-\pi \tlambda/2}}{\sqrt{-2k\tau\,}\,} 
W_{\!\frac{1}{2} + \ii\tlambda,\ii \tnu}
(\ii\hp 2k\tau)\,, 
\\
v_+( \tau,k) &=\ii \frac{e^{+\pi \tlambda/2}}{\sqrt{-2k\tau\,}\,} 
W_{\!\frac{1}{2} - \ii\tlambda,\ii \tnu}
(\ii\hp 2k\tau)\,,
\\
v_-( \tau,k) &=  \frac{e^{-\pi \tlambda/2}}{\sqrt{-2k\tau\,}\,} \tilde{M}_{\!R}\hp
W_{\!-\frac{1}{2} + \ii\tlambda,\ii \tnu}
(\ii\hp 2k\tau)\,,
\end{aligned}
\end{equation}
where $W_{\kappa,\mu}(z)$ denotes the Whittaker function, and
\begin{equation}
\label{eq:tm-tlambda-tnu}
\tilde{M}_{\!R} = \frac{M_{\!R}}{\hp H\,}\hp , 
\quad~~ 
\tlambda = \frac{\lambda}{\hp H\,}\hp , 
\quad~
\lambda = \frac{\,\dot{\phi}_0^{}\,}{\Lambda}\hp, 
\quad~
\tnu = \!\sqrt{\tilde{M}_{\!R}^2 \!+\!\tlambda^2\,}\,.
\end{equation}
On the other hand, the Whittaker function $W_{\kappa,\mu}(z)$ can be expressed 
in terms of Kummer's confluent hypergeometric function ${}_1\mathrm{F}_1(a;b;z)$ as follows:
\begin{equation}
\label{WtoM}
W_{\kappa,\mu}(z)=
\frac{\Gamma(-2\mu)\hp e^{-z/2}z^{\mu+\frac{1}{2}}\,}
{\,\Gamma(\frac{1}{2}\!-\!\mu\!-\!\kappa)\,}
{}_1\mathrm{F}_1\!\!
\(\frac{1}{2}\!+\!\mu\!-\!\kappa;1\!+\!2\mu;z\)
\!+\hsm (\mu\!\ito\! -\mu) \,,
\end{equation}
where the function ${}_1\mathrm{F}_1(a;b;z)$
is defined as 
\begin{equation}
{}_1\mathrm{F}_1(a;b;z)=
\sum_{n=0}^{\infty}\!
\frac{~(a)_n^{}z^n~}{~(b)_n^{}\hp n!~}\,, 
\end{equation}
with $(a)_n^{}$ being the rising Pochhammer symbol,
\begin{equation}
(a)_n\equiv\frac{\,\Gamma(a\!+\!n)\,}{\Gamma(a)}\,.
\end{equation}
Accordingly, the spinor coefficients $u_s$ and $v_s$ can be rewritten as follows:
{\small
\begin{align}
\hspace*{-3mm}
u_+( \tau,k) &= 
\frac{\,e^{-\ii \pi /4 +\pi \tlambda/2} \tilde{M}_{\!R}e^{\pi\tnu/2} \Gamma(-\ii\hp 2\tnu)\,}{\Gamma(1 \!+\hsm \ii \tlambda \!-\hsm \ii\hp \tnu)} (-2k\tau)^{\ii \tnu}e^{-\ii k \tau}{}_1\mathrm{F}_1(1\!+\!\ii\tlambda\!+\!\ii\tnu;
1\!+\!\ii\hp 2\tnu;\ii\hp 2 k\tau) 
\!+\! (\tnu \ito -\tnu)\,,
\nonumber
\\
\hspace*{-3mm}
u_-( \tau,k)&= \ii\frac{e^{-\ii \pi /4 - \pi\tilde{\lambda}/2}e^{\pi\tnu/2} \Gamma(-\ii\hp 2\tnu)}{\Gamma(-\ii \tlambda - \ii \tnu)} (-2k\tau)^{\ii \tnu}e^{-\ii k \tau}{}_1\mathrm{F}_1(-\ii\tlambda+\ii\tnu;1+\ii\hp 2\tnu;\ii\hp 2k\tau) + (\tnu \rightarrow -\tnu)\,, 
\nn
\\
\hspace*{-3mm}
v_+( \tau,k) &= \ii\frac{e^{-\ii \pi /4 + \pi\tilde{\lambda}/2}e^{\pi\tnu/2} \Gamma(-\ii\hp 2\tnu)}{\Gamma(\ii \tlambda - \ii \tnu)} (-2k\tau)^{\ii \tnu}e^{-\ii k \tau}{}_1\mathrm{F}_1(\ii\tlambda+\ii\tnu;1+\ii\hp 2\tnu;\ii\hp 2k\tau) + (\tnu \rightarrow -\tnu)\,,
\label{mode_function_M}
\\
\hspace*{-3mm}
v_-( \tau,k) &= 
\frac{e^{-\ii \pi /4 - \pi\tilde{\lambda}/2} \tilde{M}_{\!R}e^{\pi\tnu/2} \Gamma(-\ii\hp 2\tnu)}{\Gamma(1\!-\!\ii\hp \tlambda \!-\! \ii\hp \tnu)} (-2k\tau)^{\ii \tnu}e^{-\ii k \tau}{}_1\mathrm{F}_1(1\!-\!\ii\tlambda\!+\!\ii\hp\tnu;1\!+\!\ii\hp 2\tnu;
\ii\hp 2k\tau) \!+\! (\tnu \ito -\tnu)\,. 
\nn
\end{align}
}
\hspace*{-2mm}
Note that the Whittaker function has a branch cut in the $z$-plane.\ As a result, performing a Wick rotation in the calculation of the correlation function is not entirely straightforward. In contrast, the function ${}_1\mathrm{F}_1(a;b;z)$ is regular in the $z$-plane, which makes the Wick rotation simpler and facilitates the evaluation of the desired correlation function.

To extract the oscillatory CC signals, it is useful to determine the late-time behavior of these spinor coefficients.\ In the soft limit or late time limit $|k\tau|\ito 0\hp$, their asymptotic forms are given by
\\[-5mm]
{\small
\begin{equation}
\label{latetime_mode}
\begin{aligned}
    u_{\text{L}+}^{}( \tau, k) 
    &\simeq e^{-\ii \pi /4 +\pi \tlambda/2} \frac{~\tilde{M}_{\!R}\hp e^{\pi\tnu/2} 
    \Gamma(-\ii\hp 2\tnu)~}
    {\,\Gamma(1 \!+\! \ii\hp\tlambda \!-\! 
    \ii\hp\tnu)} (-2k\tau)^{\ii\hp\tnu} 
    \!+\! (\tnu \ito -\tnu)\,,
    \\
    u_{\text{L}-}( \tau, k) &\simeq \ii e^{-\ii \pi /4 - \pi\tlambda/2} \frac{e^{\pi\tnu/2} \Gamma(-\ii\hp 2\tnu)}{\Gamma(-\ii \tlambda - \ii \tnu)} (-2k\tau)^{\ii \tnu} + (\tnu \rightarrow -\tnu), 
    \\
    v_{\text{L}+}( \tau,k) &\simeq \ii e^{-\ii \pi /4 + \pi\tlambda/2} \frac{e^{\pi\tnu/2} \Gamma(-\ii\hp 2\tnu)}{\Gamma( \ii \tlambda - \ii \tnu)} (-2k\tau)^{\ii \tnu} + (\tnu \rightarrow -\tnu),
    \\
    v_{\text{L}-}( \tau, k) &\simeq e^{-\ii \pi /4 - \pi\tlambda/2} \tilde{M}_{\!R}\frac{e^{\pi\tnu/2} \Gamma(-\ii\hp 2\tnu)}{\Gamma(1-\ii \tlambda - \ii \tnu)} (-2k\tau)^{\ii \tnu} + (\tnu \rightarrow -\tnu)\,, \\
\end{aligned}
\end{equation}
}
\hspace*{-2.5mm}
with the $\displaystyle u_{\rm{L}\pm}( \tau, k)\equiv \lim_{|k\tau|\to 0}\!u_{\pm}^{}
(\tau,k)$ and
$\displaystyle v_{\rm{L}\pm}( \tau, k)\equiv\lim_{|k\tau|\to 0}\!v_{\pm}^{}
(\tau,k)$.

\vs 

In the next section, we will introduce the propagators of the Weyl spinor, 
$D(\tau_1,\tau_2)$, 
which are constructed from products of 
two spinor coefficients.\ The late time limit of a ``propagator'' 
contributes to the CC signal.\ For instance, for the propagator  
$D(\tau_1,\tau_2)\simeq u_{+}( \tau_1, k) u^\dagger_{+}( \tau, k)$, it has the late time limit,
{\small 
\begin{equation}
\begin{aligned}
u_{\text{L}+}( \tau_1, k) u^\dagger_{\text{L}+}( \tau_2, k)
&=\frac{-e^{\pi\tlambda}
\big[\Gamma(-\ii\hp 2\tnu)\big]^2}
{~\Gamma(\ii \tlambda\!-\!\ii\hp \tnu)
\Gamma(-\ii \tlambda\!-\!\ii\hp \tnu)~}(4k^2\tau_1\tau_2)^{\ii \hp \tnu}+(\tnu \ito -\tnu)
\\
& \quad +\frac{-e^{\pi\tlambda+\pi\tnu}
\hp\tilde{M}_{\!R}^2\hp\Gamma(-\ii\hp 2\tnu)
\Gamma(\ii\hp 2\tnu)}
{~\Gamma(1\!+\!\ii \tlambda-\ii\hp \tnu)
\Gamma(1\!-\!\ii \tlambda\!+\!\ii\hp\tnu)~}
\!\!\(\!\frac{\tau_1}{\,\tau_2\,}\!\)^{\!\!\ii\hp\tnu}
\!\hsm +\hsm (\tnu \ito -\tnu)
\\
&=\left[u_+ (\tau_1,k) u_+^\dagger (\tau_2,k)\right]_\text{(NLoc)}
\!\!+\!\left[u_+ (\tau_1,k) u_+^\dagger (\tau_2,k)\right]_{\rm{(Loc)}}.
\end{aligned}    
\end{equation}
}

Similar to the case of the scalar field, the late time limit of a fermion propagator 
could also be divided into the nonlocal part proportional to $(\tau_1\tau_2)^{\ii \hp\tnu}$ 
(accompanied by a non-analytic dependence on $k$) and the local part proportional to  
$\left(\! \frac{\,\tau_1\,}{\,\tau_2\,} \!\right)^{\!\ii\hp \tnu}$.\ 
In this work, we are interested in the nonlocal part of the cosmological collider signal 
generated by the right-handed neutrino, so we extract the nonlocal contribution 
by retaining only the terms proportional to 
$(\tau_1\tau_2)^{\ii \hp\tnu}$ and $(\tau_1\tau_2)^{-\ii \hp\tnu}$.\ 
For instance, for the product 
$u_+(\tau_1,k)u_+^\dagger(\tau_2,k)$, 
the nonlocal part (denoted by the abbreviation ``NLoc'') is given by
{\small
\begin{equation}
\begin{aligned}
\left[u_+ (\tau_1,k) u_+^\dagger (\tau_2,k)\right]_\text{(NLoc)}
& =\frac{-e^{\pi\tlambda}
\big[\Gamma(-\ii\hp2\tnu)\big]^2}
{~\Gamma(\ii\hp \tlambda\!-\!\ii\hp\tnu)
\Gamma(-\ii\hp\tlambda\!-\!\ii\hp\tnu)~}(4k^2\tau_1\tau_2)^{\ii \hp \tnu}
\!+\!(\tnu \ito -\tnu)
\\[1mm]
&=\sum_{d=\pm}\!\left[u_+ (\tau_1,k) u_+^\dagger (\tau_2,k)\right]_{\!\rm{(NLoc)},d},
\end{aligned}
\end{equation}
}
\hspace*{-3mm}
where {\small$\Big[u_+(\tau_1,k)u_+^\dagger(\tau_2,k)\Big]_{\rm{(NLoc)},d}$} denotes the component proportional to 
$(\tau_1\tau_2)^{\ii\hp d\hp\tnu}$, 
with $d\!=\!\pm\hp$.\ 
In the following, we summarize the nonlocal parts for all relevant products.\ 
As shown above, we obtain the nonlocal contribution, denoted by ``$\text{(NLoc)}$'', by summing over the two components ``$\rm{(NLoc)},d$'' with $d=\pm\hp$,
{\small 
\beqs 
\label{mode_function_nonlocal}
\begin{align}
\left[u_+ (\tau_1,k) u_+^\dagger (\tau_2,k)\right]_{\text{(NLoc)},d}
&= \dis\frac{-e^{\pi\tlambda}
\big[\Gamma(-\ii\hp 2d\tnu)\big]^2}{~\Gamma(\ii\hp \tlambda\!-\!\ii\hp d\tnu)\Gamma(-\ii\hp \tlambda\!-\!\ii\hp d\tnu)~}(4k^2\tau_1\tau_2)^{\ii\hp d\tnu} ,
\\[1.5mm]
\left[u_+ (\tau_1,k) v_+^\dagger (\tau_2,k)\right]_{\text{(NLoc)},d}
&=-\ii\dis\frac{\tilde{M}_{\!R}\hp e^{\pi\tlambda}
\big[\Gamma(-\ii\hp 2d\tnu)\big]^2}{~\Gamma(1\!+\!\ii\hp \tlambda-\ii\hp d\tnu)\Gamma(-\ii\hp \tlambda-\ii\hp d\tnu)~}(4k^2\tau_1\tau_2)^{\ii\hp d\tnu} ,
\\[1.5mm]
\left[v_+ (\tau_1,k) u_+^\dagger (\tau_2,k)\right]_{\text{(NLoc)},d}
&=\ii\dis\frac{\tilde{M}_{\!R}\hp e^{\pi\tlambda}
\big[\Gamma(-\ii\hp 2d\tnu)\big]^2}
{~\Gamma(1-\ii\hp \tlambda\!-\!\ii\hp d\tnu)\Gamma(\ii\hp \tlambda\!-\!\ii\hp d\tnu)~}(4k^2\tau_1\tau_2)^{\ii\hp d\tnu}, 
\\[1.5mm]
\left[u_- (\tau_1,k) u_-^\dagger (\tau_2,k)\right]_{\text{(NLoc)},d}
&=\dis\frac{e^{-\pi\tlambda}
\big[\Gamma(-\ii\hp 2d\tnu)\big]^2}
{~\Gamma(\ii\hp \tlambda\!-\!\ii \hp d\tnu)
\Gamma(-\ii\hp \tlambda\!-\!\ii\hp d\tnu)~}(4k^2\tau_1\tau_2)^{\ii\hp d\tnu},
\\[1.5mm]
\left[u_- (\tau_1,k) v_-^\dagger (\tau_2,k)\right]_{\rm{(NLoc)},d}
&=\ii\dis\frac{\tilde{M}_{\!R}\hp 
e^{-\pi\tlambda}
\big[\Gamma(-\ii\hp 2d\tnu)\big]^2}
{~\Gamma(1\!+\!\ii\hp\tlambda-\ii \hp d\tnu)\Gamma(-\ii\hp \tlambda\!-\!\ii\hp d\tnu)~}(4k^2\tau_1\tau_2)^{\ii\hp d\tnu},
\\[1.5mm]
\left[v_- (\tau_1,k) u_-^\dagger (\tau_2,k)\right]_{\rm{(NLoc)},d}
&=-\ii\dis\frac{\tilde{M}_{\!R}\hp e^{-\pi\tlambda}
\big[\Gamma(-\ii\hp 2d\tnu)\big]^2}
{~\Gamma(1\!-\!\ii \hp\tlambda\!-\!\ii\hp d\tnu)
\Gamma(\ii\hp \tlambda\!-\!\ii\hp d\tnu)~}
(4k^2\tau_1\tau_2)^{\ii\hp d\tnu}.
\end{align}
\eeqs 
}
\hspace*{-2.5mm}
Using the properties of the nonlocal part, 
we can express the remaining cases 
in terms of the above results
(for $s\!=\!\pm$), 
{\small
\beqs 
\label{mode_function_nonlocal_othercase}
\begin{align}
\left[v_s (\tau_1,k) v_s^\dagger (\tau_2,k)\right]_{\text{(NLoc)},d}
&=-\left[u_s (\tau_2,k) u_s^\dagger (\tau_1,k)\right]_{\text{(NLoc)},d},
\\
\left[v_s^\dagger (\tau_1,k) u_s(\tau_2,k)\right]_{\text{(NLoc)},d}
&=\left[u_s (\tau_1,k) v_s^\dagger (\tau_2,k)\right]_{\text{(NLoc)},d},
\\
\left[u_s^\dagger (\tau_1,k) v_s (\tau_2,k)\right]_{\text{(NLoc)},d}
& =\left[v_s (\tau_1,k) u_s^\dagger (\tau_2,k)\right]_{\text{(NLoc)},d}. 
\end{align}
\eeqs 
}
\hspace*{-2.5mm}

\vspace*{-5mm}
\subsection{\hspace*{-2.5mm}Fermion Propagators during Inflation in the SK Formalism}
\label{sec:2.3}
\label{Propagators of fermion}

The Schwinger–Keldysh (SK) path integral formalism is used to compute expectation values of operators at a finite time in a time-dependent background, as required for deriving cosmological correlators.\ It evolves the quantum state along a closed time contour, going forward and then backward in time, which ensures a consistent treatment of real-time dynamics.\ In this framework, fields are doubled into forward $(\oplus)$ and backward $(\ominus)$ branches, and correlators are obtained from a generating functional defined on this contour.\ This method naturally incorporates interactions and initial conditions, and is commonly used to compute inflationary correlation functions 
and primordial non-Gaussianity\,\cite{Chen170310166,Chen:2018xck,Hook:2019zxa}.\ 
To compute the inflationary correlators of $\delta \phi(t,\mathbf{x})$ generated 
by the right-handed neutrino loop, we first present the propagators 
for Weyl spinors in the SK formalism.

\vs 

As in flat spacetime, there are two types of propagators for a Majorana fermion 
in terms of Weyl spinors.\ We begin with the propagators of the 
$\langle \tilde{N}\tilde{N}^\dagger \rangle$ type, denoted by $D_{ab\,\alpha\dot{\beta}}\hp$, 
where the SK indices $(a,b)$ take the values $\oplus,\ominus$ and have four combinations
$(\oplus\oplus,\oplus\ominus,\ominus\oplus,\ominus\ominus)$ 
corresponding to the four possible SK contour assignments 
and related by time ordering.\ 
They are expressed as follows:
{\small 
\begin{align}
\label{propagator1} 
\hspace*{-6mm}
{D_{\ominus\oplus}}_{\alpha\dot{\beta}}(\mathbf{k}; \tau_1, \tau_2)
&=\sum_{s}\!\xi_{s,\alpha}(\tau_1, \mathbf{k})\,\xi^\dagger_{s,\dot{\beta}}(\tau_2, \mathbf{k})
=\sum_{s}\!u_{s}(\tau_1,k)\,u_{s}^\dagger(\tau_2,k)\,h_{s,\alpha}(\mathbf{k})h_{s,\dot{\beta}}^\dagger(\mathbf{k})\nn\,,
\\
\hspace*{-6mm}
{D_{\oplus \ominus}}_{\alpha\dot{\beta}}(\mathbf{k}; \tau_1, \tau_2)
&=\sum_{s}\!-\chi^{\dagger}_{s,\dot{\beta}}(\tau_2, -\mathbf{k})\,\chi_{s,\alpha}(\tau_1, -\mathbf{k})
=\sum_{s}\!-v_{s}(\tau_2,k)\,v_{s}^\dagger(\tau_1,k)\,h^\dagger_{-s,\dot{\beta}}(-\mathbf{k})h_{-s,\alpha}(-\mathbf{k})\nn\,,
\\
\hspace*{-6mm}
{D_{\oplus\oplus}}_{\alpha\dot{\beta}}(\mathbf{k}; \tau_1, \tau_2)
&={D_{\ominus \oplus}}_{\alpha\dot{\beta}}(\mathbf{k}; \tau_1, \tau_2)\hp
\theta(\tau_1 \!-\! \tau_2)
\hsm +\hsm {D_{\oplus \ominus}}_{\alpha\dot{\beta}}(\mathbf{k}; \tau_1, \tau_2)\hp
\theta(\tau_2 \!-\hsm \tau_1)\,,
\\[1mm]
\hspace*{-6mm}
{D_{\ominus\ominus}}_{\alpha\dot{\beta}}(\mathbf{k}; \tau_1, \tau_2)
&={D_{\oplus \ominus}}_{\alpha\dot{\beta}}(\mathbf{k}; \tau_1, \tau_2)
\hp\theta(\tau_1 \!-\hsm \tau_2)
\hsm +\hsm {D_{\ominus\oplus}}_{\alpha\dot{\beta}}(\mathbf{k}; \tau_1, \tau_2)\hp
\theta(\tau_2 \!-\hsm \tau_1)\nn\,,
\end{align}
}
\hspace{-2mm}
where $\theta(\tau)$ is the step function,
\\[-5mm]
\begin{equation}
\theta(\tau)=
\left\{
\begin{matrix}
0\hp, &~~~~ (\tau < 0) \hp, 
\\[1mm]
1\hp, &~~~~ (\tau\geqq 0)\hp,
\end{matrix}
\right.
\end{equation}
and $\sum_s$ denotes the sum over 
the two helicity states ($s\hsm =\hsm\pm)$.

\vs 

We call these four propagators the ``type-1'' propagators and give their diagrammatic representation as follows:
\\[-7mm]
\begin{equation}
\raisebox{-9mm}{\hbox{\includegraphics[height=2cm]{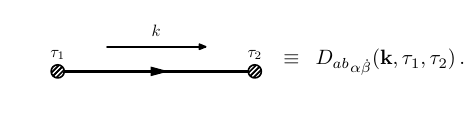}}}
\end{equation}
\\[-3mm]
In addition, there are four propagators of the $\langle \tilde{N}\tilde{N} \rangle$ type, denoted by $\hat{D}_{ab\,\alpha\beta}$, 
which are expressed as follows:
{\small 
\begin{align}
\label{propagator2}
\hspace*{-6mm}
\hat{D}_{\ominus\oplus\,\alpha \beta} (\mathbf{k}; \tau_1, \tau_2)
&= \sum_{s} \!\xi_{s,\alpha} (\tau_1, \mathbf{k}) \chi_{s,\beta} (\tau_2, \mathbf{k})
=\sum_{s}\!s\,u_{s} (\tau_1,k) v_{s}^\dagger (\tau_2,k) h_{s,\alpha}(\mathbf{k}) h_{-s,\beta}(\mathbf{k})\nn\,,  
\\
\hat{D}_{\oplus\ominus\,\alpha \beta} (\mathbf{k}; \tau_1, \tau_2)
&= \sum_{s}\! -\xi_{s,\beta} (\tau_2, -\mathbf{k}) \chi_{s,\alpha} (\tau_1, -\mathbf{k})
=\sum_{s}\!-s\,u_{s} (\tau_2,k) v_{s}^\dagger (\tau_1,k) h_{s,\beta}(-\mathbf{k}) h_{-s,\alpha}(-\mathbf{k})\nn\,,  
\\
\hat{D}_{\oplus\oplus\,\alpha \beta} (\mathbf{k}; \tau_1, \tau_2)
&= \hat{D}_{\ominus\oplus\,\alpha \beta} (\mathbf{k}; \tau_1, \tau_2)\,\theta (\tau_1 \!-\! \tau_2)
\hsm+\hsm \hat{D}_{\oplus\ominus\,\alpha \beta} (\mathbf{k}; \tau_1, \tau_2)\,\theta (\tau_2 \!-\! \tau_1)\,,  
\\[1mm]
\hat{D}_{\ominus\ominus\,\alpha \beta} (\mathbf{k}; \tau_1, \tau_2)
&= \hat{D}_{\oplus\ominus\,\alpha \beta} (\mathbf{k}; \tau_1, \tau_2)\,\theta (\tau_1 \!-\! \tau_2)
\hsm+\hsm \hat{D}_{\ominus\oplus\,\alpha \beta} (\mathbf{k}; \tau_1, \tau_2)\,\theta (\tau_2 \!-\! \tau_1)\nn\,.
\end{align}
}
\hspace*{-2.5mm}
\vspace*{-3mm}

We refer to these four propagators as the ``type-2'' propagators. Their diagrammatic representation is
\\[-6mm]
\begin{equation}
\raisebox{-9mm}{\hbox{\includegraphics[height=2cm]{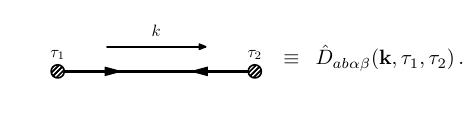}}}
\end{equation}
\\[-3mm]
The two oppositely oriented arrows indicate that this propagator connects two identical Weyl fields,  
and thus represents a Majorana contraction rather than an $N$-number-conserving propagator.

The indices can be raised by contracting with the antisymmetric tensor $\epsilon^{\alpha\beta}$:
\begin{equation}
    \hat{D}_{ab\alpha}^{\,\,\,\,\,\,\,\,\,\beta}(\mathbf{k},\tau_{1},\tau_{2})
    \equiv \epsilon^{\beta\gamma}\hat{D}_{ab\alpha\gamma}(\mathbf{k},\tau_{1},\tau_{2})
    =-\hat{D}_{ab\alpha\gamma}(\mathbf{k},\tau_{1},\tau_{2})\epsilon^{\gamma\beta}.
\end{equation}

Moreover, the Hermitian conjugates of the above propagators, namely those of the $\langle \tilde{N}^\dagger \tilde{N}^\dagger \rangle$ type, are denoted by $\check{D}_{ab\dot{\alpha}\dot{\beta}}\hp$.\  For clarity, we refer to them as the ``type-3'' propagators and express them as follows:
{\small 
\begin{align}
\label{propagator3}
\hspace*{-7mm}
\check{D}_{\ominus\oplus\,\dot{\alpha} \dot{\beta}} (\mathbf{k}; \tau_1, \tau_2)
&= \sum_{s}\! \chi^\dagger_{s,\dot{\alpha}} (\tau_1, \mathbf{k}) \xi^\dagger_{s,\dot{\beta}} (\tau_2, \mathbf{k})
=\sum_{s}\!s\,v_{s} (\tau_1,k) u_{s}^\dagger (\tau_2,k) h_{-s,\dot{\alpha}}^\dagger(\mathbf{k}) h_{s,\dot{\beta}}^\dagger(\mathbf{k})\nn\,,  
\\
\hspace*{-7mm}
\check{D}_{\oplus\ominus\,\dot{\alpha} \dot{\beta}} (\mathbf{k}; \tau_1, \tau_2)
&= \sum_{s}\! -\chi^\dagger_{s,\dot{\beta}} (\tau_2, -\mathbf{k}) \xi^\dagger_{s,\dot{\alpha}} (\tau_1, -\mathbf{k})
=\sum_{s}\!-s\,v_{s} (\tau_2,k) u_{s}^\dagger (\tau_1,k) h_{-s,\dot{\beta}}^\dagger (-\mathbf{k}) h_{s,\dot{\alpha}}^\dagger (-\mathbf{k})\nn\,,  
\\
\hspace*{-7mm}
\check{D}_{\oplus\oplus\,\dot{\alpha} \dot{\beta}} (\mathbf{k}; \tau_1, \tau_2)
&= \check{D}_{\ominus\oplus\,\dot{\alpha} \dot{\beta}} (\mathbf{k}; \tau_1, \tau_2)\,\theta (\tau_1 \!-\! \tau_2)
\hsm+\hsm \check{D}_{\oplus\ominus\,\dot{\alpha} \dot{\beta}} (\mathbf{k}; \tau_1, \tau_2)\,\theta (\tau_2 \!-\! \tau_1)\,,  
\\[1mm]
\hspace*{-7mm}
\check{D}_{\ominus\ominus\,\dot{\alpha} \dot{\beta}} (\mathbf{k}; \tau_1, \tau_2)
&= \check{D}_{\oplus\ominus\,\dot{\alpha} \dot{\beta}} (\mathbf{k}; \tau_1, \tau_2)\,\theta (\tau_1 \!-\! \tau_2)
\hsm+\hsm \check{D}_{\ominus\oplus\,\dot{\alpha} \dot{\beta}} (\mathbf{k}; \tau_1, \tau_2)\,\theta (\tau_2 \!-\! \tau_1)\nn\,.
\end{align}
}
\hspace*{-3mm}
The dotted spinor indices can be raised 
likewise by contracting with the antisymmetric tensor $\epsilon^{\dot{\alpha}\dot{\beta}}$:
\\[-6mm]
\begin{equation}
\check{D}_{ab\,\,\,\dot{\beta}}^{\,\,\,\,\,\dot{\alpha}}(\mathbf{k},\tau_{1},\tau_{2})
  \equiv \epsilon^{\dot{\alpha}\dot{\gamma}}\check{D}_{ab\dot{\gamma}\dot{\beta}}(\mathbf{k},\tau_{1},\tau_{2})
  = -\epsilon^{\dot{\gamma}\dot{\alpha}}\check{D}_{ab\dot{\gamma}\dot{\beta}}(\mathbf{k},\tau_{1},\tau_{2})\,.
\end{equation}
Their diagrammatic representation is given by
\\[-6mm]
\begin{equation}
\raisebox{-9mm}{\hbox{\includegraphics[height=2cm]{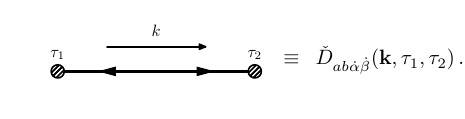}}}
\end{equation}
\\[-5mm]
As discussed in Section\,\ref{A Brief Review of Cosmological Collider Physics}, the imprint of the cosmological collider signals can be read off directly from the late-time limit of the massive propagators.\ 
For instance, we can extract the nonlocal parts of the spinor propagators in the 
late-time limit:
\begin{adjustwidth}{-12.5mm}{0cm}
\beqs 
\vspace*{-0.5cm}
{\small 
\begin{align}
\label{late time limit propagator}
& \left[D_{\ominus\oplus\alpha\dbeta}(\mathbf{k},\tau_1,\tau_2)\right]_\text{(NLoc)}\nn
\\
&=\left[u_+ (\tau_1,k) u_+^\dagger (\tau_2,k)\right]_\text{(NLoc)}\!\! h_{+,\alpha}(\mathbf{k})h_{+,\dbeta}^\dagger(\mathbf{k})+\left[u_- (\tau_1,k) u_-^\dagger (\tau_2,k)\right]_\text{(NLoc)}\!\! h_{-,\alpha}(\mathbf{k})h_{-,\dbeta}^\dagger(\mathbf{k})\nn
\\
& = \left\{\!\left[\frac{-e^{\pi\tlambda}\big[\Gamma(-\ii\hp2\tnu)\big]^2}{\,\Gamma(\ii\hp \tlambda\!-\!\ii\hp \tnu)\Gamma(-\ii\hp\tlambda\!-\!\ii\hp \tnu)}h_+(\mathbf{k})h_+^\dagger(\mathbf{k})
+\frac{e^{-\pi\tlambda}\big[\Gamma(-\ii\hp2\tnu)\big]^2}{\,\Gamma(\ii\hp \tlambda\!-\!\ii\hp \tnu)\Gamma(-\ii\hp\tlambda\!-\!\ii\hp \tnu)}
h_-(\mathbf{k})h_-^\dagger(\mathbf{k}) \right]\right. 
\\
 & \quad\times(4k^2\tau_1\tau_2)^{\ii \hp \tnu}\biggr\}_{\alpha\dbeta}
 \!+ (\tnu\ito-\tnu)\nn\,,
\end{align}
\vspace*{-0.6cm}
\begin{align}
\label{late time limit propagator hatD}
& \left[\hat{D}_{\ominus\oplus\alpha\beta}(\mathbf{k},\tau_1,\tau_2)\right]_\text{(NLoc)}\nn
\\
&=\left[u_+ (\tau_1,k) v_+^\dagger (\tau_2,k)\right]_\text{(NLoc)}\!\! h_{+,\alpha}(\mathbf{k})h_{-,\beta}(\mathbf{k})
-\!
\left[u_- (\tau_1,k) v_-^\dagger(\tau_2,k)\right]_\text{(NLoc)}\!\! h_{-,\alpha}(\mathbf{k})h_{+,\beta}(\mathbf{k})\nn
\\
& = \left\{\!\left[\frac{-\ii\tilde{M}_{\!R}\hp e^{\pi\tlambda}\big[\Gamma(-\ii\hp2\tnu)\big]^2}{\Gamma(1\!+\!\ii\hp \tlambda\!-\!\ii\hp \tnu)\Gamma(-\ii\hp \tlambda\!-\!\ii\hp \tnu)\,}h_{+}(\mathbf{k})h_{-}(\mathbf{k})
\!-\!\frac{\ii\tilde{M}_{\!R}\hp e^{-\pi\tlambda}\big[\Gamma(-\ii\hp2\tnu)\big]^2}{\Gamma(1\!+\!\ii\hp \tlambda\!-\!\ii\hp \tnu)\Gamma(-\ii\hp \tlambda\!-\!\ii\hp \tnu)}h_{-}(\mathbf{k})h_{+}(\mathbf{k})\right]\right. 
\\
& \quad\times(4k^2\tau_1\tau_2)^{\ii \hp \tnu}\biggr\}_{\alpha\beta} 
\!+ (\tnu\ito-\tnu)\nn\,,
\end{align}
\vspace*{-0.35cm}
\begin{align}
\label{late time limit propagator hatDCon}
&\left[\check{D}_{\ominus\oplus\dalpha\dbeta}(\mathbf{k},\tau_1,\tau_2)\right]_\text{(NLoc)} \nn
\\
&=\left[v_+ (\tau_1,k) u_+^\dagger (\tau_2,k)\right]_\text{(NLoc)}\!\! h_{-,\dalpha}^\dagger(\mathbf{k})h_{+,\dbeta}^\dagger(\mathbf{k})
-\!
\left[v_- (\tau_1,k) u_-^\dagger(\tau_2,k)\right]_\text{(NLoc)}\!\! h_{+,\dalpha}^\dagger(\mathbf{k})h_{-,\dbeta}^\dagger(\mathbf{k})\nn
\\
& = \left\{\!\left[\ii\frac{\tilde{M}_{\!R}\hp e^{\pi\tlambda}\big[\Gamma(-\ii\hp2\tnu)\big]^2}{\Gamma(1\!-\!\ii\hp \tlambda\!-\!\ii\hp \tnu)\Gamma(\ii\hp \tlambda\!-\!\ii\hp \tnu)}h_-^\dagger(\mathbf{k})h_+^\dagger(\mathbf{k})-\frac{-\ii\tilde{M}_{\!R}\hp e^{-\pi\tlambda}\big[\Gamma(-\ii\hp2\tnu)\big]^2}{\Gamma(1\!-\!\ii\hp \tlambda\!-\!\ii\hp \tnu)\Gamma(\ii\hp \tlambda\!-\!\ii\hp \tnu)}h_+^\dagger(\mathbf{k})h_-^\dagger(\mathbf{k})\right.\right] 
\\
& \quad\times(4k^2\tau_1\tau_2)^{\ii \hp \tnu}\biggr\}_{\dalpha\dbeta}+ (\tnu\to-\tnu)\,\nn ,
\end{align}
}
\eeqs 
\end{adjustwidth}
In the above expressions, the first term in the nonlocal part arises from the positive-helicity spinor coefficients, namely $u_+u_+^\dagger$, $u_+v_+^\dagger$, and $v_+u_+^\dagger$, whereas the second term arises from the negative-helicity spinor coefficients, namely $u_-u_-^\dagger$, $u_-v_-^\dagger$, and $v_-u_-^\dagger$.

\vs 

Without loss of generality, 
we choose a positive chemical potential $\lambda\!>\!0\hp$.\ 
Then, we readily find\footnote{%
For the Gamma functions appearing in the above expressions, one may use the asymptotic formula, $\Gamma(x\!+\!\ii\hp y)
\!\sim\!\sqrt{2\pi\,}|y|^{x-1/2}e^{-\pi|y|/2}$, 
in the regime $\lambda\!\gg\! m\!\gg\! H\hp$.}
that the positive-helicity contribution, namely $u_+u_+^\dagger$, $u_+v_+^\dagger$, 
and $v_+u_+^\dagger$, is proportional to the modified Boltzmann factor 
$e^{+\pi \tlambda-\pi\tnu}\!\thicksim\! 
e^{-\pi\tilde{M}_{\!R}^2/2\tlambda}$ when the chemical potential $\lambda\hp$ is sufficiently large.\ In contrast, in the absence of a chemical potential
($\lambda\!=\!0$), the usual Boltzmann suppression for a heavy fermion is 
$e^{-\pi \tilde{M}_{\!R}}$.\ This shows that a positive chemical potential enhances the production of fermions with positive helicity and delays the Boltzmann suppression associated with a large mass\,\cite{Chen:2018xck,Adshead:2015kza,Adshead:2018oaa,Wang:2019gbi,Wang:2020ioa}.

\vs 

In contrast, the negative-helicity contribution, namely $u_-u_-^\dagger$, $u_-v_-^\dagger$, and $v_-u_-^\dagger$, is proportional to $e^{-\pi \tlambda-\pi\tnu}$, indicating that a positive chemical potential suppresses the production of fermions with negative helicity.\ Similarly, a negative chemical potential would instead enhance the production of fermions with negative helicity.\
In either case, one helicity component of the Weyl spinor is enhanced in the presence of a sufficiently large chemical potential.\ 
Thus, it suffices for us to consider 
the case of a positive chemical potential $\lambda\!>\!0\,$.

\subsection{\hspace*{-2.5mm}Seed Integrals for Fermion Propagators}
\label{sec:2.4}
\label{Seed Integrals for Fermion Propagators}

In Section\,\ref{sec:3}, 
we will evaluate the inflaton two-point correlator $I_{ab}(\mathbf{k})$ containing a single fermion propagator,
\begin{equation}
\label{seed integral form}
\begin{aligned}
I_{ab}(\mathbf{k})
\thicksim
\int_{-\infty}^{0^-}\!\!\td \tau_1 \td \tau_2 \,
G_{a}(k,\tau_1)\hp 
D_{ab}(\mathbf{k},\tau_1,\tau_2)\hp 
G_{b}(k,\tau_2)\,,
\end{aligned}
\end{equation}
where $G_{a,b}$ is the bulk-to-boundary propagator of the inflaton $\delta\phi$.\ 
Such a two-point correlator $I_{ab}^{}(\mathbf{k})$ 
will serve as an ingredient in the evaluation of the three-point inflaton correlator 
induced by the fermion loop, which can be expressed  by a set of seed integrals.\ 
In the following, we introduce these seed integrals and organize them in a 
convenient form for the subsequent calculations.

The bulk-to-boundary propagators of the inflaton $\delta\phi$, defined in Eq.\eqref{inflaton_bulk-to-boundary}, can always be written as the product of a function of $\tau$ and an oscillatory exponential,
\begin{equation}
G_{a}(k,\tau)=P(\tau)e^{\ii\hp a k\tau},
\qquad
P(\tau)=\sum_{n=0}^{2}\! c(n)\hp\tau^n,
\end{equation}
where $P(\tau)$ is a finite polynomial 
of the conformal time $\tau$.\ Thus, substituting 
$G_{a}(\mathbf{k},\tau)$ into 
the two-point correlator $I_{ab}$ 
in Eq.\eqref{seed integral form},
we derive the following seed time integral:
\\[-2mm]
\begin{equation}
\begin{aligned}
\mathcal{I}^{p_1 p_2}_{ab}
&=
k^{1+p_1}k^{1+p_2}\!\! 
\int_{-\infty}^{0^-}\!\!\td \tau_1 \td \tau_2 \,
e^{\ii\hp a k\tau_1}e^{\ii\hp b k\tau_2}
(-\tau_1)^{p_1}(-\tau_2)^{p_2}
D_{ab}(\mathbf{k},\tau_1,\tau_2)
\\
&=
\int_0^{\infty}\!\!\td z_1 \td z_2 \,
e^{-\ii\hp a z_1}e^{-\ii\hp b z_2}
z_1^{p_1}z_2^{p_2}
D_{ab}(\mathbf{k},z_1,z_2)\, ,
\end{aligned}
\end{equation}
where we have introduced the dimensionless variables
\begin{equation}
z_i^{} = -k\hp\tau_i^{}\,,
    \qquad (i=1,2)\hp .
\end{equation}

The fermion propagators $D_{ab}(\mathbf{k},\tau_1,\tau_2)$ are constructed from the two types of spinor coefficients, $u_s(k,\tau)$ and $v_s(k,\tau)$, together with their conjugates.\ As a result, the relevant time integrals reduce to two basic forms.\  Taking $u_s(z_1) v_s^\dagger(z_2)$ 
as an example, the first form is 
the factorized time integral,
\begin{equation}
\label{factorized_integral_left}
\int_0^\infty\!\!\!\td z_1^{}\! 
\int_0^\infty\!\!\! \td z_2^{}\,
e^{\ii\hp a z_1} e^{-\ii\hp a z_2}
z_1^{p_1} z_2^{p_2}
u_s(z_1) v_s^\dagger(z_2)\, ,
\end{equation}
and the second form is the 
nested time integral,
\begin{equation}
\label{nested_integral_left}
\int_0^\infty \!\!\!\td z_1\! 
\int_0^\infty \!\!\!\td z_2\,
e^{\ii\hp a z_1} e^{\ii\hp a z_2}
z_1^{p_1} z_2^{p_2}
u_s(z_1) v_s^\dagger(z_2)
\!\times\!
\begin{cases}
\,\theta(z_1\!-\!z_2)\hp ,
\\
\,\theta(z_2\!-\!z_1)\hp , 
\end{cases}
\end{equation}
where $a\!=\!\oplus,\ominus\hp$,  
and $p_1^{}$ and $p_2^{}$ are the relevant powers associated with $z_1^{}$ and $z_2^{}$, respectively.
Hence, there are two basic classes of seed integrals: the factorized integrals in Eq.\eqref{factorized_integral_left} (corresponding to $a\!=\!-b\!=\!\oplus,\ominus$), 
and the nested integrals in Eq.\eqref{nested_integral_left} (corresponding to $a\!=\!b\!=\!\oplus,\ominus$).\ 
In what follows, we define the seed integrals for both classes. 

\vs 

All seed integrals introduced below can be evaluated by means of a Wick rotation\,\cite{Wang:2013zva,Chen170310166,Chen:2009zp,Chen:2012ge},
\begin{equation}
    e^{\pm \ii\hp z_1},\, e^{\pm \ii\hp z_2}
    \;\longrightarrow\;
    e^{-z_1},\, e^{-z_2}\, ,
\end{equation}
where the rotation directions are chosen such that the integrals remain convergent.\ The validity of this procedure is ensured by the usual $\ii\hp\epsilon$ prescription of the interacting quantum field theory 
for de Sitter spacetime.

\vs 

In the following 
Subsections\,\ref{sec:2.4.1}-\ref{sec:2.4.3}, 
we first present the seed integrals in the special frame where the spatial momentum is aligned with $z$-axis,
\begin{equation}
\mathbf{k}=k\hat{z}=(0,0,k)\hp .
\end{equation}
Then, in Section\,\ref{Rotation of Spinor Seed Integrals}, we will explain how to transform these seed integrals to the general case that has the momentum $\mathbf{k}$
in an arbitrary direction.\

\subsubsection{\hspace*{-2.5mm}Seed Integrals for \texorpdfstring{${D}_{ab\,{\alpha\dot{\beta}}}(\mathbf{k},z_1,z_2)$}{}}
\label{sec:2.4.1}
\vspace*{1.5mm}

As discussed above, the seed integrals are classified according to the type of the fermion propagator.\ We define the seed integrals $\hp\mathcal{I}^{p_1 p_2}_{ab\hspace{3mm}\alpha\dot{\beta}}$ associated with the propagator 
${D}_{\!ab\,{\alpha\dot{\beta}}}(\mathbf{k},z_1,z_2)$ as follows:
{\small 
\begin{equation}
\begin{aligned}
\mathcal{I}^{p_1 p_2}_{ab\hspace{2mm}\alpha\dot{\beta}}
&=\, 
k_1^{1+p_1}k_2^{1+p_2}\!\!
\int_{-\infty}^{0^-}\!\!\td \tau_1 \td \tau_2 \,
e^{\ii\hp a k\tau_1}e^{\ii\hp b k\tau_2}
(-\tau_1)^{p_1}(-\tau_2)^{p_2}
{D}_{ab\,\alpha\dot{\beta}}(\mathbf{k},\tau_1,\tau_2)
\\
&=
\int_0^{\infty}\!\!\td z_1 \td z_2 \,
e^{-\ii\hp a z_1}e^{-\ii\hp b z_2}
z_1^{p_1}z_2^{p_2}
{D}_{ab\,\alpha\dot{\beta}}(\mathbf{k},z_1,z_2)\,.
\end{aligned}
\end{equation}
}

\hspace*{-2.5mm}
Since ${D}_{\!ab\,{\alpha\dot{\beta}}}(\mathbf{k},z_1,z_2)$ is a rank-2 
spinor-tensor which is constructed by the direct product of two spinors, the seed integral $\mathcal{I}^{p_1 p_2}_{ab\hspace{3mm}\alpha\dot{\beta}}$ 
could be expressed by a $2\!\times\!2$ matrix 
as follows:
{\small 
\beqs
\label{eq:I-p1p2-albedot}
\begin{align}
\mathcal{I}_{\oplus\oplus\hspace{1.5mm}\alpha\dot{\beta} }^{p_1 p_2}
&=\!
\begin{pmatrix}
A_1^{p_1 p_2} \!\!-\! B_1^{p_1 p_2}
&
0
\\[1mm] 
0
&
A_2^{p_1 p_2} \!-\! B_2^{p_1 p_2}
\!\end{pmatrix}_{\!\!\alpha\dot{\beta}},
\\[1.5mm]
\label{Imm}
\mathcal{I}_{\ominus\ominus\hspace{1.5mm}\alpha\dot{\beta} }^{p_1 p_2}
&=\!\(\mathcal{I}_{\oplus\oplus\hspace{1.5mm}\beta\dot{\alpha} }^{p_2^{*} p_1^{*}}\)^{\!\dagger}
=
\!\begin{pmatrix}
(A_1^{p_2^{*} p_1^{*}} \!\!-\! B_1^{p_2^{*} p_1^{*}})^*
& 0
\\[1mm]
0 &
(A_2^{p_2^{*} p_1^{*}} \!\!-\! B_2^{p_2^{*} p_1^{*}})^*
\!\end{pmatrix}_{\!\!\alpha\dot{\beta}},
\\[1mm]
\mathcal{I}_{ \ominus\oplus\hspace{1.5mm}\alpha\dot{\beta}}^{p_1 p_2}
&=
\!\begin{pmatrix}
U_1^{p_1 p_2} & 0
\\[1mm]
0 & U_2^{p_1 p_2}
\end{pmatrix}_{\!\!\alpha\dot{\beta}},
\hspace*{8mm}
\mathcal{I}_{\oplus\ominus \hspace{1.5mm}\alpha\dot{\beta}}^{p_1 p_2}
=
\!\begin{pmatrix}
-V_1^{p_1 p_2}
&
\!\!\!0
\\[1mm] 
0 & \!\!-V_2^{p_1 p_2}
\!\end{pmatrix}_{\!\!\alpha\dot{\beta}}.
\end{align}
\eeqs
}
\hspace*{-2.5mm}
In Eq.\eqref{Imm}, the seed integral 
$\mathcal{I}_{\ominus\ominus\hspace{1.5mm}\alpha\dot{\beta} }^{p_1 p_2}$  is connected to $\mathcal{I}_{\oplus\oplus\hspace{1.5mm}\alpha\dot{\beta} }^{p_1 p_2}$ by the identity $\mathcal{I}_{\ominus\ominus\hspace{1.5mm}\alpha\dot{\beta} }^{p_1 p_2}\hsm\!=\!\hsm \(\mathcal{I}_{\oplus\oplus\hspace{1.5mm}\beta\dot{\alpha} }^{p_2^{*} p_1^{*}}\)^{\!\dagger}$.\ 
This can be proved in the following way: 
{\small 
\begin{equation}
\begin{aligned}
\mathcal{I}_{\ominus\ominus\hspace{1.5mm}\alpha\dot{\beta} }^{p_1 p_2}&=
\int_0^{\infty}\!\!\td z_1 \td z_2 \,
e^{\ii\hp z_1}e^{\ii\hp z_2}
z_1^{p_1}z_2^{p_2}
{D}_{\ominus\ominus\,\alpha\dot{\beta}}(\mathbf{k},z_1,z_2)
\\
&=\bigg[\!\int_0^{\infty}\!\!\td z_1 \td z_2 \,
e^{-\ii\hp z_1}e^{-\ii\hp z_2}
z_1^{p_1^{*}}z_2^{p_2^{*}}
\!\Big(\!{D}_{\ominus\ominus\,\alpha\dot{\beta}}(\mathbf{k},z_1,z_2) \!\Big)^{\!\!\dagger}\bigg]^{\!\dagger}
\\
&=\bigg[\!\int_0^{\infty}\!\!\td z_1 \td z_2 \,
e^{-\ii\hp z_1}e^{-\ii\hp z_2}
z_1^{p_1^{*}}z_2^{p_2^{*}}
{D}_{\oplus\oplus\,\beta\dot{\alpha}}(\mathbf{k},z_2,z_1) \bigg]^{\!\dagger}
=\(\mathcal{I}_{\oplus\oplus\hspace{1.5mm}\beta\dot{\alpha} }^{p_2^{*} p_1^{*}}\)^{\!\dagger}.
\end{aligned}
\end{equation}
}
\hspace*{-2.8mm}
As discussed above, we choose $\mathbf{k}$ 
to be aligned with the $z$-axis, i.e., $\mathbf{k}\!=\!k\hat{z}$ in the seed integral.\ 
In Eq.\eqref{eq:I-p1p2-albedot}, the matrix elements $(A_i^{p_1 p_2}, B_i^{p_1 p_2})$ ($i\!=\!1,2$) 
are defined as follows:
\\[-5mm]
{\small 
\beqs
\begin{align}
A_1^{p_1 p_2}
&= \int_0^\infty \!\!\td z_1 
\!\int_0^\infty \!\!\td z_2\,
e^{-\ii\hp z_1} e^{-\ii\hp z_2}
z_1^{p_1} z_2^{p_2}
u_+(z_1) u_+^\dagger(z_2)
\theta(z_2 \!-\! z_1)\nn
\\[1mm]
&=\tilde{M}_{\!R}^2e^{\pi \tlambda}
  I^{p_1 p_2}(1\!+\!\ii\hp\tlambda\!+\!\ii\hp\tnu,1\!+\!\ii\hp\tlambda\!-\!\ii\hp\tnu,\ii\hp\tlambda\!+\!\ii\hp\tnu)\,,
\\[1mm]
A_2^{p_1 p_2}
&= \int_0^\infty \!\!\td z_1 
\!\int_0^\infty \!\!\td z_2\,
e^{\ii\hp z_1} e^{\ii\hp z_2}
z_1^{p_1} z_2^{p_2}
u_-(z_1) u_-^\dagger(z_2)
\theta(z_2 \!-\! z_1)\nn
\\[1mm]
&=e^{-\pi \tlambda}I^{p_1 p_2}(-\ii\hp\tlambda\!+\!\ii\hp\tnu,-\ii\hp\tlambda\!-\!\ii\hp\tnu,1\!-\!\ii\hp\tlambda\!+\!\ii\hp\tnu)\,,
\\[1mm]
B_1^{p_1 p_2}
&= \int_0^\infty \!\!\td z_1 
\!\int_0^\infty \!\!\td z_2\,
e^{-\ii\hp z_1} e^{-\ii\hp z_2}
z_1^{p_1} z_2^{p_2}
v_+(z_2) v_+^\dagger(z_1)
\theta(z_1 \!-\! z_2)\nn
\\[1mm]
&= e^{\pi \tlambda}I^{p_2 p_1}(\ii\hp\tlambda\!+\!\ii\hp\tnu,\ii\hp\tlambda\!-\!\ii\hp\tnu,1\!+\!\ii\hp\tlambda\!+\!\ii\hp\tnu)\,,
\\[1mm]
B_2^{p_1 p_2}
&= \int_0^\infty \!\!\td z_1 
\!\int_0^\infty \!\!\td z_2\,
e^{-\ii\hp z_1} e^{-\ii\hp z_2}
z_1^{p_1} z_2^{p_2}
v_-(z_2) v_-^\dagger(z_1)
\theta(z_1 \!-\! z_2)\nn
\\[1mm]
&=\tilde{M}_{\!R}^2\hp e^{-\pi \tlambda} 
I^{p_2 p_1}\hsm (1\!-\!\ii\hp\tlambda\!+\!\ii\hp\tnu,1\!-\!\ii\hp\tlambda\!-\!\ii\hp\tnu,-\ii\hp\tlambda\!+\!\ii\hp\tnu)\,.
\end{align}
\eeqs
}
\hspace*{-2.5mm}
In the above, $I^{p_1 p_2}(x,y,z)$ is given by
{\small 
\begin{equation}
\begin{aligned}
\label{functionI}
I^{p_1 p_2}(x,y,z)
= &~ 
\dis\ii\hp 2^{-2-p_1-p_2}
 e^{-\frac{\ii\hp\pi}{2}\(p_1+p_2\)} \pi\,\csch(2\pi\tnu)\hp
 \Gamma(2\!+\!p_1\!+\!p_2)
 \\
&\times\!\hsm 
\Bigg[
\frac{\,e^{-\pi \tnu}\,}{\,\Gamma(x)\,}
\Gamma(1\!+\!p_1\!-\!\ii\hp\tnu)\,
\Gamma(2\!+\!p_1\!+\!p_2\!-\!\ii\hp2\tnu)
 \\
&\quad\,\,\times\!
{}_4\tilde F_3\!\hsm
\(\begin{matrix}
2\!+\!p_1\!+\!p_2,\;
1\!+\!p_1\!-\!\ii\hp\tnu,\;
y,\;
2\!+\!p_1\!+\!p_2\!-\!\ii\hp2\tnu\hp 
\\
2\!+\!p_1\!-\!\ii\hp\tnu,\;
3\!+\!p_1\!+\!p_2\!-\!z,\;
1\!-\!\ii\hp2\tnu
\end{matrix}
\bigg|\!-\!1\!\)\!-\!(\tnu\ito -\tnu)\Bigg],
\end{aligned}
\end{equation}
}
\hspace*{-2.5mm}
where ${}_4\tilde F_3$ is the regularized hypergeometric function as defined in Section\,\ref{introduction}.\ 
The other matrix elements 
$(U_i^{p_1 p_2}, V_i^{p_1 p_2})$ ($i\!=\!1,2$) 
are defined by 
{\small
\beqs
\begin{align}
U_1^{p_1 p_2} &= \int_0^\infty \!\!\td z_1 \!\int_0^\infty \!\!\td z_2\,
e^{\ii\hp z_1} e^{-\ii\hp z_2}
z_1^{p_1} z_2^{p_2}
u_+(z_1) u_+^\dagger(z_2) \nn
\\[1mm]
&= \tilde{M}_{\!R}^2 e^{\pi \tlambda} K^{p_1p_2}(2\!+\!\ii\hp \tlambda,2\!-\!\ii\hp\tlambda)\,,
\\[1mm]
U_2^{p_1 p_2}
&= \int_0^\infty \!\!\td z_1 
\!\int_0^\infty \!\!\td z_2\,
e^{\ii\hp z_1} e^{-\ii\hp z_2}
z_1^{p_1} z_2^{p_2}
u_-(z_1) u_-^\dagger(z_2)\nn
\\[1mm]
&=e^{-\pi \tlambda}K^{p_1p_2}(1\!-\!\ii\hp \tlambda,1\!+\!\ii\hp\tlambda)\,,
\\[1mm]
V_1^{p_1 p_2}
&= \int_0^\infty \!\!\td z_1 
\!\int_0^\infty \!\!\td z_2\,
e^{-\ii\hp z_1} e^{\ii\hp z_2}
z_1^{p_1} z_2^{p_2}
v_+(z_2) v_+^\dagger(z_1)\nn
\\
&=e^{\pi \tlambda}e^{-\ii\hp \pi(p_1-p_2)}K^{p_1p_2}(1\!-\!\ii\hp \tlambda,1\!+\!\ii\hp\tlambda)\,,
\\[1mm]
V_2^{p_1 p_2}
&= \int_0^\infty \!\!\td z_1 
\!\int_0^\infty \!\!\td z_2\,
e^{-\ii\hp z_1} e^{\ii\hp z_2}
z_1^{p_1} z_2^{p_2}
v_-(z_2) v_-^\dagger(z_1)\nonumber \nn
\\[1mm]
&=\tilde{M}_{\!R}^2 e^{-\pi \tlambda} e^{-\ii\hp \pi(p_1-p_2)} K^{p_1p_2}(2\!+\!\ii\hp \tlambda,2\!-\!\ii\hp\tlambda)\,,
\end{align}
\eeqs
}
\hspace*{-2.5mm}
where the function $K^{p_1p_2}(x,y)$ 
takes the following form,
{\small 
\begin{equation}
\begin{aligned}
\label{functionK}
\hspace*{-3mm}
K^{p_1p_2}(x,y)= \dis 
e^{\frac{\ii\hp \pi}{2}(p_1-p_2)}
\frac{~\Gamma(1\!+\!p_1\!-\!\ii\hp\tnu)\Gamma(1\!+\!p_1\!+\!\ii\hp\tnu)
\Gamma(1\!+\!p_2\!-\!\ii\hp\tnu)\Gamma(1\!+\!p_2\!+\!\ii\hp\tnu)~
}{2^{2+p_1+p_2}\Gamma(p_1\!+\!x)\Gamma(p_2\!+\!y)}
\,.
\end{aligned}
\end{equation}
}
\hspace*{-2mm}
The derivations of the above seed integral matrices are provided in Appendix\,\ref{representative derivations of the seed integrals}.

\vspace*{1.5mm}
\subsubsection{\hspace*{-2.5mm}Seed Integrals for \texorpdfstring{$\hat{D}_{ab\,\alpha\beta}(\mathbf{k},z_1,z_2)$}{}}
\label{sec:2.4.2}
\vspace*{1.5mm}

Then, we consider the propagator $\hat{D}_{ab\,\alpha\beta}(\mathbf{k},z_1,z_2)$
and define its associated seed integrals $\hat{\mathcal{I}}_{ab\hspace{3mm}\alpha\beta}^{p_1 p_2}$ as follows:
{\small
\begin{equation}
\label{seedDhat}
\begin{aligned}
\hat{\mathcal{I}}_{ab\hspace{2mm}\alpha\beta}^{p_1 p_2}
&=
k_1^{1+p_1}k_2^{1+p_2}\!\!
\int_{-\infty}^{0^-}\!\!\td \tau_1 \td \tau_2 \,
e^{\ii\hp a k\tau_1}e^{\ii\hp b k\tau_2}
(-\tau_1)^{p_1}(-\tau_2)^{p_2}
\hat{D}_{ab\,\alpha\beta}(\mathbf{k},\tau_1,\tau_2)
\\
&=
\int_0^{\infty}\!\!\td z_1 \td z_2 \,
e^{-\ii\hp a z_1}e^{-\ii\hp b z_2}
z_1^{p_1}z_2^{p_2}
\hat{D}_{ab\,\alpha\beta}(\mathbf{k},z_1,z_2)\,.
\end{aligned}
\end{equation}
}
\hspace*{-2.5mm} 
The seed integral $\hat{\mathcal{I}}_{ab\hspace{3mm}\alpha\beta}^{p_1 p_2}$ can be expressed as the following  $2\!\times\!2$ matrix:
{\small
\beqs
\begin{align}
\hat{\mathcal{I}}_{\oplus\oplus\hspace{1.5mm}{\alpha\beta}}^{p_1 p_2}
&=
\!\begin{pmatrix}
0 & S_1^{p_1 p_2} \!+\! S_1^{p_2 p_1} 
\\[1mm]
-\(S_2^{p_2 p_1} \!+\! S_2^{p_1 p_2}\) & 0
\end{pmatrix}_{\!\!\alpha\beta},
\\[1.5mm]
\label{Ihatmmdefine}
\hat{\mathcal{I}}_{\ominus\ominus\hspace{1.5mm}{\alpha\beta}}^{p_1 p_2}
&=
\!\begin{pmatrix}
0 & S_3^{p_1 p_2} \!+\! S_3^{p_2 p_1} 
\\[1mm]
-\(S_4^{p_2 p_1} \!+\! S_4^{p_1 p_2}\) & 0
\end{pmatrix}_{\!\!\alpha\beta},
\\
\hat{\mathcal{I}}_{\oplus\ominus\hspace{1.5mm}{\alpha\beta}}^{p_1 p_2}
&=
\!\begin{pmatrix}
0 & W_1^{p_1 p_2} 
\\[1mm]
- W_2^{p_1 p_2} & 0
\end{pmatrix}_{\!\!\alpha\beta},
\hspace*{8mm}
\hat{\mathcal{I}}_{\ominus\oplus\hspace{1.5mm}{\alpha\beta}}^{p_1 p_2}
=
\!\begin{pmatrix}
0 & W_1^{p_2 p_1} 
\\[1mm]
- W_2^{p_2 p_1} & 0
\end{pmatrix}_{\!\!\alpha\beta}.
\end{align}
\label{eq:I-p1p2+-/-+-W12}
\eeqs
}
\hspace*{-2.5mm}
In the above seed integrals, we define 
$S_i^{p_1 p_2}$ ($i=1,2,3,4$) as follows:
\vspace{-1mm}
{\small
\beqs
\begin{align}
S_1^{p_1 p_2}&= \int_0^\infty \!\!\td z_1 \!\int_0^\infty \!\!\td z_2\,
e^{-\ii\hp z_1} e^{-\ii\hp z_2}
z_1^{p_1} z_2^{p_2}
u_+(z_1) v_+^\dagger(z_2)\,
\theta(z_2 \!-\! z_1)\nn
\\[1mm]
&=-\ii\tilde{M}_{\!R}\hp e^{\pi \tlambda}
  I^{p_1 p_2}(1\!+\!\ii\hp\tlambda\!+\!\ii\hp\tnu,1\!+\!\ii\hp\tlambda\!-\!\ii\hp\tnu,1\!+\!\ii\hp\tlambda\!+\!\ii\hp\tnu)\,,
\\[1mm]
S_2^{p_1 p_2} &= \int_0^\infty \!\!\td z_1 \!\int_0^\infty \!\!\td z_2\,
e^{-\ii\hp z_1} e^{-\ii\hp z_2}
z_1^{p_1} z_2^{p_2}
u_-(z_1) v_-^\dagger(z_2)\,
\theta(z_2 \!-\! z_1)\nn
\\[1mm]
&=\ii\tilde{M}_{\!R}\hp e^{-\pi \tlambda}I^{p_1 p_2}(-\ii\hp\tlambda\!+\!\ii\hp\tnu,-\ii\hp\tlambda\!-\!\ii\hp\tnu,-\ii\hp\tlambda\!+\!\ii\hp\tnu)\,,
\\[1mm]
\label{S3define}
S_3^{p_1 p_2} &= \int_0^\infty \!\!\td z_1 \!\int_0^\infty \!\!\td z_2\,
e^{\ii\hp z_1} e^{\ii\hp z_2}
z_1^{p_1} z_2^{p_2}
u_+(z_1) v_+^\dagger(z_2)\,
\theta(z_1 \!-\! z_2)\nn
\\[1mm]
&= -\ii\tilde{M}_{\!R}\hp e^{\pi \tlambda} J^{p_2 p_1}(-\ii\hp\tlambda\!+\!\ii\hp\tnu,-\ii\hp\tlambda\!-\!\ii\hp\tnu,-\ii\hp\tlambda\!+\!\ii\hp\tnu)\,,
\\[1mm]
S_4^{p_1 p_2} &= \int_0^\infty \!\!\td z_1 \!\int_0^\infty \!\!\td z_2\,
e^{\ii\hp z_1} e^{\ii\hp z_2}
z_1^{p_1} z_2^{p_2}
u_-(z_1) v_-^\dagger(z_2)\,
\theta(z_1 \!-\! z_2)
\nn
\\[1mm]
&=\ii\tilde{M}_{\!R}\hp e^{-\pi \tlambda} J^{p_2 p_1}(1\!+\!\ii\hp\tlambda\!+\!\ii\hp\tnu,1\!+\!\ii\hp\tlambda\!-\!\ii\hp\tnu,1\!+\!\ii\hp\tlambda\!+\!\ii\hp\tnu)
\,,
\end{align}
\eeqs
}
\hspace*{-2.5mm}
where the function $I^{p_1 p_2}(x,y,z)$ 
is defined in Eq.\eqref{functionI}, 
and the function $J^{p_1 p_2}(x,y,z)$ 
takes the following form:
{\small
\begin{equation}
\begin{aligned}
J^{p_1 p_2}(x,y,z)=&~
\ii\hp2^{-2-p_1-p_2}
 e^{\frac{\ii\hp\pi}{2}\(p_1+p_2\)} \pi\,\csch(2\pi\tnu)\, \Gamma(2\!+\!p_1\!+\!p_2)
 \\
&\times\!
\Bigg[
\frac{\,e^{\pi \tnu}\,}{\,\Gamma(x)\,}\Gamma(1\!+\!p_1\!-\!\ii\hp\tnu)\,
\Gamma(2\!+\!p_1\!+\!p_2\!-\!\ii\hp2\tnu)
 \\
&\quad\,\,\times\!
{}_4\tilde F_3\!\(
\begin{matrix}
2\!+\!p_1\!+\!p_2,\;
1\!+\!p_1\!-\!\ii\hp\tnu,\;
y,\;
2\!+\!p_1\!+\!p_2\!-\!\ii\hp2\tnu\hp
\\
2\!+\!p_1\!-\!\ii\hp\tnu,\;
3\!+\!p_1\!+\!p_2\!-\!z,\;
1\!-\!\ii\hp2\tnu
\end{matrix}
\bigg|-\!\!1\!\)\!-\!(\tnu\ito -\tnu)\Bigg]\,.
\end{aligned}
\end{equation}
}
\hspace*{-2.5mm}
In Eq.\eqref{eq:I-p1p2+-/-+-W12}, the matrix 
elements $W_i^{p_1 p_2}$ ($i=1,2$) 
are defined as follows:
{\small
\beqs
\begin{align}
W_1^{p_1 p_2}
&= \int_0^\infty \!\!\td z_1 
\!\int_0^\infty \!\!\td z_2\,
e^{\ii\hp z_1} e^{-\ii\hp z_2}
z_1^{p_1} z_2^{p_2}
u_+(z_1) v_+^\dagger(z_2)
=-\ii\tilde{M}_{\!R}\hp e^{\pi \tlambda}K^{p_1p_2}(2\!+\!\ii\hp \tlambda,1\!-\!\ii\hp\tlambda)\,,
\\[1mm]
W_2^{p_1 p_2}
 &= \int_0^\infty \!\!\td z_1 
\!\int_0^\infty \!\!\td z_2\,
e^{-\ii\hp z_1} e^{\ii\hp z_2}
z_1^{p_1} z_2^{p_2}
u_-(z_1) v_-^\dagger(z_2)
=\ii\tilde{M}_{\!R}\hp e^{-\pi \tlambda}K^{p_1p_2}(1\!-\!\ii\hp \tlambda,2\!+\!\ii\hp\tlambda)\,,
\end{align}
\eeqs
}
\hspace*{-2.5mm}
where the function $K^{p_1 p_2}(x,y)$ 
is defined in Eq.\eqref{functionK}.

\vspace*{1.5mm}
\subsubsection{\hspace*{-2.5mm}Seed Integrals for \texorpdfstring{$\check{D}_{ab\,\dalpha\dbeta}(\mathbf{k},z_1,z_2)$}{}}
\label{sec:2.4.3}
\vspace*{1.5mm}

Similarly, we consider the propagator $\check{D}_{ab\,\alpha\beta}(\mathbf{k},\tau_1,\tau_2)$ and 
define its associated seed integrals $\check{\mathcal{I}}_{ab\hspace{3mm}\alpha\beta}^{p_1 p_2}$ as follows: 
{\small
\begin{equation}
\begin{aligned}
\check{\mathcal{I}}_{ab\hspace{3mm}\dalpha\dbeta}^{p_1 p_2}
&=
k_1^{1+p_1}k_2^{1+p_2}\!\!
\int_{-\infty}^{0^-}\!\!\td \tau_1 \td \tau_2 \,
e^{\ii\hp a k\tau_1}e^{\ii\hp b k\tau_2}
(-\tau_1)^{p_1}(-\tau_2)^{p_2}
\check{D}_{ab\,\dalpha\dbeta}(\mathbf{k},\tau_1,\tau_2)
\\
&=
\int_0^{\infty}\!\!\td z_1 \td z_2 \,
e^{-\ii\hp a z_1}e^{-\ii\hp b z_2}
z_1^{p_1}z_2^{p_2}
\check{D}_{ab\,\dalpha\dbeta}(\mathbf{k},z_1,z_2)\,.
\end{aligned}
\end{equation}
}
\hspace*{-2.5mm}
The seed integrals $\check{\mathcal{I}}_{ab\hspace{3mm}\dalpha\dbeta}^{p_1 p_2}$ are related to the hatted counterparts $\hat{\mathcal{I}}_{ab\hspace{3mm}\alpha\beta}^{p_1 p_2}$ through the relation, 
\\[-6mm]
{\small
\begin{equation}
\label{eq:check{I}=-hat{I}*}
\check{\mathcal{I}}_{ab\hspace{2mm}\dalpha\dbeta}^{p_1p_2}
=\!\(\hat{\mathcal{I}}_{-b\,-a\hspace{0.5mm}\beta\alpha}^{p_2^{*} p_1^{*}}\)^{\!*},
\end{equation}
}
which could be derived through Eq.\eqref{conjugation_propagator}.

\vspace*{1.5mm}
\subsubsection{\hspace*{-2.5mm}Rotation of Spinor Seed Integrals}
\label{Rotation of Spinor Seed Integrals}
\label{sec:2.4.4}
\vspace*{1.5mm}

In the above subsections, we have computed all these seed integrals by choosing the spatial momentum to be along the direction of the $z$-axis  ($\mathbf{k}\!=\!k\hat z\hp$).\ 
Since the seed integrals are spinor-tensors, 
we can derive them for a general momentum direction by a proper $SU(2)$ rotation.\  

Consider the spatial momentum in a general direction,
\begin{equation}
\hat{\,\mathbf k}=(\sin\theta\cos\varphi,\sin\theta\sin\varphi,\cos\theta) \hp . 
\end{equation}
The helicity spinor $h_\alpha(\hat{\,\mathbf k})$ is 
in the $(\frac{1}{2},0)$ representation of the Lorentz group.\  
The corresponding rotation matrix for the helicity spinor $h_\alpha(\mathbf{k})$ 
from $h_\alpha(\hat z)$ is given by 
{\small
\begin{equation}
U(\theta,\varphi) =
e^{-\frac{\ii}{2}\varphi\sigma^3}
e^{-\frac{\ii}{2}\theta\sigma^2}=
\begin{pmatrix} 
e^{-\frac{\ii}{2}\varphi}\cos\frac{\theta}{2} & 
~-e^{-\frac{\ii}{2}\varphi}\sin\frac{\theta}{2}
\\[1.5mm]
e^{\frac{\ii}{2}\varphi}\sin\frac{\theta}{2} 
& ~e^{\frac{\ii}{2}\varphi}\cos\frac{\theta}{2}
\end{pmatrix} \!,
\end{equation}
}%
which is a unitary matrix obeying $U^\dagger(\theta,\varphi)\!=\! 
U^{-1}(\theta,\varphi)\hp$.
Similarly, we can derive the rotation matrix for the spinor 
$h^\dagger_\dalpha(\hat{\,\mathbf k})$, which belongs to the $(0,\frac{1}{2})$ 
representation of the Lorentz group.\ 
For a spinor tensor as a direct product of two spinors (such as $M_{\alpha\beta}$), 
we apply the rotation $U(\theta,\varphi)$ 
to each of the spinor-tensor indices.\ 
We summarize in Table\,\ref{tab:1} the rotation rules 
for six types of the spinor-tensors considered in the present analysis.

\tabcolsep 1pt
\begin{table}[H]
\centering
\renewcommand{\arraystretch}{1.5}
\begin{tabular}{c||c|c}
\hline\hline
Tensor Type\, & General Rotation Rule & Matrix Form \\
\hline\hline
{\small
$M_{\alpha\beta}$ 
}
&
{\small
$M_{\alpha\beta}(\mathbf{k})
\!=\!
U_{\alpha}{}^{\gamma}
U_{\beta}{}^{\delta}
M^{(z)}_{\gamma\delta}(k)$
}
&
{\small
$M(\mathbf{k})\!=\!UM^{(z)}U^T$
}
\\
\hline
{\small
$M_{\dot\alpha\dot\beta}$ 
}
&
{\small
$M_{\dot\alpha\dot\beta}(\mathbf{k})
\!=\!
U^*_{\dot\alpha}{}^{\dot\gamma}
U^*_{\dot\beta}{}^{\dot\delta}
M^{(z)}_{\dot\gamma\dot\delta}(k)$
}
&
{\small
$M(\mathbf{k})\!=\!U^*M^{(z)}U^\dagger$
}
\\
\hline
{\small
$M_{\alpha\dot\beta}$ 
}
&
{\small
$M_{\alpha\dot\beta}(\mathbf{k})
\!=\!
U_{\alpha}{}^{\gamma}
U^*_{\dot\beta}{}^{\dot\delta}
M^{(z)}_{\gamma\dot\delta}(k)$
}
&
{\small
$M(\mathbf{k})\!=\!UM^{(z)}U^\dagger$
}
\\
\hline
{\small
$M_{\alpha}{}^{\beta}$ 
}
&
{\small
$M_{\alpha}{}^{\beta}(\mathbf{k})
\!=\!
U_{\alpha}{}^{\gamma}
M^{(z)}_{\gamma}{}^{\delta}(k)
(U^{-1})_{\delta}{}^{\beta}$
}
&
{\small
$M(\mathbf{k})\!=\!UM^{(z)}U^{-1}\!=\!UM^{(z)}U^{\dagger}$
}
\\
\hline
{\small
$M^{\dot\alpha}{}_{\dot\beta}$ 
}
&
{\small
$M^{\dot\alpha}{}_{\dot\beta}(\mathbf{k})
\!=\!
\big((U^\dagger)^{-1}\big)^{\dot\alpha}{}_{\dot\gamma}
M^{(z)\dot\gamma}{}_{\dot\delta}(k)
U^*_{\dot\beta}{}^{\dot\delta}$
}
&
{\small
$M(\mathbf{k})\!=\!(U^\dagger)^{-1}M^{(z)}U^\dagger\!=\!UM^{(z)}U^{\dagger}$
}
\\
\hline
{\small
$M^{\dot\alpha\beta}$ 
}
&
{\small
$M^{\dot\alpha\beta}(\mathbf{k})
\!=\!
\big((U^\dagger)^{-1}\big)^{\dot\alpha}{}_{\dot\gamma}
M^{(z)\dot\gamma\delta}(k)
(U^{-1})_{\delta}{}^{\beta}$
}
&
{\small
$M(\mathbf{k})\!=\!(U^\dagger)^{-1}M^{(z)}U^{-1}\!=\!UM^{(z)}U^{\dagger}$
}
\\
\hline\hline
\end{tabular}
\caption{\small 
Rotation rules for six types of the spinor tensors.\ The symbol $M^{(z)}$ denotes the relevant result evaluated for the spatial momentum in the direction of $z$-axis.}
\label{tab:spinor_rotation_general}
\label{tab:1}
\end{table}

Consider a special case $\varphi\!=\!0\hp$, 
where $\mathbf{k}$ lies in the $x$-$z$ plane and has an angle $\theta$ relative to the $z$-axis,
namely, 
$\hat{\,\mathbf k}\!=\!(\sin\theta,0,\cos\theta)$.\ 
Thus, the rotation matrix reduces to
\begin{equation}
U(\theta,0)
=
e^{-\frac{\ii}{2}\theta\sigma^2}
=
\begin{pmatrix}
\cos\frac{\theta}{2} & -\sin\frac{\theta}{2}
\\[1.5mm]
\sin\frac{\theta}{2} & \cos\frac{\theta}{2}
\end{pmatrix}\!.
\end{equation}
Since $U(\theta,0)$ is real, we have 
$U^*(\theta,0)\!=\!U(\theta,0)$ and  $U^\dagger(\theta,0)\!=\!U^T(\theta,0)\!=\!U^{-1}(\theta,0)\hp$.\ 
Thus, the rotation for all types of spinor-tensors in Table\,\ref{tab:1} can be written in a single matrix form,
\begin{equation}
\label{rotation_no_phi}
    M(\mathbf{k})=U(\theta,0)\,M^{(z)}\,U(\theta,0)^T.
\end{equation}

\section{\hspace*{-2.5mm}Cosmological Signatures of Right-Handed Neutrino during Inflation}
\label{sec:3}
\label{sec:4new}

After presenting the spinor coefficients, the Majorana fermion propagators, and the seed integrals within the Schwinger-Keldysh formalism in Section\,\ref{sec:2}, we then turn to the 
loop-induced primordial signatures generated by the right-handed neutrino.\ The derivative interaction in Eq.\eqref{dimension5_operator} not only induces an effective chemical potential through the slow-roll inflaton background, but also couples the inflaton fluctuation $\delta\phi$ directly to the right-handed neutrinos.\ In consequence, the heavy-neutrino loop leaves imprints on the inflaton bispectrum $\langle \delta\phi(\mathbf{k}_1)\delta\phi(\mathbf{k}_2)\delta\phi(\mathbf{k}_3)\rangle'$ and can generate the non-analytic cosmological collider signals, as shown in Fig.\,\ref{illustration}.

\vs 

The rest of this section is organized as follows.\ In Section\,\ref{The Inflaton Three-Point Function from the Right-Handed Neutrino Loop}, we present the inflaton three-point correlator generated by the right-handed neutrino triangle loop and classify the eight distinct contractions.\ 
In Section\,\ref{Calculation of the Three-Point Correlator}, we factorize the loop diagram into three subdiagrams containing the left part, the right part, and the middle part (bubble) in the nonlocal sector, which will be computed separately.\ In Section\,\ref{Final Result of the Three-Point Correlator}, we derive the final bispectrum, identify the dominant contributions, and analyze their angular dependence.

\subsection{\hspace*{-2.5mm}Three-Point Inflaton Correlator from Right-Handed Neutrino Loop}
\label{The Inflaton Three-Point Function from the Right-Handed Neutrino Loop}

We discussed in Section\,\ref{Fermions in an Inflationary Slow-roll Background} how the inflaton background $\phi_0^{}$ modifies the dynamics of the right-handed neutrino through the dimension-5 operator in Eq.\eqref{dimension5_operator}.\ 
In addition, the inflaton fluctuation $\delta\phi$ also couples to the right-handed neutrino through the same dimension-5 operator,
\begin{equation}
    \Delta\LG =\! \sqrt{-g\,}\frac{-1}{a(\tau)\Lambda} \pd_\mu \delta\phi \, N^\dagger \bar{\sigma}^\mu N
    = \frac{\,-1\,}{\Lambda} \pd_\mu \delta\phi \, \tilde{N}^\dagger \bar{\sigma}^\mu \tilde{N} \,,
\end{equation}
where $\tilde{N}\!=\!a^{3/2}{N}\hp$.\ 
Through this interaction, the quantum fluctuation $\delta\phi$ receives a backreaction from the right-handed neutrino $N$.\ This effect is ultimately encoded in the primordial non-Gaussianity through the three-point correlator of $\delta\phi$ generated by the right-handed neutrino triangle loop.\ The right-handed neutrino has a nonzero Majorana mass 
$m\!\gtrsim\! H$ and a large chemical potential $\lambda\hsm\thicksim\hsm 60H$.\ This loop contribution can generate a cosmological collider signal in the three-point correlator of $\delta\phi\hp$,
\begin{equation}
\begin{aligned}
\langle\delta\phi(\mathbf{k}_1)\delta\phi(\mathbf{k}_2)\delta\phi(\mathbf{k}_3)\rangle^{\prime}
\,\thicksim\, A \!\(\frac{k_1}{k_3}\)^{\!\!\ii\hp \tnu}\!\!+\!\! \hc\!\hsm ,
\end{aligned}
\end{equation}
where we have defined
$\tilde{M}_{\!R} \!=\!{M_{\!R}}/{H}$, 
$\tlambda \!=\!{\lambda}/{H}$, 
and $\tnu \!=\! \sqrt{\tilde{M}_{\!R}^2 \!+\! \tlambda^2\,}$.\ 

\vs 

\begin{figure}[t]
\centering
\resizebox{0.9\linewidth}{!}{%
\begin{minipage}{\linewidth}
\centering
\captionsetup[subfigure]{skip=1pt} 
\begin{subfigure}[b]{0.24\textwidth}
\centering
\hspace*{-0.2\textwidth}
\vspace*{-3mm}
\includegraphics[width=1.3\linewidth]{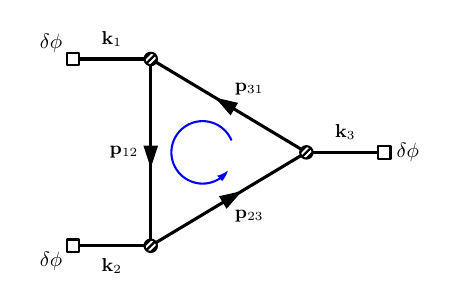}
\caption*{(1)}
\label{fig:feyn1}
\end{subfigure}\hfill
\begin{subfigure}[b]{0.24\textwidth}
\centering
  \hspace*{-0.2\textwidth}
  \vspace*{-3mm}
\includegraphics[width=1.3\linewidth]{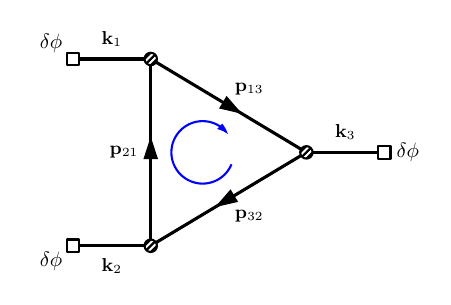}
\caption*{(2)}
\label{fig:feyn2}
\end{subfigure}\hfill
\begin{subfigure}[b]{0.24\textwidth}
\centering
  \hspace*{-0.2\textwidth}
  \vspace*{-3mm}
\includegraphics[width=1.3\linewidth]{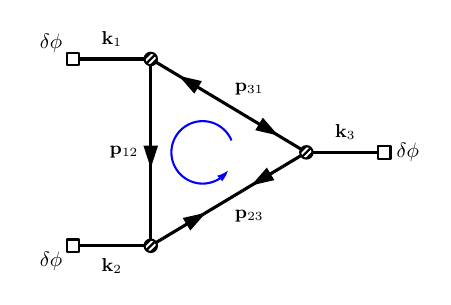}
\caption*{(3)}
\label{fig:feyn3}
\end{subfigure}\hfill
\begin{subfigure}[b]{0.24\textwidth}
\centering
  \hspace*{-0.2\textwidth}
  \vspace*{-3mm}
\includegraphics[width=1.3\linewidth]{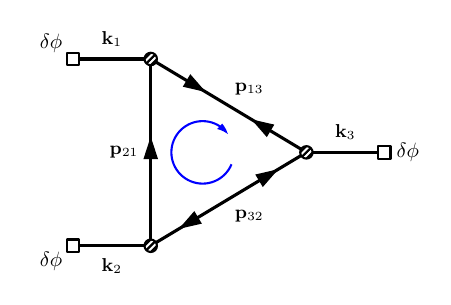}
\caption*{(4)}
\label{fig:feyn4}
\end{subfigure}
\\[-2ex]
\begin{subfigure}[b]{0.24\textwidth}
\centering
  \hspace*{-0.2\textwidth}
  \vspace*{-3mm}
\includegraphics[width=1.3\linewidth]{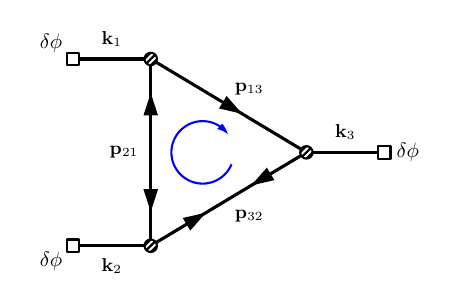}
\caption*{(5)}
\label{fig:feyn5}
\end{subfigure}\hfill
\begin{subfigure}[b]{0.24\textwidth}
\centering
  \hspace*{-0.2\textwidth}
  \vspace*{-3mm}
\includegraphics[width=1.3\linewidth]{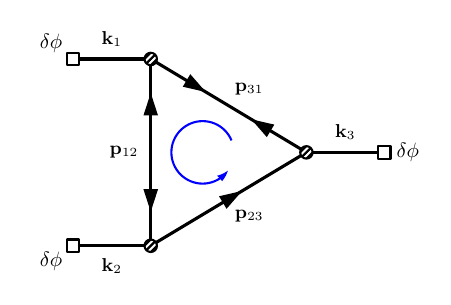}
\caption*{(6)}
\label{fig:feyn6}
\end{subfigure}\hfill
\begin{subfigure}[b]{0.24\textwidth}
\centering
  \hspace*{-0.2\textwidth}
  \vspace*{-3mm}
\includegraphics[width=1.3\linewidth]{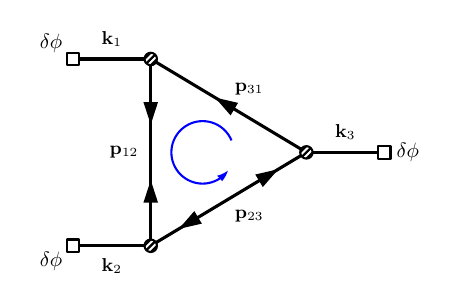}
\caption*{(7)}
\label{fig:feyn7}
\end{subfigure}\hfill
\begin{subfigure}[b]{0.24\textwidth}
\centering
  \hspace*{-0.2\textwidth}
  \vspace*{-3mm}
\includegraphics[width=1.3\linewidth]{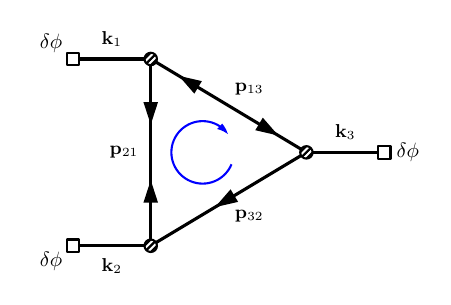}
\caption*{(8)}
\label{fig:feyn8}
\end{subfigure}
\end{minipage}%
}
\vspace*{-2mm}
\caption{\small Relevant diagrams arising from the right-handed neutrino loop.\ In this figure, each  propagator of the right-handed neutrino, 
such as $\langle\tilde N \tilde N^\dagger\rangle$
and $\langle\tilde N \tilde N\rangle$, 
is represented by a solid line with a single arrow or two reverse arrows.\ The line connecting a dot and a square without an arrow denotes a bulk-to-boundary propagator of the inflaton 
$\langle\delta \phi^2\rangle$.\ 
The square at one end marks the boundary point $(\tau_{\!f}\!\to\! 0^-)\,$.\ The shaded dot at the vertex indicates the two cases of time ordering in this vertex, including the time-ordered case ``$\oplus$'' and reverse-time-ordered case ``$\ominus$''.}
\label{fig:feyn_all_2x4}
\label{fig:2}
\end{figure}

In the SK formalism, the three-point correlator generated by the right-handed neutrino loop 
can be represented diagrammatically.\ Using the identity of the sigma matrices,
$\bar{\sigma}^{\mu\dot{\alpha}\alpha}
\!=\!
\epsilon^{\alpha\beta}\epsilon^{\dot{\alpha}\dot{\beta}}\sigma^{\mu}_{\beta\dot{\beta}}\hp$,
the inflaton-right-handed-neutrino interaction
term can be expressed as follows:
\begin{equation}
    -\frac{1}{\Lambda} \pd_\mu \delta\phi \, \tilde{N}^\dagger \bar{\sigma}^\mu \tilde{N}
    = \frac{1}{\Lambda} \pd_\mu \delta\phi \, \tilde{N} \sigma^\mu \tilde{N}^\dagger .
\end{equation}
As a result, there are eight distinct contractions of the neutrino fields in the triangle loop.\ 
This means that for a fixed momentum configuration $(\mathbf{k}_1, \mathbf{k}_2, \mathbf{k}_3)$, 
the three-point correlator receives contributions from eight diagrams, as shown in  Fig.\,\ref{fig:feyn_all_2x4}. 

\vs 

In the above diagrams, the external lines terminating at squares represent the inflaton propagator $\langle \delta\phi^2 \rangle$, whereas the internal arrow-lines represent the neutrino propagators.\ The square at one end denotes the boundary point $(\tau_{\!f} \ito 0^-)\,$.\ 
The shaded dot at each vertex indicates the two possible time-ordering prescriptions at that vertex: the time-ordered case ``$\oplus$'' and the reverse-time-ordered case ``$\ominus$''.\ 
Using the SK path-integral formalism, the 
three-point correlator of the inflaton,
$\langle\delta\phi(\mathbf{k}_1)\delta\phi(\mathbf{k}_2)\delta\phi(\mathbf{k}_3)\rangle^{\prime}$\,,
obtained by summing over the eight diagrams, can be derived as follows:
{\small
\begin{equation}
\label{3pt_form}
\begin{aligned}
\hspace*{-5mm}
&\langle\delta\phi(\mathbf{k}_1)\delta\phi(\mathbf{k}_2)\delta\phi(\mathbf{k}_3)\rangle^{\prime}
=\sum_{j=1}^{8}\langle\delta\phi(\mathbf{k}_1)\delta\phi(\mathbf{k}_2)\delta\phi(\mathbf{k}_3)\rangle^{\prime(j)}
\\
\hspace*{-5mm}
&=\!\!\sum_{a,b,c=\pm} \!\!\!a\hp b\hp c \(\!\frac{\,-\ii\,}{\,\Lambda\,}\!\)^{\!\!3} 
\!\!\!\int_{-\infty}^{0^-}\!\!\!
\mathrm{d}\tau_1 \mathrm{d}\tau_2 \mathrm{d}\tau_3 \,
 \mathcal{F}_{\mu a}(\mathbf{k}_1,\tau_1)
 \mathcal{F}_{\nu b}(\mathbf{k}_2,\tau_2)
 \mathcal{F}_{\lambda c}(\mathbf{k}_3,\tau_3)
\!\!\int\!\!
\frac{\mathrm{d}^3\mathbf{q}}{\hp (2\pi)^3\,}
\!\sum_{j=1}^{8}\!\mathcal{Y}_{abc}^{\mu\nu\lambda}(j)\,,
\end{aligned}
\end{equation}
}
\hspace*{-2.5mm}
where 
$\langle\delta\phi(\mathbf{k}_1)\delta\phi(\mathbf{k}_2)\delta\phi(\mathbf{k}_3)\rangle^{\prime(j)}$ denotes the contribution from the diagram-$(j)$ in Fig.\,\ref{fig:feyn_all_2x4}, 
$a,b,c\!=\!\pm$ are the SK labels, and the factor $(-\ii\hp/\Lambda)^3$ arises from the three interaction vertices.\ The vector function $\mathcal{F}_{\mu a}(\mathbf{k},\tau)$ comes from the external leg of the inflaton field $\delta\phi(\mathbf{k})$ and is given by
{\small
\begin{equation}
\label{def_external_{(d)}erivative}
\mathcal{F}_{\mu a}(\mathbf{k},\tau) 
= \!
\begin{pmatrix}
\partial_{\tau}G_a(k;\tau) 
\\[1mm]
\ii\hp \mathbf{k}G_a(k;\tau)
\end{pmatrix}
=
\frac{\,H^2\,}{\,2k^3\,}\!\!
\begin{pmatrix}
k^2\tau 
\\[1mm]
\ii\hp \mathbf{k}(1-\ii\hp ak\tau)
\end{pmatrix}
\!e^{+\ii\hp ak\tau},
\end{equation}
}
\hspace*{-2.5mm}
where $G_a(k;\tau)$ denotes the boundary-to-bulk propagator of the inflaton field treated 
as an approximately massless scalar field, shown in Eq.\eqref{inflaton_bulk-to-boundary} in the Appendix\,\ref{A Brief Review of Cosmological Collider Physics}
\begin{equation}
\label{boundarypropagator}
G_{a}(k; \tau) = 
\frac{\,H^2\,}{2k^3} ( 1- \ii\hp a \hp k \tau) 
e^{ \ii\hp a\hp k \tau}\,.
\end{equation}
In Eq.\eqref{3pt_form}, 
the functions $\mathcal{Y}_{abc}^{\mu\nu\lambda}(j)$ (with $j=1,2,\cdots,8)$ represent the neutrino triangle loop associated with the $j$-th diagram in Fig.\,\ref{fig:feyn_all_2x4} and 
take the following forms,
{\small
\begin{align}
\label{all_triangleloopform}
\hspace*{-8mm}
   \mathcal{Y}_{abc}^{\mu\nu\lambda}(1) =& -\bar{\sigma}^{\mu\dalpha\alpha}D_{ab\alpha\dbeta}(\mathbf{p}_{12},\tau_{1},\tau_{2})\bar{\sigma}^{\nu\dbeta\beta}D_{bc\beta\dgamma}(\mathbf{p}_{23},\tau_{2},\tau_{3})\bar{\sigma}^{\lambda\dgamma\gamma}D_{ca\gamma\dalpha}(\mathbf{p}_{31},\tau_{3},\tau_{1})\,,\nn
\\
\hspace*{-8mm}
   \mathcal{Y}_{abc}^{\mu\nu\lambda}(2) =& -\bar{\sigma}^{\nu\dalpha\alpha}D_{ba\alpha\dbeta}(\mathbf{p}_{21},\tau_{2},\tau_{1})\bar{\sigma}^{\mu\dbeta\beta}D_{ac\beta\dgamma}(\mathbf{p}_{13},\tau_{1},\tau_{3})\bar{\sigma}^{\lambda\dgamma\gamma}D_{cb\gamma\dalpha}(\mathbf{p}_{32},\tau_{3},\tau_{2})\,,\nn
\\
\hspace*{-8mm}
  \mathcal{Y}_{abc}^{\mu\nu\lambda}(3) =& \,\, \bar{\sigma}^{\mu\dalpha\alpha}D_{ab\alpha\dbeta}(\mathbf{p}_{12},\tau_{1},\tau_{2})\bar{\sigma}^{\nu\dbeta\beta}\hat{D}_{bc\beta}^{\,\,\,\,\,\,\,\,\gamma}(\mathbf{p}_{23},\tau_{2},\tau_{3})\sigma^{\lambda}_{\,\,\,\gamma\dgamma}\check{D}_{ca\,\,\dalpha}^{\,\,\,\,\,\dgamma}(\mathbf{p}_{31},\tau_{3},\tau_{1})\,,\nn
\\
\hspace*{-8mm}
   \mathcal{Y}_{abc}^{\mu\nu\lambda}(4) =&\,\, \bar{\sigma}^{\nu\dalpha\alpha}D_{ba\alpha\dbeta}(\mathbf{p}_{21},\tau_{2},\tau_{1})\bar{\sigma}^{\mu\dbeta\beta}\hat{D}_{ac\beta}^{\,\,\,\,\,\,\,\,\,\gamma}(\mathbf{p}_{13},\tau_{1},\tau_{3})\sigma^{\lambda}_{\,\,\,\gamma\dgamma}\check{D}_{cb\,\,\dalpha}^{\,\,\,\,\dgamma}(\mathbf{p}_{32},\tau_{3},\tau_{2})\,,\nn
\\
\hspace*{-8mm}
   \mathcal{Y}_{abc}^{\mu\nu\lambda}(5) =& \,\,\sigma^{\nu}_{\,\,\,\alpha\dalpha}\check{D}_{ba\,\,\,\dbeta}^{\,\,\,\,\dalpha}(\mathbf{p}_{21},\tau_{2},\tau_{1})
   \bar{\sigma}^{\mu\dbeta\beta}D_{ac\beta\dgamma}(\mathbf{p}_{13},\tau_{1},\tau_{3})\bar{\sigma}^{\lambda\dgamma\gamma}\hat{D}_{cb\gamma}^{\,\,\,\,\,\,\,\,\,\alpha}(\mathbf{p}_{32},\tau_{3},\tau_{2})\,,
\\
\hspace*{-8mm}
   \mathcal{Y}_{abc}^{\mu\nu\lambda}(6) =& \,\,\sigma^{\mu}_{\,\,\,\alpha\dalpha}\check{D}_{ab\,\,\,\dbeta}^{\,\,\,\,\,\dalpha}(\mathbf{p}_{12},\tau_{1},\tau_{2})
   \bar{\sigma}^{\nu\dbeta\beta}D_{bc\beta\dgamma}(\mathbf{p}_{23},\tau_{2},\tau_{3})\bar{\sigma}^{\lambda\dgamma\gamma}\hat{D}_{ca\gamma}^{\,\,\,\,\,\,\,\,\,\alpha}(\mathbf{p}_{31},\tau_{3},\tau_{1})\,,\nn
\\
   \hspace*{-8mm}
   \mathcal{Y}_{abc}^{\mu\nu\lambda}(7) =& \,\,\bar{\sigma}^{\mu\dalpha\alpha}\hat{D}_{ab\alpha}^{\,\,\,\,\,\,\,\,\,\beta}(\mathbf{p}_{12},\tau_{1},\tau_{2})\sigma^{\nu}_{\,\,\,\beta\dbeta}\check{D}_{bc\,\,\,\dgamma}^{\,\,\,\,\,\dbeta}(\mathbf{p}_{23},\tau_{2},\tau_{3})\bar{\sigma}^{\lambda\dgamma\gamma}D_{ca\gamma\dalpha}(\mathbf{p}_{31},\tau_{3},\tau_{1})\,,\nn
\\
   \hspace*{-8mm}
   \mathcal{Y}_{abc}^{\mu\nu\lambda}(8) =& \,\,\bar{\sigma}^{\nu\dalpha\alpha}\hat{D}_{ba\alpha}^{\,\,\,\,\,\,\,\,\,\beta}(\mathbf{p}_{21},\tau_{2},\tau_{1})\sigma^{\mu}_{\,\,\,\beta\dbeta}\check{D}_{ac\,\,\,\dgamma}^{\,\,\,\,\,\dbeta}(\mathbf{p}_{13},\tau_{1},\tau_{3})\bar{\sigma}^{\lambda\dgamma\gamma}D_{cb\gamma\dalpha}(\mathbf{p}_{32},\tau_{3},\tau_{2})\,, \nn  
\end{align}
}
\hspace*{-3.5mm}
where $ \mathbf{p}_{12}^{} \!=\! -\mathbf{p}_{21}^{} \!=\! \mathbf{q}\!+\!\mathbf{k}_1\hp$, 
$\mathbf{p}_{23}^{} \!=\!-\mathbf{p}_{32} \!=\! \mathbf{q} \!+\! \mathbf{k}_s\hp$,\, $\mathbf{p}_{31} \!=\!-\mathbf{p}_{13} \!=\! \mathbf{q}$\,, and $ \mathbf{q} $ is the loop momentum.\ The propagators with upper indices including $\hat{D}_{ab\alpha}^{\,\,\,\,\,\,\,\,\,\beta}$ and $\check{D}_{ab\,\,\,\dalpha}^{\,\,\,\,\,\,\dbeta}$ are defined as follows: 
{\small
\begin{equation}
\label{raiseindex_propagator}
\begin{aligned}
    \hat{D}_{ab\alpha}^{\,\,\,\,\,\,\,\,\,\beta}(\mathbf{k},\tau_{1},\tau_{2})&\equiv \epsilon^{\beta\gamma}\hat{D}_{ab\alpha\gamma}(\mathbf{k},\tau_{1},\tau_{2})=-\hat{D}_{ab\alpha\gamma}(\mathbf{k},\tau_{1},\tau_{2})\epsilon^{\gamma\beta}\,,
\\[1mm]    \check{D}_{ab\,\,\,\dbeta}^{\,\,\,\,\,\dalpha}(\mathbf{k},\tau_{1},\tau_{2})&\equiv\epsilon^{\dalpha\dgamma}\check{D}_{ab\dgamma\dbeta}(\mathbf{k},\tau_{1},\tau_{2})=-\epsilon^{\dgamma\dalpha}\check{D}_{ab\dgamma\dbeta}(\mathbf{k},\tau_{1},\tau_{2})\,.
\end{aligned}
\end{equation}
}
\hspace*{-2.5mm}
We note that in the literature\,\cite{Chen:2018xck,Hook:2019zxa,Aoki:2026olh} only the first two of the eight diagrams 
in Fig.\,\ref{fig:feyn_all_2x4}
were considered.\ 
In these works\,\cite{Chen:2018xck,Hook:2019zxa,Aoki:2026olh}, the hard internal line $|\mathbf{p}_{12}|$ was estimated using the saddle-point approximation, and the loop momentum was effectively fixed around a characteristic value.\  
The saddle-point approximation may capture the leading exponential dependence, 
but could miss important fluctuation prefactors around the saddle [as will be shown in Fig.\,\ref{fig:4a}].
In this work, we include all eight diagrams shown in 
Fig.\,\ref{fig:feyn_all_2x4} and perform a complete analysis of these diagrams.

\subsection{\hspace*{-2.5mm}Calculation of Three-Point Inflaton Correlator}
\label{Calculation of the Three-Point Correlator}

In this subsection, we calculate the three-point correlator associated with the triangle loop shown in Fig.\,\ref{fig:feyn_all_2x4}.\ 
We will extract the cosmological collider signal from the three-point correlator in the squeezed limit $k_3 \!\ll\! k_1, k_2\,$.

\vs 

We consider a specific momentum configuration in which $\mathbf{k}_1$, $\mathbf{k}_2$, and $\mathbf{k}_3$ are nearly collinear along the $z$-axis:
\begin{equation}
\mathbf{k}_1 \!\simeq\! -\mathbf{k}_2\,, \qquad 
\mathbf{k}_1\!=\!(0,0,k_1)\hp , \qquad 
\mathbf{k}_3\!=\!(0,0,k_3)\hp.
\end{equation}
Then, we can consider the case with the angle $\Theta$ between $\mathbf{k}_1$ and $\mathbf{k}_3$ being arbitrary and $\mathbf{k}_3$ being along the $z$-axis.\ 
In this case, we have 
\begin{equation}
\mathbf{k}_1 \!\simeq\!-\mathbf{k}_2\,, \qquad 
\mathbf{k}_1\!=\!k_1(\sin \Theta,0,\cos \Theta)\hp, \qquad 
\mathbf{k}_3\!=\!(0,0,k_3)\hp .  
\end{equation}

As discussed in Section\,\ref{The Local and the nonlocal Signal}, the cosmological collider signal contains both local and nonlocal contributions, where  
these two contributions have different physical origins and can be distinguished more clearly in the four-point correlator.\  
Although their physical origins are different, both the local and nonlocal signals exhibit the same scaling behavior $\(\!\frac{k_1}{k_3}\!\)^{\!\hp\ii \tnu}$ 
in the three-point correlator.\ For the present analysis, 
we will focus on the nonlocal contribution and extract the nonlocal part 
of the fermion propagator.

\vs

In this subsection, we take the diagram-(7) in Fig.\,\ref{fig:feyn_all_2x4} 
as a sample for illustration and present a complete calculation for it.\ 
The same procedure can be applied to the other seven diagrams 
in Fig.\,\ref{fig:feyn_all_2x4}. For simplicity, we set the Hubble parameter during inflation, $H$, to unity ($H\!=\!1$) in this subsection.

\subsubsection{\hspace*{-2.5mm}Factorization of Fermion Loop Diagrams}
\label{Factorization of the correlator loop diagrams}

To extract nonlocal cosmological collider signals from the fermion loop, 
it is useful to isolate the kinematic region in which two internal lines simultaneously become soft.\ 
In analogy with cosmological collider signals from scalar loops, one expects 
that the non-analytic part of the diagram should arise precisely from this soft region 
and admit a factorized description in terms of left, right, and bubble subdiagrams.\

For scalar loops, this factorization can be implemented directly.\ It has been shown that the non-analytic behavior of a loop diagram originates from the kinematic region in which two specific internal momenta simultaneously become soft\,\cite{Xianyu230413295}.\ In this limit, the corresponding internal lines can be effectively cut, and the diagram factorizes into three parts: a left tree-level subdiagram $\mathcal{T}^{(\mathsf{L})}$, a right tree-level subdiagram $\mathcal{T}^{(\mathsf{R})}$, and a bubble loop subdiagram $\mathcal{B}$ in the middle.\ Accordingly, the nonlocal cosmological collider signal generated by a scalar loop takes the following schematic form\,\cite{Xianyu230413295}:
{\small
\begin{equation}
\lim_{k_s\!\to\hp 0}
\langle \delta\phi^n(\mathbf{k}_i)\rangle^{\prime}_{\mathrm{(NLoc)}}
\sim
\mathcal{T}^{(\mathsf{L})}\!\big(\hsm\{\mathbf{k}^{(\mathsf{L})}\}\hsm\big)
\!\times\!
\mathcal{B}(k_s)  
\!\times\!
\mathcal{T}^{(\mathsf{R})}\!\big(\hsm\{\mathbf{k}^{(\mathsf{R})}\}\hsm\big) \hp.
\end{equation}
}
\hspace*{-2.5mm}
Following the same logic, for the fermionic triangle diagrams in Fig.\,\ref{fig:feyn_all_2x4}, the internal momenta $\mathbf{p}_{31}^{}$ and $\mathbf{p}_{23}^{}$ can be taken to the soft limit, $\mathbf{p}_{31}^{}\,,\mathbf{p}_{23}^{}\!\to\! {O}(k_s)$, which corresponds to the late-time limit described in Eq.\eqref{latetime_mode}.\ For the nonlocal contribution, we expect the three-point correlator in the squeezed limit to have the following factorized form:

\begin{equation}
\label{factorization_3pt}
\begin{aligned}
\lim_{k_s\!\to\hp 0}
\langle
\delta\phi(\mathbf{k}_1)\delta\phi(\mathbf{k}_2)\delta\phi(\mathbf{k}_3)
\rangle^{\prime}_{\mathrm{(NLoc)}}
\simeq
\sum_{d=\pm}\!
\mathcal{T}^{(\mathsf{L})}_{(d)}\!(\mathbf{k}_1,\mathbf{k}_2)
\hp\mathcal{B}_{(d)}(\mathbf{k}_s)\hp 
\mathcal{T}^{(\mathsf{R})}_{(d)}\!(\mathbf{k}_3) \hp ,
\end{aligned}
\end{equation}
where $\mathbf{k}_s\!=\!\mathbf{k}_1\!+\!\mathbf{k}_2\!=\!-\mathbf{k}_3$
for the three-point correlator.

\vs 

As far as the nonlocal signal is concerned, we apply the cutting rule for the bubble loop subdiagram $\mathcal{B}$, so that the two soft lines $D_{ab}$ in the 
bubble loop subdiagram $\mathcal{B}$ 
could only take the form of Eqs.\eqref{late time limit propagator}-\eqref{late time limit propagator hatDCon} 
without theta-function dependence:\footnote{For the soft lines, the nested-time part in $D_{\oplus\oplus}$ and $D_{\ominus\ominus}$ contains no nonlocal cosmological collider signals \cite{Tong:2021wai,Xianyu230413295}, and the nested time-integrals over 
$\tau_1$ and $\tau_2$ become factorized, $\iint \!\td\tau_1 \td\tau_2
\!\longrightarrow\!\!
\int \!\td\tau_1 \!\int \!\td\tau_2\,$.}
\begin{equation}
D_{ab}(\mathbf{k},\tau_1,\tau_2)\rightarrow \left[D_{\ominus\oplus}(\mathbf{k},\tau_1,\tau_2)\right]_\text{(NLoc)}\,.
\end{equation}

\begin{figure}[t]
\centering
\includegraphics[width=0.85\linewidth]{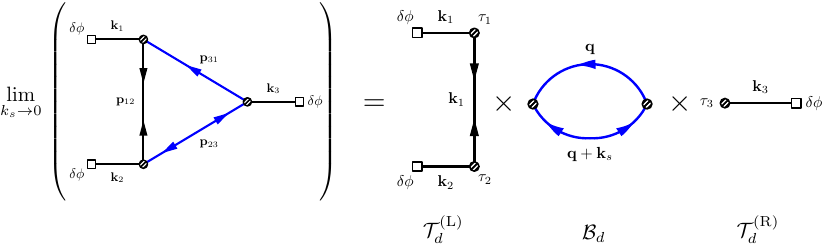}
\vspace*{-2mm}
\caption{\small After applying the factorization method, the diagram-(7) in Fig.\,\ref{fig:feyn_all_2x4} 
can be decomposed into a left subdiagram, a bubble loop, and a right subdiagram.}
\label{fig:3}
\end{figure}

With the above prescription, the factorization of the fermion loop can be implemented diagram by diagram.\ 
As an example, we illustrate in Fig.\,\ref{fig:3} the decomposition of diagram-(7) in Fig.\,\ref{fig:feyn_all_2x4}.\ 
In this decomposition, all time integrals are assigned to the left and right parts.\ 
The left subdiagram $\mathcal{T}^{(\mathsf{L})}_{(d)}$ is a two-point object containing the time integral 
$\iint \!\td\tau_1 \td\tau_2$ together with the time-dependent factors $(-\tau_1)^{\ii\hp d\tnu}$ and $(-\tau_2)^{\ii\hp d\tnu}$ 
inherited from the two soft lines.\ The right subdiagram $\mathcal{T}^{(\mathsf{R})}_{(d)}$ contains only the time integral 
$\int\! \td\tau_3$ and the corresponding factor $(-\tau_3)^{\ii\hp 2d\tnu}$.\ 
In contrast, the bubble factor $\mathcal{B}_{(d)}$ contains only the loop integral over 
$\d\mathbf{q}$, together with the residual Gamma-function coefficients in the two soft propagators.\ 
All non-analytic dependence arises from evaluating this bubble factor in the soft limit 
$\mathbf{p}_{31},\mathbf{p}_{23}\!\to\! {O}(k_s)\hp$.

We specify the SK ordering of the three vertices explicitly and  
denote the three parts by
$\mathcal{T}^{(\mathsf{L})}_{(d),ab}$, $\mathcal{B}_{(d)}$,  
and $\mathcal{T}^{(\mathsf{R})}_{(d),c}$, respectively.\ 
Thus, we express the nonlocal contribution to the squeezed limit three-point correlator 
as follows: 
{\small
\begin{equation}
\begin{aligned}
\lim_{k_s\!\to0}
\langle\delta\phi^3\rangle^{\prime}_{\mathrm{(NLoc)}}
\simeq
\sum_{\substack{a,b,c=\pm \\ d=\pm}}
\!\!\!\!a\hp b\hp c\,
\mathcal{T}^{(\mathsf{L})}_{(d),ab}(\mathbf{k}_1,\mathbf{k}_2)\hp 
\mathcal{B}_{(d)}(\mathbf{k}_s)\hp 
\mathcal{T}^{(\mathsf{R})}_{(d),c}(\mathbf{k}_3) \, .
\end{aligned}
\end{equation}
}
\hspace*{-2.5mm}
For the diagram-(7), the left subdiagram is a two-point correlator containing two external inflaton legs, 
$\mathcal{F}_{\mu a}(\mathbf{k}_1,\tau_1)$ and $\mathcal{F}_{\nu b}(\mathbf{k}_2,\tau_2)$, 
together with one hard internal fermion propagator $\hat{D}_{ab}(\mathbf{p}_{12},\tau_1,\tau_2)$,
which takes the following form:
{\small
\begin{equation}
\mathcal{T}^{(\mathsf{L})}_{(d),ab}
=\!
\(\!\frac{\,-\ii\,}{\,\Lambda\,}\!\)^{\!\!2}
\!\!\!\int \!\td\tau_1 \td\tau_2\,
\mathcal{F}_{\mu a}(\mathbf{k}_1,\tau_1)\hp\bar{\sigma}^{\mu}
\hat{D}_{ab}(\mathbf{p}_{12},\tau_1,\tau_2)
\hp\sigma^{\nu}\mathcal{F}_{\nu b}(\mathbf{k}_2,\tau_2)
(-\tau_1)^{\ii\hp d\tnu}(-\tau_2)^{\ii\hp d\tnu} \, ,
\end{equation}
}
\hspace*{-2.5mm}
where the factors $(-\tau_1)^{\ii\hp d\tnu}(-\tau_2)^{\ii\hp d\tnu}$ arise from the late-time behavior of the two soft lines in Eq.\eqref{late time limit propagator hatD}, and the momenta satisfy $\mathbf{k}_1\!=\!-\mathbf{k}_2^{}\hsm =\hsm \mathbf{p}_{12}^{}\hp$.
The right subdiagram contains no internal fermion propagator and it can be expressed as follows:
\begin{equation}
\mathcal{T}^{(\mathsf{R})}_{(d),c}
=
\(\!\frac{\,-\ii\,}{\,\Lambda\,}\!\)
\!\int \!\td\tau_3\,
\mathcal{F}_{\lambda c}(\mathbf{k}_3,\tau_3)\bar{\sigma}^{\lambda}
(-\tau_3)^{\ii\hp 2d\tnu} \, ,
\end{equation}
where the factor $(-\tau_3)^{\ii\hp 2d\tnu}$ also comes from the two soft lines.

\vs 

We note that the bubble factor depends only on the loop momentum and 
the residual Gamma-function coefficients from the two soft propagators.\
Thus, we can express it in the following form: 
\begin{equation}
\label{bubbleterm_form}
\mathcal{B}_{(d)}
=\sum_{n}C_n
\!\int\!\!\frac{\td^3\mathbf{q}}{(2\pi)^3}(\cos{\theta})^n
q^{\ii\hp 2d\tnu}
\Big|\mathbf{q}\!+\!\mathbf{k}_s\Big|^{\ii\hp 2d\tnu}
\{\Gamma\}_{(d),abc} \, ,
\end{equation}
where $\{\Gamma\}_{(d),abc}$ denotes the remaining Gamma-function factors associated with the two soft lines, and $(\cos{\theta})^n$ is the angle dependence from the spinor structure in the two soft lines and the interaction form, with $\theta$ defined as the angle between the loop momentum $\mathbf{q}$ and the momentum $\mathbf{k}_s$.

In summary, as shown in Fig.\,\ref{fig:3},
the factorization of the triangle diagram in the squeezed limit 
$\mathbf{k}_3\!\ll\! \mathbf{k}_1\!\simeq\! \mathbf{k}_2$ 
originates from the soft momentum region in which the momenta of two internal propagators 
$\mathbf{p}_{31}^{}\!=\!\mathbf{q}$ and 
$\mathbf{p}_{23}^{}\!=\!\mathbf{q}\hsm +\hsm\mathbf{k}_3$ are both soft and of $O(\mathbf{k}_3^{})$, 
whereas the momentum of the third propagator $\mathbf{p}_{12}^{}\!=\!\mathbf{k}_1^{}\hsm -\mathbf{q}$ 
is hard and of $O(\mathbf{k}_1^{})$.\  
In this region, the two soft propagators describe the long-distance propagation of the massive particle 
and generate the non-analytic clock dependence on $\mathbf{k}_3^{}/\mathbf{k}_1^{}$, 
whereas the hard part reduces to local form factors.\  
An auxiliary scale $\Lambda$ satisfying $\mathbf{k}_3\!<\!\Lambda\!\ll\! \mathbf{k}_1$ 
is introduced to explicitly separate the soft and hard parts of the triangle diagram.\  
The soft-cutoff integral itself contains both the cutoff-independent non-analytic term 
and the cutoff-dependent analytic endpoint-tail terms.\  
We find that, once these analytic terms are separated, the non-analytic coefficient extracted 
from the soft-cutoff integral agrees with the coefficient obtained from the full loop-momentum integration  
after analytic continuation, as discussed in Appendix\,\ref{Soft Cutoff and Extraction of the Non-analytic Coefficient}.\  
Hence the cutoff construction provides a physically intuitive prescription for defining the signal and exposing  
the factorized origin of the nonlocal clock term.\ 
In both treatments (either the factorization method or the full loop-momentum integration) under the squeezed limit,  
the physical cosmological collider signal is given by the same cutoff-independent non-analytic coefficient.\  
In the following subsections, we explicitly compute the diagram-(7) as a representative example.\ 
The computation of the remaining diagrams of Fig.\,\ref{fig:2} can be performed in the same fashion.\

\vspace*{1.5mm}
\subsubsection{\hspace*{-2.5mm}Left Part of the Diagram}
\label{sec: the left part}
\label{sec:3.2.2}
\vspace*{1.5mm}

We begin with the left subdiagram of diagram-(7) as shown in Fig.\,\ref{fig:3}.\ 
This two-point subdiagram contributes to $\mathcal{T}_{(d),ab}^{(\mathsf{L})}$
and consists of two massless external inflaton lines, $\mathcal{F}_{\mu a}(\mathbf{k}_1,\tau_1)$ 
and $\mathcal{F}_{\nu b}(\mathbf{k}_2,\tau_2)$, together with one massive internal fermion propagator 
$\hat{D}_{ab}(\mathbf{p}_{12},\tau_1,\tau_2)$.\ 
It carries two spinor indices and takes the following form:
\begin{equation}
\label{left_expression_fig7}
\begin{aligned}
\mathcal{T}^{(\mathsf{L},7)\,\,\dalpha}_{(d),ab\quad\!\!\!\dbeta}(\mathbf{k}_1)
&=\(\!\frac{\,-\ii\,}{\,\Lambda\,}\!\)^{\!\!2}\!\!
\int_{-\infty}^{0^-} \!\!\td\tau_1\td\tau_2 \,
(-\tau_1)^{\ii\hp d\tnu} (-\tau_2)^{\ii\hp d\tnu}
\\
&\quad \times
\mathcal{F}_{\mu a}(\mathbf{k}_1,\tau_1)\,
\bar{\sigma}^{\mu\dalpha\alpha}\,
\hat{D}_{ab\alpha}^{\,\,\,\,\,\,\,\,\,\beta}(\mathbf{p}_{12},\tau_{1},\tau_{2})\,
\sigma^{\nu}_{\beta\dbeta}\,
\mathcal{F}_{\nu b}(\mathbf{k}_2,\tau_2)\, ,
\end{aligned}
\end{equation}
where the factor $(-\tau_1)^{\ii\hp d\tnu} (-\tau_2)^{\ii\hp d\tnu}$ with $d=\pm$ comes from the time dependence 
of the soft lines in the bubble subdiagram $\mathcal{B}_{(d)}$, and the momentum satisfies 
$\mathbf{k}_1\!=\!-\mathbf{k}_2\!=\!\mathbf{p}_{12}$.\ 
For simplicity, we choose $\mathbf{k}_1$ to be along the direction of the $z$-axis, 
$\mathbf{k}_1\!=\!(0,0,k_1)\hp$.\ 
The result for a general orientation of $\mathbf{k}_1$ can be obtained 
by the rotation rule derived in Section\,\ref{Rotation of Spinor Seed Integrals}.\ 

\vs 

In Refs.\,\cite{Chen:2018xck,Hook:2019zxa}, the hard internal propagator $\hat{D}_{ab}(\mathbf{p}_{12},\tau_{1},\tau_{2})$ was evaluated 
using a saddle-point approximation around,
$|2k_1\tau_1|\!=\!|2k_1\tau_2|\simeq \tilde{M}_{\!R}\hp$,
in the regime $\tlambda\gg \tilde{M}_{\!R}$.\ 
In contrast, we introduced in Section\,\ref{Seed Integrals for Fermion Propagators} 
a set of seed integrals that allow one to evaluate the left subdiagram directly.\ 
To express Eq.\eqref{left_expression_fig7} in terms of the seed integrals $\hat{\mathcal{I}}_{ab\hspace{2mm}\alpha\beta}^{p_1p_2}$, 
we substitute the external lines defined in Eq.\eqref{def_external_{(d)}erivative} 
into Eq.\eqref{left_expression_fig7}, and contract the Lorentz indices between 
$\mathcal{F}_{\mu}$ and $\bar{\sigma}^{\mu}$ (or $\sigma^\nu$).\ 
For instance, we can derive:
\begin{equation}
\begin{aligned}
\mathcal{F}_{\mu\oplus}(\mathbf{k}_1,\tau_1)\bar{\sigma}^{\mu}
&=
\mathcal{F}_{0\oplus}(\mathbf{k}_1,\tau_1)\bar{\sigma}^{0}
\!+\!\mathcal{F}_{3\oplus}(\mathbf{k}_1,\tau_1)\bar{\sigma}^{3}
\\
&=
\frac{\,\tau_1 e^{\ii\hp k_1 \tau_1}\,}{2k_1}\bar{\sigma}^{0}
\!+\!\frac{\,\ii\hp e^{\ii k_1\tau_1}(1\!-\!\ii\hp k_1 \tau_1)\,}{2k_1^2}\bar{\sigma}^{3}\, .
\end{aligned}
\end{equation}
In this way, $\mathcal{F}_{\mu\oplus}(\mathbf{k}_1,\tau_1)\bar{\sigma}^{\mu}$ is rewritten as 
a sum of terms with definite power-law dependence on $\tau_1$. Matching the resulting expression 
to the definition of the seed integrals in Eq.\eqref{seedDhat}, we decompose the left subdiagram 
into a linear combination of seed integrals with different powers of $\tau_1$ and $\tau_2\hp$.

\vs 

As an illustration, we consider the case of $a\hsm =\hsm b\hsm =\hsm\oplus\hp$ 
and derive the following:
{\small
\begin{align}
\label{leftpp_fig7}
&\mathcal{T}^{(\mathsf{L},7)\quad\!\!\!\dalpha}_{(d),\oplus\oplus\quad\!\!\!\!\dbeta}(\mathbf{k}_1)\nonumber
\\
&= \(\!\frac{\,-\ii\,}{\,\Lambda\,}\!\)^{\!\!2}\!\!\int_{-\infty}^{0^-} \!\!\td\tau_1\td\tau_2 (-\tau_1)^{\ii\hp d\tnu} (-\tau_2)^{\ii\hp d\tnu}
\mathcal{F}_{\mu \oplus}(\mathbf{k}_1,\tau_1)\bar{\sigma}^{\mu\dalpha\alpha}
\hat{D}_{\oplus\oplus\alpha}^{\quad\,\,\,\,\,\,\beta}(\mathbf{p}_{12},\tau_{1},\tau_{2})
\sigma^{\nu}_{\beta\dbeta}\mathcal{F}_{\nu \oplus}(\mathbf{k}_2,\tau_2)\nonumber
\\
&=\(\!\frac{\,-\ii\,}{\,\Lambda\,}\!\)^{\!\!2}\!\!\frac{1}{\,4(k_1k_2)^{3+\ii\hp d \tnu}\,}
\Big[(\bar{\sigma}^{0}\!+\!\bar{\sigma}^{3})^{\dalpha\alpha}
(\hat{\mathcal{I}}_{\oplus\oplus}^{1+\ii\hp d\tnu,1+\ii\hp d\tnu})^{\,\,\,\,\beta}_{\alpha}
(\sigma^{0}\!-\!\sigma^{3})_{\!\beta\dbeta}
\!-\!\ii\hp\bar{\sigma}^{3\dalpha\alpha}
(\hat{\mathcal{I}}_{\oplus\oplus}^{\ii\hp d\tnu,1+\ii\hp d\tnu})^{\,\,\,\,\beta}_{\alpha}
(\sigma^{0}\!-\!\sigma^{3})_{\!\beta\dbeta}\nonumber
\\
&\qquad\qquad\qquad\qquad\qquad
\!+\!\ii\hp(\bar{\sigma}^{0}+\bar{\sigma}^{3})^{\dalpha\alpha}
(\hat{\mathcal{I}}_{\oplus\oplus}^{1+\ii\hp d\tnu,\ii\hp d\tnu})^{\,\,\,\,\beta}_{\alpha}
\sigma^{3}_{\,\,\beta\dbeta}
\!+\!\bar{\sigma}^{3\dalpha\alpha}
(\hat{\mathcal{I}}_{\oplus\oplus}^{\ii\hp d\tnu,\ii\hp d\tnu})^{\,\,\,\,\beta}_{\alpha}\,
\sigma^{3}_{\,\,\beta\dbeta}\Big]\nonumber
\\
&=\(\!\frac{\,-\ii\,}{\,\Lambda\,}\!\)^{\!\!2}\!\!\frac{1}{\,2(k_1k_2)^{3+\ii\hp d \tnu}\,}\!\!
\Bigg[\!-\!S_1^{\ii\hp d\tnu,\ii\hp d\tnu}\delta^\dalpha_{\,\,\,\,1}\delta_\dbeta^{\,\,\,1}\nonumber
\\
&\quad+\!\(-S_2^{\ii\hp d\tnu,\ii\hp d\tnu} \!-\!\ii\hp2 S_2^{\ii\hp d\tnu,1+\ii\hp d\tnu}\!-\!\ii\hp2 S_2^{1+\ii\hp d\tnu,\ii\hp d\tnu}\!+\!4S_2^{1+\ii\hp d\tnu,1+\ii\hp d\tnu}\)\delta^\dalpha_{\,\,\,2}\delta_\dbeta^{\,\,\,2}\Bigg]\, ,
\end{align}
}
\hspace*{-2.5mm}
where 
$\hat{\mathcal{I}}_{ab\quad\alpha}^{p_1,p_2\,\,\,\beta}
=
-\hat{\mathcal{I}}_{ab\quad\alpha\gamma}^{p_1,p_2}\,\epsilon^{\gamma\beta}\hp$.\ 
We also use the following matrix notation,
\begin{equation}
\begin{aligned} 
\delta^\dalpha_{\,\,\,1}\delta_\dbeta^{\,\,\,1}
=
\bigg(
\begin{matrix}
\,1\,&\,0\,\,
\\
\,0\,&\,0\,\,
\end{matrix}\bigg)^{\!\!\dalpha}_{\,\,\,\dbeta},
\qquad
\delta^\dalpha_{\,\,\,2}\delta_\dbeta^{\,\,\,2}
=
\bigg(
\begin{matrix}
\,0\,&\,0\,\,
\\
\,0\,&\,1\,\,
\end{matrix}\bigg)^{\!\!\dalpha}_{\,\,\,\dbeta}\, .
\end{aligned}
\end{equation}
Substituting the seed integrals $S_1^{p_1,p_2}$ and $S_2^{p_1,p_2}$ 
(with $p_1,p_2\!=\!\ii\hp d\tnu,\,1\!+\!\ii\hp d\tnu$) 
into Eq.\eqref{leftpp_fig7}, we obtain the explicit result for the left subdiagram 
$\mathcal{T}^{(\mathsf{L},7)\quad\!\!\!\!\dalpha}_{(d),\oplus\oplus\quad\!\!\!\!\dbeta}$.

The case $a\!=\!b\!=\!\ominus$ can be treated in the same manner.\ 
The corresponding subdiagram, $\mathcal{T}^{(\mathsf{L},7)\quad\!\!\!\dalpha}_{(d),\ominus\ominus\quad\!\!\!\!\dbeta}$, 
can be expressed in terms of the seed integrals $S_3^{p_1,p_2}$ and $S_4^{p_1,p_2}$ as follows:
{\small
\label{leftmm_fig7}
\begin{align}
&\mathcal{T}^{(\mathsf{L},7)\quad\!\!\!\dalpha}_{(d),\ominus\ominus\quad\!\!\!\!\dbeta}(\mathbf{k}_1)\nn
\\
&= \(\!\frac{\,-\ii\,}{\,\Lambda\,}\!\)^{\!\!2}\!\!\int_{-\infty}^{0^-} \!\!\td\tau_1\td\tau_2 (-\tau_1)^{\ii\hp d\tnu} (-\tau_2)^{\ii\hp d\tnu}\,
\mathcal{F}_{\mu \ominus}(\mathbf{k}_1,\tau_1)\bar{\sigma}^{\mu\dalpha\alpha}
\hat{D}_{\ominus\ominus\alpha}^{\quad\,\,\,\,\,\,\beta}(\mathbf{p}_{12},\tau_{1},\tau_{2})
\sigma^{\nu}_{\beta\dbeta}\mathcal{F}_{\nu \ominus}(\mathbf{k}_2,\tau_2)\nn
\\
&=\(\!\frac{\,-\ii\,}{\,\Lambda\,}\!\)^{\!\!2}\!\frac{1}{
\,4(k_1k_2)^{3+\ii\hp d \tnu}\,}
\Big[(\bar{\sigma}^{0}\!-\!\bar{\sigma}^{3})^{\dalpha\alpha}\,
(\hat{\mathcal{I}}_{\ominus\ominus}^{1-\ii\hp d \tnu,1-\ii\hp d \tnu})^{\,\,\,\,\beta}_{\alpha}
(\sigma^{0}\!+\!\sigma^{3})_{\beta\dbeta}
\!-\!\ii\hp\bar{\sigma}^{3\dalpha\alpha}\,
(\hat{\mathcal{I}}_{\ominus\ominus}^{-\ii\hp d \tnu,1-\ii\hp d \tnu})^{\,\,\,\,\beta}_{\alpha}
(\sigma^{0}\!+\!\sigma^{3})_{\beta\dbeta}\nn
\\
&\qquad\qquad\qquad\qquad\qquad
+\!\ii\hp(\bar{\sigma}^{0}\!-\!\bar{\sigma}^{3})\,
(\hat{\mathcal{I}}_{\ominus\ominus}^{1-\ii\hp d \tnu,-\ii\hp d \tnu})^{\,\,\,\,\beta}_{\alpha}\,
\sigma^{3}_{\,\,\beta\dbeta}
\!+\!\bar{\sigma}^{3\dalpha\alpha}\,
(\hat{\mathcal{I}}_{\ominus\ominus}^{-\ii\hp d \tnu,-\ii\hp d \tnu})^{\,\,\,\,\beta}_{\alpha}
\sigma^{3}_{\,\,\beta\dbeta}\Big]\nn
\\
&=\(\!\frac{\,-\ii\,}{\,\Lambda\,}\!\)^{\!\!2}\!\frac{1}{\,2(k_1k_2)^{3+\ii\hp d \tnu}\,}
\Bigg[\!-\!S_4^{\ii\hp d\tnu,\ii\hp d\tnu}\delta^\dalpha_{\,\,\,2}\delta_\dbeta^{\,\,\,2}\nn
\\
&\quad+\!\(-S_3^{\ii\hp d\tnu,\ii\hp d\tnu} \!\!+\!\ii\hp 2 S_3^{\ii\hp d\tnu,1+\ii\hp d\tnu}\!+\!\ii\hp 2S_3^{1+\ii\hp d\tnu,\ii\hp d\tnu}\!+\!4S_3^{1+\ii\hp d\tnu,1+\ii\hp d\tnu}\)\delta^\dalpha_{\,\,\,1}\delta_\dbeta^{\,\,\,1}\Bigg]\, .
\end{align}
}
\hspace*{-2.5mm}
We next consider diagram-(8) in Fig.\,\ref{fig:feyn_all_2x4}, which is related to diagram-(7) by reversing the fermion flow.\ Thus the same factorization can be applied.\ 
Its left subdiagram is given by 
{\small
\begin{equation}
\label{left_expression_fig8}
\begin{aligned}
\mathcal{T}^{(\mathsf{L},8)\dalpha}_{(d),ab\,\,\dbeta}(\mathbf{k}_1)
&= \(\!\frac{\,-\ii\,}{\,\Lambda\,}\!\)^{\!\!2}\!\int_{-\infty}^{0^-} \!\!\td\tau_1\td\tau_2 \,
(-\tau_1)^{\ii\hp d\tnu} (-\tau_2)^{\ii\hp d\tnu}
\\[1.5mm]
&\quad\times
\mathcal{F}_{\mu b}(\mathbf{k}_2,\tau_2)\bar{\sigma}^{\mu\dalpha\alpha}
\hat{D}_{ba\alpha}^{\,\,\,\,\,\,\,\,\,\beta}(\mathbf{p}_{21},\tau_{2},\tau_{1})
\sigma^{\nu}_{\beta\dbeta}\mathcal{F}_{\nu a}(\mathbf{k}_1,\tau_1)\, ,
\end{aligned}
\end{equation}
}
\hspace*{-2.5mm}
where $\mathbf{p}_{21}\!=\!-\mathbf{p}_{12}\!=\!-\mathbf{k}_1$.\ 
Because the fermion flow of diagram-(8) is reversed relative to diagram-(7), 
its internal momentum is given by $\mathbf{p}_{21}\!=\!-\mathbf{k}_1\!=\!(0,0,-k_1)$, 
and its time-integral variables are interchanged ($\tau_1^{}\!\leftrightarrow\!\tau_2^{}$).\ 
In Eq.\eqref{left_expression_fig8}, the propagator 
$\hat{D}_{ba\alpha}^{\,\,\,\,\,\,\,\,\,\beta}(-\mathbf{k}_1^{},\tau_{2},\tau_{1})$ 
can be obtained from the corresponding propagator
$\hat{D}_{ba\,\gamma}^{\hspace{5mm}\delta}(\mathbf{k}_1^{},\tau_{2},\tau_{1})$ 
through a rotation with $\theta\!=\!\varphi\!=\!\pi$ as follows:
\begin{equation}
    \hat{D}_{ba\,\alpha}^{\hspace{5mm}\beta}(-\mathbf{k}_1^{},\tau_{2},\tau_{1})
    =
    U(\pi,\pi)_{\alpha}^{\,\,\,\gamma}
    \hat{D}_{ba\,\gamma}^{\hspace{5mm}\delta}(\mathbf{k}_1^{},\tau_{2},\tau_{1})
    U(\pi,\pi)_{\,\,\delta}^{\dagger\,\,\,\beta}\, .
\end{equation}
Accordingly, the nested-integral cases $\mathcal{T}^{(\mathsf{L},8)}_{(d),\oplus\oplus}$ and 
$\mathcal{T}^{(\mathsf{L},8)}_{(d),\ominus\ominus}$ are obtained from the results of diagram-(7) 
by exchanging $\tau_1^{}$ and $\tau_2^{}$ and performing a rotation $U(\pi,\pi)$:
{\small
\beqs 
\begin{align}
\mathcal{T}^{(\mathsf{L},8)\hspace{2mm}\dalpha}_{(d),\oplus\oplus\hspace{2mm}\dbeta}(\mathbf{k}_1)
&=
U(\pi,\pi)^{\dalpha}_{\,\,\,\,\,\dgamma}\,
\mathcal{T}^{(\mathsf{L},7)\quad\!\!\!\dgamma}_{(d),\oplus\oplus\quad\!\!\dot{\delta}}\bigg|_{\tau_1^{}\leftrightarrow\tau_2^{}}\!\!(\mathbf{k}_1)\,
U(\pi,\pi)^{\dagger\,\dot{\delta}}_{\,\,\,\,\,\,\,\dbeta}\nn
\\
&=
\! \(\!\frac{\,-\ii\,}{\,\Lambda\,}\!\)^{\!\!2}\!\frac{1}{\,2(k_1k_2)^{3+\ii\hp d \tnu}\,}
\!\Bigg[\!-\!S_1^{\ii\hp d\tnu,\ii\hp d\tnu}\delta^\dalpha_{\,\,\,2}\delta_\dbeta^{\,\,\,2}\nn
\\
&\quad+\!\(\!-S_2^{\ii\hp d\tnu,\ii\hp d\tnu} \!-\!\ii\hp 2 S_2^{\ii\hp d\tnu,1+\ii\hp d\tnu}\!-\!\ii\hp 2 S_2^{1+\ii\hp d\tnu,\ii\hp d\tnu}\!+\!4S_2^{1+\ii\hp d\tnu,1+\ii\hp d\tnu}\)\delta^\dalpha_{\,\,\,1}\delta_\dbeta^{\,\,\,1}\Bigg]\! ,
\label{leftpp_fig8}
\\
\mathcal{T}^{(\mathsf{L},8)\hspace{2mm}\dalpha}_{(d),\ominus\ominus\hspace{2mm}\dbeta}(\mathbf{k}_1)
&=
U(\pi,\pi)^{\dalpha}_{\,\,\,\,\,\dgamma}\,
\mathcal{T}^{(\mathsf{L},7)\quad\!\!\!\dgamma}_{(d),\ominus\ominus\quad\!\!\dot{\delta}}\bigg|_{\tau_1^{}\leftrightarrow\tau_2^{}}\!\!(\mathbf{k}_1)\,
U(\pi,\pi)^{\dagger\,\dot{\delta}}_{\,\,\,\,\,\,\,\dbeta}\nn
\\
&=
\(\!\frac{\,-\ii\,}{\,\Lambda\,}\!\)^{\!\!2}\!\frac{1}{\,2(k_1k_2)^{3+\ii\hp d \tnu}\,}
\!\Bigg[\!-\!S_4^{\ii\hp d\tnu,\ii\hp d\tnu}\delta^\dalpha_{\,\,\,1}\delta_\dbeta^{\,\,\,1}\nn
\\
&\quad+\(\!-S_3^{\ii\hp d\tnu,\ii\hp d\tnu} \!+\!\ii\hp2 S_3^{\ii\hp d\tnu,1+\ii\hp d\tnu}\!+\!\ii\hp2 S_3^{1+\ii\hp d\tnu,\ii\hp d\tnu}\!+\!4S_3^{1+\ii\hp d\tnu,1+\ii\hp d\tnu}\)\delta^\dalpha_{\,\,\,2}\delta_\dbeta^{\,\,\,2}\Bigg]\! .
\label{leftmm_fig8}
\end{align}
\eeqs 
}
\hspace*{-2.5mm}
Here the notation  {\small$\left[ 
\mathcal{T}^{(\mathsf{L},7)}\big|_{\tau_1\!\leftrightarrow\tau_2}
\right]$}
means that the roles of $\tau_1$ and $\tau_2$ are exchanged in the corresponding time integrals.

An important observation is that the dominant contribution to $\mathcal{T}^{(\mathsf{L})}_{(d),ab}$ 
corresponds to the cases with $a\hp d,b\hp d\!>\!0\hp$.\ This can be understood by inspecting 
the following integral (which is contained in the time integral of the left subdiagram):
\begin{equation}
\begin{aligned}
    \int_0^\infty\!\!\td z \, e^{-\ii\hp a z} z^{\ii\hp d\hp \tnu}
    =
    -\ii\hp a\! \int_0^\infty\!\!\td z \, e^{- z} (-\ii\hp a z)^{\ii\hp d\hp \tnu}
    =
    -\ii\hp a\, e^{\frac{\pi(ad)\tnu}{2}}\,\Gamma(1\!+\!\ii\hp d\hp\tnu)\, .
\end{aligned}
\end{equation}
From the above, we see that when $a\hp d,b\hp d\!<\!0\hp$, 
the integral is exponentially suppressed.\ 
In consequence, the left subdiagram is dominated by contributions with
$a\hp d,b\hp d\!>\!0\hp$.\

Finally, when the momentum of the propagator in the left subdiagram 
is not aligned with the $z$-axis, we define $\theta$ as the polar angle between $\mathbf{p}_{12}^{}\!=\!\mathbf{k}_1$ and the $z$-axis, namely 
$\mathbf{k}_1^{}\!=\!k_1^{}(\sin \Theta,0,\cos \Theta)$ and 
$\mathbf{k}_1^{}\!\simeq\!-\mathbf{k}_2\hp$.\ 
Thus, the corresponding left subdiagram for nonzero $\theta$ 
can be obtained from the $\theta\!=\!0$ case by the rotation rule 
given in Section\,\ref{Rotation of Spinor Seed Integrals}.\

\vspace*{1mm}
\subsubsection{\hspace*{-2.5mm}Right Part of the Diagram}
\vspace*{1mm}

The right subdiagram is considerably simpler than the left one.\ 
It can be viewed as a degenerate ``one-point'' function, which is 
a $2\!\times\!2$ spinor matrix with no internal fermion propagator.\ 
As an explicit example, we consider the right subdiagram of diagram-(7)
and it is given by the following:
\begin{equation}
\begin{aligned}
\mathcal{T}_{(d),c}^{(\mathsf{R},7)\dgamma\gamma} =
\(\!\frac{\,-\ii\,}{\,\Lambda\,}\!\)\!\!\int_{-\infty}^{0^-} \!\!\td\tau_3 \,
    \mathcal{F}_{\lambda c}(\mathbf{k}_3,\tau_3)\,
    \bar{\sigma}^{\lambda\dgamma\gamma}
    (-\tau_3)^{\ii\hp 2 d\hp\tnu}\, ,
\end{aligned}
\end{equation}
where the factor $(-\tau_3^{})^{\ii\hp 2d\hp\tnu}$ with $d\!=\!\pm$ comes from 
the time dependence of the two soft lines.\
The corresponding expressions for the other diagrams can be derived in the same way.

Since the right subdiagram contains only one massless external line 
$\mathcal{F}_{\lambda c}(\mathbf{k}_3,\tau_3)$, 
the time integral can be directly evaluated by using the integral formula,
\begin{equation}
\int_0^\infty\!\!\td t\, t^z e^{-\ii\hp\omega\hp t} =
(\ii\hp\omega)^{-1-z}\hp\Gamma(1\!+\!z)\, . 
\end{equation}
Thus, we derive the contribution to the right subdiagram as follows: 
\begin{equation}
\label{right_fig7}
\begin{aligned}
\mathcal{T}_{(d),\oplus}^{(\mathsf{R},7)\dgamma\gamma}
&=
\(\!\frac{-\ii}{\,2\Lambda\,}\!\)\!
k_3^{-3-\ii\hp 2d\tnu}
e^{d\pi\tnu}\Gamma(1\!+\!\ii\hp 2 d  \tnu)
\begin{pmatrix}
-1 & 0 \\
0 & 3 \!+\! \ii\hp 4d  \tnu
\end{pmatrix}^{\!\!\!\dgamma\gamma},
\\[2mm]
\mathcal{T}_{(d),\ominus}^{(\mathsf{R},7)\dgamma\gamma}
&=
\(\!\frac{-\ii}{\,2\Lambda\,}\!\)\!
k_3^{-3-\ii\hp 2d\tnu}
e^{-d\pi\tnu}\Gamma(1\!+\!\ii\hp 2 d  \tnu)
\begin{pmatrix}
3 \!+\! \ii\hp 4d  \tnu & 0 \\
0 & -1
\end{pmatrix}^{\!\!\!\dgamma\gamma}.
\end{aligned}
\end{equation}
Eq.\eqref{right_fig7} shows that the right subdiagram could avoid the additional 
exponential suppression factor $e^{-2\pi\tnu}$ only when 
$d\!=\!c\!=\!\pm$.\ Accordingly, in the following sections 
we will only consider the cases with $d\!=\!c\hp$.\ 

Among the eight diagrams in Fig.\,\ref{fig:feyn_all_2x4}, there are six diagrams
(with numbers 1,2,5,6,7,8) in which their right subdiagrams take the same form.\ 
The only exceptions are diagrams\,(3) and (4) which take a different form and
are given by
\begin{equation}
\begin{aligned}
\mathcal{T}_{(d),c\,\,\gamma\dgamma}^{(\mathsf{R},3)}
=
\mathcal{T}_{(d),c\,\,\gamma\dgamma}^{(\mathsf{R},4)}
=
\frac{\,-\ii\,}{\,\Lambda\,}
\!\int_{-\infty}^{0^-} \!\!\!\td\tau_3 \,
\mathcal{F}_{\lambda c}(\mathbf{k}_3,\tau_3)\,
\sigma^{\lambda}_{\gamma\dgamma}
(-\tau_3)^{\ii\hp 2\hp d\hp\tnu}\, .
\end{aligned}
\end{equation}
These can be obtained from the counterparts of diagram-(7) by exchanging
the diagonal elements of the $2\!\times\!2$ spinor matrix.\ 
Thus, using the formula for diagram-(7) we derive the following: 
\\[-7mm]
\beqs 
\begin{align}
\mathcal{T}_{(d),\oplus\,\,\gamma\dgamma}^{(\mathsf{R},3)}
&=
\mathcal{T}_{(d),\oplus\,\,\gamma\dgamma}^{(\mathsf{R},4)}
=
\(\!\frac{-\ii}{\,2\Lambda\,}\!\)\!
k_3^{-3-\ii\hp 2d\tnu}
e^{d\pi\tnu}\Gamma(1\!+\!\ii\hp 2 d\tnu)
\!\!\begin{pmatrix}
3 \!+\!\ii\hp 4d\hp\tnu & 0 \\
0 & -1
\end{pmatrix}_{\!\!\gamma\dgamma},
\\
\mathcal{T}_{(d),\ominus\,\,\gamma\dgamma}^{(\mathsf{R},3)}
&=
\mathcal{T}_{(d),\ominus\,\,\gamma\dgamma}^{(\mathsf{R},4)}
=
\(\!\frac{-\ii}{\,2\Lambda\,}\!\)\!
k_3^{-3-\ii\hp 2d\tnu}
e^{-d\pi\tnu}\Gamma(1\!+\!\ii\hp 2 d  \tnu)
\!\!\begin{pmatrix}
-1 & 0 \\
0 & 3 + \ii\hp 4d\hp \tnu
\end{pmatrix}_{\!\!\gamma\dgamma} .
\end{align}
\eeqs

\vspace*{1mm}
\subsubsection{\hspace*{-2.5mm}Central Part: Fermion Bubble Contribution}
\label{sec: the bubble contribution}
\vspace*{1mm}

In this subsection, we analyze the bubble contribution $\mathcal{B}_{(d)}$, 
which contains the loop-momentum integral {\small $\dis \int \!\td^3\mathbf{q}\hp$}, 
the dependence on the soft external momentum $\mathbf{k}_s$, and the remaining Gamma-function factors from the nonlocal parts of the two soft internal lines.\

As shown in the previous two subsections, the left subdiagram is dominated by the sector $d=a=b$, while the right subdiagram is dominated by $d=c$. It then follows that the leading contribution to the full diagram comes from the sector $d=a=b=c$, whereas the remaining choices of $a$, $b$, and $c$ are suppressed.

In the following, we use the two representative cases
\begin{equation}
\begin{cases}
    d=-,\quad a=b=c=\ominus\, ,\\
    d=+,\quad a=b=c=\oplus\, ,
\end{cases}
\end{equation}
to illustrate the evaluation of the bubble contribution.\ 
In diagram-(7), the central part of the bubble loop integral 
contains the product of two propagators with the spinor structure:
{\small $\check{D}_{\ominus\oplus\,\,\,\,\dgamma}^{\quad\,\dbeta} D_{\ominus\oplus\gamma\dalpha}
\to 
\mathcal{B}^{(7)\,\dbeta}_{(d)\,\,\,\,\,\dgamma\gamma\dalpha}\hp$}.\ 
\hspace*{-2.5mm}
For fixed $d$, the bubble loop contribution $\mathcal{B}^{(7)\,\dbeta}_{(d)\,\,\,\,\,\dgamma\gamma\dalpha}$ can be expressed as follows:
{\small
\begin{align}
\hspace*{-2mm}
&\mathcal{B}^{(7)\,\dbeta}_{(d)\,\,\,\,\,\dgamma\gamma\dalpha}
=
\int\!\!\frac{\,\mathrm{d}^3\mathbf{q}\,}{(2\pi)^3}
\frac{\,\left[\hsm\check{D}^{\quad\,\dbeta}_{\ominus\oplus\,\,\,\,\rm{\dgamma}}(\mathbf{p}_{23};\tau_2,\tau_3)\hsm\right]_{\!\rm{(NLoc)},d}\,}{(\tau_2\tau_3)^{\ii\hp d\tnu}}
\frac{\,\left[\hsm D_{\ominus\oplus\gamma\dalpha}(\mathbf{p}_{31}; \tau_3, \tau_1)\hsm\right]_{\!\rm{(NLoc)},d}~}{(\tau_3\tau_1)^{\ii\hp d\tnu}}
\nn
\\
\hspace*{-2mm}
&=
\int\!\!\frac{\,\mathrm{d}^3\mathbf{q}\,}{(2\pi)^3} \sum_{s_1,s_2}
\frac{\,\dis\!\hsm {s_1}\left[ v_{s_1} (\tau_2,\mathbf{p}_{23}) u_{s_1}^\dagger (\tau_3,\mathbf{p}_{23})\hsm\right]_{\!\rm{(NLoc)},d}\,}
{(\tau_2\tau_3)^{\ii\hp d\tnu}}
\frac{\,\dis\!\hsm\left[ u_{s_2} (\tau_3,\mathbf{p}_{31}) u_{s_2}^\dagger (\tau_1,\mathbf{p}_{31})\hsm\right]_{\!\rm{(NLoc)},d}~}
{(\tau_3\tau_1)^{\ii\hp d\tnu}}\nn
\\ &\hspace*{20mm}\times\mathcal{S}^{(7)\dbeta}_{\quad\,\,\,\,\,\dgamma\gamma\dalpha}(\hat{\mathbf{p}}_{23},\hat{\mathbf{p}}_{31},s_1,s_2)
\hp , \hspace*{2mm}
\label{bubble_fig7_pre}
\end{align}
}
\hspace*{-2.5mm}
where 
$\mathbf{p}_{23}\! =\! \mathbf{q} \hsm +\hsm \mathbf{k}_s\hp$, 
$\mathbf{p}_{31} \! =\! \mathbf{q}\hp$, 
and $\mathcal{S}^{(7)\dbeta}_{\quad\,\,\,\,\,\dgamma\gamma\dalpha}(\hat{\mathbf{p}}_{23},\hat{\mathbf{p}}_{31},s_1,s_2)$ 
is the spinor kinematic function given by
\begin{equation}
\mathcal{S}^{(7)\dbeta}_{\quad\,\,\,\,\,\dgamma\gamma\dalpha}(\hat{\mathbf{p}}_{23},\hat{\mathbf{p}}_{31},s_1,s_2)=h^{\dagger\quad\!\!\dbeta}_{-s_1}(\hat{\mathbf{p}}_{23})h^{\dagger}_{s_1,\dgamma}(\hat{\mathbf{p}}_{23})h_{s_2,\gamma}(\hat{\mathbf{p}}_{31})h^{\dagger}_{s_2,\dalpha}(\hat{\mathbf{p}}_{31})
\!
\end{equation}
In the above formula, the nonlocal mode-function combinations
{\small$\big[ v_{s_1} (\tau_2,\mathbf{p}_{23}) u_{s_1}^\dagger 
(\tau_3,\mathbf{p}_{23})\big]_{\!\rm{(NLoc)},d}$}
and 
{\small$\big[ u_{s_2} (\tau_3,\mathbf{p}_{31}) u_{s_2}^\dagger 
(\tau_1,\mathbf{p}_{31})\big]_{\!\rm{(NLoc)},d}\hp$}
(with $s\!=\!\pm$) are defined in 
Eq.\,\eqref{mode_function_nonlocal}.

\vs 

In Eq.\eqref{bubble_fig7_pre}, the time-dependent factors 
$(\tau_2\tau_3)^{\ii\hp d\hp\tnu}$ and $(\tau_3\tau_1)^{\ii\hp d\hp\tnu}$ 
associated with the two soft internal lines have already been included  
in the time integrals of the left and right subdiagrams.\ 
Hence, they do not appear in the bubble loop contribution.\  
Because the Gamma-function factors contained in the nonlocal parts 
are independent of the internal momenta $\mathbf{p}_{23}^{}$ and 
$\mathbf{p}_{31}^{}$, we factorize the loop integral as follows:
\begin{equation}
\label{bubble_fig7}
\begin{aligned}
\mathcal{B}^{(7)\,\dbeta}_{(d)\,\,\,\,\,\dgamma\gamma\dalpha}
&=
\sum_{s_1,s_2} \!\!\left\{\Gamma\right\}_{(d),s_1s_2}\!\!\int\!\!\frac{\,\mathrm{d}^3\mathbf{q}\,}{\,(2\pi)^3\,}
q^{\ii\hp 2d\tnu}|\mathbf{q}\!+\!\mathbf{k}_s|^{\ii\hp 2d\tnu}\mathcal{S}^{(7)\dbeta}_{\quad\,\,\,\,\,\dgamma\gamma\dalpha}(\hat{\mathbf{q}}\!+\!\hat{\mathbf{k}}_s,\hat{\mathbf{q}},s_1,s_2)\, ,
\end{aligned}
\end{equation}
where $\left\{\Gamma\right\}_{(d),s_1s_2}$ (with $d\!=\!\pm$) denotes the Gamma-function 
from the two nonlocal soft propagators.\ 
The spinor kinematic structure $\mathcal{S}^{(7)\dbeta}_{\quad\,\,\,\,\,\dgamma\gamma\dalpha}$ could generate additional angle dependence $\cos^n\theta$, as shown in Eq.\eqref{bubbleterm_form} in Section\,\ref{Factorization of the correlator loop diagrams}. Since a nonzero chemical potential could enhance the positive helicity mode and suppress the negative helicity mode, we will omit the nonlocal contribution from the negative helicity mode 
and only take the positive helicity mode ($s_1=s_2=+$), so the remaining Gamma function is given by 
{\small
\begin{align}
\left\{\Gamma\right\}_{(d),++}
&=
\frac{\Big[ v_{+} (\tau_2,\mathbf{p}_{23}) u_{+}^\dagger (\tau_3,\mathbf{p}_{23})\Big]_{\text{(NLoc)},d}\,}
{(\hp p_{23}^2\tau_2\tau_3)^{\ii\hp d\tnu}}
\frac{\Big[ u_{+} (\tau_3,\mathbf{p}_{31}) u_{+}^\dagger (\tau_1,\mathbf{p}_{31})\Big]_{\text{(NLoc)},d}\,}
{(\hp p_{31}^2\tau_3\tau_1)^{\ii\hp d\tnu}}\nn
\\
&=
\frac{\,\ii\hp 2^{\ii\hp 4d\tnu}
 e^{2\pi\tlambda} \tilde{M}_{\!R}^3
\big[\Gamma(-\ii\hp 2d\tnu)\big]^{4}
\,}{~\Gamma\!\big(\ii\hp\tlambda\!-\!\ii\hp d\tnu\big)
\Gamma\!\big(1\!+\!\ii\hp\tlambda\!-\!\ii\hp d\tnu\big)
\big[\Gamma\!\big(1\!-\!\ii\hp\tlambda\!-\!\ii\hp d\tnu\big)\!\big]^{2}
~}
\,\, ,
\end{align}
}%

For a three-point correlator $k_s\!=\!k_3$, we can define the product of 
the bubble term and the right subdiagram {\small$\mathcal{BR}^{(7)\,\,\,\,\,\dbeta}_{(d),a\,\,\,\dalpha}\hsm(\mathbf{k}_3)$}:
{\small
\begin{equation}
\mathcal{BR}^{(7)\,\,\,\,\,\dbeta}_{(d),a\,\,\,\dalpha}\hsm(\mathbf{k}_3)
\,\equiv\,
\mathcal{B}^{(7)\,\dbeta}_{(d)\,\,\,\,\,\dgamma\gamma\dalpha}\hsm(\mathbf{k}_s)
\,\mathcal{T}_{(d),a}^{(\mathsf{R},7)\dgamma\gamma}\hsm(\mathbf{k}_3)\,.
\end{equation}
}
\hspace*{-2.5mm}
Substituting Eqs.\eqref{right_fig7} and \eqref{bubble_fig7}
into the above formula, we derive the product as follows.\ 
Combining the spinor structure in the  bubble term and the right subdiagram, this product {\small$\mathcal{BR}^{(7)\,\,\,\,\,\dbeta}_{(d),a\,\,\,\dalpha}\hsm(\mathbf{k}_3)$} 
takes the following form (with $a=d$)
{\small
\label{eq:BR(7)}
\begin{align}
\hspace*{-4mm}\mathcal{BR}^{(7)\,\,\,\,\,\dbeta}_{(d),d\,\,\,\dalpha}\hsm(\mathbf{k}_3)=& \(\!\frac{\,-\ii\,}{\,\Lambda\,}\!\)k_3^{-3-\ii\hp 2d\tnu}\frac{\, e^{-d\pi \tnu}\Gamma(1\!+\!\ii\hp 2d\tnu)
\,}{8}\!\left\{\Gamma\right\}_{(d),++}\nn
\\
&\times\int\!\!\frac{\,\sin\theta\mathrm{d}\theta\mathrm{d}q\,}{\,(2\pi)^2\,}
q^{2+\ii\hp 2d\tnu}\,
|\mathbf{q}\!+\!\mathbf{k}_s|^{-\!1+\ii\hp 2d\tnu}\mathcal{A}^{\quad\dbeta}_{(d)\,\,\,\dalpha}(\mathbf{q},\mathbf{q}\!+\!\mathbf{k}_s)\,,
\end{align}   
}
\hspace{-2.5mm}
where we have already performed the integral about the azimuth angle $\varphi$. $\mathcal{A}^{\quad\dbeta}_{(d)\,\,\,\dalpha}$ is a $2\!\times\!2$ matrix and depends on the choice of $d\hp$ 
\begin{equation}
\label{A_matrix}
\mathcal{A}^{\quad\,\dbeta}_{(-)\,\,\,\dalpha}=\begin{pmatrix}
\mathcal{A}_{(-)1}
&
0
\\[1ex]
0
&
\mathcal{A}_{(-)2}
\end{pmatrix}^{\!\!\dbeta}_{\,\,\,\,\,\dalpha},\hspace*{5mm}
\mathcal{A}^{\quad\,\dbeta}_{(+)\,\,\,\dalpha}=\begin{pmatrix}
\mathcal{A}_{(+)1}
&
0
\\[1ex]
0
&
\mathcal{A}_{(+)2}
\end{pmatrix}^{\!\!\dbeta}_{\,\,\,\,\,\dalpha},
\end{equation}
with the component $\mathcal{A}_{(-)1}$,  $\mathcal{A}_{(-)2}$, $\mathcal{A}_{(+)1}$ and $\mathcal{A}_{(+)2}$ defined as
\beqs
\label{A_matrix_elements}
\begin{align}
\mathcal{A}_{(-)1} &= 
(1\!+\hsm\cos\theta)\Big[k_s(3\!-\!\ii\hp 4\tnu)
\!+\!q(-1\!+\!4\cos\theta\!-\!\ii\hp4 \tnu\cos\theta )\!+\!|\mathbf{q}\!+\!\mathbf{k}_s|\,(3\!-\!\ii\hp 4\tnu)
\Big],
\\
\mathcal{A}_{(-)2} &=
-(1\!-\hsm\cos\theta)\Big[\!
-\!k_s\!+\!q(-3\!-\!4\cos\theta\!+\!\ii\hp4\tnu\!+\!\ii\hp 4  \tnu\cos\theta)\!+\!|\mathbf{q}\!+\!\mathbf{k}_s|
\Big],
\\
\mathcal{A}_{(+)1} &=
-(1\!+\hsm\cos\theta)\Big[
k_s\!+\!q(-3\!+\!4\cos\theta\!-\!\ii\hp 4\tnu\!+\!\ii\hp 4 \tnu\cos\theta)\!+\!|\mathbf{q}\!+\!\mathbf{k}_s|
\Big],
\\
\mathcal{A}_{(+)2} &=
-(1\!-\hsm\cos\theta)\Big[
k_s(3\!+\! \ii\hp 4\tnu)\!+\!q(1\!+\!4\cos\theta\!+\!\ii\hp 4\tnu\cos\theta)
\!+\!|\mathbf{q}\!+\!\mathbf{k}_s|(-3\!-\!\ii\hp4\tnu)
\Big].
\end{align}
\eeqs
In the above equations, $\theta$ is defined as the angle between the loop momentum $\mathbf{q}$ and 
the momentum $\mathbf{k}_s$.\ According to the power of $\cos\theta$, the full integral can be split into 
three types of contributions, denoted by $\mathcal{J}_{0}^{a,b}, \mathcal{J}_{1}^{a,b}$, 
and $\mathcal{J}_{2}^{a,b}$, as shown in Appendix\,\ref{loop integral through Feynman parameter approach}.\ 
Then, the complete loop-momentum integral can be represented as follows: 
{\small
\begin{align}
\label{eq:BR}
\mathcal{BR}^{(7)\,\,\,\,\,\,\dbeta}_{(d),d\,\,\,\,\dalpha}\hsm(\mathbf{k}_3)
&= \(\!\frac{\,-\ii\,}{\,\Lambda\,}\!\)k_3^{-3-\ii\hp 2d\tnu}\frac{\, e^{\pi \tnu}\Gamma(1\!+\!\ii\hp 2d\tnu)
\,}{8}\!\left\{\Gamma\right\}_{(d),++} \mathcal{C}^{\quad\dbeta}_{(d)\,\,\,\dalpha}\,,
\end{align}   
}
\hspace*{-2.5mm}
In the above, $\mathcal{C}^{\quad\dbeta}_{(d)\,\,\,\dalpha}$ 
is a $2\!\times\!2$ matrix and depends on the choice of $d$\,,
\\[-4mm]
\begin{equation}
\label{cfactor_matrix}
\mathcal{C}^{\quad\dbeta}_{(-)\,\,\,\dalpha}\!=\!
\begin{pmatrix}
\mathcal{C}_{(-)1}
&
0
\\[1ex]
0
&
\mathcal{C}_{(-)2}
\end{pmatrix}^{\!\!\dbeta}_{\,\,\,\dalpha},
\hspace*{8mm}
\mathcal{C}^{\quad\dbeta}_{(+)\,\,\,\dalpha}\!=\!
\begin{pmatrix}
\mathcal{C}_{(+)1}
&
0
\\[1ex]
0
&
\mathcal{C}_{(+)2}
\end{pmatrix}^{\!\!\dbeta}_{\,\,\,\dalpha},
\end{equation}
with the component $\mathcal{C}_{(-)1}$,  $\mathcal{C}_{(-)2}$, $\mathcal{C}_{(+)1}$ and $\mathcal{C}_{(+)2}$ defined as
{\small
\beqs
\begin{align}
\mathcal{C}_{(-)1}&=
(3 \!-\! \ii\hp 4\tnu)
\Bigl(
\mathcal{J}_{0}^{\,\ii\hp \tnu,\,\ii\hp\tnu}
\!+\!k_s\mathcal{J}_{0}^{\,\ii\hp \tnu,\,\frac12+\ii\hp\tnu}\!+\!
\mathcal{J}_{1}^{\,\ii\hp \tnu,\,\ii\hp\tnu}\!+\!k_s\,\mathcal{J}_{1}^{\,\ii\hp \tnu,\,\frac12+\ii\hp\tnu}\!+\!
\mathcal{J}_{1}^{\,-\frac12+\ii\hp \tnu,\,\frac12+\ii\hp\tnu}
\Bigr)\nn
\\[1mm]
&\quad\!-\!
\mathcal{J}_{0}^{\,-\frac12+\ii\hp \tnu,\,\frac12+\ii\hp\tnu}
\!+\!
4(1 \!-\! \ii\hp\tnu)\,
\mathcal{J}_{2}^{\,-\frac12+\ii\hp \tnu,\,\frac12+\ii\hp\tnu}
\\[1mm]
&=\frac{\,(\ii\!+\!2\tnu)[\csch(\pi\tnu)]^2\sech(2\pi\tnu)k_s^{3-\ii\hp 4\tnu}\,}{4(\ii\hp2\tnu\!-\!3)\Gamma(5\!-\!\ii\hp4\tnu)\big[\Gamma(\ii\hp2\tnu)\big]^2}\Big(32\tnu^3\!+\!\ii\hp 96\tnu^2-92\tnu\!-\!\ii\hp29\!+\!\big(\ii\hp 8\tnu^2\!-\!20\tnu\!-\!\ii\hp11\big)\cosh(2\pi\tnu)\Big)\,,\nn
\\[3mm]
\mathcal{C}_{(-)2}&=k_s\,\mathcal{J}_{0}^{\,\ii\hp \tnu,\,\frac12+\ii\hp\tnu}
\!-\!
\mathcal{J}_{0}^{\,\ii\hp \tnu,\,\ii\hp\tnu}
\!+\!
(3 \!-\! \ii\hp 4\tnu)\,
\mathcal{J}_{0}^{\,-\frac12+\ii\hp \tnu,\,\frac12+\ii\hp\tnu}\nn
\\[1mm]
&\quad
\!-\!
k_s\,\mathcal{J}_{1}^{\,\ii\hp \tnu,\,\frac12+\ii\hp\tnu}
\!+\!
\mathcal{J}_{1}^{\,\ii\hp \tnu,\,\ii\hp\tnu}
\!+\!
\mathcal{J}_{1}^{\,-\frac12+\ii\hp \tnu,\,\frac12+\ii\hp\tnu}
\!+\!
\ii\hp 4(\ii \!+\! \tnu)\,
\mathcal{J}_{2}^{\,-\frac12+\ii\hp \tnu,\,\frac12+\ii\hp\tnu}
\\
&=
-\frac{
(1\!-\!\ii\hp2 \tnu )
(5\!-\! \ii\hp4 \tnu)
\sech(2\pi\tnu)k_s^{3-\ii\hp4 \tnu}
}{
4\,
\Gamma(7\!-\!\ii\hp4 \tnu )\,
[\Gamma(\ii\hp2 \tnu)]^{2}
}\Big(
4(1\!-\! \ii\hp\tnu)^{2}[\csch(\pi\tnu)]^{2}
\!+\!
(1 \!-\! \ii\hp 2 \tnu)
(3 \!-\! \ii\hp2 \tnu)
[\sech(\pi\tnu)]^{2}
\Big)\,,\nn
\\[3mm]
\mathcal{C}_{(+)1}&=
-\mathcal{J}_{0}^{\,-\ii\hp \tnu,\,-\ii\hp\tnu}
\!+\!
(3 \!+\! \ii\hp 4\tnu)\,
\mathcal{J}_{0}^{\,-\frac12-\ii\hp \tnu,\,\frac12-\ii\hp\tnu}
\!-\!
k_s
\mathcal{J}_{0}^{\,-\ii\hp \tnu,\,\frac12-\ii\hp\tnu}
\!-\!
k_s\mathcal{J}_{1}^{\,-\ii\hp \tnu,\,\frac12-\ii\hp\tnu}\nn
\\[1mm]
&\qquad
\!-\!
\mathcal{J}_{1}^{\,-\ii\hp \tnu,\,-\ii\hp\tnu}
\!-\!
\mathcal{J}_{1}^{\,-\frac12-\ii\hp \tnu,\,\frac12-\ii\hp\tnu}
\!-\!
4(1 \!+\! \ii\hp\tnu)\,
\mathcal{J}_{2}^{\,-\frac12-\ii\hp \tnu,\,\frac12-\ii\hp\tnu}
\\
&=
-\frac{
\left(
8 \tnu^{2}
\!-\!
\ii\hp14 \tnu
\!-\!
5
\right)
\left(
8 \tnu^{2}
\!-\!
\ii\hp16 \tnu
\!-\!
7
\!-\!
\sech(2\pi\tnu)
\right)
[\csch(2\pi\tnu)]^{2}
k_s^{3+\ii\hp4 \tnu}
}{
2\,
\Gamma(7\!+\!\ii\hp4 \tnu)\,
[\Gamma(-\ii\hp2 \tnu)]^{2}
},\nn
\\[3mm]
\mathcal{C}_{(+)2}&=
(3 \!+\! \ii\hp 4\tnu)
\Bigl(-k_s
\mathcal{J}_{0}^{\,-\ii\hp \tnu,\,\frac12-\ii\hp\tnu}\!+\!
\mathcal{J}_{0}^{\,-\ii\hp \tnu,\,-\ii\hp\tnu}\!+\!
k_s\,\mathcal{J}_{1}^{\,-\ii\hp \tnu,\,\frac12-\ii\hp\tnu}
\!-\!
\mathcal{J}_{1}^{\,-\ii\hp \tnu,\,-\ii\hp\tnu}
\!-\!
\mathcal{J}_{1}^{\,-\frac12-\ii\hp \tnu,\,\frac12-\ii\hp\tnu}
\Bigr)\nn
\\[1mm]
&\quad
\!-\!
\mathcal{J}_{0}^{\,-\frac12-\ii\hp \tnu,\,\frac12-\ii\hp\tnu}\!+\!
4(1 \!+\! \ii\hp\tnu)\,
\mathcal{J}_{2}^{\,-\frac12-\ii\hp \tnu,\,\frac12-\ii\hp\tnu}
\\
&=
\frac{
\sech(2\pi\tnu)\,
k_s^{\,3+\ii\hp4 \tnu}
}{
128
(1\!+\!\ii\hp\tnu)\,
[\Gamma(-2 \!-\! \ii\hp2 \tnu)]^{2}\,
\Gamma(3\!+\!\ii\hp4 \tnu)
}
\Bigg(
\frac{
[\sech(\pi\tnu)]^{2}
}{
(1\!+\!\ii\hp\tnu)^{2}
}
\!-\!
\frac{
4(5 \!+\! \ii\hp4 \tnu)
[\csch(\pi\tnu)]^{2}
}{
(1 \!+\! \ii\hp2 \tnu)
(3 \!+\! \ii\hp2 \tnu)
(3 \!+\! \ii\hp4 \tnu)
}
\Bigg).\nn
\end{align}
\eeqs
}
\hspace*{-2.5mm}
We note that in the large $\tnu$ limit the factor $\mathcal{C}_{(d)j}$ of 
Eq.\eqref{cfactor_matrix} has the following asymptotic behavior:
\begin{equation}
\begin{aligned}
&|\mathcal{C}_{(d)j}|\!\simeq\!2^{-13/2}\pi^{-3/2}\tnu^{-3/2} k_s^3\,,
\hspace*{5mm} (d=\pm,\,j=1,2)\,,
\end{aligned}
\end{equation}
which arises from the loop-momentum integral together with the right diagram.\ 
We note that the coefficient $1\!+\hsm\cos\theta=2\cos^2\hsm\frac{\theta}{2}$ 
or $1\!-\hsm\cos\theta=2\sin^2\hsm\frac{\theta}{2}$ of Eq.\eqref{A_matrix_elements}
originates from the helicity overlap of spin-$\frac{1}{2}$ states [or equivalently, from the Wigner
function $d^{1/2}_{1/2,1/2}(\theta)$] and 
leads to the helicity-induced suppression\,\cite{Jacob:1959at,Weinberg:1995mt}.\ 
In our case, this helicity zero occurs precisely at the endpoint that dominates the large-$\tnu$
asymptotics.\ In consequence, the leading boundary contribution of $O(\tnu^{-1/2})$
is canceled and the asymptotic expansion starts at the next-to-leading order of $O(\tnu^{-3/2})$, 
causing an additional suppression by a factor of $1/\tnu\hp$.\

For clarity, in the large-chemical-potential limit ($\tlambda\!\!\gg\!\! 1\hp$), 
the absolute value for $\mathcal{BR}^{(7)\,\,\,\,\,\,\dbeta}_{(d),d\,\,\,\,\dalpha}$ 
can be simplified into the following expression:
{\small
\beqs
\begin{align}
|\mathcal{BR}^{(7)\,\,\,\,\,\,\,\,\dbeta}_{(-),\ominus\,\,\,\dalpha}|
&= \frac{\, e^{-3\pi (\tnu-\tlambda)}\tilde{M}_{\!R}^3 P_\zeta^{\frac{\,1\,}{\,2\,}}\pi
\,}{256\sqrt{2}\tlambda^2}\delta^{\,\dbeta}_{\,\,\,\,\dalpha}\,,
\\
|\mathcal{BR}^{(7)\,\,\,\,\,\,\,\,\dbeta}_{(+),\oplus\,\,\,\dalpha}|
&= \frac{\, e^{-3\pi (\tnu-\tlambda)}\tilde{M}_{\!R}^5 P_\zeta^{\frac{\,1\,}{\,2\,}}\pi
\,}{1024\sqrt{2}\tlambda^4}\delta^{\,\dbeta}_{\,\,\,\,\dalpha}\,,
\end{align}
\eeqs
}
\hspace*{-2.5mm}
which is the coefficient of $k_3^{\ii\hp d\hp 2\tnu}$ in $\mathcal{BR}^{(7)\,\,\,\,\,\,\dbeta}_{(d),d\,\,\,\,\dalpha}\,$.\
\hspace*{-2.5mm}

\hspace*{-2.5mm}
The bubble contribution depends only on the types of the two soft propagators in the loop.\ 
Moreover, in the case of $d\!=\!a\!=\!b\!=\!c\,$, 
we find that the product of the bubble term and the right subdiagram of diagram-(8) (denoted as $\mathcal{BR}^{(8)}_{(d),a}$ in Fig.\,\ref{fig:feyn_all_2x4}) 
is identical to that of diagram-(7), because of the same types of the two soft propagators 
in the bubble contribution and the same spinor structure of the right subdiagram.\  
In general, for the case of $d\!=\!a\!=\!b\!=\!c\,$, 
we derive the following relations between these products 
($\mathcal{BR}_{(d),a}$) for the different diagrams:
\begin{equation}
\begin{aligned}
\mathcal{BR}^{(1)}_{(d),a}\!=\!\mathcal{BR}^{(2)}_{(d),a}\hp ,\qquad
\mathcal{BR}^{(3)}_{(d),a}\!=\!\mathcal{BR}^{(4)}_{(d),a}\hp ,\\
\mathcal{BR}^{(5)}_{(d),a}\!=\!\mathcal{BR}^{(6)}_{(d),a}\hp ,\qquad
\mathcal{BR}^{(7)}_{(d),a}\!=\!\mathcal{BR}^{(8)}_{(d),a}\hp .
\end{aligned}
\end{equation} 

\vspace*{2mm}
\subsection{\hspace*{-2.5mm}Full Result of the Three-Point Correlator}
\label{Final Result of the Three-Point Correlator}
\vspace*{1.5mm}

In this subsection, we are ready to derive the full result of the nonlocal contribution 
to the three-point correlator from summing up all diagrams of Fig.\,\ref{fig:feyn_all_2x4}.\ 
For each given diagram-$(j)$ of Fig.\,\ref{fig:feyn_all_2x4}, the nonlocal contribution 
takes the following form:
{\small
\begin{equation}
\lim_{k_s\!\to0}\langle\delta\phi^3\rangle^{\prime(j)}_\text{(NLoc)}
\,\simeq\!
\!\sum_{\substack{a,b,c=\pm \\ d=\pm}}\!\!\!\!\!
a\hp b\hp c\,
\mathcal{T}_{(d),ab}^{(\mathsf{L},j)}\hsm (\mathbf{k}_1,\mathbf{k}_2)\hp 
\mathcal{B}^{(j)}_{(d),abc}\hsm (\mathbf{k}_s)\hp  
\mathcal{T}_{(d),c}^{(\mathsf{R},j)}\hsm (\mathbf{k}_3)\hp .
\end{equation}
}
\hspace*{-2.5mm}
As discussed above, the dominant contribution comes from the cases with 
$d\!=\!a\!=\!b\!=\!c$.\ 
Thus, the above formula reduces to
{\small
\begin{equation}
\lim_{k_s\!\to0}\langle\delta\phi^3\rangle^{\prime(j)}_\text{(NLoc)}
\,\simeq\!\!
\sum_{\substack{a=d=\pm}}\!\!\!\!
d\,
\mathcal{T}_{(d),aa}^{(\mathsf{L},j)}\hsm (\mathbf{k}_1,\mathbf{k}_2) 
\hp\mathcal{B}^{(j)}_{(d),aaa}\hsm(\mathbf{k}_s)\hp 
\mathcal{T}_{(d),a}^{(\mathsf{R},j)}\hsm (\mathbf{k}_3)\hp .
\end{equation}
}
\hspace*{-2.5mm}
\vspace*{-3.5mm}

Then, we present the results for diagrams (7) and (8).\ 
Since these two diagrams share the same bubble contribution and the same right subdiagram, 
the corresponding three-point correlator can be expressed as follows:
{\small
\begin{equation}
\begin{aligned}
\hspace*{-5mm}
\lim_{k_s\hsm\to\hp 0}
\!\hsm\left[\langle\delta\phi^3\rangle^{\prime(7)}_\text{(NLoc)}\!+\!\langle\delta\phi^3\rangle^{\prime(8)}_\text{(NLoc)}\right]
&\simeq
\sum_{\substack{a=d=\pm}}
\!\!\!d
\!\left[\hsm\mathcal{T}_{(d),aa}^{(\mathsf{L},7)}\hsm(\mathbf{k}_1)
\hsm\!+\!\mathcal{T}_{(d),aa}^{(\mathsf{L},8)}\!(\mathbf{k}_1)\hsm\right]
\!\mathcal{B}^{(7)}_{(d),aaa}\!(\mathbf{k}_s)\hp 
\mathcal{T}_{(d),a}^{(\mathsf{R},7)}\hsmx (\mathbf{k}_3)\hp .\\
&=\sum_{a=d=\pm}
\!\!\!d
\!\left[\mathcal{T}^{(\mathsf{L},7)\,\dalpha}_{(d),aa\quad\!\!\!\!\dbeta}\hsm(\mathbf{k}_1)\!+\!\mathcal{T}^{(\mathsf{L},8)\,\dalpha}_{(d),aa\quad\!\!\!\!\dbeta}\hsm(\mathbf{k}_1)\right]\!
\mathcal{BR}^{(7)\,\,\,\,\,\dbeta}_{(d),a\,\,\,\dalpha}\hsm(\mathbf{k}_3)\,,
\end{aligned}
\label{eq:dphi7+dphi8}
\end{equation}
}
\hspace*{-2.5mm}
\hspace*{-2.5mm}
where the term $\mathcal{T}^{(\mathsf{L},7)\,\dalpha}_{(d),aa\quad\!\!\!\!\!\dbeta}\hsm(\mathbf{k}_1)$ 
is given by Eqs.\eqref{leftpp_fig7} and \eqref{leftmm_fig7},  the term  $\mathcal{T}^{(\mathsf{L},8)\,\dalpha}_{(d),aa\quad\!\!\!\!\!\dbeta}\hsm(\mathbf{k}_1)$ 
is given by Eqs.\eqref{leftpp_fig8} and \eqref{leftmm_fig8},
and the term $\mathcal{BR}^{(7)\,\,\,\,\,\dbeta}_{(d),a\,\,\,\dalpha}$ is given by Eq.\eqref{eq:BR(7)}.

\vs

\begin{figure}[t]
    \centering
    \begin{subfigure}[b]{0.485\textwidth}
        \centering
        \includegraphics[width=\linewidth]{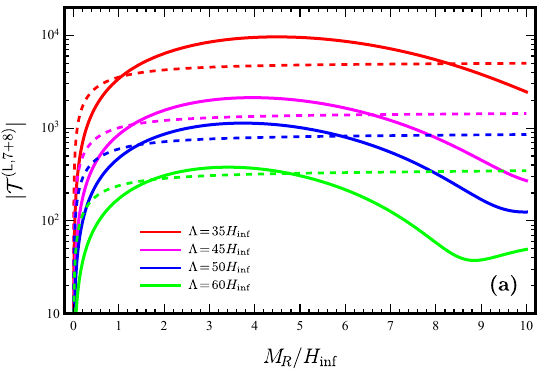}
        \phantomcaption
        \label{fig:4a}
    \end{subfigure}
        \hfill\hspace*{-3mm}
        \begin{subfigure}[b]{0.5\textwidth}
        \centering
        \includegraphics[width=\linewidth]{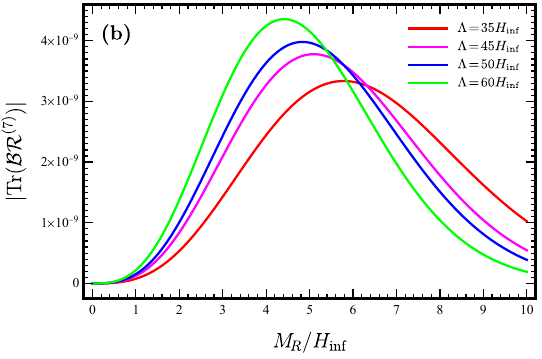}
        \phantomcaption
        \label{fig:4b}
\end{subfigure}
\vspace*{-0.8cm}
\caption{\small Contributions of diagrams\,(7)-(8) of Fig.\,\ref{fig:2}
to the three-point correlator as a function of right-handed neutrino mass $M_{\!R}$ for a set of sample $\Lambda$ values.\ 
The plot\,(a) shows the contribution of the sum of the left subdiagrams,
the plot\,(b) depicts the contribution of the trace of the term $\mathcal{BR}^{(7)\,\,\,\,\,\dbeta}_{(d),a\,\,\,\dalpha}$ given by Eq.\eqref{eq:BR(7)}.\ 
For comparison, the dashed curves of plot\,(a) also show the results as derived 
by using the saddle point approximation\,\cite{Chen:2018xck,Hook:2019zxa,Aoki:2026olh}. 
}
\label{fig:4} 
\label{function_of_m}
\end{figure}

Finally, the nonlocal CC signal of the full three-point correlator is given 
by the following sum of the eight diagrams of Fig.\,\ref{fig:2}:
\begin{equation} 
\lim_{k_s\!\to\hp 0}\langle\delta\phi^3\rangle^{\prime}_\text{(NLoc)}
\,=\,\lim_{k_s\!\to\hp 0}\sum_{j=1}^8 \langle\delta\phi^3\rangle^{\prime(j)}_\text{(NLoc)}\,.
\end{equation}
The calculations for the remaining six diagrams are similar and we do not list 
their explicit expressions here.\ 
Moreover, as proved in Appendix\,\ref{app:C1}, we have the following
conjugate relations between two diagrams: 
\begin{equation}
\label{eq:conjugationrelation}
\begin{aligned}
\langle\delta\phi^3\rangle^{\prime(1)}_\text{(NLoc)}=
\left[\hsm\langle\delta\phi^3\rangle^{\prime(2)}_\text{(NLoc)}\hsm\right]^*, 
\hspace*{12mm}
\langle\delta\phi^3\rangle^{\prime(3)}_\text{(NLoc)}=
\left[\hsm\langle\delta\phi^3\rangle^{\prime(4)}_\text{(NLoc)}\hsm\right]^*,
\\[0.6mm]
\langle\delta\phi^3\rangle^{\prime(5)}_\text{(NLoc)}=
\left[\hsm\langle\delta\phi^3\rangle^{\prime(7)}_\text{(NLoc)}\hsm\right]^*, 
\hspace*{12mm}
\langle\delta\phi^3\rangle^{\prime(6)}_\text{(NLoc)}=
\left[\hsm\langle\delta\phi^3\rangle^{\prime(8)}_\text{(NLoc)}\hsm\right]^*.
\end{aligned}
\end{equation}

In the above analysis, we have chosen the small momentum $\mathbf{k}_3$ 
to be on the $z$-axis and the large momentum $\mathbf{k}_1$ to be approximately 
aligned with the $z$-axis.\ 
For a more general configuration with $\mathbf{k}_1$ having an angle $\theta$ 
relative to the $z$-axis, we find that the total nonlocal cosmological collider signal 
of the three-point correlator is independent of $\theta$.\ Hence the above analysis
applies to the general case.\

In Fig.\,\ref{function_of_m}, we present the 
contributions of diagrams\,(7)-(8) of Fig.\,\ref{fig:2}
to the three-point correlator as a function of right-handed neutrino mass parameter 
$M_{\!R}$ for a set of sample values of the UV cutoff,
$\Lambda\!=\!(35,\,45,\,50,\,60)H_{\rm{inf}}$ 
[corresponding to
$\lambda\!\simeq\!(99,\,77,\,69,\,58)H_{\rm{inf}}\hp$]\footnote{%
Our analysis is based on the $\phi\hsm -\! N$ interaction through the effective dimension-5
operator \eqref{dimension5_operator}.\ 
The next most relevant operator that contributes to the
$\phi\hsm -\!N$ interaction related to the chemical potential and respects shift symmetry 
is given by a dimension-9 operator, 
$\frac{1}{\Lambda^5}(\partial_{\nu}\phi )^2
 (\partial_{\mu}^{}\phi N^\dagger\bar{\sigma}^\mu N)$.\ 
This shows that the relevant perturbative expansion parameter for this analysis
is given by the dimensionless ratio $(\dot{\phi}_0^2/\Lambda^4)$ up to a ratio of
the two coefficients [which are naturally $O(0.1\!-\!1)$] of this dimension-9 operator and 
our dimension-5 operator \eqref{dimension5_operator}.\  
The validity of the perturbative expansion requires the expansion parameter 
to be smaller than one, $(\dot{\phi}_0^2/\Lambda^4)\!\lesssim\! 1$.\ 
After considering the ranges of the two naturally $O(0.1\!-\!1)$ coefficients 
of the dimension-5 and dimension-9 operators, 
this perturbation condition gives a conservative lower bound on the cutoff scale,
$\Lambda \!\gtrsim\!\dot{\phi}_0^{1/2}\!/\hsm 10^{1/4}\hsm\simeq\hsm 34\Hinf\,$ 
for $\dot{\phi}\!\simeq\!(60\Hinf)^2$ as in Eq.\eqref{eq:dot(phi)-value}.}
which are shown respectively by the (red,\,pink,\,blue,\,green) curves.\  

\vs

The left subdiagram shows the strongest dependence on the chemical potential $\lambda\hp$.\ 
Its magnitude is significantly enhanced for larger chemical potential, 
and for each benchmark choice it reaches its largest value around the intermediate mass region.\ 
We also show in Fig.\,\ref{fig:4}(a)
the results obtained by using the saddle-point approximation of 
Refs.\,\cite{Chen:2018xck,Hook:2019zxa,Aoki:2026olh} shown by the dashed curves.\ 
The deviation of the saddle-point approximation 
from the complete left-subdiagram result 
depends on both the right-handed neutrino mass $M_{\!R}$ 
and the chemical potential $\lambda\hp$.

In Fig.\,\ref{fig:4}(b), 
the bubble contribution combined with the right subdiagram peaks around the region of 
$M_{\!R}\!\sim\!(3\!-\!5)H_{\rm{inf}}$, but is relatively insensitive to 
the chemical potential $\tilde\lambda\hp$.\
Since the contribution to the CC signal is determined by the product 
of the three parts, the enhancement of the left 
subdiagrams controls the behavior of the CC signal as the chemical potential increases.\ 
Consequently, the full oscillatory signal becomes stronger 
for larger chemical potential and is most significant 
around the intermediate mass region.

\section{\hspace*{-2.5mm}Amplitude and Shape of Cosmological Collider Signatures}
\label{sec:4}

In the previous section, we derived the nonlocal oscillatory contribution to the inflaton three-point correlator generated by the right-handed neutrino loop.\ 
In this section, we study its phenomenological application.\ 
First, we analyze the cosmological collider (CC) signal (including its magnitude) 
generated by the three-point inflaton correlator 
as contributed by all the eight diagrams in Fig.\,\ref{fig:feyn_all_2x4}.\ 
Second, we compare our result with that of the previous work\,\cite{Chen:2018xck} 
and discuss the difference between our study and the previous one. 

\subsection{\hspace*{-2.5mm}Analyzing the Cosmological Collider Signatures}
\label{sec:4.1}

Our goal in this subsection is twofold.\ We first express the squeezed limit 
bispectrum in the conventional non-Gaussianity normalization and extract the corresponding oscillatory amplitude parameter $\fNLCC$.\ 
Then, we analyze the dependence of the CC signals 
on the heavy right-handed neutrino mass and the cutoff scale, 
and display the characteristic oscillatory shape 
of the CC signals.\  
We will focus on the nonlocal contribution as computed 
in Section\,\ref{sec:3}, 
which captures the clock signals associated with 
the right-handed neutrino propagation.\

\vs 

According to the convention of Planck collaboration\,\cite{Planck9}, 
the bispectrum of the Bardeen potential $\Phi$ is defined as follows:
{\small
\begin{equation}
\langle \Phi_{\mathbf{k}_1}\!\Phi_{\mathbf{k}_2}\!\Phi_{\mathbf{k}_3}\rangle'
=
\frac{\,6A^2 \fNL\,}{\,(k_1k_2k_3)^2\,}\,
\mathcal{C}(k_1,k_2,k_3)\,,
\end{equation}
}
\hspace*{-2.5mm} 
where $\fNL$ characterizes the magnitude of the primordial 
non-Gaussianity, and $\mathcal{C}(k_1,k_2,k_3)$ denotes the corresponding 
shape template.\ The normalization constant $A$ is defined through 
the power spectrum of 
{\small $\Phi$, $P_\Phi\hsm\!=\!\hsm
\langle \Phi_{\mathbf{k}_1}\Phi_{\mathbf{k}_2}\rangle'
\hsm\!=\hsm\!{A}/{k^3}\,$},
where $k\!=\!|\mathbf{k}_1^{}|\hp$.\ 
On superhorizon scales, the Bardeen potential 
is connected to the comoving curvature perturbation ($\zeta$) by the relation 
{\small$\Phi\!=\!\frac{\,3\,}{\,5\,}\zeta\hp$}, 
and thus {\small$A\!=\!2\pi^2\!\(\!\frac{\,3\,}{\,5\,}\!\)^{\!2}\!P_\zeta\hp$}, 
where {\small$P_\zeta^{}\!=\!\frac{k^3}{\,2\pi^2\,}\langle \zeta_{\mathbf{k}_1}\zeta_{\mathbf{k}_2}\rangle' \simeq 2\!\times\! 10^{-9}$} 
is the observed dimensionless scalar power spectrum.

\vs 

For the oscillatory cosmological collider signal, the bispectrum of $\hp\zeta$ 
may be written as
{\small
\begin{equation}
\langle \zeta_{\mathbf{k}_1}\zeta_{\mathbf{k}_2}\zeta_{\mathbf{k}_3}\rangle'
\hp =\hp  
\fNL
\frac{~72\hp\pi^4 P_\zeta^2~}{~5\hp k_1^2k_2^2k_3^2~}\hp 
\mathcal{C}(k_1,k_2,k_3) \,.
\end{equation}
}
\hspace*{-2.5mm}
In the above, the function $\mathcal{C}(k_1,k_2,k_3)$ is 
the oscillatory template and takes the following form in the squeezed limit,
{\small
\begin{equation}
\mathcal{C}(k_1,k_2,k_3)
= 
e^{\ii\hp\varphi_0}\hsm\!\(\hsm\!\frac{\,k_1\,}{\,k_3\,}\hsm\!\)^{\!\!\ii\hp 2\tilde\nu-2}
\!\!+\rm{c.c.}
=
2\!\(\hsm\!\frac{\,k_1\,}{\,k_3\,}\hsm\!\)^{\!\!\!-2}\!
\cos\!\hsm\(\hsm\! 
2\tilde\nu\ln\!\frac{\,k_1\,}{\,k_3\,}
\!+\!\varphi_0 \!\)\!,
\end{equation}
}
\hspace*{-2.5mm}
where $\varphi_0$ is an overall phase.\ 
Then, we derive the corresponding oscillatory non-Gaussian amplitude $\fNLCC$ as follows: 
{\small
\begin{align}
\label{eq:fNLCC}
\fNLCC
&=
\frac{5}{\,72\hp\pi^4P_\zeta^2\,}
\frac{k_1^2k_2^2k_3^2}{\,\mathcal{C}(k_1,k_2,k_3)\,}
\langle \zeta_{\mathbf{k}_1}\zeta_{\mathbf{k}_2}\zeta_{\mathbf{k}_3}\rangle'_{\text{squeezed}}\nn
\\
&=
\frac{5}{\,72\hp\pi^4P_\zeta^2\,}\!\!
\(\!\!\frac{\,H_{\rm{inf}}\,}{\dot\phi_0}\!\!\)^{\!\!3}\!\! 
\frac{k_1^2k_2^2k_3^2}{\,\mathcal{C}(k_1,k_2,k_3)\,}
\langle \delta\phi_{\mathbf{k}_1}\delta\phi_{\mathbf{k}_2}
\delta\phi_{\mathbf{k}_3}\rangle'_{\text{squeezed}}
\nn\\
&
\simeq
\Bigg|\frac{\,5\pi^3 \tilde{M}_{\!R} P_\zeta\,\tlambda^3 e^{\pi(2\tlambda+\tnu)} 
[\csch(2\pi\tnu)]^3
\sech(2\pi\tnu)\Gamma(\ii\hp2\tnu)\,S_3^{1-\ii\hp\tilde\nu,1-\ii\hp\tilde\nu}\,}
{\,9(\tnu\!-\!\lambda)\,
\Gamma(5\!-\!\ii\hp4\tnu)\,
[\Gamma(\ii\hp\tnu\!-\!\ii\hp\lambda)]_{}^2\,
[\Gamma(\ii\hp\lambda\!+\!\ii\hp\tnu)]_{}^2\,}\\
&\,\,\quad\times(\ii\!+\!2\tnu)\left(8\tnu^2\!+\!\ii\hp12\tnu\!-\!5\!+\!(\ii\hp4\tnu\!-\!3)\cosh(2\pi\tnu)\right)
\Bigg|\,,\nn
\end{align}
}
\hspace*{-2.5mm}
where in the second line above we have used 
$\zeta\!=\!-\frac{H}{\,\dot\phi_0\,}\delta\phi\,$.\  
The above quantity $S_3^{1-\ii\tilde\nu,1-\ii\tilde\nu}$ is the dominant entry  
in the seed integral matrix 
$\hat{\mathcal I}^{p_1p_2}_{\ominus\ominus\,\alpha\beta}$ 
(with $p_1\!=\!p_2\!=\!1\hsm -\hsm\ii\hp\tilde\nu$)
as defined in Eqs.\eqref{Ihatmmdefine} and \eqref{S3define}, 
{\small
\begin{align}
\label{eq:s3_1-inu}
\hspace*{-7mm}
S_3^{1-\ii\hp\tilde\nu,1-\ii\hp\tilde\nu}
=&
\frac{\,-\ii\hp\pi\tilde{M}_{\!R}\, 2^{-4+\ii\hp2\tilde\nu}
e^{\pi\tilde\lambda+\pi\tilde\nu}\csch(2\pi\tnu)
\Gamma(4\!-\!\ii\hp2\tilde\nu)\,}{\,\Gamma(\ii\hp\tnu\!-\!\ii\hp\tlambda)\Gamma(-\ii\hp\tnu\!-\!\ii\hp\tlambda)\,}
\nn
\\
&\times\Bigg[\!\!-\!6\,e^{-\pi\tnu}\Gamma(\ii\hp\tnu\!-\!\ii\hp\tlambda){}_4\tilde F_3\!\Bigg(\!
\begin{matrix}
4\!-\!\ii\hp2\tnu,\,
2,\,
\ii\hp\tnu\!-\!\ii\hp\tlambda,\,
4\,
\\
3,\,
5\!+\!\ii\hp\tilde\lambda\!-\!\ii\hp\tilde\nu,\,
1\!+\!\ii\hp2\tilde\nu
\end{matrix}
\bigg|\!-\!1\!
\Bigg)
\\
&\,\,\,\quad\!+\!e^{\pi\tnu}\Gamma(2\!-\!\ii\hp2\tnu)\Gamma(4\!-\!\ii\hp4\tnu)\Gamma(-\ii\hp\tnu\!-\!\ii\hp\tlambda){}_4\tilde F_3\!
\Bigg(\!
\begin{matrix}
4\!-\!\ii\hp2\tnu,\,
2\!-\!\ii\hp2\tnu,\,
-\ii\hp\tnu\!-\!\ii\hp\tlambda,\,
4\!-\!\ii\hp4\tnu\,\\
3\!-\!\ii\hp2\tnu,\,
5\!+\!\ii\hp\tilde\lambda\!-\!\ii\hp3\tilde\nu,\,
1\!-\!\ii\hp2\tilde\nu
\end{matrix}
\bigg|\!-\!1\!
\!\Bigg)
\!\Bigg]\hsm .
\nn
\end{align}
}
\vspace*{-3mm}

\begin{figure}[t]
\centering
\includegraphics[width=0.75\textwidth]{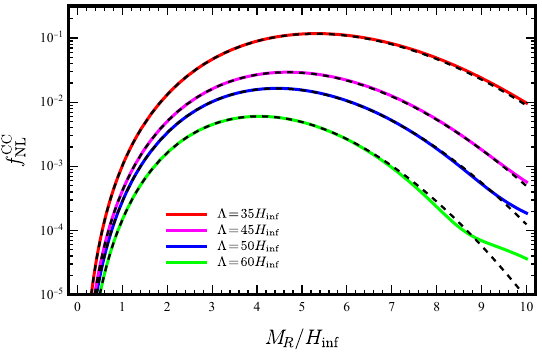}
\vspace*{-3mm}
\caption{\small Amplitude of the oscillatory non-Gaussianity $\fNLCC$ as a function of the 
right-handed neutrino mass $M_{\!R}$ (in the unit of $\Hinf$).\ 
The (red,\,pink,\,blue,\,green) curves correspond to $\Lambda\!=\!(35,\,45,\,50,\,60)H_{\rm{inf}}^{}$, 
which in turn correspond to $\lambda\!\simeq\!(99,\,77,\,69,\,58)H_{\rm{inf}}^{}\hp$.\ 
Each black dashed curve shows the corresponding analytic result 
under the large-chemical-potential limit ($\tlambda\!\!\gg\!\! 1$) 
as in Eq.\eqref{fnllimit}.
}
\label{fig:5}
\end{figure}

We further simplify the above result for the non-Gaussian amplitude $\fNLCC$ 
by taking the large-chemical-potential limit $\tlambda\!\gg\! 1\hp$, 
which gives the following expression:
\begin{equation}
\label{fnllimit}
\fNLCC\simeq\frac{\,5\hp \pi_{}^{\frac{\,5\,}{\,2\,}}\,}{\,2^7\hp 9\sqrt{2\,}\,}
P_\zeta^{}\tilde{M}_{\!R}^4 \hp\tlambda^{\!\frac{\,7\,}{\,2\,}} 
e^{-\frac{\hp 9\pi\hp}{2}(\tnu -\tlambda)} \,. 
\end{equation}
This clearly shows that the usual exponential suppression factor 
$e^{-\pi m/\Hinf}$ is softened by the chemical potential 
$e^{-\frac{\hp 9\pi\hp}{2}(\tnu -\tlambda)}\sim 
e^{-\frac{\hp 9\pi\hp}{4}(\tilde{M}_{\!R}^2/\tlambda)}$.  

In Fig.\,\ref{fig:5}, we show the amplitude of the oscillatory non-Gaussianity $\fNLCC$ as a function of the right-handed neutrino mass $M_{\!R}$ 
for a set of benchmark values of the cutoff scale $\Lambda\hp$.\ 
The (red,\,pink,\,blue,\,green) curves correspond to 
$\Lambda\!=\!(35,\,45,\,50,\,60)\,H_{\rm{inf}}$ respectively, 
which in turn give $\lambda\!\simeq\!(99,\,77,\,69,\,58)\,H_{\rm{inf}}^{}\hp$.\ 
In the same plot, we show analytic results 
under the large-chemical-potential limit ($\tlambda\!\gg\! 1$) 
as given by Eq.\eqref{fnllimit},
which are depicted by the black dashed curves for each case.

As shown in Fig.\,\ref{fig:5}, the oscillatory non-Gaussianity depends sensitively on both the cutoff scale and the right-handed neutrino mass.\ 
When the cutoff scale $\Lambda \sim 35 H_{\rm {inf}}$, the cosmological collider signal 
can reach $\fNLCC\!=\! {O}(0.1)$ for
$M_{\!R}\!=\! (3\!-\!6)\,H_{\rm{inf}}\hp$.\ 
Such signals may be probed by the future searches through the measurements  
of the CMB, the large scale structure, and especially the 21cm tomography.\ 
Even for a moderately larger cutoff, such as $\Lambda\!=\!50\,H_{\rm{inf}}$, 
part of the parameter space with $M_{\!R}\!=\! (2\!-\!6)\,H_{\rm{inf}}$ 
can still give $\fNLCC\!\gtrsim\! {O}(0.01)$,
which may be probed by the future 21cm tomography surveys.
Fig.\,\ref{fig:5} also shows that the cosmological collider signal 
becomes weaker as $\Lambda$ increases, or equivalently, as the chemical potential $\lambda\!=\!\dot\phi_0/\Lambda$ decreases.\
This trend can be understood from the factorized structure 
discussed in Section\,\ref{Calculation of the Three-Point Correlator}.\ 
As shown in Fig.\,\ref{function_of_m}, the left subdiagram 
is enhanced by a larger chemical potential, 
whereas the combination of bubble contribution and right diagram 
mildly depends on the chemical potential.\ 
Consequently, the full oscillatory signal becomes stronger when the 
chemical potential increases.\
This feature highlights the central role of the chemical potential.\  
In the absence of a chemical potential, the cosmological collider signal 
from heavy fermions is strongly suppressed by the usual Boltzmann factor, 
and an additional suppression arises from the loop level.\ 
In contrast, a nonzero chemical potential softens the Boltzmann suppression 
and significantly enhances the dominant helicity mode.\ This opens up 
a window for observing the cosmological collider signals 
generated by heavy fermions such as the right-handed neutrinos.

\vs 

\begin{figure}[t]
    \centering
    \begin{subfigure}[b]{0.7\textwidth}
        \centering
        \hspace{-0.23cm}\includegraphics[width=\linewidth]{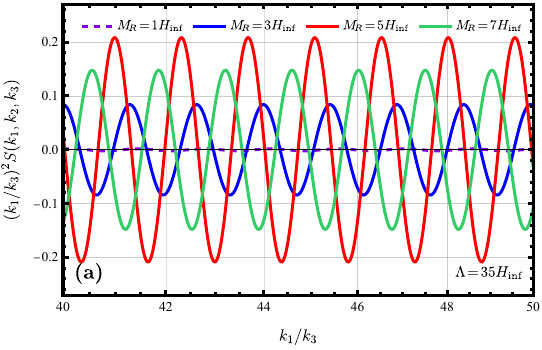}
    \end{subfigure}
    \begin{subfigure}[b]{0.7\textwidth}
        \centering
        \hspace{-0.1cm}\includegraphics[width=\linewidth]{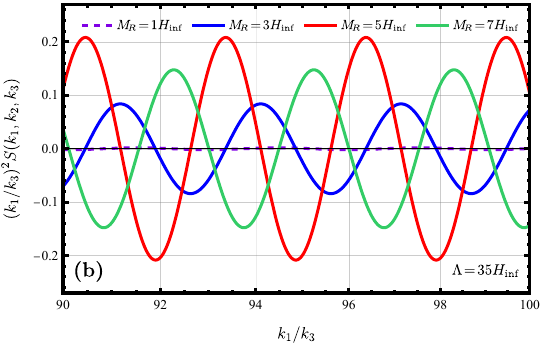}
    \end{subfigure}
\centering
\vspace*{-3mm}
\caption{\small Dimensionless shape function $(k_1/k_3)^2S(k_1,k_2,k_3)$ plotted as a function of the ratio $k_1/k_3$ on a logarithmic horizontal axis.\ 
The (purple-dashed,\,purple,\,blue,\,red,\,green) curves correspond to 
$M_{\!R}\!=\!(1,\,3,\,5,\,7)\,H_{\rm{inf}}$ respectively, 
with the cutoff scale  
$\Lambda \!=\!35\Hinf\hp$.\ 
The upper and lower plots correspond to different ranges of $k_1/k_3$, 
illustrating the characteristic logarithmic scaling behavior 
along the horizontal axis.}
\label{fig:6}
\end{figure}

Figure\,\ref{fig:5} also shows that the non-Gaussianity $\fNLCC$ 
is suppressed in both the large-mass and small-mass regions.\ 
In the large-mass region, the suppression is controlled by the Boltzmann factor.\ 
In the small-mass region, the suppression is determined by 
two complementary factors.\ First, certain left subdiagrams in 
Fig.\,\ref{fig:3}, especially the diagrams\,(3)–(7), 
are effectively controlled by mass-insertion-type couplings.\ 
As the Majorana mass decreases, the corresponding mass-insertions 
become less important, and the oscillatory signals become reduced.\ 
Second, the suppression is also encoded in the soft-fermion propagators: 
the nonlocal part of the type-1 propagator in Eq.\eqref{propagator1} 
scales as $M_{\!R}^2$ in the small-mass limit, whereas the corresponding 
nonlocal parts of the type-2 and type-3 propagators in Eqs.\eqref{propagator2} 
and \eqref{propagator3} scale linearly with $M_{\!R}$ at the leading order.\ 
Combining these factors, we deduce the following:
\\[-6mm]
\begin{equation}
\lim_{\tilde{M}_{\!R}\to\hp 0} \!f_\rm{NL}^{\text{CC}} \sim \tilde{M}_{\!R}^4\,,
\vspace{-2mm}
\end{equation}
which shows that the neutrino-loop signal is quickly reduced via $\tilde{M}_{\!R}^4$
as the heavy fermion mass decreases.

To further characterize the structure of the cosmological collider signal, 
one can introduce a dimensionless shape function $S(k_1,k_2,k_3)$ 
through the following relation:
{\small
\begin{equation}
\langle \zeta_{\mathbf{k}_1}\zeta_{\mathbf{k}_2}\zeta_{\mathbf{k}_3}\rangle'
=
\frac{(2\pi)^4P_\zeta^2}{\,(k_1k_2k_3)^2\,}S(k_1,k_2,k_3)\,.
\end{equation}
}
\hspace*{-2.5mm}
Unlike the template function $\mathcal{C}(k_1,k_2,k_3)$ defined above, the shape function $S(k_1,k_2,k_3)$ also contains the overall amplitude information.\ 
Using the squeezed limit form of the bispectrum, it can be derived as follows:
{\small
\begin{equation}
\begin{aligned}
S(k_1,k_2,k_3)
=
\langle \zeta_{\mathbf{k}_1}\zeta_{\mathbf{k}_2}\zeta_{\mathbf{k}_3}\rangle'
\frac{\,k_1^2k_2^2k_3^2\,}{\,(2\pi)^4P_\zeta^2~}
=
\frac{2\fNLCC}{~(2\pi)^4P_\zeta^2~}
\!\!\(\!\!\frac{\,H_{\rm{inf}}\,}{\,\dot\phi_0\,}\!\hsm\)^{\hsm\!\!3}\!\!
\(\!\frac{\,k_3\,}{k_1}\!\)^{\hsm\!\!2}
\!\cos\!\hsm\(\!\hsm 
2\tilde\nu\ln\!\frac{\,k_1\,}{\,k_3\,}\!+\!\varphi_0
\!\) \!.
\end{aligned}
\end{equation}
}
\hspace*{-2.5mm}

In Fig.\,\ref{fig:6}, we plot the dimensionless quantity 
$(k_1/k_3)^2S(k_1,k_2,k_3)$ as a function of the ratio $k_1/k_3$ 
on a logarithmic horizontal axis.\ 
The (purple-dashed,\,purple,\,blue,\,red,\,green) curves correspond 
to the benchmark masses
$M_{\!R}\!=\!(1,\,3,\,5,\,7)\,H_{\rm{inf}}$ 
respectively, and the cutoff scale takes a sample value    
$\Lambda \!=\!35\Hinf\hp.$
This figure demonstrates the characteristic oscillatory behavior of 
the cosmological collider signals in the squeezed limit.\ 
Moreover, because the horizontal axis is logarithmic, 
the oscillation frequency can be read off directly from the period 
of each curve and is determined by
\begin{equation}
\omega=2\tilde\nu=2\sqrt{\tilde{M}_{\!R}^2\!+\!\tilde\lambda^2\,}\,.
\end{equation}
Hence, the shape of the cosmological collider signal directly 
encodes the right-handed neutrino mass scale and the chemical potential.

\vs

\subsection{\hspace*{-2.5mm}Comparison with the Previous Work}
\label{comparison with previous work}
\label{sec:4.2}

In this subsection, we briefly compare our result 
for the inflaton bispectrum with the estimate   
of Ref.\,\cite{Chen:2018xck}, which gave a qualitative estimate 
on the characteristic oscillatory signal generated by heavy fermions during inflation.\ 
In contrast, our work uses
a new method to quantitatively compute the cosmological collider signals generated by the
heavy right-handed neutrino loop.\ In the following, we will explain why the overall amplitude 
derived in the present work is numerically smaller than that in Ref.\,\cite{Chen:2018xck}, and clarify
how this difference arises in our calculation.

We note that the previous analysis\,\cite{Chen:2018xck} adopted two simplifying approximations: {\bf (i)}\,the hard propagator was treated by a saddle-point approximation; 
and {\bf (ii)}\,the loop momentum was replaced by a characteristic value instead of being integrated over explicitly.\ By contrast, our calculation is based on the factorized loop treatment developed in Section\,\ref{sec:4new} and Appendix\,\ref{app:C}, 
in which the diagram is decomposed into a left part, a right part, 
and a bubble loop part in which the loop-momentum integration is explicitly performed 
within this framework.

\begin{figure}[t]
\centering
\includegraphics[width=0.62\textwidth]{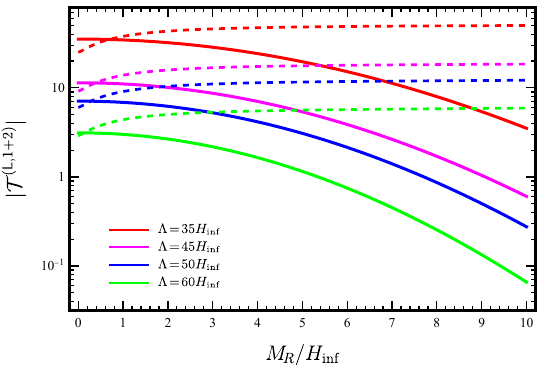}
\vspace*{-3mm}
\caption{\small Comparison for magnitude of the left subdiagrams of the diagrams (1)+(2) in Fig.\,\ref{fig:2}.\  
The solid curves present the precise result of the left subdiagrams from the diagrams-(1)+(2), 
and the dashed curves present the result of the same left subdiagrams using the saddle point approximation adopted by Refs.\,\cite{Chen:2018xck,Hook:2019zxa,Aoki:2026olh}.}
\label{fig:7}
\end{figure}

\begin{figure}[t]
\centering
\includegraphics[width=0.62\textwidth]{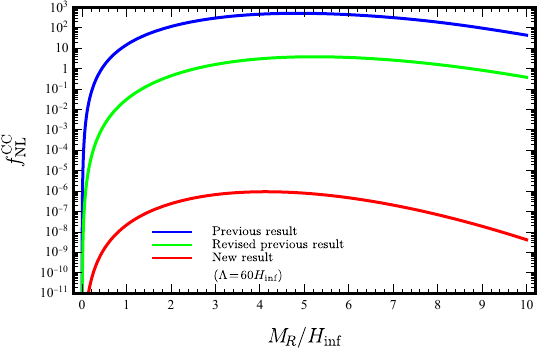}
\vspace*{-2mm}
\caption{\small Comparison among the previous result of Ref.\,\cite{Chen:2018xck}
(shown as blue curve), 
the revised previous result after correcting the propagators 
(shown as green curve), and the new result (shown as red curve), 
for a sample cutoff value $\Lambda \!=\! 60H_{\rm{inf}}$, where only 
the diagrams\,(1) and (2) of Fig.\,\ref{fig:2} are included 
for this comparison.}
\label{fig:8}
\end{figure}

The numerical difference between our analysis and the previous one\,\cite{Chen:2018xck} 
mainly arises from the following sources.\ 
First, the propagator structure of Ref.\,\cite{Chen:2018xck} improperly 
included all four helicity combinations ($++,\,+-,\,-+,\,--$), 
whereas in the present formalism only two independent helicity contributions 
($++,\,--$) enter the final result.\
This is because the anti-commutator \eqref{anticommutation relation} vanishes
for the mixed helicity-combinations ($+-,\,-+$).\  
Second, Ref.\,\cite{Chen:2018xck} adopted the saddle-point approximation for 
the hard propagator and included the extra helicity channels ($+-,\,-+$) 
[which should not exist due to the vanishing anti-commutator 
\eqref{anticommutation relation}].\  
These two factors enlarged the result of Ref.\,\cite{Chen:2018xck} as compared with 
our result, and caused a large discrepancy of order ${O}(10\!-\!100)$ 
in the intermediate-mass range.\  In contrast, if one keeps only the two helicity channels ($++,\,--$), the saddle-point-approximation-based result for the left subdiagram remains close to the precise result for intermediate mass range, but can deviate by about two orders of magnitude in the large-mass range, as illustrated in Fig.\,\ref{fig:7}.\
Third, and more importantly, the loop integration measure is treated very 
differently.\ Ref.\,\cite{Chen:2018xck} effectively approximates the loop integral 
as the following oversimplified form,
\begin{equation}
\frac{1}{\,(2\pi)^3\,}\!\!\int \!\mathrm d\phi
=
\frac{1}{\,(2\pi)^2\,},
\label{old_loop_measure}
\end{equation}
whereas in the present analysis we compute the full bubble-loop integral explicitly, 
as given in Appendix\,\ref{loop integral through Feynman parameter approach}.\
Numerically, this leads to an additional $O(10^{6})$ difference 
in the intermediate- and large-mass ranges, 
as shown in Fig.\,\ref{fig:8}, and constitutes our major new improvements.\

This main numerical improvement can be traced to the $\mathcal{C}_{(d)j}$ factor 
from the bubble loop integration combined with the right subdiagram 
which can be computed in analogy with Eq.\,\eqref{cfactor_matrix} for diagram-(7).\ 
Interestingly, we find the $\mathcal{C}_{(d)j}$ factors 
for diagrams (1) and (2) are identical to that of diagram-(7).\  
In the large $\tnu$ limit, the factor $\mathcal{C}_{(d)j}$ in our new result 
has the following asymptotic behavior:
\begin{align}
|\mathcal{C}_{(d)j}|\!\simeq\!2^{-13/2}\pi^{-3/2}\tnu^{-3/2} k_s^3\,.
\label{eq:cfactor}
\end{align}
This factor arises from the explicit bubble loop-momentum integration 
in which the helicity conservation provides a suppression $\tnu^{-1}$ 
such that the final $\tnu$ dependence becomes $\tnu^{-3/2}$.\ 
The above behavior \eqref{eq:cfactor} is not captured 
by the estimate based on pure dimensional analysis of the bubble integral 
in the literature\,\cite{Chen:2018xck}, 
where the corresponding factor can be expressed as follows:
\\[-8mm]
\begin{align}
\label{eq:cfactor_old}
\Big|\mathcal{C}_{(d)j}\text{\cite{Chen:2018xck}}\hsm\Big|
\!\simeq\!\pi^{-2}\tnu k_s^3\,.
\end{align}
The ratio of the corresponding factors between the two results is given by
\begin{align}
\frac{|\mathcal{C}_{(d)j}|}{\,|\mathcal{C}_{(d)j}\text{\cite{Chen:2018xck}}|\,}
=2^{-13/2}\pi^{1/2}\tnu^{-5/2}\,.
\end{align}
For a sample input $\hp\tilde\nu\! =\! 60$
(corresponding to $\Lambda\!\simeq\!60\Hinf$), the ratio of this factor is of 
$O(10^{-6})\hp$.\ 
Thus, the previous result\,\cite{Chen:2018xck} has  
an overestimate of the resultant loop amplitude 
by about six orders of magnitude at this frequency value.

\begin{figure}[t]
    \centering
    \begin{subfigure}[b]{0.62\textwidth}
        \centering
        \hspace{-0.23cm}\includegraphics[width=\linewidth]{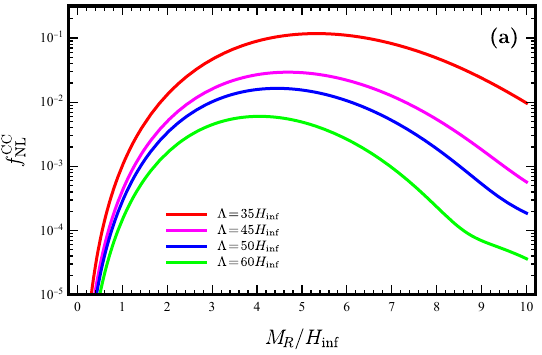}
         \label{fig:compare_12_only}

    \end{subfigure}
    \begin{subfigure}[b]{0.62\textwidth}
        \centering
        \hspace{-0.1cm}\includegraphics[width=\linewidth]{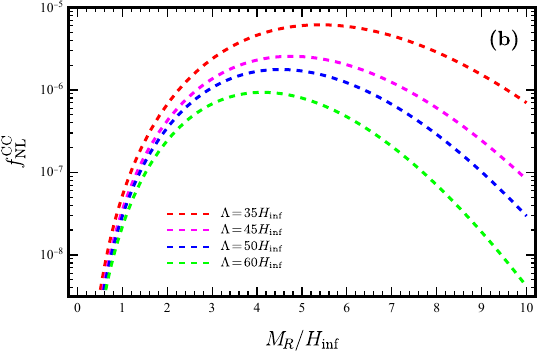}
\label{fig:compare_all}
\end{subfigure}
\vspace*{-8mm}
\caption{\small Comparison of the correlation amplitude of the cosmological collider non-Gaussianity $\fNLCC$ 
as a function of the heavy neutrino mass scale $M_{\!R}$, 
which is contributed by all diagrams of Fig.\,\ref{fig:2} 
as shown in the present plot\,(a) and contributed by the 
diagrams (1)+(2) only as shown in the present plot\,(b).\  
The (red,\,pink,\,blue,\,green) curves correspond to a set of 
sample values of the cutoff scale $\Lambda \!=\!(35,\,45,\,50,\,60)\Hinf\hp$, 
which in turn correspond to $\lambda\!\simeq\!(99,\,77,\,69,\,58)H_{\rm{inf}}^{}\hp$.}
\label{fig:compare}
\label{fig:9}
\end{figure}

We can make a direct comparison in Fig.\,\ref{fig:8} for the same cutoff 
choice $\Lambda \!=\! 60H_{\rm{inf}}^{}$ and for including only 
the diagrams\,(1) and (2) of Fig.\,\ref{fig:2}.\ 
In Fig.\,\ref{fig:8}, we plot the previous result of 
Ref.\,\cite{Chen:2018xck} (shown as blue curve), the result obtained after 
correcting the propagator treatment (i.e., the correct helicity contributions 
and the precise evaluation of the hard propagator in the left subdiagram)
(shown as green curve), and the full result 
obtained in the present work (shown as red curve).\ 
As shown in Fig.\,\ref{fig:8}, the previous result of Ref.\,\cite{Chen:2018xck} 
(the blue curve) is about two orders of magnitude larger than the result 
obtained after correcting the propagators (the green curve).\ 
The propagator-corrected result (the green curve) is still 
larger than the present result (the red curve) 
by about six orders of magnitude.\  
This substantial difference between the green curve and the red curve 
is mainly due to the different treatments of the bubble loop
integration combined with the right subdiagram contribution  
as given by our Eq.\eqref{eq:cfactor} 
and by the previous approximation in Eq.\eqref{eq:cfactor_old}.\

\vs

Moreover,  using the present method
we compare the difference between the contribution 
of the diagrams\,(1)+(2) only (as in Ref.\,\cite{Chen:2018xck}) 
and the contribution of all eight diagrams
of Fig.\,\ref{fig:2} (as in the present study).\ 
We find that the contribution of the diagrams\,(1)+(2) only
can cause a substantial underestimate of the CC signal 
by about a factor of $O(10^{-4})$.\ 
For an explicit comparison, we show in Fig.\,\ref{fig:compare}(b)
the non-Gaussian amplitude $\fNLCC$ by including only diagrams\,(1)+(2)    
of Fig.\,\ref{fig:2} in our approach.\ 
In this special case, the resultant non-Gaussianity $\fNLCC$ 
[shown in the present Fig.\,\ref{fig:9}(b)] is substantially 
smaller than the previous qualitative estimate of Ref.\,\cite{Chen:2018xck}
and than our full result [shown in the present Fig.\,\ref{fig:9}(a)].

\begin{figure}[t]
\centering
\includegraphics[width=0.62\textwidth]{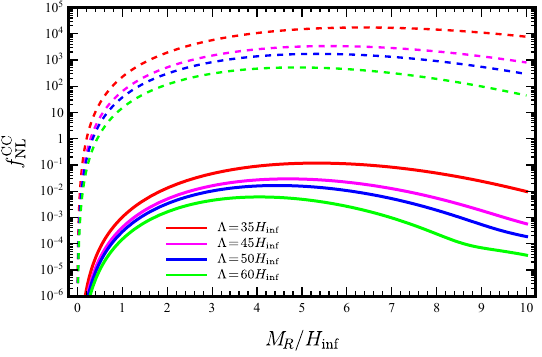}
\vspace*{-2mm}
\caption{\small Comparison of the correlation amplitude of the cosmological collider non-Gaussianity 
$\fNLCC$ between our new result and the estimate of Ref.\,\cite{Chen:2018xck}.\ 
The (red,\,pink,\,blue,\,green) solid curves denote our results for 
$\Lambda\!=\!(35,\,45,\,50,\,60)H_{\rm{inf}}$ respectively, 
whereas the dashed curves denote the corresponding estimates 
obtained using the method of Ref.\,\cite{Chen:2018xck}.\ 
In each case, the (red,\,pink,\,blue,\,green) curves correspond to 
$\lambda\!\simeq\!(99,\,77,\,69,\,58)H_{\rm{inf}}^{}\hp$ respectively.
}
\label{fig:10}
\end{figure}

Finally, in Fig.\,\ref{fig:10}, 
we present a comparison of the correlation amplitude of the cosmological collider 
non-Gaussianity $\fNLCC$ between our new result \eqref{eq:fNLCC}
and the estimate of Ref.\,\cite{Chen:2018xck}, 
as a function of the heavy neutrino mass scale $M_{\!R}\hp$.\ 
For the cutoff scales
$\Lambda=(35,\,45,\,50,\,60)H_{\rm{inf}}$, the previous estimate is larger
than our result by roughly $O(10^5)$ in the intermediate and large mass ranges.\ 
This difference is mainly caused by the combined effects of the corrected
fermion propagators, the complete helicity and Schwinger-Keldysh structure,
the inclusion of the full set of relevant loop diagrams, and the explicit
loop-momentum integration performed in the present calculation.\ 
Fig.\,\ref{fig:10} shows that our newly improved results (solid curves) become
significantly smaller than the previous results (dashed curves).\ 
This reduction would pose a challenge to the observability of such CC signals 
and highlights the importance of computing the full result (including the 
local contribution) in future work.

\vs 

In summary, by replacing the oversimplified approximations made in \cite{Chen:2018xck}
with more reliable methods, we have obtained a more precise estimate of the nonlocal 
heavy fermion loop contribution to the cosmological collider signal, 
which turns out to be a substantial improvement over the previous result.\ 
In our present approach, the loop integration is consistently factorized 
in the squeezed limit, and the full set of relevant diagrams in Fig.\,\ref{fig:2}
is computed in our analyses.\ We have presented the comparisons 
in Figs.\,\ref{fig:7}-\ref{fig:10}.\  
Due to these reasons, our present method and analyses
have made important progress for more reliably predicting 
cosmological collider signals from the heavy fermion loop contributions.

\section{\hspace*{-2.5mm}Conclusions}
\label{Conclusion}

Cosmological collider (CC) physics\,\cite{Chen:2009we,Chen:2009zp,Chen:2012ge,Baumann:2011nk,Pi:2012gf,Noumi:2012vr,Gong:2013sma,Arkani-Hamed2015,Chen:2015lza,Baumann160703735,Chen:2016uwp,Kumar:2017ecc,Kumar:2019ebj}
provides a unique means of probing heavy particles present during inflation 
through their contributions to the primordial non-Gaussianity.\ So far most of the existing studies 
have focused on tree-level signals or scalar-loop effects, whereas signatures of heavy fermion loops are 
technically more challenging and remain largely unexplored.\ However, heavy fermions (such as the 
right-handed neutrinos) are strongly motivated\,\cite{You:2024hit,Han:2024qbw} 
from the structure of the Standard Model of particle physics
and from the light neutrino mass generations through the canonical seesaw 
mechanism\,\cite{Minkowski:1977sc,GellMann:1980vs,Yanagida:1979as,Glashow:1979nm}.\ 
Hence, it is important to develop a systematic method for computing the contributions of heavy fermion loops 
to the inflaton correlators and to predict the fermion-loop signatures through the primordial non-Gaussianity.\

In this work, we studied cosmological collider signals generated by heavy fermion loops in an inflationary setup 
including the derivative interaction between the inflaton and right-handed neutrinos.\ 
A central ingredient of our analysis is the effective chemical potential induced by the slow-roll inflaton background.\ 
This chemical potential leads to helicity-dependent fermion production and can substantially enhance one of the helicity modes, 
thereby weakening the usual Boltzmann suppression of heavy-particle production during inflation.\ 
In this way, heavy fermions such as the heavy right-handed neutrinos 
(that would otherwise generate extremely small signals) 
may become significantly more visible in the primordial non-Gaussianity.\ 
For the computational side, we systematically analyzed the spinor structure, the factorization of fermion triangle loop, 
the loop-momentum integrals and the time seed integrals,  
which give a controlled computation of the nonlocal contribution 
to the CC signals of right-handed neutrinos.\

In Section\,\ref{sec:2}, we formulated the dynamics of the right-handed Majorana neutrinos in the slow-roll inflationary background.\ 
With the unique derivative interaction between the inflaton and right-handed neutrinos as given in Eq.\eqref{Total_Lagrangian_weyl}, 
we showed how the rolling inflaton background induces an effective chemical potential \eqref{Lagrangian_weyl_free} 
for the right-handed neutrinos.\ 
We then derived the corresponding Weyl spinor mode functions \eqref{modefunction}, constructed the Schwinger-Keldysh propagators 
in Eqs.\eqref{propagator1}, \eqref{propagator2}, and \eqref{propagator3}, and extracted their late-time nonlocal components 
in Eqs.\eqref{late time limit propagator}--\eqref{late time limit propagator hatDCon}.\
These late-time components encode the characteristic oscillatory clock signals and provide 
the basic building blocks for the fermion-loop calculation.

In Section\,\ref{sec:3}, we computed the three-point inflaton correlator generated 
by the right-handed neutrino triangle loop.\ 
We classified the eight nonequivalent contractions of the Majorana fermion loop 
[shown in Fig.\,\ref{fig:2} and Eqs.\eqref{3pt_form}\eqref{all_triangleloopform}]
and organized their contributions in terms of the factorized left subdiagram, 
central bubble loop, and right subdiagram 
[shown in Fig.\,\ref{fig:3} and Eq.\eqref{factorization_3pt}].\ 
In this treatment, we precisely computed the left subdiagram and 
retained the loop-momentum integration in the central bubble loop, 
thereby improving the calculation of nonlocal signals 
as compared with the previous estimates based on a saddle-point approximation 
of the hard internal propagator.\ 

\vs 

In Section\,\ref{sec:4}, we analyzed the size and shape of the predicted cosmological collider signatures.\  
In subsection\,\ref{sec:4.1}, we expressed the squeezed limit bispectrum 
in terms of the oscillatory non-Gaussianity amplitude $\fNLCC$ in Eq.\eqref{eq:fNLCC}
and studied its dependence on the right-handed neutrino mass, the induced chemical potential, 
and the UV cutoff scale of the inflaton-right-handed-neutrino interaction.\  
We found that the chemical potential enhances one of the helicity modes and can substantially soften 
the usual Boltzmann suppression of heavy fermion production.\  
In consequence, as shown in Fig.\,\ref{fig:5}, 
the loop-induced signals can become sizable in the intermediate-mass range, 
whereas they are suppressed both in the large-mass range by the heavy-particle Boltzmann factor 
and in the small-mass range by the mass-dependence of the fermion propagators 
and the mass-insertion structure.\  
We demonstrated in Fig.\,\ref{fig:5} and Figs.\,\ref{fig:9}-\ref{fig:10}   
that {\it the CC signal $\fNL$ is enhanced by the increase of the chemical potential $\tlambda\hp$.}\ 
We further compared in subsection\,\ref{sec:4.2}
our results with the previous estimates as presented in Figs.\,\ref{fig:7}-\ref{fig:10}.\
The main difference arises from a more precise treatment of the full loop structure 
as shown in Fig.\,\ref{fig:8}.\  
In earlier analysis, the hard propagator was estimated by a saddle-point approximation and 
the magnitude of loop momentum was assumed to be a cutoff value.\  
Moreover, not all fermion loop contractions were included, 
so part of the loop-diagram contributions was effectively omitted.\  
In contrast, we keep the hard propagator inside the left subdiagram, 
include the complete set of relevant fermion loop contractions, 
and perform the loop-momentum integration precisely in the bubble subdiagram.\ 
This treatment precisely retains the momentum dependence, 
helicity dependence, and Schwinger-Keldysh structure 
of the fermion propagators.\  
In consequence, we found that the resultant signal amplitude is substantially smaller than 
the previous estimate, showing that the saddle-point approximation, the oversimplification 
of the loop integral and the incomplete inclusion of Feynman diagrams 
cause large overestimates of the loop-induced CC signals.\

Overall, our analysis provides a systematic treatment of the nonlocal contribution 
of the fermion loop to the cosmological collider signals.\  
We have shown that heavy fermion derivative interaction with the inflaton may generate potentially observable 
oscillatory non-Gaussian signatures even when their masses are significantly above the Hubble scale.\ 
This opens up a new window to sensitively probe the high-scale new physics involving heavy fermions, 
such as the right-handed neutrinos which are responsible for the light neutrino mass generation 
through the seesaw mechanism.\  
In such cosmological collider signals, the oscillation frequency is mainly controlled by 
the effective chemical potential (which is set by the cutoff scale $\Lambda$), 
whereas the signal amplitude is sensitive to the mass scale of the right-handed neutrino.\  
We also note that our current results include only the nonlocal leading-pole contribution to the CC signal (similar to the literature\,\cite{Chen:2018xck,Hook:2019zxa,Hook:2019vcn,Aoki:2026olh}).\ 
It would be important to further reliably compute the contributions from other poles and the local contribution.\ 
It is expected that very different methods might be required 
to achieve a complete result and we leave this challenge for future work.\ 

Moreover, we would stress the complementarity between the present study of probing the neutrino seesaw scale 
using cosmological collider signals 
and the previous study\,\cite{You:2024hit,Han:2024qbw} of probing the neutrino seesaw mechanism 
using the analytic non-Gaussianity\footnote{%
For a nonzero three-point correlator, the dependence on the momentum ($k$) can be analytic 
only for the contributions of the equilateral, orthogonal, and local non-Gaussianity templates 
as shown in Ref.\,\cite{Planck9}.}.\ 
In Refs.\,\cite{You:2024hit,Han:2024qbw}, the analytic non-Gaussian signals arise from 
the Higgs-modulated reheating and depend on the seesaw parameters 
(including the right-handed neutrino mass and the neutrino Yukawa coupling) and   
the Higgs self-coupling during inflation.\  
Hence, the two types of observables can probe different combinations of the seesaw and inflationary parameters: 
the non-analytic cosmological collider signals are primarily sensitive to the heavy right-handed neutrino mass 
and the chemical-potential, whereas the analytic non-Gaussianity constrains the reheating dynamics and 
the seesaw parameters (connected to the light neutrino mass as well).\ 
Combining these two types of primordial non-Gaussian signatures will give complementary probes   
of the full seesaw parameter space.\

\vspace*{6mm}
\noindent 
{\large\bf Acknowledgments}
\\[1mm] 
We thank Shuntaro Aoki, Alessandro Strumia, Zhehan Qin and Yuhang Zhu for discussions.\  
The research of H.\,J.\,H., L.\,S.\ and J.\,Y.\ was supported in part by the National Natural Science Foundation of China (Grant Nos.\,12175136, 12435005, and 11835005) and by the Shenzhen Science and Technology Program (Grant No.\ JCYJ20240813150911015).\ C.\,H.\ acknowledges support from the National Key R{\&}D Program of China under Grant 2023YFA1606100, the National Natural Science Foundation of China under Grants No.\ 12435005, and  the Key Laboratory of Particle Astrophysics and Cosmology (MOE) 
of Shanghai Jiao Tong University.\

\vspace*{6mm}

\appendix

\noindent
{\Large\bf Appendix}	

\vspace*{-2mm}

\appendix

\section{\hspace*{-1.5mm}Weyl Spinors in the Inflationary Background}
\label{app:A}
\label{details: helicity and eom}

\noindent 

In this Appendix, we retain the spinor indices explicitly and 
set up the spinor-helicity conventions.\ Then, we derive a self-consistent set 
of equations of motion for the Weyl spinor mode functions.\ 

\vs 

Consider the curved spacetime with a metric tensor $g_{\mu\nu}^{}\hp$,
\begin{equation}
\d s^2=g_{\mu\nu}\d x^\mu \d x^\nu \,. 
\end{equation}
Using the vierbein $e^a_{\ \mu}\!=\hsm\partial\xi^a\hsmx \big/ \partial x^\mu$ with $\xi^a$ 
being the coordinate of the Minkowski metric $\eta_{ab}^{}$, 
the metric tensor can be expressed as 
$g_{\mu\nu}^{} \!=\! e^a_{\ \mu} e^b_{\ \nu} \eta_{ab}^{}\hp$,
where $\eta_{ab}\!=\hsm\rm{diag}(-1,1,1,1)$ is the Minkowski metric tensor. 

\vs 

In curved spacetime, spinors are defined with respect to a local Lorentz frame and the covariant derivative includes the spin connection\,\cite{Birrell:1982ix,Parker:2009uva,Chen160407841}.\ 
The covariant derivative acting on a Weyl spinor $\psi_\alpha^{}$ is defined as
\begin{equation}
\label{aeq:Dpsi}
\text{D}_\mu \psi_\alpha
= \partial_\mu \psi_\alpha
\!+\! \frac{1}{2}\,\omega_{\mu ab}
\big(\Sigma^{ab}\big)_{\!\alpha}^{\,\,\beta}\,\psi_\beta\,,
\end{equation}
where $\omega_{\mu ab}$ is the spin connection, 
\begin{equation}
\omega_{\mu ab}= e^{a}_{\ \nu} \mathrm{D}_\mu e^{b\nu}
= e^{a}_{\ \nu}\!\(\partial_\mu e^{b\nu}
\!+\! \Gamma^\nu_{\mu\rho} e^{b\rho} \),
\end{equation}
with $\Gamma^\nu_{\mu\rho}$ being the Christoffel symbol.\ 
In Eq.\eqref{aeq:Dpsi}, the quantity 
$\(\Sigma^{ab}\)_\alpha^{\,\,\,\,\beta}$ 
denotes the Lorentz generators in the Weyl representation,
\begin{equation}
\big(\Sigma^{ab}\big)_{\!\alpha}^{\,\,\beta}
= \frac{1}{\hs 4\hs}( \sigma^a \bar{\sigma}^b 
\!-\! \sigma^b \bar{\sigma}^a )_{\!\alpha}^{\,\,\beta}\,,
\end{equation}
where $\sigma^a \!=\! (1, \vec{\sigma})$ and 
$\bar{\sigma}^a \!=\! (1,\hsm -\vec{\sigma})$ 
are the sigma matrices in the flat spacetime.

\vs 

Then, we consider the spatially flat FLRW metric in the conformal coordinates,
\begin{equation}
\td s^2 = a^2(\tau)\big(\!-\!\d\tau^2 \!+\hsm \d\vec{x}^{\hp 2}\big).
\end{equation}
The vierbein is given by 
$e^a_{\ \mu} \!= a(\tau)\,\delta^a_{\ \mu}$  
and 
$e_a^{\ \mu} \!= a^{-1}(\tau)\,\delta_a^{\ \mu}$.\ 
The non-vanishing Christoffel symbols $\Gamma^\nu_{\mu\rho}$ are derived as follows:
\begin{equation}
\Gamma^0_{00} \hsm =\hsm \mathcal{H} \hp, 
\quad
\Gamma^0_{ij} \hsm = \mathcal{H}\hp\delta_{ij} \hp, 
\quad
\Gamma^i_{0j} \!=\hsm \mathcal{H}\hp\delta^i_j \hp, 
\end{equation}
where $\mathcal{H} \!=\! \frac{\,a'}{a}$ with 
$a'\!=\!\d a(\tau)/\d\tau\hp$.\ 
Thus, the non-vanishing spin connection 
components are given by
\begin{equation}
\omega_{ij0}^{} =- \omega_{i0j}^{}=\mathcal{H}\hp\delta_{ij}^{}\hp, 
\quad (i,j\!=\!1,2,3), 
\end{equation}
and all other components vanish.\ Using the formula 
\begin{equation}
\(\Sigma^{0i}\)_\alpha^{\,\,\,\beta}=\!-\!\(\Sigma^{i0}\)_\alpha^{\,\,\,\beta}
=\frac{1}{4}
\!\( \sigma^0 \bar{\sigma}^i \!-\! \sigma^i \bar{\sigma}^0 \)_\alpha^{\,\,\,\beta}=\frac{1}{2}\!\(\sigma^0 \bar{\sigma}^i\)_\alpha^{\ \beta} ,
\end{equation}
we simplify the covariant derivative (acting on the Weyl spinor) as follows:
\begin{equation}
\label{covariant derivative}
\rm{D}_0^{} \psi_\alpha^{} \!= \partial_0^{}\psi_\alpha^{}\hp, 
\quad 
\rm{D}_i^{} \psi_\alpha^{}
= \partial_i^{} \psi_\alpha^{}
\!-\! \frac{1}{2}\mathcal{H}\big(\sigma^0 \bar{\sigma}^i\big)_\alpha^{\ \beta} \psi_\beta\,, 
\end{equation}
where $\sigma^a $ and $\bar{\sigma}^a$ are the sigma matrices in the flat spacetime.

\vs 

Consider curved spacetime with a metric tensor,
Consider free Lagrangian of a Weyl spinor in the curved spacetime,
\begin{equation}
\Delta\LG= \sqrt{-g\,}\;
\ii\hp \psi_{\dalpha}^\dagger \bar{\varsigma}^{\hp\mu\dalpha\alpha} 
\rm{D}_\mu \psi_\alpha\,,
\end{equation}
where the Pauli sigma matrices $\varsigma_{\mu}^{}$ 
and $\bar{\varsigma}^{\mu}$ are defined 
from the flat-space sigma matrices $\sigma^a$ 
through the vierbein $e_a^{\ \mu}$,
\begin{equation}
\begin{aligned}
\varsigma^{\mu} = e_b^{\ \mu} \sigma^{b} ,\,
\hspace{6mm}
\bar{\varsigma}^{\mu} = e_b^{\ \mu} \bar{\sigma}^{b}\,.
\end{aligned}
\end{equation}
For the FLRW universe, the Pauli sigma matrices $\varsigma^{\mu}$ 
and $\bar{\varsigma}^{\mu}$ are expressed as follows:
\begin{equation}
\begin{aligned}
\varsigma^{\mu} = a(\tau)^{-1}\sigma^{\mu},\,
\hspace*{6mm}
\bar{\varsigma}^{\mu} = a(\tau)^{-1}\bar{\sigma}^{\mu}\,.
\end{aligned}
\end{equation}
Then, the kinematic term of the free Lagrangian for a Weyl spinor in FLRW spacetime 
takes the following form:
{\small
\begin{equation}
\Delta\LG= \sqrt{-g\,}
\frac{\,\ii\,}{\,a\,} \psi_{\dalpha}^\dagger \bar{\sigma}^{\mu\dalpha\alpha} \text{D}_\mu \psi_\alpha\,.
\end{equation}
}
\hspace*{-3mm}
Using $\sqrt{-g\,} \!=\! a(\tau)^4$ and the covariant derivative formula
\eqref{covariant derivative}, we derive the above Lagrangian as follows:
{\small
\begin{equation}
\label{aeq:DeltaL-FLRW}
\Delta\LG = \ii\hp a^3
 \psi^\dagger\!\hsm 
\(\!
\bar{\sigma}^{0} \partial_0
\!+\! \bar{\sigma}^{i} \partial_i
\!+\! \frac{3}{2}\mathcal{H} \bar{\sigma}^{0}\!\hsm 
\)\hsm\!\psi\,,
\end{equation}
}
\hspace*{-2.5mm}
where we have used the identity,
\begin{equation}
\sum_i \hsm ( \bar{\sigma}^i \sigma^0 \bar{\sigma}^i)^{\dalpha\beta}
=-3\bar{\sigma}^{0\dalpha\beta}\,.
\end{equation}
With the field redefinition 
\begin{equation}
\tilde{\psi} = a^{3/2} \psi\,,
\end{equation}
we can simplify the Lagrangian to the flat-space form:
{\small
\begin{equation}
\Delta\LG = 
\ii\hp \tilde{\psi}^\dagger
(\bar{\sigma}^0 \partial_0
+ \bar{\sigma}^i \partial_i)
\tilde{\psi}=
\ii\hp \tilde{\psi}^\dagger
\bar{\sigma}^\mu \partial_\mu
\tilde{\psi}\,.
\end{equation}
}
\hspace*{-3.5mm}
Thus, using the field redefinition for the right-handed neutrino field 
$\tilde{N} \!\!=\hsm\! a^{3/2} N$, we can simplify the free Lagrangian 
\eqref{Lagrangian_weyl_free} of the right-handed neutrinos 
with a chemical potential as follows:
{\small
\begin{equation}
\label{free_lagrangian_app}
\Delta\LG = \ii\hp {\tilde{N}^\dagger_\dalpha} \bar{\sigma}^{\mu\dalpha\alpha} \partial_\mu \tilde{N}_\alpha \!-\! a\lambda {\tilde{N}^\dagger_\dalpha} \bar{\sigma}^{0 \dalpha\alpha} \tilde{N}_\alpha \!-\! \frac{1}{2} a M_{\!R} (\tilde{N}^\alpha \tilde{N}_\alpha \!+\!\tilde{N}^{\dagger\dalpha}{\tilde{N}^\dagger_\dalpha})\,.
\end{equation}
}
Then, we derive the equations of motion for $\tilde{N}^\dagger_\dalpha$, 
\begin{equation}
\ii\hp \bar{\sigma}^{\mu\dalpha\alpha} \partial_\mu \tilde{N}_\alpha = a  \lambda \bar{\sigma}^{0 \dalpha\alpha} \tilde{N}_\alpha + a M_{\!R} \tilde{N}^{\dagger\dalpha}\,.
\end{equation}
\hspace*{-2.5mm}
With this, we define the quantized right-handed neutrino fields:
{\small
\begin{equation}
\label{mode_expansion_app}
\begin{aligned}
\tilde{N}_\alpha(\tau, \mathbf{x}) &= 
\int\!\! \frac{\td^3 \mathbf{k}}{(2\pi)^3} 
\sum_{s=\pm}\! \left[ \xi_{ s,\alpha} (\tau, \mathbf{k}) b_s (\mathbf{k}) e^{\ii\hp \mathbf{k} \cdot \mathbf{x}} \!+\hsm \chi_{ s,\alpha} (\tau, \mathbf{k}) b_s^{\dagger} (\mathbf{k}) e^{-\ii\hp \mathbf{k} \cdot \mathbf{x}} \right]\!,
\\
\tilde{N}^\dagger_\dalpha(\tau, \mathbf{x}) &= 
\int\!\! \frac{\td^3 \mathbf{k}}{(2\pi)^3} 
\sum_{s=\pm}\! \left[ \xi^\dagger_{ s,\dalpha} (\tau, \mathbf{k}) b_s^{\dagger} (\mathbf{k}) e^{-\ii\hp \mathbf{k} \cdot \mathbf{x}} \!+\hsm  \chi^\dagger_{s,\dalpha} (\tau, \mathbf{k}) b_s (\mathbf{k}) e^{\ii\hp \mathbf{k} \cdot \mathbf{x}} \right]\!,
\end{aligned}
\end{equation}
}
\hspace*{-2.5mm}
where $ b_s $ and $ b_s^\dagger $ are annihilation and creation operators satisfying
\begin{equation}
    \{b_s(\mathbf{k}), b_s^\dagger(\mathbf{k'})\} = (2\pi)^3 \delta_{ss'}\delta^3(\mathbf{k} \!-\! \mathbf{k'})\,.
\end{equation}
In the above expansions \eqref{mode_expansion_app},
$\xi_{s,\alpha}(\tau,\mathbf{k})$ and $\chi_{s,\alpha}(\tau,\mathbf{k})$ are the mode functions.\  
Substituting Eq.\eqref{mode_expansion_app} into the equation of motion 
\eqref{free_lagrangian_app}, we derive the equations of motion for the mode functions:
\beqs
\begin{align}
\label{eom1}
\ii\hp ( \bar{\sigma}^{0\,\dalpha\beta} \xi'_{s,\beta}
\!+\hsm \ii\hp \bar{\sigma}^{i\,\dot{\alpha}\beta} k_i \xi_{s,\beta} )
&= a\lambda \bar{\sigma}^{0\,\dalpha\beta} \xi_{s,\beta}
\!+\! a M_{\!R} \chi^{\dagger \dot{\alpha}}_s\,,
\\
\label{eom2}
\ii\hp ( \bar{\sigma}^{0\,\dalpha\beta} \chi'_{s,\beta}
\!-\! \ii\hp \bar{\sigma}^{i\,\dot{\alpha}\beta} k_i \chi_{s,\beta} )
&= a\lambda \bar{\sigma}^{0\,\dalpha\beta} \chi_{s,\beta}
\!+\! a M_{\!R} \xi^{\dagger \dot{\alpha}}_s\,,
\end{align}
\eeqs 
where we have used the notation $f'(\tau)\!=\hsm \d f(\tau)/\d\tau\hp$
with $f(\tau)\!=\!\xi^{}_{s,\beta}(\tau)$ or
$f(\tau)\!=\!\chi^{}_{s,\beta}(\tau)$.\ 
We can re-express the above second equation and have the following coupled equations of motion 
for the two mode functions $\xi_{s,\alpha}(\mathbf{k})$ and 
$\chi_{s}^{\dagger\dot{\alpha}}(\mathbf{k})$:
\beqs 
\label{eom_modes}
\begin{align}
\ii\hp ( \bar{\sigma}^{0\,\dalpha\beta} \xi'_{s,\beta}
\!+\! \ii\hp \bar{\sigma}^{i\,\dot{\alpha}\beta} k_i \xi_{s,\beta} )
&= a\lambda \bar{\sigma}^{0\,\dalpha\beta} \xi_{s,\beta}
\!+\! a M_{\!R} \chi^{\dagger \dot{\alpha}}_s\,,
\\
- \ii\hp (
- \sigma^0_{\beta\dot{\alpha}} \chi^{\dagger \dot{\alpha}\,\prime}_s
\!-\! \ii\hp k_i \sigma^i_{\beta\dot{\alpha}} \chi^{\dagger \dot{\alpha}}_s
)
&= - a\lambda \sigma^0_{\beta\dot{\alpha}} \chi^{\dagger \dot{\alpha}}_s
\!+\! a M_{\!R} \xi_{s,\beta}\,.
\end{align}
\eeqs

We further factorize the helicity spinors and define the spinor coefficients 
$u_s^{}$ and $v_s^{}$ as follows:
\begin{equation}
\label{helicity_expansion}
\xi_{s,\alpha}(\tau, \mathbf{k}) = u_s(\tau,k) h_{s,\alpha}(\mathbf{k})\hp, 
\quad~ 
\chi^{\dagger\dalpha}_{s}(\tau, \mathbf{k}) = v_s(\tau,k)\hp s\,h^{\dagger\dalpha}_{-s}(\mathbf{k})\hp,
\end{equation}
where 
$s\,h^{\dagger\dalpha}_{-s}(\mathbf{k})
 =\pm h^{\dagger\dalpha}_{\mp}(\mathbf{k})$ 
for $s\!=\!\pm\,$.\ 
The unit helicity eigenspinors $h_s$ for momentum $\mathbf{k}$ in the direction
\begin{equation}
\hat{\mathbf{\,k}}= (\sin \theta \cos \varphi,\sin \theta \sin \varphi, \cos \theta)
\end{equation}
are chosen to take the following forms:
\beqs 
\begin{align}
\hspace*{-3.5mm}h_{+,\alpha}(\mathbf{k}) &=
h_{+,\alpha}(\theta,\varphi)=
\begin{pmatrix}
e^{-\ii\hp \frac{\varphi}{2}}\!\cos\!\frac{\theta}{2} 
\\[2mm]
e^{+\ii\hp \frac{\varphi}{2}}\!\sin\!\frac{\theta}{2}
\end{pmatrix}_{\!\!\!\alpha}\!,
\\
h_{-,\alpha}(\mathbf{k}) &=
h_{-,\alpha}(\theta,\varphi)=
\begin{pmatrix}
-e^{-\ii\hp \frac{\varphi}{2}}\!\sin\!\frac{\theta}{2}
\\[2mm]
e^{\ii\hp \frac{\varphi}{2}}\!\cos\!\frac{\theta}{2}
\end{pmatrix}_{\!\!\!\alpha}\!.
\end{align}
\eeqs 
We also have the relation
$h_{s,\alpha}^{}\!=\! s\hp h^{\dagger\dalpha}_{-s}$\,,
which leads to the following: 
\begin{equation}
h^{\dagger\dalpha}_{-}(\mathbf{k})=
\begin{pmatrix}
e^{-\ii\hp \frac{\varphi}{2}}\!\cos\!\frac{\theta}{2} 
\\[2mm]
e^{\ii\hp \frac{\varphi}{2}}\!\sin\!\frac{\theta}{2}
\end{pmatrix}^{\!\!\!\dalpha}\!,\quad 
- h^{\dagger\dalpha}_{+}(\mathbf{k})=
\begin{pmatrix}
\!-e^{-\ii\hp \frac{\varphi}{2}}\!\sin\!\frac{\theta}{2} 
\\[2mm]
\!e^{\ii\hp \frac{\varphi}{2}}\!\cos\!\frac{\theta}{2}
\end{pmatrix}^{\!\!\!\dalpha}\!.
\end{equation}
The helicity eigenspinors $h_{s,\alpha}^{}$ satisfy the following relations:
{\small
\beqs
\begin{align}
\big(\!-\!\vec{\bar{\sigma}} \cdot \hat{\mathbf{k}}\big)^{\!\dalpha\beta} h_{s,\beta}(\mathbf{k}) 
& = h_{-s}^{\dagger\dalpha}(\mathbf{k}) \hp, 
\hspace*{7mm}
\bar{\sigma}^{0\,\dalpha\beta} h_{s,\beta}(\mathbf{k})=s h_{-s}^{\dagger\dalpha}(\mathbf{k})\hp,
\\
\big(\vec{\sigma} \cdot \hat{\mathbf{k}} \big)_{\!\alpha\dbeta} h^{\dagger\dbeta}_{s}(\mathbf{k}) & = h_{-s,\alpha}(\mathbf{k})\hp,
\hspace*{7.4mm}
\sigma^0_{\alpha\dbeta} h^{\dagger\dbeta}_{s}(\mathbf{k})=-s h_{-s,\alpha}(\mathbf{k})\hp,
\end{align}
\eeqs 
}
\hspace*{-3.5mm}
which is consistent with Eq.\eqref{property_helicity}.\ 
In the above equations, we express the spinor indices explicitly, 
with $\alpha$ denoting left-handed $(\frac{\,1\,}{\,2\,},0)$ indices 
and $\dalpha$ denoting right-handed $(0,\frac{\,1\,}{\,2\,})$ indices.\ 

\vs 

Substituting Eq.\eqref{helicity_expansion} into the equations of motion \eqref{eom_modes}, 
we derive the following coupled equations for the spinor coefficients 
$u_s^{}$ and $v_s^{}\hp$:
\beqs 
\begin{align}
\ii\hp u'_{\pm} \hsm \pm\hsm k u_{\pm}
&= a\lambda u_{\pm} \hsm +\hsm a M_{\!R} v_{\pm}\,,
\\
\ii\hp v'_{\pm} \hsm\mp\hsm k v_{\pm}
&= - a\lambda v_{\pm} \hsm +\hsm a M_{\!R} u_{\pm}\,.
\end{align}
\eeqs
Then, we recast these two equations into the diagonal form:
\beqs 
\begin{align}
u''_{\pm} \!-\! a H u'_{\pm}
\!+\! \left[ (\pm k \!-\! a\lambda)^2 \!+\! a^2 M_{\!R}^2 
\hsm\pm\hsm \ii\hp a H k \right]\! u_{\pm}
&= 0\,,
\\
v''_{\pm} \!-\! a H v'_{\pm}
\!+\! \left[ (\mp k \!+\! a\lambda)^2 \!+\! a^2 M_{\!R}^2 
\hsm\mp\hsm \ii\hp a H k \right]\! v_{\pm}
&= 0\,.
\end{align}
\eeqs
The above second-order differential equations precisely give Eqs.\eqref{Eq:2.18} 
in Section\,\ref{sec:2.2}, which can be solved to give the solutions
in Eq.\eqref{modefunction}.\ 

\section{\hspace*{-1.5mm}Schwinger-Keldysh Propagators of Weyl Spinors}
\label{app:B}
\label{SK path integral}

In this Appendix, we briefly summarize the propagators of two-component Weyl spinors 
in the Schwinger-Keldysh (SK) path-integral formalism.\ We present the generating functional and the fermion propagators for Weyl spinors 
in the presence of a nonzero chemical potential.\ 

Since the path-integral variables for fermion fields are anticommuting Grassmann quantities, 
some care is required for constructing the doubled path integral in the in-in formalism.\ 
In particular, the reverse-time-ordered contour should be placed to the left-hand side 
of the time-ordered contour.\ Let $\psi_{+}^{}$ and $\psi_{-}^{}$ denote the path-integral variables on the time-ordered and reverse-time-ordered contours respectively, 
and let $I_{+}$ and $I_{-}$ be the corresponding external sources.\ 
Then, the generating functional of the in-in correlators can be written as follows: 
{\small 
\begin{equation}
\label{aeq:Z-psi+psi-}
\begin{aligned}
Z[I_{-},I_{-}^{\dagger};I_{+},I_{+}^{\dagger}] =&
\left.\int\!\!\mathcal{D}\psi_-^\dagger\mathcal{D}\psi_-
\exp\!\left[-\ii\hp S[\psi_-,\psi_-^\dagger]
\hsm -\hsm
\ii\!  \int\!\!\mathrm{d}^4x\Big(I_-^\alpha\psi_{-\alpha}\!+\!I_{-\dalpha}^\dagger\psi_-^{\dagger\dalpha}\Big)\!\right]\right.  
\\
 & \times\!\int\!\!\mathcal{D}\psi_+^\dagger\mathcal{D}\psi_+\exp\!\left[\ii\hp S[\psi_+,\psi_+^\dagger]
 \hsm +\hsm 
 \ii\!  \int\!\!\mathrm{d}^4x\Big(I_+^\alpha\psi_{+\alpha}
 \!+\!I_{+\dalpha}^\dagger\psi_+^{\dagger\dalpha}\Big)\!\right]\!. 
\end{aligned}
\end{equation}
}
\hspace{-2mm}
In the above path integral, we treat the left-handed spinors $\psi_{\pm\alpha}^{}$ and the 
right-handed spinors $\psi_{\pm}^{\dagger a}$ as independent variables.\ 
It is understood that the boundary conditions $\psi_+^{} \!=\hsm \psi_-$ 
and $\psi_+^\dagger \!= \psi_-^\dagger$ at future infinity $\tau \!=\! 0$ are imposed.

From the generating functional \eqref{aeq:Z-psi+psi-}, 
we derive three types of two-point correlators (propagators)
of the Weyl spinors $\psi$ and $\psi^\dagger$ 
by taking two functional derivatives with respect to 
the corresponding external sources:
{\small
\beqs
\begin{align}
\hat{D}_{a b \alpha \beta}^{}\!\( \tau_{1},\mathbf{x}; \tau_{2},\mathbf{y}\) 
& =\!\int\!\! \mathrm{d}^{3} \mathbf{x}\, e^{-\ii\hp \mathbf{k} \cdot \mathbf{x}} \frac{\delta}{\,\ii\hp a\delta I_{a}^{\alpha}(\tau_{1}, \mathbf{x})\,} 
\frac{\delta}{\,\ii\hp b\hp\delta I_{b}^{\beta}(\tau_{2}, \mathbf{y})\,} 
Z\big[I_{-}, I_{-}^{\dagger} ; I_{+}, I_{+}^{\dagger}\big]\hsm\bigg|_{I=0}, 
\\
\check{D}_{a b \dalpha \dbeta}^{}( \tau_{1},\mathbf{x}; \tau_{2},\mathbf{y}) 
& =\!\int\!\!\mathrm{d}^{3}\,\mathbf{x}\, e^{-\ii\hp \mathbf{k} 
\cdot \mathbf{x}} \frac{\delta}{\,\ii\hp   a \delta I_{a }^{\dagger\dalpha}
(\tau_{1}, \mathbf{x})\,} \frac{\delta}{\,\ii\hp b\hp\delta I_{b }^{\dagger\dbeta}(\tau_{2}, \mathbf{y})\,} Z\big[I_{-}, I_{-}^{\dagger} ; I_{+}, I_{+}^{\dagger}\big]\bigg|_{I=0}, 
\\
D_{a b \alpha\dbeta}\( \tau_{1},\mathbf{x}; \tau_{2},\mathbf{y}\) 
& =\!\int\!\!\mathrm{d}^{3} \mathbf{x}\,e^{-\ii\hp\mathbf{k} \cdot \mathbf{x}} 
\frac{\delta}{\,\ii\hp   a \delta I_{a}^{\alpha}\(\tau_{1}, \mathbf{x}\)\,} \frac{\delta}{\,\ii\hp   b\hp \delta I_{b }^{\dagger\dbeta}\(\tau_{2}, \mathbf{y}\)\,} Z\big[I_{-}, I_{-}^{\dagger} ; I_{+}, I_{+}^{\dagger}\big]\bigg|_{I=0} .
\end{align}
\eeqs
}
\hspace*{-2.5mm}
These objects correspond to the two-point vacuum expectation values in the in-in formalism.\ 
Then the propagators of the $D_{a b \alpha\dbeta}$ type (type-1) are given by 
{\small
\beqs
\begin{align}
D_{\ominus\oplus \alpha \dbeta}\(\mathbf{k} ; \tau_{1}, \tau_{2}\) 
& =\int\!\! \mathrm{d}^{3} \mathbf{x}\, e^{-\ii\hp \mathbf{k} \cdot \mathbf{x}}  \langle 0 | \psi_\alpha(\tau_1,\mathbf{x})\psi^\dagger_{\dbeta} (\tau_2,\mathbf{0})  | 0 \rangle\,,
\\
D_{\oplus\ominus \alpha \dbeta}\(\mathbf{k} ; \tau_{1}, \tau_{2}\) 
& =\!-\!\!\int\!\!\mathrm{d}^{3} \mathbf{x}\,e^{-\ii\hp \mathbf{k} \cdot \mathbf{x}}  
\langle 0 | \psi^\dagger_{\dbeta} (\tau_2,\mathbf{0})
\psi_\alpha(\tau_1,\mathbf{x})   | 0 \rangle\,,
\\
D_{\oplus\oplus \alpha \dbeta}\(\mathbf{k} ; \tau_{1}, \tau_{2}\) 
& =\int\!\!\mathrm{d}^{3} \mathbf{x}\,e^{-\ii\hp \mathbf{k} \cdot \mathbf{x}}  
\langle 0 | \mathbf{T}\Big\{\psi_\alpha(\tau_1,\mathbf{x})
\psi^\dagger_{\dbeta} (\tau_2,\mathbf{0})\!\Big\}  | 0 \rangle\,,
\\
D_{\ominus\ominus \alpha \dbeta}\(\mathbf{k} ; \tau_{1}, \tau_{2}\) 
& =\int\!\! \mathrm{d}^{3} \mathbf{x} \,e^{-\ii\hp \mathbf{k} \cdot \mathbf{x}}  
\langle 0 | \bar{\mathbf{T}}\Big\{\!\psi_\alpha(\tau_1,\mathbf{x})\psi^\dagger_{\dbeta} (\tau_2,\mathbf{0})\!\Big\}  | 0 \rangle\,,
\end{align}
\eeqs
}
\hspace{-2.5mm}
where $\hp\mathbf{T}\hp$ denotes the time-ordered product and $\bar{\mathbf{T}}$ 
the reverse-time-ordered product.\ The extra minus sign in $D_{\oplus\ominus \alpha \dbeta}$ 
arises from exchanging the fermionic fields when the reverse-time-ordered field is placed 
to the left of the time-ordered field. 

The propagators of the $\hat{D}_{a b \alpha\beta}$ type (type-2) are given by 
{\small
\beqs
\begin{align}
\hat{D}_{\ominus\oplus \alpha \beta}\(\mathbf{k} ; \tau_{1}, \tau_{2}\) 
& =\int\!\! \mathrm{d}^{3} \mathbf{x}\, e^{-\ii\hp \mathbf{k} \cdot \mathbf{x}}  \langle 0 | \psi_\alpha(\tau_1,\mathbf{x})\psi_{\beta} (\tau_2,\mathbf{0})  | 0 \rangle\,,
\\
\hat{D}_{\oplus\ominus \alpha \beta}\(\mathbf{k} ; \tau_{1}, \tau_{2}\) 
& =\!-\!\!\int\!\!\mathrm{d}^{3} \mathbf{x}\,e^{-\ii\hp \mathbf{k} \cdot \mathbf{x}}  
\langle 0 | \psi_{\beta} (\tau_2,\mathbf{0})
\psi_\alpha(\tau_1,\mathbf{x})   | 0 \rangle\,,
\\
\hat{D}_{\oplus\oplus \alpha \beta}\(\mathbf{k} ; \tau_{1}, \tau_{2}\) 
& =\int\!\!\mathrm{d}^{3} \mathbf{x}\,e^{-\ii\hp \mathbf{k} \cdot \mathbf{x}}  
\langle 0 | \mathbf{T}\Big\{\psi_\alpha(\tau_1,\mathbf{x})
\psi_{\beta} (\tau_2,\mathbf{0})\!\Big\}  | 0 \rangle\,,
\\
\hat{D}_{\ominus\ominus \alpha \beta}\(\mathbf{k} ; \tau_{1}, \tau_{2}\) 
& =\int\!\! \mathrm{d}^{3} \mathbf{x} \,e^{-\ii\hp \mathbf{k} \cdot \mathbf{x}}  
\langle 0 | \bar{\mathbf{T}}\Big\{\!\psi_\alpha(\tau_1,\mathbf{x})\psi_{\beta} (\tau_2,\mathbf{0})\!\Big\}  | 0 \rangle\,,
\end{align}
\eeqs
}

The propagators of the $\check{D}_{a b \dalpha\dbeta}$ type (type-3) are given by 
{\small
\beqs
\begin{align}
\check{D}_{\ominus\oplus \dalpha \dbeta}\(\mathbf{k} ; \tau_{1}, \tau_{2}\) 
& =\int\!\! \mathrm{d}^{3} \mathbf{x}\, e^{-\ii\hp \mathbf{k} \cdot \mathbf{x}}  \langle 0 | \psi^\dagger_\dalpha(\tau_1,\mathbf{x})\psi^\dagger_{\dbeta} (\tau_2,\mathbf{0})  | 0 \rangle\,,
\\
\check{D}_{\oplus\ominus \dalpha \dbeta}\(\mathbf{k} ; \tau_{1}, \tau_{2}\) 
& =\!-\!\!\int\!\!\mathrm{d}^{3} \mathbf{x}\,e^{-\ii\hp \mathbf{k} \cdot \mathbf{x}}  
\langle 0 | \psi^\dagger_{\dbeta} (\tau_2,\mathbf{0})
\psi^\dagger_\dalpha(\tau_1,\mathbf{x})   | 0 \rangle\,,
\\
\check{D}_{\oplus\oplus \dalpha \dbeta}\(\mathbf{k} ; \tau_{1}, \tau_{2}\) 
& =\int\!\!\mathrm{d}^{3} \mathbf{x}\,e^{-\ii\hp \mathbf{k} \cdot \mathbf{x}}  
\langle 0 | \mathbf{T}\Big\{\psi^\dagger_\dalpha(\tau_1,\mathbf{x})
\psi^\dagger_{\dbeta} (\tau_2,\mathbf{0})\!\Big\}  | 0 \rangle\,,
\\
\check{D}_{\ominus\ominus \dalpha \dbeta}\(\mathbf{k} ; \tau_{1}, \tau_{2}\) 
& =\int\!\! \mathrm{d}^{3} \mathbf{x} \,e^{-\ii\hp \mathbf{k} \cdot \mathbf{x}}  
\langle 0 | \bar{\mathbf{T}}\Big\{\!\psi^\dagger_\dalpha(\tau_1,\mathbf{x})\psi^\dagger_{\dbeta} (\tau_2,\mathbf{0})\!\Big\}  | 0 \rangle\,,
\end{align}
\eeqs
}

\vs

Substituting the mode expansion of $\psi_\alpha$ and $\psi^\dagger_{\dbeta}$ of  Eq.\eqref{mode_expansion} into the above definitions, 
we find that only the term containing $b_s(\mathbf{k})b_{s}^\dagger(\mathbf{k})$ survives because
\begin{equation}
\{b_s(\mathbf{k}), b_{s'}^\dagger(\mathbf{k'})\} = 
(2\pi)^3 \delta_{ss'} \delta^3(\mathbf{k} \!-\! \mathbf{k'})\,, 
\end{equation}
with which the fermion propagators can be derived as in 
Section\,\ref{Propagators of fermion}\,.

\section{\hspace*{-1.5mm}Partial Mellin--Barnes Representation of the Spinor Propagator}
\label{PMB representation}

In this subsection, we introduce the partial Mellin--Barnes (PMB) representation, which is useful for deriving the late-time limit of the propagators. In this representation, the late-time behavior can be identified with the contributions from the leading poles. More details on the use of PMB representations in the calculation of cosmological correlators can be found in Refs.~\cite{Xianyu220501692,Qin:2022fbv,Xianyu230107047,Xianyu230413295}~.

We first list the PMB representations for the Hankel and Whittaker functions:
\begin{align}
\mathrm{H}_{\nu}^{(j)}(a z)&=\int_{-\mathrm{i} \infty}^{\mathrm{i} \infty} \frac{\mathrm{~d} s}{2 \pi \mathrm{i}} \frac{(a z / 2)^{-s}}{2 \pi} e^{(-1)^{j+1}(s-\nu-1) \pi \mathrm{i} / 2} \Gamma\left[\frac{s-\nu}{2}, \frac{s+\nu}{2}\right], \quad(j=1,2)~,\\
\label{PMB_whittaker}
\mathrm{W}_{\mu,\nu}(az)&=e^{-az/2}\int_{-\ii  \infty}^{\ii  \infty}\frac{\mathrm{d}s}{2\pi\ii  }\left(az\right)^{-s}\Gamma
\begin{bmatrix}
\frac{1}{2}+\nu+s,\frac{1}{2}-\nu+s,-\mu-s \\
\frac{1}{2}+\nu-\mu,\frac{1}{2}-\nu-\mu
\end{bmatrix}~.
\end{align}

For the Whittaker function, it is convenient to use another equivalent representation obtained by shifting the integration variable $s\rightarrow s-\frac{1}{2}$ and deforming the integration contour:
\begin{equation}
\mathrm{W}_{\mu,\nu}(az)=e^{-az/2}\int_{-\ii  \infty}^{\ii  \infty}\frac{\mathrm{d}s}{2\pi\ii  }\left(az\right)^{-s+\frac{1}{2}}\Gamma
\begin{bmatrix}
\nu+s,-\nu+s,\frac{1}{2}-\mu-s \\
\frac{1}{2}+\nu-\mu,\frac{1}{2}-\nu-\mu
\end{bmatrix}.
\end{equation}

Before deriving the PMB representation of the right-handed neutrino propagators, 
we rewrite the above formula with $z\equiv-k\tau$:
\begin{equation}
\begin{cases}
\mathrm{W}_{-\frac{1}{2}-\ii  \tlambda,\ii  \widetilde{\nu}}(-2\ii z)=e^{-\ii  \pi/4+\ii z}\int_{-\ii  \infty}^{\ii  \infty}\frac{\mathrm{d}s}{2\pi\ii  }e^{\ii  \pi s/2}(2z)^{1/2-s}\Gamma
\begin{bmatrix}
s+\ii  \widetilde{\nu},s-\ii  \widetilde{\nu},-s+1+\ii  \tlambda \\
1+\ii  \widetilde{\nu}+\ii  \tlambda,1-\ii  \widetilde{\nu}+\ii  \tlambda
\end{bmatrix}, \\
\\
\mathrm{W}_{\frac{1}{2}-\ii  \tlambda,\ii  \widetilde{\nu}}(-2\ii z)=e^{-\ii  \pi/4+\ii z}\int_{-\ii  \infty}^{\ii  \infty}\frac{\mathrm{d}s}{2\pi\ii  }e^{\ii  \pi s/2}(2z)^{1/2-s}\Gamma
\begin{bmatrix}
s+\ii  \widetilde{\nu},s-\ii  \widetilde{\nu},-s+\ii  \tlambda \\
\ii  \widetilde{\nu}+\ii  \tlambda,-\ii  \widetilde{\nu}+\ii  \tlambda
\end{bmatrix}~.
\end{cases}
\end{equation}

For the neutrino propagator $D_{\ominus\oplus\alpha\dbeta} (k,\tau_1,\tau_2)$,
\begin{equation}
\begin{aligned}
        D_{\ominus\oplus\alpha\dbeta} (k,\tau_1,\tau_2) &=\sum_{s}u_s (\tau_1,\mathbf{k}) u_{s}^\dagger (\tau_2,\mathbf{k}) h_{s\alpha}(\mathbf{k})h_{s\dbeta}^\dagger(\mathbf{k})~,
\end{aligned}
\end{equation}
we present the PMB representation of the term $u_+(\tau_1,\mathbf{k})u_+^\dagger(\tau_2,\mathbf{k})$ as an example:
\begin{equation}
    \begin{aligned}
    u_+(\tau_1,\mathbf{k})u_+^\dagger(\tau_2,\mathbf{k})&=\\
    &=\tilde{M}_{\!R}^2 \frac{e^{\pi\tilde{\lambda}}}{2k}(-\tau_1)^{-1/2}(-\tau_2)^{-1/2}W_{-\frac{1}{2}-\ii  \tlambda,\ii  \tnu}(2\ii   k\tau_1)W_{-\frac{1}{2}+\ii  \tlambda,-\ii  \tnu}(-2\ii   k\tau_2)\\
    &=\tilde{M}_{\!R}^2e^{\pi\tilde{\lambda}}e^{-\ii   k\tau_1+\ii   k\tau_2}\int_{-\ii  \infty}^{\ii  \infty}\frac{\rm{d}s_1}{2\pi\ii  }\frac{\rm{d}s_2}{2\pi\ii  }(2k)^{-s_{12}}e^{\ii  \pi(s_1-s_2)/2}(-\tau_1)^{-s_1}(-\tau_2)^{-s_2}\\
    &\,\,\quad\Gamma\begin{bmatrix}
    s_1-\ii  \tnu,s_1+\ii  \tnu,-s_1+1+\ii  \tlambda,s_2-\ii  \tnu,s_2+\ii  \tnu,-s_2+1-\ii  \tlambda \\
    1+\ii  \tlambda-\ii  \tnu,1+\ii  \tlambda+\ii  \tnu,1-\ii  \tlambda-\ii  \tnu,1-\ii  \tlambda+\ii  \tnu
    \end{bmatrix}
    \end{aligned}
\end{equation}

Similarly, for the propagator $\hat{D}_{\ominus\oplus\alpha\beta} (k,\tau_1,\tau_2)$,
\begin{equation}
\begin{aligned}
\hat{D}_{\ominus\oplus\,\alpha \beta} (\mathbf{k}; \tau_1, \tau_2)
=\sum_{s}s\,u_{s} (\tau_1,\mathbf{k}) v_{s}^\dagger (\tau_2,\mathbf{k}) h_{s,\alpha}(\mathbf{k}) h_{\beta,-s}(\mathbf{k})
\end{aligned}~,
\end{equation}
we present the PMB representation of the term $u_+(\tau_1,\mathbf{k})v_+^\dagger(\tau_2,\mathbf{k})$ as an example:
\begin{equation}
    \begin{aligned}
        u_+(\tau_1,\mathbf{k})v_+^\dagger(\tau_2,\mathbf{k})&=\\
        &=\tilde{M}_{\!R} \frac{e^{\pi\tilde{\lambda}}}{2k}(-\tau_1)^{-1/2}(-\tau_2)^{-1/2}W_{-\frac{1}{2}-\ii  \tlambda,\ii  \tnu}(2\ii   k\tau_1)W_{\frac{1}{2}+\ii  \tlambda,-\ii  \tnu}(-2\ii   k\tau_2)\\
        &=\tilde{M}_{\!R}e^{\pi\tilde{\lambda}}e^{-\ii   k\tau_1+\ii   k\tau_2}\int_{-\ii  \infty}^{\ii  \infty}\frac{\rm{d}s_1}{2\pi\ii  }\frac{\rm{d}s_2}{2\pi\ii  }(2k)^{-s_{12}}e^{\ii  \pi(s_1-s_2)/2}(-\tau_1)^{-s_1}(-\tau_2)^{-s_2}\\
    &\,\,\quad\Gamma\begin{bmatrix}
    s_1-\ii  \tnu,s_1+\ii  \tnu,-s_1+1+\ii  \tlambda,s_2-\ii  \tnu,s_2+\ii  \tnu,-s_2-\ii  \tlambda \\
    1+\ii  \tlambda+\ii  \tnu,1+\ii  \tlambda-\ii  \tnu,-\ii  \tlambda-\ii  \tnu,-\ii  \tlambda+\ii  \tnu
    \end{bmatrix}
    \end{aligned}
\end{equation}
The poles in the PMB representation can be divided into two sets:
\begin{equation}
    s_1=-n_1\pm\ii\tilde{\nu},\,\,s_2=-n_2\pm\ii\tilde{\nu}~,
\end{equation}
with $n_1,n_2=0,1,2,\dots$. The late-time limit, which is equivalent to the soft limit $|k\tau|\ll1$, can be obtained by taking the leading poles with $n_1=n_2=0$.

\vs 

In the late-time limit, the nonlocal part is obtained by choosing the leading poles,
\begin{equation}
s_1^{} = s_2^{} =\pm\ii\hp\tilde{\nu} \,,
\end{equation}
whereas the local part is obtained by choosing
\begin{equation}
s_1^{} = -s_2^{} =\pm\ii\hp\tilde{\nu} \,.
\end{equation}

\section{\hspace*{-1.5mm}Derivations of the Relevant Formulas}
\label{details of calculation}
\label{app:C}

In this Appendix, we present several derivations for computing the loop contribution of right-handed neutrinos  
to the three-point inflaton correlation function.\ Our purpose is to make 
the calculation strategy more transparent and to separate the analytic derivations from the physical analysis.

\vs 

We proceed in the following order.\ First, we summarize several general relations that are 
repeatedly used in the loop calculation, including the conjugation properties of the external 
lines and loop diagrams.\ Then, we present sample derivations of the seed integrals introduced in Section\,\ref{Seed Integrals for Fermion Propagators}.\ These seed integrals form the basic 
building blocks of the factorized loop calculation.\ We further explain how 
to use these seed integrals to form the three-point inflaton correlators and clarify 
the origin of the dominant contribution in the squeezed limit.\  
After that, we discuss the cutting rule, 
which provides a useful structural interpretation of the nonlocal part of the signal.\ 
Finally, we derive the bubble loop integral 
by using the Feynman diagram approach.

\subsection{\hspace*{-1.5mm}Conjugation Relations between Three-Point Correlators}
\label{app:C1}

In this subsection, 
we analyze the relations between the three-point loop correlators.\ 
Specifically, we can classify the eight diagrams of Fig.\,\ref{fig:2}
into four pairs, where the two diagrams within each pair are 
related by complex conjugation.
A particularly useful relation is the conjugation property of the external function
\eqref{def_external_{(d)}erivative}, 
\begin{equation}
\mathcal{F}_{\mu a}(\mathbf{k},\tau)^*
=
\mathcal{F}_{\mu,\text{-}a}(-\mathbf{k},\tau)\,,
\end{equation}
which follows directly from the definition of the SK bulk-to-boundary propagator \eqref{boundarypropagator}.\ 
In consequence, the pairs of correlators related by reversing the fermion flow in the loop  
are complex conjugates of each other.\ This relation can also be checked numerically.

\vs

As an illustration, consider diagram-(7) of Fig.\,\ref{fig:2}, whose contribution takes 
the following form: 
{\small
\begin{align}
\label{aeq:C2}
\langle\delta\phi(\mathbf{k}_1)\delta\phi(\mathbf{k}_2)\delta\phi(\mathbf{k}_3)\rangle^{\prime(7)}
 =& \sum_{a,b,c=\pm}\!\! a\hp b\hp c \(\!\frac{\,-\ii\,}{\,\Lambda\,}\!\)^{\!\!3}
 \!\!\int_{-\infty}^{0^-}{\d}\tau_1{\d}\tau_2{\d}\tau_3
\\
 & 
 \hspace*{4mm}
 \mathcal{F}_{\mu a}^{}(\mathbf{k}_1,\tau_1)
 \mathcal{F}_{\nu b}(\mathbf{k}_2,\tau_2)
 \mathcal{F}_{\lambda c}(\mathbf{k}_3,\tau_3)
 \!\!\int\!\!\frac{\,\mathrm{d}^3\mathbf{q}\,}{(2\pi)^3}
 \mathcal{Y}_{abc}^{\mu\nu\lambda}(7)\,,
\nn 
\end{align}
}
\hspace{-2.5mm}
where the function $\mathcal{Y}_{abc}^{\mu\nu\lambda}(7)$ is given by
{\small
\begin{equation}
\mathcal{Y}_{abc}^{\mu\nu\lambda}(7)
=
\bar{\sigma}^{\mu\dalpha\alpha}\hat{D}_{ab\alpha}^{\,\,\,\,\,\,\,\,\,\beta}(\mathbf{p}_{12},\tau_{1},\tau_{2})
\sigma^{\nu}_{\,\,\,\beta\dbeta}
\check{D}_{bc\,\,\,\dgamma}^{\,\,\,\,\,\dbeta}(\mathbf{p}_{23},\tau_{2},\tau_{3})
\bar{\sigma}^{\lambda\dgamma\gamma}
D_{ca\gamma\dalpha}(\mathbf{p}_{31},\tau_{3},\tau_{1})\,,
\end{equation}
}
\hspace{-2.5mm}
with $\mathbf{p}_{12}^{}\!=\! \mathbf{q}\!+\!\mathbf{k}_1$, 
$\mathbf{p}_{23}^{}  \!=\! \mathbf{q} \!+\! \mathbf{k}_s$, 
and $\mathbf{p}_{31}^{} \hsm\!=\hsm \mathbf{q}\hp$.

\vs 

Taking the complex conjugate of Eq.\eqref{aeq:C2}, we derive the following: 
{\small
\begin{equation}
\begin{aligned}
\left[\langle\delta\phi(\mathbf{k}_1)\delta\phi(\mathbf{k}_2)\delta\phi(\mathbf{k}_3)\rangle^{\prime(7)}\right]^*
 = & \sum_{a,b,c=\pm}\! a\hp b\hp c \(\!\frac{\,-\ii\,}{\,\Lambda\,}\!\)
^{\!\!3}\!\! \int_{-\infty}^{0^-}\mathrm{d}\tau_1 \mathrm{d}\tau_2 \mathrm{d}\tau_3
 \\
 & 
 \hspace*{4mm}
 \mathcal{F}_{\mu a}(\mathbf{k}_1,\tau_1)^*
 \mathcal{F}_{\nu b}(\mathbf{k}_2,\tau_2)^*
 \mathcal{F}_{\lambda c}(\mathbf{k}_3,\tau_3)^*
 \int\frac{\,\mathrm{d}^3\mathbf{q}}{(2\pi)^3\,}
 \mathcal{Y}_{abc}^{\mu\nu\lambda}(7)^*\,.
\end{aligned}
\end{equation}
}
\hspace{-2.5mm}
Using
{\small
$\mathcal{F}_{\mu a}(\mathbf{k},\tau)^*=\mathcal{F}_{\mu, \text{-}a}(-\mathbf{k},\tau)$}\,, 
we deduce the following:
{\small
\begin{equation}
\label{aeq:3dphi-(7)}
\begin{aligned}
\left[\langle\delta\phi(\mathbf{k}_1)\delta\phi(\mathbf{k}_2)\delta\phi(\mathbf{k}_3)\rangle^{\prime(7)}\right]^*
  =& \sum_{a,b,c=\pm}\! a\hp b\hp c \(\!\frac{\,-\ii\,}{\,\Lambda\,}\!\)
^{\!\!3}\!\! \int_{-\infty}^{0^-}\!\mathrm{d}\tau_1 \mathrm{d}\tau_2 \mathrm{d}\tau_3 
 \\
 &
 \mathcal{F}_{\mu a}(-\mathbf{k}_1,\tau_1)
 \mathcal{F}_{\nu b}(-\mathbf{k}_2,\tau_2)
 \mathcal{F}_{\lambda c}(-\mathbf{k}_3,\tau_3)
 \!\int\!\!\frac{\,\mathrm{d}^3\mathbf{q}}{(2\pi)^3\,}
 \mathcal{Y}_{\text{-}a,\text{-}b,\text{-}c}^{\mu\nu\lambda}(7)^*\,.
\end{aligned}
\end{equation}
}
\hspace{-2.5mm}

To evaluate $\mathcal{Y}_{\text{-}a,\text{-}b,\text{-}c}^{\mu\nu\lambda}(7)^{\!*}$, 
it is convenient to suppress spinor indices and use the following trace notation:
{\small
\begin{equation}
\label{aeq:Y(7)*}
\begin{aligned}
\mathcal{Y}_{\text{-}a,\text{-}b,\text{-}c}^{\mu\nu\lambda}(7)^*
&=
\left[
\text{Tr}\!\(\hsm 
\bar{\sigma}^{\mu}\hsm\hat{D}_{\text{-}a,\text{-}b}(\mathbf{p}_{12},\tau_{1},\tau_{2})
\sigma^{\nu}\check{D}_{\text{-}b,\text{-}c}(\mathbf{p}_{23},\tau_{2},\tau_{3})
\bar{\sigma}^{\lambda}D_{\text{-}c,\text{-}a}(\mathbf{p}_{31},\tau_{3},\tau_{1})
\hsm\)\hsmx 
\right]^* 
\\
&=
\rm{Tr}\left\{\!
\sigma^{\nu}\!
\left[\hat{D}_{\text{-}a,\text{-}b}(\mathbf{p}_{12},\tau_{1},\tau_{2})\right]^\dagger
\!\!\bar{\sigma}^{\mu}\!
\left[D_{\text{-}c,\text{-}a}(\mathbf{p}_{31},\tau_{3},\tau_{1})\right]^\dagger
\!\bar{\sigma}^{\lambda}\!
\left[\check{D}_{\text{-}b,\text{-}c}(\mathbf{p}_{23},\tau_{2},\tau_{3})\right]^{\!\dagger}\!
\right\}\!.
\end{aligned}
\end{equation}
}
\hspace{-2.5mm}
The Hermitian conjugates of the propagators satisfy the relations: 
{\small
\begin{align}
\label{conjugation_propagator}
\hspace*{-7mm}
\left[\hat{D}_{\text{-}a,\text{-}b\,\alpha\beta}
(\mathbf{p},\tau_{1},\tau_{2})\right]^\dagger
&\hsm\!=\!
\check{D}_{ba\dbeta\dalpha}
(\mathbf{p},\tau_{2},\tau_{1})\,,
&
\left[\hat{D}_{\text{-}a,\text{-}b\,\alpha}^{\quad\quad\beta}
(\mathbf{p},\tau_{1},\tau_{2})\right]^\dagger
&\hsm\!=\!
-\check{D}_{ba\quad\!\!\!\!\dalpha}^{\quad\!\!\dbeta}
(\mathbf{p},\tau_{2},\tau_{1})\,,\nn
\\
\hspace*{-7mm}
\left[\check{D}_{\text{-}a,\text{-}b\,\dalpha\dbeta}
(\mathbf{p},\tau_{1},\tau_{2})\right]^\dagger
&\hsm\!=\!
\hat{D}_{ba\beta\alpha}
(\mathbf{p},\tau_{2},\tau_{1})\,,
&
\left[\check{D}_{\text{-}a,\text{-}b\quad\!\!\!\!\!\dbeta}^{\quad\,\,\,\dalpha}
(\mathbf{p},\tau_{1},\tau_{2})\right]^\dagger
&\hsm\!=\!
-\hat{D}_{ba\,\beta}^{\quad\,\,\alpha}
(\mathbf{p},\tau_{2},\tau_{1})\,,
\\
\hspace*{-7mm}
\left[D_{\text{-}a,\text{-}b\,\alpha\dbeta}
(\mathbf{p},\tau_{1},\tau_{2})\right]^\dagger
&\hsm\!=\!
D_{ba\beta\dalpha}
(\mathbf{p},\tau_{2},\tau_{1})\,.\nn
\end{align}
}
Thus, using the above relations, we further derive Eq.\eqref{aeq:Y(7)*} as follows:
\begin{equation}
\label{aeq:Y(7)*-2}
\begin{aligned}
\mathcal{Y}_{\text{-}a,\text{-}b,\text{-}c}^{\mu\nu\lambda}(7)^*
&=
\text{Tr}\left\{
\sigma^{\nu}\check{D}_{ba}(\mathbf{p}_{12},\tau_{2},\tau_{1})
\bar{\sigma}^{\mu}D_{ac}(\mathbf{p}_{31},\tau_{1},\tau_{3})
\bar{\sigma}^{\lambda}\hat{D}_{cb}(\mathbf{p}_{23},\tau_{3},\tau_{2})
\right\}
\\
&=
\sigma^{\nu}_{\,\,\,\alpha\dalpha}
\check{D}_{ba\,\,\,\dbeta}^{\,\,\,\,\,\dalpha}(\mathbf{p}_{12},\tau_{2},\tau_{1})
\bar{\sigma}^{\mu\dbeta\beta}
D_{ac\beta\dgamma}(\mathbf{p}_{31},\tau_{1},\tau_{3})
\bar{\sigma}^{\lambda\dgamma\gamma}
\hat{D}_{cb\gamma}^{\,\,\,\,\,\,\,\,\alpha}(\mathbf{p}_{23},\tau_{3},\tau_{2})\,.
\end{aligned}
\end{equation}

Then, we reverse the direction of each external momentum as $\mathbf{k}_j\rightarrow-\mathbf{k}_j$, 
and further reverse the direction of the loop momentum $\mathbf{q}\rightarrow-\mathbf{q}$.\ 
With these and using Eq.\eqref{aeq:Y(7)*-2}, 
we further express the three-point correlator \eqref{aeq:3dphi-(7)} as follows:
{\small
\begin{equation}
\begin{aligned}
&\left[\langle\delta\phi(-\mathbf{k}_1)\delta\phi(-\mathbf{k}_2)
\delta\phi(-\mathbf{k}_3)\rangle^{\prime(7)}\right]^{\hsm *}
\\
&= \sum_{a,b,c=\pm}\! a\hp b\hp c \(\!\frac{\,-\ii\,}{\,\Lambda\,}\!\)
^{\!\!3} 
\!\!\int_{-\infty}^{0^-}\mathrm{d}\tau_1 \mathrm{d}\tau_2 \mathrm{d}\tau_3 \,
\mathcal{F}_{\mu a}(\mathbf{k}_1,\tau_1)
\mathcal{F}_{\nu b}(\mathbf{k}_2,\tau_2)
\mathcal{F}_{\lambda c}(\mathbf{k}_3,\tau_3)
\\
&\quad\times\!\!
\int\!\!\frac{\,\mathrm{d}^3\mathbf{q}\,}{\,(2\pi)^3\,}
\sigma^{\nu}_{\,\,\,\alpha\dalpha}
\check{D}_{ba\,\,\,\dbeta}^{\,\,\,\,\,\dalpha}(\mathbf{p}_{21},\tau_{2},\tau_{1})
\bar{\sigma}^{\mu\dbeta\beta}
D_{ac\beta\dgamma}(\mathbf{p}_{13},\tau_{1},\tau_{3})
\bar{\sigma}^{\lambda\dgamma\gamma}
\hat{D}_{cb\gamma}^{\,\,\,\,\,\,\,\,\alpha}(\mathbf{p}_{32},\tau_{3},\tau_{2})\,,
\end{aligned}
\end{equation}
}
\hspace{-2.5mm}
which just describes the contribution from the diagram-(5) of Fig.\,\ref{fig:2}.\ 
Hence, we deduce the following relation:
\begin{equation}
\label{aeq:3dphi(7)*=(5)}
\left[\hsm\langle\delta\phi(-\mathbf{k}_1)\delta\phi(-\mathbf{k}_2)
\delta\phi(-\mathbf{k}_3)\rangle^{\prime(7)}\hsm\right]^{\hsm *}
=
\langle\delta\phi(\mathbf{k}_1)\delta\phi(\mathbf{k}_2)\delta\phi(\mathbf{k}_3)\rangle^{\prime(5)}\,.
\end{equation}

Note that the cosmological collider signal takes the schematic form in the squeezed limit,
{\small
\begin{equation}
\lim_{k_s\!\to 0}
\langle\delta\phi(\mathbf{k}_1)\delta\phi(\mathbf{k}_2)\delta\phi(\mathbf{k}_3)\rangle^{\prime}_\text{(NLoc)}
\thicksim
\frac{A}{\,k_1^3k_3^3\,}\!
\(\!\frac{\,k_1\,}{\,k_3\,}\!\)^{\!\!n+\ii\hp 2\tnu}
\!f(\hat{\,\mathbf{k}}_1\!\cdot\!\hat{\,\mathbf{k}}_3)
+\!\!\hc\!,
\end{equation}
}
\hspace*{-2.5mm}
where $n$ is a certain real number and the function $f(\hat{\,\mathbf{k}}_1\!\cdot\!\hat{\,\mathbf{k}}_3)$ 
depends only on the angle between the two external momenta.\ Thus, the signal should
be invariant under the transformation $\mathbf{k}_j\ito -\mathbf{k}_j$, 
with which we derive the following:
{\small
\begin{align}
\left[\lim_{k_s\!\to0}
\langle\delta\phi(\mathbf{k}_1)\delta\phi(\mathbf{k}_2)\delta\phi(\mathbf{k}_3)\rangle^{\prime(7)}_\text{(NLoc)}\right]^*
&=
\left[\lim_{k_s\!\to0}
\langle\delta\phi(-\mathbf{k}_1)\delta\phi(-\mathbf{k}_2)\delta\phi(-\mathbf{k}_3)\rangle^{\prime(7)}_\text{(NLoc)}\right]^*
\nn\\
&=
\lim_{k_s\!\to0}
\langle\delta\phi(\mathbf{k}_1)\delta\phi(\mathbf{k}_2)\delta\phi(\mathbf{k}_3)\rangle^{\prime(5)}_\text{(NLoc)}\,,
\end{align}
}
\hspace*{-2.5mm}
where in the last step we have used Eq.\eqref{aeq:3dphi(7)*=(5)}.\ 
Moreover, we can prove that the other pairs of diagrams in Fig.\,\ref{fig:2}, 
including each pair of diagrams-(1)(2), (3)(4), and (6)(8), also obey the same type of relations 
as derived in \eqref{eq:conjugationrelation}.

\subsection{\hspace*{-1.5mm}Derivation of Seed Integrals for Fermion Propagators}
\label{representative derivations of the seed integrals}

In Section\,\ref{Seed Integrals for Fermion Propagators}, we introduced the seed integrals 
associated with the three types of fermion propagators and listed the entries of 
the corresponding seed integral matrices, such as $A_i^{p_1,p_2}$\!, $S_i^{p_1,p_2}$\!, 
$U_i^{p_1,p_2}$\!, and $W_i^{p_1,p_2}$ (with $i\!=\!1,2$).\ 
In this subsection, we derive these entries explicitly.\ 
As representative examples, we present the derivation of the time-ordered integral 
$S_1^{p_1,p_2}$ and the non-time-ordered integral $W_1^{p_1,p_2}$.\ 
The remaining entries can be obtained analogously.

\vspace*{4mm}
\noindent 
{\bf $\blacklozenge$\,Time-Ordered Seed Integral:}
\\[3mm]
We first consider the following time-ordered integral $S_1^{p_1,p_2}$,
{\small
\begin{align}
S_1^{p_1p_2}
=& \int_{0}^{\infty}\!\!\!\d z_1\!\int_{0}^{\infty}\!\!\!\d z_2\,
e^{-\ii\hp z_1}e^{-\ii\hp z_2}z_1^{p_1}z_2^{p_2}\,
u_+(z_1)v_+^{\dagger}(z_2)\theta(z_2\!-\!z_1)
\nn\\
=& \int_{0}^{\infty}\!\!\!\d z_1\!\int_{z_1}^{\infty}\!\!\!\d z_2\,
e^{-\ii\hp z_1}e^{-\ii\hp z_2}z_1^{p_1}z_2^{p_2}
u_+(z_1)v_+^{\dagger}(z_2)
\nn\\
=&
 \int_{0}^{\infty}\!\!\!\d z_1\!\int_{0}^{\infty}\!\!\!\d z\,
e^{-\ii\hp z_1}e^{-\ii\hp (z+z_1)}z_1^{p_1}(z\!+\!z_1)^{p_2}\,
u_+(z_1)v_+^{\dagger}(z\!+\!z_1)\,.
\label{aeq:S1-p1p2} 
\end{align}
}
\hspace{-2.5mm} 
where $(p_1^{},\hp p_2^{})$ are the relevant powers associated with $(z_1^{},\hp z_2^{})$.\  
In the last step of the above derivation, we have changed the integral variable $z_2^{}$
by $z\!=\! z_2^{}\!-\!z_1^{}$, whereas the spinor coefficients 
$u_+^{}$ and $v_+^{}$ are given by
\beqs
\begin{align}
\label{aeq:u+}
u_+^{}(z) &=  \frac{\,e^{+\pi\tlambda/2}\,}{\sqrt{2z\,}\,} \tilde{M}_{\!R}^{}\hp 
W_{\!-\frac{1}{2} - \ii\tlambda,\ii \tnu}
(-\ii\hp 2z)\,, 
\\
\label{aeq:v+}
v_+^{}(z) &= \ii\frac{\,e^{+\pi \tlambda/2}\,}{\sqrt{2 z\,}\,} 
W_{\!\frac{1}{2} - \ii\tlambda,\ii \tnu}
(-\ii\hp 2z)\,.
\end{align}
\eeqs 
The power factors $p_1$ and $p_2$ are complex in general and 
all such complex powers are defined on the principal branch,
\begin{equation}
w^p \equiv e^{p\log w},
\qquad
\arg (w)\in(-\pi,\pi].
\end{equation}
In particular, for $x\!>\!0\hp$, we have 
$(-\ii\hp x)^p \!=\! e^{-\ii\hp\pi p/2}x^p$.\ 

\vs 

For Eq.\eqref{aeq:S1-p1p2}, we first make Wick rotation $z\!\ito\! -\ii\hp z$
for the integral contour of $z\hp$.\ 
Since $z_1\!-\!\ii\hp z$ stays in the fourth quadrant for $z_1\!>\!0$ and $z\!>\!0$, 
the contour deformation does not cross the branch cut:
{\small
\begin{align}
S_1^{p_1p_2}
= 
\!\int_{0}^{\infty}\!\!\!\d z_1\,e^{-\ii\hp z_1}z_1^{p_1}u_+(z_1)\!\!
\int_{0}^{\infty}\!\!\!\d z\,(-\ii)\,
e^{-\ii(-\ii\hp z+z_1)}(-\ii\hp z\!+\!z_1)^{p_2}\,
v_+^{\dagger}(-\ii\hp z\!+\!z_1)\,.
\end{align}
}
\hspace{-2.5mm}
The subsequent Wick rotation of the $z_1$ contour is more delicate, because $u_+(z_1)$ contains the Whittaker function
according to the expression \eqref{aeq:u+},  
which is a function of $z_1$ and has a branch cut along the negative imaginary axis.\ 
Hence, the contour deformation should be understood as follows: 
\begin{equation}
z_1 \to e^{-\ii\hp(\pi/2-\epsilon)}z_1 \hp,
\qquad (\epsilon\!>\!0),
\end{equation}
and the limit $\epsilon\ito 0^+$ is taken after the deformation.

\begin{figure}[t]
  \centering
  \begin{subfigure}[b]{0.32\textwidth}
    \centering
    \includegraphics[width=\linewidth]{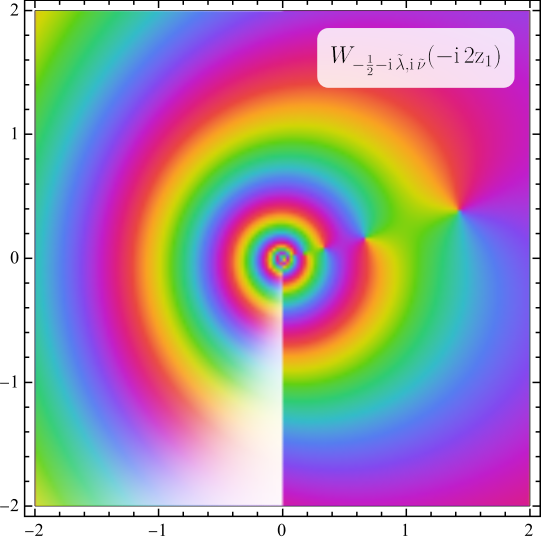}
    \caption{}
  \end{subfigure}\hfill
  \begin{subfigure}[b]{0.32\textwidth}
    \centering
    \includegraphics[width=\linewidth]{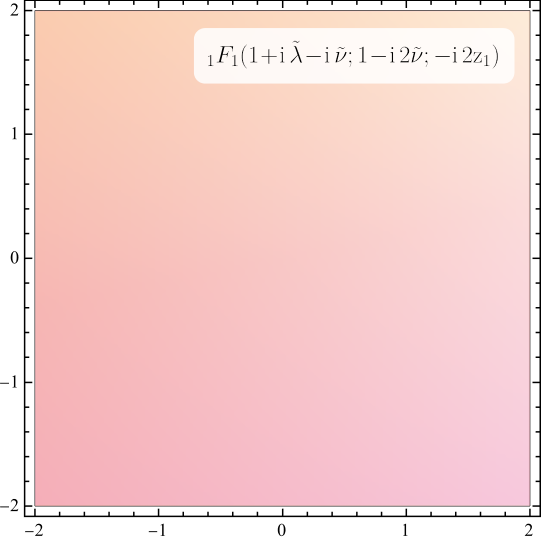}
    \caption{}
  \end{subfigure}\hfill
  \begin{subfigure}[b]{0.32\textwidth}
    \centering
    \includegraphics[width=\linewidth]{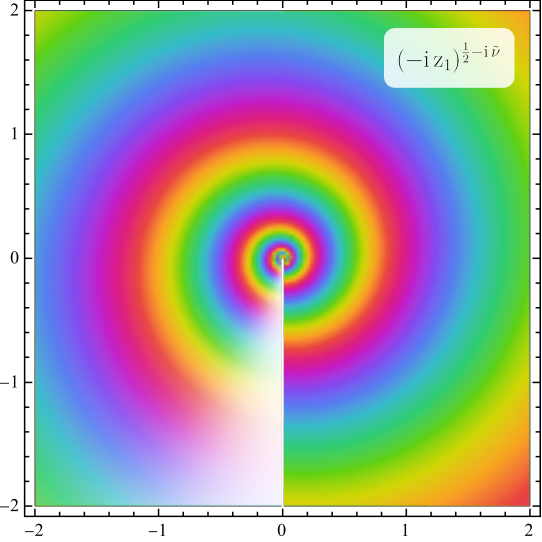}
    \caption{}
\end{subfigure}
\vspace*{-1mm}
\caption{\small 
Domain-coloring plots illustrating the origin of the branch cut in the $z_1$ plane:
(a).\ $W_{-\frac{1}{2}\!-\!\ii\hp\tlambda,\ii\hp\tnu}(-\ii\hp 2 z_1)$,
(b).\ ${}_1F_1(1\!\hp+\!\ii\hp\tlambda\hsm-\hsm\ii\hp\tnu,1\!-\!\ii\hp 2\tnu;-\ii\hp 2 z_1)$, and
(c).\ $(-\ii\hp z_1)^{1/2\!-\!\ii\hp\tnu}$.\ 
The horizontal and vertical axes denote $\mathrm{Re}\hp z_1$ and $\mathrm{Im}\hp z_1$, respectively.\ 
The color indicates the complex phase, whereas the brightness indicates the magnitude.\ 
Plot\,(b) is regular in the $z_1$ plane, whereas the non-analyticity in plot\,(a) 
is associated with the fractional-power behavior shown in plot\,(c), 
whose principal-branch cut lies along the negative imaginary axis of the $z_1$ plane.
}
\label{fig:branch}
\label{fig:11}
\end{figure}

\vs 

To identify the origin of the branch cut, we rewrite the Whittaker function 
in terms of the confluent hypergeometric function ${}_1F_1$ by using Eq.\eqref{WtoM}:
{\small
\begin{equation}
\begin{aligned}
\hspace*{-3.5mm}W_{\!-\frac{1}{2}- \ii\hp\tlambda,\ii\hp \tnu}(-\ii\hp 2 z_1)=
\frac{\,2^{\frac{1}{2}-\ii\hp \tnu}e^{\ii\hp z_1}(-\ii\hp z_1)^{\frac{1}{2}-\ii\hp \tnu}
\Gamma(\ii\hp 2\tnu)\,}
{\Gamma(1\!+\!\ii\hp \tlambda\!+\!\ii\hp\tnu)}
{}_1F_1(1\!+\!\ii\hp\tlambda\!-\!\ii\hp\tnu,1\!-\!\ii\hp 2\tnu;-\ii\hp 2 z_1)
\!+\!(\tnu\!\rightarrow\!-\tnu)\hp.
\label{HGfunction}
\end{aligned}
\end{equation}
}
\hspace{-3.5mm}
Since ${}_1F_1(a;b;x)$ is analytic in $x$ and does not generate branch cuts, 
the non-analyticity in the $z_1$ plane originates from the fractional-power factor 
$(-\ii\hp z_1)^{1/2\pm\ii\hp\tnu}$ in Eq.\eqref{HGfunction}, 
as illustrated in Fig.\,\ref{fig:branch}.\ 
With the principal-branch convention adopted above, this factor has a branch cut 
along the negative imaginary axis of the $z_1$ plane.\ 
Hence, the Wick rotation of the $z_1$ contour should be understood as approaching this axis 
from the fourth quadrant, namely, through the regulated deformation 
$z_1\ito e^{-\ii(\pi/2-\epsilon)}z_1$, with $\epsilon\ito 0^+$.\ 
After the principal branch has been fixed by this prescription, 
the Wick rotation of each fractional-power factor appearing in Eq.\eqref{HGfunction} 
can be evaluated as 
$(-\ii\hp z_1)^\alpha\ito\left(e^{-\ii\hp\pi}z_1\right)^\alpha
\!=\!e^{-\ii\hp\pi\alpha}z_1^\alpha\hp$.\ 
With this prescription and making the Wick rotation $z_1\!\ito\! -\ii\hp z_1$
for the integral contour of $z_1\hp$, we arrive at 
{\small
\begin{align}
\quad S_1^{p_1p_2}
&= 
 \int_{0}^{\infty}\!\!\!\d z_1(-\ii)\,e^{-z_1}(-\ii\hp z_1)^{p_1}u_+(-\ii\hp z_1)
\!\int_{0}^{\infty}\!\!\!\d z\,(-\ii)\,
e^{-(z+z_1)}(-\ii\hp z\!-\!\ii\hp z_1)^{p_2}\,
v_+^{\dagger}(-\ii\hp z\!-\!\ii\hp z_1)
\nn\\
&= -\!\int_{0}^{\infty}\!\!\!\d z_1\,e^{-z_1}(-\ii\hp z_1)^{p_1}u_+(-\ii\hp z_1)
\!\int_{0}^{\infty}\!\!\!\d z\,
e^{-(z+z_1)}(-\ii\hp z\!-\!\ii\hp z_1)^{p_2}\,
v_+^{\dagger}(-\ii\hp z\!-\!\ii\hp z_1) \hp, 
\end{align}
}
\hspace{-2.5mm}
where the overall minus sign comes from the two Jacobian factors, $(-\ii)^2\!=\!-1$.

\vs 

Finally, defining $z'\!=\!z\!+\!z_1$ and $\d z'\!=\!\d z$, we further derive 
{\small
\begin{align}
S_1^{p_1p_2}
&= -\!\int_{0}^{\infty}\!\!\!\d z_1\,e^{-z_1}(-\ii\hp z_1)^{p_1}u_+(-\ii\hp z_1)
\!\int_{z_1}^{\infty}\!\!\!\d z'\,e^{-z'}(-\ii\hp z')^{p_2}\,
v_+^{\dagger}(-\ii\hp z')
\nn\\
&= -\!\!\int_{0}^{\infty}\!\!\!\d z_1\,e^{-z_1}(-\ii\hp z_1)^{p_1}u_+(-\ii\hp z_1)
\!\int_{z_1}^{\infty}\!\!\!\d z_2\,e^{-z_2}(-\ii\hp z_2)^{p_2}\,
v_+^{\dagger}(-\ii\hp z_2) \hp.
\end{align}
}
\hspace{-2.5mm}
Equivalently, using the principal-branch relation 
$(-\ii\hp x)^p\!=\!e^{-\ii\hp\pi p/2}x^p$ (for $x\!>\!0$), we obtain the following:
{\small
\begin{equation}
\label{s1calcu}
S_1^{p_1p_2}
=
-e^{-\frac{\ii}{2}\hp\pi(p_1+p_2)}
\!\!\int_{0}^{\infty}\!\!\!\d z_1\,e^{-z_1}z_1^{p_1}u_+(-\ii\hp z_1)
\!\int_{z_1}^{\infty}\!\!\!\d z_2\,e^{-z_2}z_2^{p_2}v_+^{\dagger}(-\ii\hp z_2) \hp.
\end{equation}
}
\hspace{-2.5mm}

Next, we proceed to explicitly evaluate the double integral over $z_1$ and $z_2$ 
in Eq.\eqref{s1calcu}.\ Since the $z_2$ integration is nested within the $z_1$ integration, 
we begin with the $z_2$ integral.\ The integral over $z_2$ can be evaluated analytically, 
yielding an indefinite integral in terms of a Meijer $G$ function.\ 
Taking the asymptotic expansion at infinity and then setting the lower limit to 
$z_2\!=\!z_1$, we obtain
{\small
\begin{align}
\mathcal{I}_2(z_1)&\equiv\int_{z_1}^{\infty}\!\!\!\d z_2\, e^{-z_2} z_2^{p_2}\,v_+^{\dagger}(-\ii\hp z_2)
\nn\\
&=-\ii\hp 2^{-\frac{\,3\,}{\,2\,}-p_2}e^{\frac{\,\pi\tlambda\,}{2}}e^{\ii\frac{\,\pi\,}{4}}z_2^{-\frac{\,1\,}{\,2\,}}
G^{2,1}_{2,3}\!\!
        \left.\left[\!\!
        \begin{array}{c}
            \frac{\,3\,}{\,2\,}, \frac{\,3\,}{\,2\,}\!+\!p_2\!-\!\ii\hp\tlambda 
            \\
            \frac{\,3\,}{\,2\,}\!+\!p_2\!-\!\ii\hp\tnu, \frac{\,3\,}{\,2\,}\!+\!p_2\!+\!\ii\hp\tnu,\frac{\,1\,}{\,2\,}
        \end{array}\!\!;2z_2
        \right]\right|_{z_1}^{\infty}
        \nn\\
&=-\bigg(\!\ii \hp 2^{-1-p_2}e^{\frac{\,\pi\tlambda\,}{2}}e^{\ii\hp\frac{\,\pi\,}{4}}\frac{\,\Gamma(1\!+\!p_2\!-\!\ii\hp\tnu)
\Gamma(1\!+\!p_2\!+\!\ii\hp\tnu)\,}{{\Gamma(1\!+\!p_2\!-\!\ii\hp\tlambda)}}\!\bigg)
\nn\\
&\quad +\bigg(\ii \hp 2^{-\frac{\,3\,}{\,2\,}-p_2}
e^{\frac{\,\pi\tlambda\,}{2}}e^{\ii\hp\frac{\,\pi\,}{4}}z_1^{-\frac{\,1\,}{\,2\,}}
G^{2,1}_{2,3}\!\hsm 
\left[
\begin{array}{c}
\frac{\,3\,}{\,2\,}, \frac{\,3\,}{\,2\,}\!+\!p_2\!-\!\ii\hp\tlambda 
\\
\frac{\,3\,}{\,2\,}\!+\!p_2\!-\!\ii\hp\tnu, \frac{\,3\,}{\,2\,}\!+\!p_2\!+\!\ii\hp\tnu,\frac{\,1\,}{\,2\,}
\end{array}\!\!;2z_1
\right]\!\hsm\bigg).
\end{align}
}
\hspace{-2.5mm}
The Meijer $G$ function is defined as follows:
{\small
\begin{align}
\hspace*{-4mm}
G^{m,n}_{p,q}\!\!\left[\!
        \begin{array}{c}
            a_1, a_2 , \cdots,a_p\\
            b_1, b_2,\cdots,b_q
        \end{array};z
\right]
\equiv
\frac{1}{\,2\pi\ii\,}\!\!\bigintsss_{C}\!\!\d s\, z^{s}
\frac{~\bigg[\dis\prod_{k=1}^{n}\!\Gamma(1\!-\!a_k\!+\!s)\bigg]\hsm\bigg/\hsm
\bigg[\!\dis\prod_{k=m+1}^{q}\!\!\!\Gamma(1\!-\!b_k\!+\!s)\bigg]~}
{\bigg[\!\dis\prod_{k=n+1}^{p}
\!\!\!\Gamma(a_k\!-\!s)\!\bigg]\hsm\bigg/\hsm
\bigg[\!\dis\prod_{k=1}^{m}\!\Gamma(b_k\!-\!s)\hsm\bigg]}. 
\end{align}
}
\hspace{-2.5mm}
In the second step, we evaluate the integral over $z_1$ from $0$ to $\infty$.\ 
By virtue of Eq.\eqref{WtoM} and after adopting the branch-cut prescription specified above, 
$u_+(-\ii\hp z_1)$ can be written as follows: 
{\small
\begin{align}
\hspace*{-6mm}
u_+(-\ii\hp z_1)
=\tilde{M}_{\!R}\hp e^{ z_1}e^{-\frac{\ii\hp \pi}{4}}e^{\frac{\,\pi \tlambda\,}{2}}
\!\Bigg[\!\frac{\,2^{-\ii\hp \tnu}e^{-\pi \tnu}z_1^{-\ii\hp \tnu}\Gamma(\ii\hp 2\tnu)\,}
{\Gamma(1\!+\!\ii\hp \tlambda\!+\!\ii\hp \tnu)}
{}_1F_1(1\!+\!\ii\hp\tlambda\!-\!\ii\hp\tnu,1\!-\!\ii\hp 2\tnu;-\ii\hp 2 z_1)\!+\!(\tnu\rightarrow\!-\tnu)\!\Bigg]\! .
\end{align}
}
\hspace{-2.5mm}
Then, we can evaluate the remaining $z_1$ integral directly, which yields the following: 
{\small
\begin{equation}
\begin{aligned}
 S_1^{p_1p_2} =\,& 
-\!e^{-\frac{\ii}{2}\hp\pi(p_1+p_2)}
\!\!\int_{0}^{\infty}\!\!\d z_1\,e^{-z_1}z_1^{p_1}u_+(-\ii\hp z_1)\mathcal{I}_2(z_1)
\\
=\,&  
\frac{\,\pi\,2^{-2-p_1-p_2}
 e^{-\frac{\,\ii\hp\pi\,}{2}(p_1+p_2)}
 e^{\pi\tlambda}
 \tilde{M}_{\!R} \,\csch(2\pi\tnu)\,
 \Gamma(2\!+\!p_1\!+\!p_2)\,}
{\Gamma(1\!+\!\ii\hp\tlambda\!-\!\ii\hp\tnu)\,
\Gamma(1\!+\!\ii\hp\tlambda\!+\!\ii\hp\tnu)}
\\
&\times
\Bigg[
e^{-\pi\tnu}
\Gamma(1\!+\!p_1\!-\!\ii\hp\tnu)\,
\Gamma(2\!+\!p_1\!+\!p_2\!-\!\ii\hp 2\tnu)\,
\Gamma(1\!+\!\ii\hp\tlambda\!-\!\ii\hp\tnu)
\\
&\times
{}_4\tilde F_3\!\(
\begin{matrix}
2\!+\!p_1\!+\!p_2,\;
1\!+\!p_1\!-\!\ii\hp\tnu,\;
1\!+\!\ii\hp\tlambda-\ii\hp\tnu,\;
2\!+\!p_1\!+\!p_2\!-\!\ii\hp 2\tnu
\\
2\!+\!p_1\!-\!\ii\hp\tnu,\;
2\!+\!p_1\!+\!p_2\!-\!\ii\hp\tlambda\!-\!\ii\hp\tnu,\;
1\!-\!\ii\hp 2\tnu
\end{matrix}
;-1
\!\hsm\)\!-\!(\tnu\rightarrow\!-\tnu)\hsm\Bigg].
\end{aligned}
\end{equation}
}
\vspace*{3mm}
\noindent 
{\bf $\blacklozenge$\,Non-Time-Ordered Seed Integral:}
\\[3mm]
Next, we consider the following non-time-ordered integral $W_1^{p_1 p_2}$, 
{\small
\begin{equation}
\begin{aligned}
W_1^{p_1 p_2}
&= \int_0^\infty \!\!\!\td z_1 \!\int_0^\infty \!\!\!\td z_2\,
e^{\ii\hp z_1} e^{-\ii\hp z_2}
z_1^{p_1} z_2^{p_2}
u_+(z_1) v_+^\dagger(z_2)
\\
&= \int_0^\infty \!\!\!\td z_1 e^{\ii\hp z_1} z_1^{p_1} u_+(z_1)
   \!\!\int_0^\infty \!\!\!\td z_2 e^{-\ii\hp z_2}  z_2^{p_2}  v_+^\dagger(z_2)
   \\
&= \int_{0}^{\infty}\!\!\!\td z_1\,e^{-z_1}(\ii\hp z_1)^{p_1}u_+(\ii\hp z_1)
\!\!\int_{0}^{\infty}\!\!\!\td z_2\,e^{-z_2}(-\ii\hp z_2  )^{p_2}\,
v_+^{\dagger}(-\ii\hp z_2)
\\
&= e^{\frac{\,\ii\hp \pi\,}{2}(p_1-p_2)}\!\!\hsm\int_{0}^{\infty}\!\!\!\d z_1\,e^{-z_1}z_1^{p_1}u_+(\ii\hp z_1)
\!\!\int_{0}^{\infty}\!\!\!\d z_2\,e^{-z_2}z_2^{p_2}\,
v_+^{\dagger}(-\ii\hp z_2)\,.
\end{aligned}
\end{equation}
}
\hspace{-2.5mm}
Unlike the time-ordered integrals, the variables $z_1$ and $z_2$ are not coupled in the non-time-ordered case, 
so the double integral factorizes into a product of two single integrals:
{\small
\beqs 
\begin{align}
&\int_{0}^{\infty}\!\!\!\d z_1\,e^{-z_1}z_1^{p_1}u_+(\ii\hp z_1)
=
(1\!-\!\ii) 2^{-\frac{\,3\,}{\,2\,} - p_1}  e^{\frac{\,\pi \tlambda\,}{2}}  \tilde{M}_{\!R}
\frac{\,\Gamma(1 \!+\! p_1 \!-\! \ii\hp \tnu)  \Gamma(1 \!+\! p_1 \!+\! \ii\hp \tnu)\,}{\,\Gamma(2 \!+\! p_1 \!+\! \ii\hp \tlambda)\,}\,,
\\
&\int_{0}^{\infty}\!\!\!\d z_2\,e^{-z_2}(-\ii\hp z_2)^{p_2}\,
v_+^{\dagger}(-\ii\hp z_2)
=
-\ii(1\!+\!\ii) 2^{-\frac{\,3\,}{\,2\,} - p_2}  e^{\frac{\,\pi \tlambda\,}{2}}
\frac{\,\Gamma(1 \!+\! p_2 \!-\! \ii\hp \tnu)  \,\Gamma(1 \!+\! p_2 \!+\! \ii\hp \tnu)\,}{\Gamma(1 \!+\! p_2 \!-\! \ii\hp \tlambda)}\,.
\end{align}
\eeqs 
}
\hspace{-2.5mm}
With these, we further derive $W_1^{p_1 p_2}$ as follows:
{\small
\begin{align}
W_1^{p_1 p_2}
=
-\ii\tilde{M}_{\!R}\hp e^{\frac{\,\ii\hp \pi\,}{2}(p_1-p_2)}2^{-2-p_1- p_2}
\frac{\,\Gamma(1 \!+\! p_1 \!-\! \ii\hp \tnu)  \Gamma(1 \!+\! p_1 \!+\! \ii\hp \tnu)\,}{\Gamma(2 \!+\! p_1 \!+\! \ii\hp \tlambda)}
\frac{\,\Gamma(1 \!+\! p_2 \!-\! \ii\hp \tnu)  \,\Gamma(1 \!+\! p_2 \!+\! \ii\hp \tnu)\,}{\Gamma(1 \!+\! p_2 \!-\! \ii\hp \tlambda)}
\hp .
\end{align}
}
\hspace{-2.5mm}

\subsection{\hspace*{-1.5mm}Folded-limit Analysis and the Cutting Rule}
\label{The cutting rule}

In this subsection, we first discuss the folded-limit analysis for the reduction 
of the four-point correlation to the three-point correlation.\ 
Then, we present the cutting rule implementation when computing the nonlocal signal of  
the correlator, and we show how the cutting rule is implemented for the fermionic case.\ 

\vspace*{4mm}
\noindent 
{\bf $\blacklozenge$\,\textcolor{black}{Folded-limit Analysis:}}
\\[3mm]
In general, a three-point correlator may be viewed as the folded limit of a four-point correlator,
\\[-5mm]
\begin{equation}
\lim_{k_4\to  0}\,\langle\delta\phi(\mathbf{k}_1)\delta\phi(\mathbf{k}_2)\delta\phi(\mathbf{k}_3)\delta\phi(\mathbf{k}_4)\rangle^{\prime}
\Longrightarrow 
\langle\delta\phi(\mathbf{k}_1)\delta\phi(\mathbf{k}_2)\delta\phi(\mathbf{k}_3)\rangle^{\prime}\,.
\end{equation}
Once the expression for the corresponding four-point correlator is obtained, one may directly infer the structure of the three-point correlator\,\cite{Arkani-Hamed2015,Xianyu230107047}.

\begin{figure}[t]
\centering
\includegraphics[width=1\linewidth]{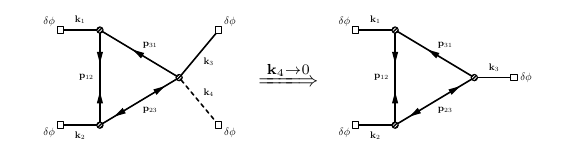}
\vspace*{-6mm}
\caption{From the four-point correlator to the three-point correlator in the folded limit.}
\label{4ptloop}
\end{figure}

To realize this, we introduce the auxiliary coupling
\begin{equation}
\label{fictitious_coupling}
    \Delta\mathcal{L}=\(\frac{\,a\,}{\Lambda}\) \delta\phi' \pd_\mu \phi \(\tilde{N}^\dagger \bar{\sigma}^\mu \tilde{N}\)\,,
\end{equation}
which allows us to construct a four-point correlator that reduces to the original three-point correlator in the folded limit, as shown in Fig.\,\ref{4ptloop}.

\vs 

This four-point correlator can be expressed as
\begin{equation}
\label{4-point_form}
\begin{aligned}
 & \quad\langle\delta\phi(\mathbf{k}_1)\delta\phi(\mathbf{k}_2)\delta\phi(\mathbf{k}_3)\delta\phi(\mathbf{k}_4)\rangle^{\prime} \\
 & =\sum_{a,b,c=\pm}\! a\hp b\hp c \(\!\frac{\,-\ii\,}{\,\Lambda\,}\!\)
^{\!\!3} 
\!\!
 \int_{-\infty}^{0^-}\!\mathrm{d}\tau_1\mathrm{d}\tau_2\mathrm{d}\tau_3\,
 \mathcal{F}_{\mu a}(\mathbf{k}_1,\tau_1)
 \mathcal{F}_{\nu b}(\mathbf{k}_2,\tau_2)
 \mathcal{F}_{\lambda c}(\mathbf{k}_3,\tau_3)\\
 &\quad\times(-H\tau_3)^{-1}G_c^{\prime}(k_4,\tau_3)
 \!\int\!\!\frac{\mathrm{d}^3\mathbf{q}}{(2\pi)^3}\mathcal{T}_{abc}^{\mu\nu\lambda}\,,
\end{aligned}
\end{equation}
where $G_c^{\prime}(k_4,\tau_3)=\frac{\,H^2\,}{2k_4}\tau_3 e^{\ii   c k_4 \tau_3}$ and
\begin{equation}
\label{loop_kernal2}
\begin{aligned}
\mathcal{T}_{abc}^{\mu\nu\lambda}= & - \left[\bar{\sigma}^{\mu\dalpha\alpha}{\hat{D}_{ab\alpha}}^{\beta}(\mathbf{p}_{12},\tau_{1},\tau_{2})\sigma^{\nu}_{\beta\dbeta}\check{D}_{bc\dgamma}^{\dbeta}(\mathbf{p}_{23},\tau_{2},\tau_{3})\bar{\sigma}^{\lambda\dgamma\gamma}D_{ca\gamma\dalpha}(\mathbf{p}_{31},\tau_{3},\tau_{1})\right] \\
 & - \left[\bar{\sigma}^{\mu\dalpha\alpha}D_{ac\alpha\dbeta}(\mathbf{p}_{13},\tau_1,\tau_3)\bar{\sigma}^{\lambda\dbeta\beta}{\hat{D}_{cb\beta}}^{\gamma}(\mathbf{p}_{32},\tau_3,\tau_2)\sigma^{\nu}_{\gamma\dgamma}\check{D}_{ba\dalpha}^{\dgamma}(\mathbf{p}_{21},\tau_2,\tau_1)\right].
\end{aligned}
\end{equation}
The new vertex in the four-point correlator introduces only an additional factor
\begin{equation}
   \lim_{k_4\to 0} k_4 (-H\tau_3)^{-1}G_c^{\prime}(k_4,\tau_3)= -\frac{\,H\,}{2}\,.
\end{equation}
Thus, upon taking the folded limit $k_4\rightarrow0$, the desired three-point correlator is given by
\begin{equation}
\label{4ptloop_to_3ptloop}
\begin{aligned}
 & \langle\delta\phi(\mathbf{k}_1)\delta\phi(\mathbf{k}_2)\delta\phi(\mathbf{k}_3)\rangle^{\prime} 
 \\
 & =\sum_{a,b,c=\pm} \! a\hp b\hp c \(\!\frac{\,-\ii\,}{\,\Lambda\,}\!\)
^{\!\!3} 
\!\!
 \int_{-\infty}^{0^-}\mathrm{d}\tau_1\mathrm{d}\tau_2\mathrm{d}\tau_3\,
 \mathcal{F}_{\mu a}(\mathbf{k}_1,\tau_1)
 \mathcal{F}_{\nu b}(\mathbf{k}_2,\tau_2)
 \mathcal{F}_{\lambda c}(\mathbf{k}_3,\tau_3)
 \!\int\!\!\frac{\mathrm{d}^3\mathbf{q}}{(2\pi)^3}\mathcal{T}_{abc}^{\mu\nu\lambda}
 \\
 &=\lim_{k_4\rightarrow0} -2\frac{k_4}{H}
 \langle\delta\phi(\mathbf{k}_1)\delta\phi(\mathbf{k}_2)\delta\phi(\mathbf{k}_3)\delta\phi(\mathbf{k}_4)\rangle^{\prime}\,.
\end{aligned}
\end{equation}

For notational convenience, we define the following momentum variables and ratios used for the four-point correlator:
\begin{equation}
\label{momentum_def}
    \begin{aligned}
    k_s &\equiv \lvert \mathbf{k}_1 \!+\! \mathbf{k}_2 \rvert, \quad
    k_{12} \equiv k_1 \!+\! k_2, \quad
    k_{34} \equiv k_3 \!+\! k_4, \\
    r_1 &\equiv k_s / k_{12}, \quad
    r_2 \equiv k_s / k_{34}.
\end{aligned}
\end{equation}
The four-point correlator has one more degree of freedom than the three-point correlator. For the three-point case, only
the two quantities 
\begin{equation}
\label{external_momentum}
k_{12}^{}=k_1\!+\hsm k_2^{}\hp, \qquad
r_1\equiv \frac{k_s}{\,k_{12}\,}\hp , 
\end{equation}
remain independent, whereas effectively $r_2\hsm\ito\hsm 1$ in the folded limit.

\vs 

Using the PMB approach introduced in Appendix\,\ref{PMB representation}, 
the nonlocal signal of the four-point correlator can be written schematically as follows: 
\beq 
\begin{aligned}
\hspace*{-3mm} 
& \langle\delta\phi(\mathbf{k}_1)\delta\phi(\mathbf{k}_2)\delta\phi(\mathbf{k}_3)\delta\phi(\mathbf{k}_4)\rangle^{\prime}
\\
\hspace*{-3mm} 
&\thicksim P(k_1,k_2,k_3,k_4) (r_1r_2)^{2\ii  \tnu+\frac{3}{2}}
\hspace*{-6mm} \sum_{n_1,n_2,n_3,n_4=0}^\infty  \hspace*{-6mm} 
f(n_1,n_2,n_3,n_4,\tilde{M}_{\!R},\tnu)
(r_1)^{n_{13}}(r_2)^{n_{24}}
\!+\!(\tnu\ito-\tnu)\hp,
\end{aligned}
\eeq 
where $P(k_1,k_2,k_3,k_4)$ is a polynomial in the external momenta, $f(n_1,n_2,n_3,n_4,\tilde{M}_{\!R},\tnu)$ 
is a function of $n_i$ and $(\tilde{M}_{\!R},\tnu)$, and $n_{ij}^{}\!\equiv\hsm n_i\!+\!n_j$.

\vs

Taking the squeezed limit $k_s\!\to\hsm 0$ means $r_1,r_2\!\to\hsm 0\hp$, 
under which only the leading term with $n_i^{}\!=\!0$ survives and gives the following form: 
\begin{equation}
\langle\delta\phi(\mathbf{k}_1)\delta\phi(\mathbf{k}_2)\delta\phi(\mathbf{k}_3)\delta\phi(\mathbf{k}_4)\rangle^{\prime}
\thicksim
C_0^{} P(k_1,k_2,k_3,k_4) (r_1r_2)^{2\ii  \tnu+\frac{3}{2}}
+(\tnu \ito -\tnu) \hp ,
\end{equation}
where $C_0^{}$ is a constant determined by the parameters of the model.
Taking the folded limit
\begin{equation}
k_4^{}\!\to\hsm 0 ~\Longrightarrow~
k_s \hsm\ito k_3^{}  ~\Longrightarrow~
r_2^{}\!=\!\frac{k_s^{}}{\,k_{34}^{}\,}\ito 1 \hp,
\end{equation}
we see that except the terms with $n_1\!=\!n_3\!=\!0\hp$, 
other terms with positive $n_1$ or $n_3$ are suppressed by powers of $r_1$.\
The sums over $n_2$ and $n_4$ are not easy to perform analytically.\ 
As a practical approximation, one may use the squeezed limit result with $r_2\!=\!0$ 
to estimate the folded-limit behavior at $r_2\!=\!1$, thereby retaining only the leading term with $n_2\!=\!n_4\!=\!0$.\ 
This provides a reasonable estimate of the order of magnitude of the signal in the folded limit.

\newpage 
\noindent 
{\bf $\blacklozenge$\,Cutting Rule Interpretation:}
\\[3mm]
In this subsection, we discuss the cutting rule interpretation of the nonlocal signal, 
which provides useful insight into the physical origin of the oscillatory contribution 
as extracted in the squeezed limit.

\vs 

As discussed in Section\,2, the nonlocal cosmological collider signal is extracted 
by taking the soft limit for the internal propagators.\ 
We refer to such internal propagators as the soft propagators.\ 
To illustrate the cutting rule, we first consider a massive scalar field $\sigma$ 
with mode function $y_k^{}(\tau)$.\ Its in-in propagators are given by 
{\small
\begin{equation}
\begin{aligned}
D_{\oplus\oplus }(\mathbf{k},\tau_1,\tau_2)&=
D_{\ominus\oplus }(\mathbf{k},\tau_1,\tau_2)\theta(\tau_1\!-\!\tau_2)
+
D_{\oplus\ominus }(\mathbf{k},\tau_1,\tau_2)\theta(\tau_2\!-\!\tau_1) \hp,
\\
D_{\ominus\ominus }(\mathbf{k},\tau_1,\tau_2)&=
D_{\oplus\ominus }(\mathbf{k},\tau_1,\tau_2)\theta(\tau_1\!-\!\tau_2)
+
D_{\ominus\oplus }(\mathbf{k},\tau_1,\tau_2)\theta(\tau_2\!-\!\tau_1)\hp,
\\
D_{\ominus\oplus }(\mathbf{k},\tau_1,\tau_2)&=
y_k(\tau_1)y_k^*(\tau_2)\hp, \qquad
D_{\oplus\ominus }(\mathbf{k},\tau_1,\tau_2)=
y_k^*(\tau_1)y_k(\tau_2)\hp.
\end{aligned}
\end{equation}
}
\hspace{-2.5mm}
The nonlocal part of a soft massive propagator satisfies the relations:
{\small
\begin{equation}
\label{condition_cutting_rule}
    D_{\oplus \ominus}(\mathbf{k}, \tau_1, \tau_2)_\text{(NLoc)}
    =
    D_{\oplus \ominus}^\dagger (\mathbf{k}, \tau_1, \tau_2)_\text{(NLoc)}
    =
    D_{\ominus\oplus }(\mathbf{k}, \tau_1, \tau_2)_\text{(NLoc)}.
\end{equation}
}
\hspace{-2.5mm}
In consequence, for the nonlocal signal the $\theta$ functions in the time-ordered and 
reverse-time-ordered propagators do not affect the time integrals.\ 
For the nonlocal signal of a four-point tree diagram with only one massive propagator 
$D_{ab}(\mathbf{k}, \tau_1, \tau_2)$ as an internal line, it can be expressed as  
the following integral:
\begin{equation}
\langle\delta\phi^4\rangle ~\thicksim~
\sum_{a,b=\pm}\!\int_{\infty}^0\!\hsm\td\tau_1 \td\tau_2 f_L^a(\tau_1) f_R^b(\tau_2)
D_{ab}\hsm (\mathbf{k}_s,\tau_1,\tau_2) \hp ,
\end{equation}
where $f_L^{}(\tau_1)$ and $f_R^{}(\tau_2)$ denote the external-line factors associated with $\tau_1$ and $\tau_2$ 
respectively, which have the typical form of $f(\tau)\!\thicksim\!\tau^n \hsm\exp(\ii\hp k\hp \tau)$
with $k\!=\!|\mathbf{k}|\hp$.\

Using the cutting rule, it could be factorized into a product of the following two separate integrals:
{\small
\begin{equation}
\label{after_cutting_example_scalar}
\begin{aligned}
\langle\delta\phi^4\rangle
&~\rightarrow~
\int_{-\infty}^{0^-}\!\!\!\!\!\td\tau_1 
\!\int_{-\infty}^{0^-}\!\!\!\!\!\td\tau_2
\Big[\!\sum_{a,b=\pm}\!\!\!f_L^a(\tau_1)f_R^b(\tau_2)\hsm\Big]\hsm
D_{\ominus\oplus }(\mathbf{k}_s,\tau_1,\tau_2) \hp \\
&~\rightarrow~
\sum_{a,b=\pm}\int_{-\infty}^{0^-}\!\!\!\!\!\td\tau_1 f_L^a(\tau_1) y_{k_s}(\tau_1)
\!\int_{-\infty}^{0^-}\!\!\!\!\!\td\tau_2
f_R^b(\tau_2)
y_{k_s}^*(\tau_2) \hp~,
\end{aligned}
\end{equation}
}

\vs 

For the fermion propagator ${D}_{ab\alpha \dbeta}(\mathbf{k},\tau_{1},\tau_{2})$, we have 
{\small
\begin{equation}
\begin{aligned}
{D_{\ominus\oplus}}_{\alpha\dbeta}(\mathbf{k}; \tau_1, \tau_2)
&=
\sum_{s}\!u_{s} (\tau_1,k) u_{s}^\dagger (\tau_2,k) h_{s,\alpha}(\mathbf{k})h_{s,\dbeta}^\dagger(\mathbf{k})\hp,
\\
{D_{\oplus \ominus}}_{\alpha\dbeta} (\mathbf{k}; \tau_1, \tau_2)
&=
\sum_{s}\!-v_{s} (\tau_2,k) v_{s}^\dagger (\tau_1,k) h^\dagger_{-s,\dbeta}(-\mathbf{k})h_{-s,\alpha}(-\mathbf{k})\hp,
\\
{D_{\oplus\oplus }}_{\alpha\dbeta} (\mathbf{k}; \tau_1, \tau_2)
&=
{D_{\ominus \oplus}}_{\alpha\dbeta}(\mathbf{k}; \tau_1, \tau_2) \theta (\tau_1 \!-\! \tau_2)
\!+\! {D_{\oplus \ominus}}_{\alpha\dbeta} (\mathbf{k}; \tau_1, \tau_2)\theta (\tau_2 \!-\! \tau_1)\hp,
\\
{D_{\ominus\ominus }}_{\alpha\dbeta} (\mathbf{k}; \tau_1, \tau_2)
&=
{D_{\oplus \ominus}}_{\alpha\dbeta} (\mathbf{k}; \tau_1, \tau_2)\theta (\tau_1 \!-\! \tau_2)
\!+\! {D_{ \ominus\oplus}}_{\alpha\dbeta}(\mathbf{k}; \tau_1, \tau_2) \theta (\tau_2 \!-\! \tau_1)\hp.
\end{aligned}
\end{equation}
}
\hspace{-2.5mm}
As shown in Eq.\eqref{mode_function_nonlocal} and Eq.\eqref{mode_function_nonlocal_othercase}, 
we have the following relations:
{\small
\beqs
\begin{align}
\label{uu_vv_nonlocal}
\hspace*{-5mm}\left[u_{+}(\tau_1,k) u_{+}^\dagger(\tau_2,k) \right]_\text{(NLoc)}\!\!&=-\left[v_{+}(\tau_2,k)  v_{+}^\dagger(\tau_1,k)\right]_\text{(NLoc)}\!, 
\\
\left[u_{-}(\tau_1,k) u_{-}^\dagger(\tau_2,k) \right]_\text{(NLoc)}\!\!&=-\left[v_{-}(\tau_2,k)  v_{-}^\dagger(\tau_1,k)\right]_\text{(NLoc)}\!.
\end{align}
\eeqs
}
\hspace{-3mm}
For the helicity eigenspinors defined in Eq.\eqref{helicity_value}, which appear in ${D_{\ominus\oplus}}_{\alpha\dbeta}(\mathbf{k}; \tau_1, \tau_2)$ and ${D_{\oplus \ominus}}_{\alpha\dbeta} (\mathbf{k}; \tau_1, \tau_2)$, the following identity holds:
\begin{equation}
    h_{s,\alpha}(\mathbf{k})h_{s,\dbeta}^\dagger(\mathbf{k})=h_{-s,\alpha}(-\mathbf{k})h^\dagger_{-s,\dbeta}(-\mathbf{k}).
\end{equation}
With the above relations, we find that the nonlocal parts of the time-ordered and reverse-time-ordered fermion propagators are also equal, 
{\small
\begin{equation}
    {D_{\ominus\oplus}}_{\alpha\dbeta}(\mathbf{k}; \tau_1, \tau_2)_\text{(NLoc)}
    =
    {D_{\oplus\ominus}}_{\alpha\dbeta}(\mathbf{k}; \tau_1, \tau_2)_\text{(NLoc)}\,.
\end{equation}
}
\hspace{-2.5mm}
Hence, the cutting rule could be implemented directly for the fermion case as well.

\vspace*{2mm}
\subsection{\hspace*{-1.5mm}Loop Integration Using Feynman Parameterization}
\label{loop integral through Feynman parameter approach}
\vspace*{2mm}

In this subsection, we evaluate the relevant loop-momentum integration 
by using Feynman parameterization.\ 
This derivation makes the separation between the local UV-sensitive part 
and the nonlocal, non-analytic part fairly transparent.

\vs 

We consider the following master integral,
{\small
\begin{equation}
\label{eq:bubble_master_appendix}
\mathcal{J}_{\rm{bubble}}^{}(k_s)
=
\!\!\int\!\! \frac{\,\mathrm d^3 \mathbf q\,}{(2\pi)^3}\,
|\mathbf q|^{-2a}\,
|\mathbf q\!+\!\mathbf k_s|^{-2b},
\end{equation}
}
\hspace{-2.5mm}
where $a$ and $b$ are complex parameters in general, and $\mathbf k_s$ denotes an external momentum.

For generic complex factors $a$ and $b$, the integral is absolutely convergent 
only if the following conditions are obeyed, 
{\small
\begin{equation}
\label{eq:bubble_convergence_domain}
\mathrm{Re}(a)\!<\!\frac{\,3\,}{\,2\,},\qquad
\mathrm{Re}(b)\!<\!\frac{\,3\,}{\,2\,},\qquad
\mathrm{Re}(a\!+\!b)\!>\!\frac{\,3\,}{\,2\,}.
\end{equation}
}
\hspace{-2.5mm}
Outside the above convergent domain, the integral can be regulated by the analytic continuation.\ 
Within the convergent domain, it can be evaluated explicitly using the Feynman parameterization, 
and the resultant expression can be analytically continued to the parameter region where  
the fermion loop integration is performed.

\vs 

Using the formula 
{\small
\begin{equation}
\frac{1}{A^a B^b}
=
\frac{\,\Gamma(a\!+\!b)\,}{\,\Gamma(a)\Gamma(b)\,}
\!\!\int_0^1\!\! \mathrm d\xi_1\, \mathrm d\xi_2\,
\delta(1\!-\!\xi_1\!-\!\xi_2)\,
\frac{\xi_1^{a-1}\xi_2^{b-1}}{\,(\xi_1 A\!+\!\xi_2 B)^{a+b}\,},
\end{equation}
}
\hspace{-2.5mm}
we can re-express Eq.\eqref{eq:bubble_master_appendix} as follows: 
{\small
\begin{equation}
\label{aeq:I-buble}
\begin{aligned}
\mathcal{J}_{\mathrm{bubble}}(k_s)
&=
\frac{\,\Gamma(a\!+\!b)\,}{\,\Gamma(a)\Gamma(b)\,}
\!\!\int_0^1 \!\!\!\mathrm d\xi_1^{}\hp \mathrm d\xi_2^{}\hp 
\delta(1\!-\!\xi_1\!-\!\xi_2)\!
\!\int \hsm\!\!\frac{\,\mathrm d^3 \mathbf q\,}{\,(2\pi)^3\,}
\frac{\xi_1^{a-1}\xi_2^{b-1}}
{\,\big[\xi_1 |\mathbf q|^2\!+\!\xi_2 |\mathbf q\!+\!\mathbf k_s|^2\big]^{a+b}\,}.
\end{aligned}
\end{equation}
}
\hspace{-2.5mm}
Then, shifting the loop momentum 
$\mathbf q \!=\! \mathbf p \hsm -\hsm \xi_2 \mathbf k_s^{}\hp$,
we derive  
$\,\xi_1 |\mathbf q|^2\!+\!\xi_2 |\mathbf q\!+\!\mathbf k_s|^2
\!=\! p^2\!+\!\Delta$  with 
$\Delta \!=\! \xi_1\xi_2 k_s^2\hp$.\ 
Thus, the integral \eqref{aeq:I-buble} becomes
{\small
\begin{equation}
\begin{aligned}
\mathcal{J}_{\rm{bubble}}(k_s)
&=
\frac{\,\Gamma(a\!+\!b)\,}{\,\Gamma(a)\Gamma(b)\,}
\!\!\int_0^1 \!\!\!\d\xi_1\hp \d\xi_2\hp 
\delta(1\!-\!\xi_1\!-\!\xi_2)\,
\xi_1^{a-1}\xi_2^{b-1}
\!\!\!\int\hsm\!\! \frac{\,\mathrm d^3 \mathbf p\,}{\,(2\pi)^3\,}
\frac{1}{\,(p^2\!+\!\Delta)^{a+b}\,}\,.
\end{aligned}
\end{equation}
}
\hspace{-2.5mm}
The remaining momentum integral is computed in the standard way,
{\small
\begin{equation}
\int\!\!\hsm \frac{\,\mathrm d^3 \mathbf p\,}{\,(2\pi)^3\,}
\frac{1}{\,(p^2\!+\!\Delta)^{a+b}\,}
=
\frac{\,\Gamma\!\(a\!+\!b\!-\!\frac{\,3\,}{\,2\,}\)\,}{\,(4\pi)^{3/2}\hp\Gamma(a\!+\!b)\,}
\Delta^{\frac{\,3\,}{\,2\,}-a-b}.
\end{equation}
}
\hspace{-2.5mm}
Substituting $\Delta\hsm =\hsm\xi_1\xi_2 k_s^2$ into the above integral, 
we derive the integral $\mathcal{J}_{\mathrm{bubble}}$ as follows:
{\small
\begin{equation}
\begin{aligned}
\mathcal{J}_{\mathrm{bubble}}(k_s)
&=
\frac{\,k_s^{\,3-2a-2b}\,}{\,(4\pi)^{3/2}\,}
\frac{\,\Gamma\!\(a\!+\!b\!-\!\frac32\)\,}{\,\Gamma(a)\Gamma(b)\,}\!\!
\int_0^1 \!\!\!\mathrm d\xi_1\, \mathrm d\xi_2\,
\delta(1\!-\!\xi_1\!-\!\xi_2)\,
\xi_1^{\frac32-b-1}\xi_2^{\frac32-a-1}.
\end{aligned}
\end{equation}
}
\hspace{-2.5mm}
We compute the Feynman-parameter integral, 
{\small
\begin{equation}
\int_0^1 \!\!\mathrm d\xi_1\, \mathrm d\xi_2\,
\delta(1\!-\!\xi_1\!-\!\xi_2)\,
\xi_1^\alpha \xi_2^\beta
=\frac{\,\Gamma(1\!+\!\alpha)\Gamma(1\!+\!\beta)\,}{\Gamma(2\!+\!\alpha\!+\!\beta)}\,,
\end{equation}
}
\hspace{-2.5mm}
which leads to 
{\small
\begin{equation}
\label{loop_integral_form}
\mathcal{J}_{\mathrm{bubble}}(k_s)
=
\frac{\,\Gamma\!\(a\!+\!b\!-\!\frac{\,3\,}{\,2\,}\)
\Gamma\!\(\frac{\,3\,}{\,2\,}\!-\!b\)
\Gamma\!\(\frac{\,3\,}{\,2\,}\!-\!a\)\,}
{(4\pi)^{3/2}\Gamma(a)\Gamma(b)\Gamma(3\!-\!a\!-\!b)}
k_s^{\,3-2a-2b}.
\end{equation}
}
\hspace{-2.5mm}
We note that the integral \eqref{loop_integral_form} is first derived in the convergence domain 
\eqref{eq:bubble_convergence_domain}, and then it can be obtained for other domains by analytic continuation.\ 

\vs 

It is useful to compare this analytic-continuation prescription with other regularization schemes.\ 
If we introduce a sharp UV cutoff, $|\mathbf q|\!<\!\Lambda_{\mathrm{UV}}$, then the large-$q$ expansion 
yields the following: 
{\small
\begin{equation}
\label{eq:bubble_cutoff_structure}
\mathcal{J}_{\Lambda}(k_s)
=
\frac{\Lambda_{\mathrm{UV}}^{\,3-2a-2b}}{\,2\pi^2(3\!-\!2a\!-\!2b)\,}
\!+\!\frac{~b(2b\!-\!1)k_s^2\Lambda_{\mathrm{UV}}^{\,1-2a-2b}\,}{\,6\pi^2(1\!-\!2a\!-\!2b)\,}
+\cdots
+\mathcal{J}_{\mathrm{NLoc}}(k_s)\,,
\end{equation}
}
\hspace{-2.5mm}
where the omitted terms are higher-order analytic functions of $k_s^2\hp$.\ 
These cutoff-dependent terms can be absorbed into counterterms as generated by local operators.\ 
Then the remaining non-analytic part, $\mathcal{J}_{\mathrm{NLoc}}(k_s)$, 
just corresponds to Eq.\eqref{loop_integral_form} and 
is independent of the regularization scheme,    
which contributes to the cosmological collider signal.

\vs 

With the dimensional regularization, we continue the spatial dimension to $D\!=\!3\!-\!\epsilon$ 
and consider the following integral:
{\small
\begin{equation}
\label{eq:bubble_DR_appendix}
\mathcal{J}_{\mathrm{bubble}}^{(D)}
=
\mu^\epsilon\hsm 
\!\!\int\!\!\! \frac{\,\mathrm d^D \mathbf q\,}{(2\pi)^D}\,
(\mathbf q^2)^{-a}
\big[(\mathbf q\!+\!\mathbf k_s)^2\big]^{-b}.
\end{equation}
}
\hspace{-2.5mm}
Using the standard massless two-point integral, we compute the above integral and deduce the following:   
{\small
\begin{equation}
\label{eq:bubble_DR_result_appendix}
\mathcal{J}_{\mathrm{bubble}}^{(D)}
= 
\frac{~\mu^\epsilon (k_s^2)^{\frac{\,D\,}{\,2\,}-a-b}\,}{\,(4\pi)^{D/2}\,}
\frac{\,
\Gamma\!\hsm\(\hsm\frac{\,D\,}{\,2\,}\!-\!a\)\hsm 
\Gamma\!\hsm\(\hsm\frac{\,D\,}{\,2\,}\!-\!b\)\hsm 
\Gamma\!\hsm\(a\!+\!b\!-\!\frac{\,D\,}{\,2\,}\hsm\)
\,}{
\Gamma(a)\Gamma(b)\Gamma(D\!-\!a\!-\!b)
}.
\end{equation}
}
\hspace{-2.5mm}
In this scheme, the power-divergent analytic terms do not appear as explicit $1/\epsilon$ poles.\ 
This is expected for the dimensional regularization: 
the power-divergent analytic contributions do not appear, 
whereas the non-analytic momentum dependence is retained.\ 

\vs 

The important point is that different regularization schemes affect only the analytic UV-sensitive part 
of the bubble integral.\ In the cutoff scheme, these appear as analytic terms such as 
those shown in Eq.\eqref{eq:bubble_cutoff_structure}, whereas in the dimensional regularization 
the corresponding power-divergent analytic terms do not exist.\ 
In contrast, the non-analytic dependence on $k_s$ is captured by the expression 
\eqref{loop_integral_form} which is obtained from the analytic continuation.

For the integrals 
{\small
\begin{equation}
\label{eq:bubble_master_appendix_other}
\mathcal{J}_{\rm{bubble}}^{}(k_s)
=
\!\!\int\!\! \frac{\,\mathrm d^3 \mathbf q\,}{(2\pi)^3}\,
|\mathbf q|^{-2a}\,
|\mathbf q\!+\!\mathbf k_s|^{-2b} \cos^n(\theta),
\end{equation}
}
one can always transform the $\cos^n(\theta)$ into 
\begin{equation}
    \cos^n\theta=\frac{1}{(qk_s)^n}\left(\mathbf{q}\cdot\mathbf{k}_s\right)^n=\frac{1}{(2qk_s)^n}\left(|\mathbf q\!+\!\mathbf k_s|^2-q^2-k_s^2\right)^n.
\end{equation}
The resulting expression can then be expanded into a finite sum of terms of the form discussed above, reducing the integral to the previously analyzed cases.

Here we give the expressions of the integrals 
for $n\!=\!0, 1, 2$ after the azimuthal angle $\varphi$ has been integrated out,
{\small
\beqs
\begin{align}
\label{eq: Jn}
\mathcal{J}_{0}^{a,b}&=\int^{\pi}_{0}\!\!\frac{\,\sin\theta\rm{d}\theta\,}{(2\pi)^2}\!\!\int^{\infty}_{0}\!\!q^2\rm{d}q\,q^{-2a}|\mathbf{q}\!+\!\mathbf{k}_s|^{-2b}(\cos\theta)^0 \nn
\\[1mm]
&=\frac{\,\Gamma\!\(a\!+\!b\!-\!\frac{\,3\,}{\,2\,}\)
\Gamma\!\(\frac{\,3\,}{\,2\,}\!-\!b\)
\Gamma\!\(\frac{\,3\,}{\,2\,}\!-\!a\)\,}
{(4\pi)^{3/2}\Gamma(a)\Gamma(b)\Gamma(3\!-\!a\!-\!b)}
k_s^{\,3-2a-2b},
\\[3mm]
\mathcal{J}_{1}^{a,b}&=\int^{\pi}_{0}\!\!\frac{\,\sin\theta\rm{d}\theta\,}{(2\pi)^2}\!\!\int^{\infty}_{0}\!\!q^2\rm{d}q\,q^{-2a}|\mathbf{q}\!+\!\mathbf{k}_s|^{-2b}(\cos\theta)^1 \nn
\\[1mm]
&=\frac{1}{\,2k_s\,}\Big(\mathcal{J}_{0}^{\frac{1}{2}+a,-1+b}\!-\!\mathcal{J}_{0}^{-\frac{1}{2}+a,b}\Big)\!-\!\frac{\,k_s\,}{2}\mathcal{J}_{0}^{\frac{1}{2}+a,b}\nn
\\[1mm]
&=
\frac{
2(a\!-\!1)\,
\Gamma(1\!-\!2a)\,
\Gamma(2\!-\!2b)\,
\Gamma(-3\!+\!2a\!+\!2b)
\,k_s^{\,3-2a-2b}}{
(2a\!+\!2b\!-\!5)\,
\Gamma\!\left(\frac{\,1\,}{\,2\,}\!-\!a\right)\,
\Gamma\!\left(\frac{\,1\,}{\,2\,}\!+\!a\right)\,
\Gamma(1\!-\!b)\,
\Gamma\!\left(\frac{\,1\,}{\,2\,}\!-\!a\!-\!b\right)\,
\Gamma(b)\,
\Gamma\!\left(\frac{\,1\,}{\,2\,}\!+\!a\!+\!b\right)
}\,,
\\[3mm]
\mathcal{J}_{2}^{a,b}&=\int^{\pi}_{0}\!\!\frac{\,\sin\theta\rm{d}\theta\,}{(2\pi)^2}\!\!\int^{\infty}_{0}\!\!q^2\rm{d}q\,q^{-2a}|\mathbf{q}\!+\!\mathbf{k}_s|^{-2b}(\cos\theta)^2 \nn
\\[1mm]
=&
\frac{1}{\,4k_s^2\,}\Big(\mathcal{J}_{0}^{-1+a,b}\!-\!2\mathcal{J}_{0}^{a,-1+b}\!+\!\mathcal{J}_{0}^{1+a,-2+b}\Big)\!+\!\frac{1}{\,2\,}\Big(\mathcal{J}_{0}^{a,b}\!-\!\mathcal{J}_{0}^{1+a,-1+b}\Big)\!+\!\frac{\,k_s^2\,}{\,4\,}\mathcal{J}_{0}^{1+a,b}\nn
\\[1mm]
=&
\frac{
\left(3\!-\!2b\!+\!2a(-3\!+\!a\!+\!b)\right)
\Gamma(2\!-\!2a)\,
\Gamma(2\!-\!2b)\,
\Gamma(2a\!+\!2b\!-\!4)\,
k_s^{\,3-2a-2b}
}{
2(a\!+\!b\!-\!3)\,
\Gamma(1\!-\!a)\,
\Gamma(1\!+\!a)\,
\Gamma(1\!-\!b)\,
\Gamma(1\!-\!a\!-\!b)\,
\Gamma(b)\,
\Gamma(a\!+\!b)
}\,,
\end{align}
\eeqs
}
\hspace*{-2.5mm}
where the superscripts $a$ and $b$ in $\mathcal{J}_{i}^{a,b}$ ($i=0,1,2$) denote the exponents of the factors $q$ and $|\mathbf{q}\!+\!\mathbf{k}_s|$, respectively, appearing as $q^{2-2a}$ and $|\mathbf{q}+\mathbf{k}_s|^{-2b}$.

\vs

\subsection{\hspace*{-1.5mm}Soft Cutoff and Extraction of Non-analytic Coefficients}
\label{Soft Cutoff and Extraction of the Non-analytic Coefficient}
\vspace*{1.5mm}

In this subsection, we explicitly explain how the non-analytic clock signal can be identified 
from a cutoff-regulated soft integral.\ Consider the bubble-type momentum integral,
{\small
\begin{equation}
\label{aeq:Icut<ks}
\mathcal{J}_{<\Lambda}(k_s)
=
\int_{|\mathbf q|<\Lambda}
\!\!\frac{\,\td^3 \mathbf{q}\,}{\,(2\pi)^3\,}
|\mathbf q|^{-2a}
|\mathbf q\!+\!\mathbf k_s|^{-2b},
\end{equation}
}
\hspace*{-2.5mm}
where $k_s \!<\! \Lambda \!\ll\! k_1 $ 
with $k_s$ denoting the soft external momentum.\  
Choosing $\mathbf k_s$ to be the $z$-axis direction, we integrate the angular part 
of the integral and derive the following:
{\small
\begin{equation}
\mathcal{J}_{<\Lambda}(k_s) =
\frac{1}{\,8\pi^2(1\!-\!b)k_s~}\!\!
\int_0^\Lambda \!\!\td q\,q^{1-2a}
\!\left[
(q\!+\!k_s)^{2-2b}\!-\!|q\!-\!k_s|^{2-2b}
\right] \hsm, 
\end{equation}
}
\hspace*{-2.5mm}
for $b\!\neq\! 1$ and with $q\!=\!|\mathbf{q}|$.\  
We define a ratio,
{\small
\begin{equation}
L\equiv \frac{\,\Lambda\,}{\,k_s\,}\!>\!1\,,
\qquad
(\beta\equiv 3\!-\!2a\!-\!2b)\hp.
\end{equation}
}
\hspace*{-2.5mm}
Splitting the radial integral at $q\!=\!k_s$ and changing variable to $q\!=\!k_s x$, 
we compute the integral \eqref{aeq:Icut<ks} as follows: 
{\small
\begin{equation}
\mathcal{J}_{<\Lambda}(k_s)
=
\frac{k_s^\beta}{\,8\pi^2(1\!-\!b)\,}
\!\left[
A_1(L)\!-\!A_2\!-\!A_3(L)
\right],
\end{equation}
}
\hspace*{-2.5mm}
which contains three parts, 
{\small
\beqs
\begin{align}
A_1(L)
&=
\!\int_0^L \!\!\!\td x\,x^{1-2a}(1\!+\!x)^{2-2b},
\\
A_2
&=
\!\int_0^1 \!\!\td x\,x^{1-2a}(1\!-\!x)^{2-2b}
=
B(2\!-\!2a,3\!-\!2b)\hp ,
\label{A2define}
\\
A_3(L)
&=
\!\int_1^L \!\!\!\td x\,x^{1-2a}(x\!-\!1)^{2-2b}.
\end{align}
\eeqs
}
\hspace*{-2.5mm}
In the above, $B(z_1,z_2)$ denotes the Euler beta function,
{\small
\begin{equation}
    B(z_1,z_2)=\!\int_0^1 \!\!\td t\, t^{z_1-1}(1\!-\!t)^{z_2-1},
\end{equation}
}
\hspace*{-2.5mm}
which can be expressed in terms of gamma functions, 
{\small
\begin{equation}
B(z_1,z_2)=\frac{~\Gamma(z_1)\Gamma(z_2)~}{\,\Gamma(z_1\!+\!z_2)\,}.
\end{equation}
}
\hspace*{-2.5mm}
\vspace*{-4mm}

The cutoff-regulated expression should not be identified directly with the physical nonlocal contribution,
because the $L$-dependent terms contain the product of the $L$-dependent non-analytic function of $L\!=\!\Lambda/k_s$
and the overall factor $k_s^\beta$ which could result in an analytic function of $k_s^{}$.\ 
After this multiplication, they become analytic powers of the soft momentum $k_s$ 
and should be assigned to contact contributions.\  
The nonlocal clock signal is the cutoff-independent term that remains proportional to $k_s^\beta$.

\vs 

To make this separation explicit, we first isolate the finite part of $A_1(L)\!-\!A_3(L)$.\  
Thus, it is useful to write
\\[-6mm]
{\small
\beqs
\begin{align}
A_1(L)
&=
A_1(\infty)
\!-\!
\int_L^\infty \!\td x\,x^{1-2a}(1\!+\!x)^{2-2b},
\\
A_3(L)
&=
A_3(\infty)
\!-\!
\int_L^\infty \!\td x\,x^{1-2a}(x\!-\!1)^{2-2b},
\end{align}
\eeqs
}
\hspace*{-2.5mm}
with which we derive the following,
{\small
\begin{equation}
A_1(L)\!-\!A_3(L)
=
S_{\rm{finite}}
\!+\!
T_\Lambda(L)\,,
\end{equation}
}
\hspace*{-2.5mm}
where the two terms on the right-hand side are given by 
{\small
\beqs 
\begin{align}
S_{\rm{finite}}
&\equiv
A_1(\infty)\!-\!A_3(\infty)\,,
\\ 
T_\Lambda(L)
&=
-\!\!\int_L^\infty \!\!\!\td x\,x^{1-2a}
\!\left[
(1\!+\!x)^{2-2b}\!-\!(x\!-\!1)^{2-2b}
\right] \!.
\label{aeq:TLambda(L)}
\end{align}
\eeqs 
}
\hspace*{-2.5mm}
Since the remaining $L$-dependent contribution is an endpoint-tail integral, it can be expanded in the large-$x$ region.
The finite part can be evaluated directly.\  
Using $t\!=\!x/(1\!+\!x)$, we evaluate $A_1(\infty)$ as follows:
{\small
\begin{equation}
A_1(\infty)
=
\int_0^\infty \hsm\!\!\td x\,x^{1-2a}(1\!+\!x)^{2-2b}
=
B(2\!-\!2a,2a\!+\!2b\!-\!4)\,.
\end{equation}
}
\hspace*{-2.5mm}
Then, changing the integration variable to $y\!=\!x\!-\!1$ and to $t\!=\!y/(1\!+\!y)$,
we compute $A_3(\infty)$ as follows: 
{\small
\begin{equation}
A_3(\infty)
=
\int_1^\infty \hsm\!\!\td x\,x^{1-2a}(x\!-\!1)^{2-2b}
=
B(3\!-\!2b,2a\!+\!2b\!-\!4)\hp .
\end{equation}
}
\hspace*{-2.5mm}
With the above, we derive the following expression for the $S_{\rm{finite}}$ term: 
{\small
\begin{equation}
S_{\rm{finite}}
=
B(2\!-\!2a,2a\!+\!2b\!-\!4)
\!-\!
B(3\!-\!2b,2a\!+\!2b\!-\!4)\hp .
\end{equation}
}
\hspace*{-2.5mm}
Since the full cutoff-regulated integral contains $A_1(L)\!-\!A_2\!-\!A_3(L)$, 
the cutoff-independent finite contribution (relevant for the nonlocal term) contains both 
$S_{\rm{finite}}$ and $A_2\hp$,
{\small
\begin{align}
S_{\rm{finite}}\!-\!A_2
=&\,
B(2\!-\!2a,2a\!+\!2b\!-\!4)
\!-\!
B(3\!-\!2b,2a\!+\!2b\!-\!4)
\!-\!
B(2\!-\!2a,3\!-\!2b)\hp .
\end{align}
}
\hspace*{-2.5mm}
\vspace*{-4mm}

Next, we show that the remaining $L$-dependent tail gives only analytic terms 
after including the multiplicative factor $k_s^\beta$.\  
Since $T_\Lambda(L)$ is an endpoint tail integral, we derive the following expanded expressions:  
{\small
\beqs
\begin{align}
(1\!+\!x)^{2-2b}
&=
x^{2-2b}
\!\left(\hsm\!1\!+\!\frac{\,1\,}{\,x\,}\!\right)^{\!\!2-2b}
\!=x^{2-2b}\sum_{n=0}^{\infty}\!
{2\!-\!2b\choose n} x^{-n},
\\
(x\!-\!1)^{2-2b}
&=
x^{2-2b}
\!\left(\hsm\!1\!-\!\frac{1}{\,x\,}\!\right)^{\!\!2-2b}
\!=x^{2-2b}\sum_{n=0}^{\infty}\!
{2\!-\!2b\choose n} \hsm (-1)^n x^{-n}.
\end{align}
\eeqs
}
\hspace*{-2.5mm}
For the difference between the above two terms, only the odd-power terms survive 
and thus we derive the following: 
{\small
\begin{equation}
x^{1-2a}
\left[
(1\!+\!x)^{2-2b}\!-\!(x\!-\!1)^{2-2b}
\right]
=
2\sum_{m=0}^{\infty}\!\!
{2\!-\!2b\choose 2m\!+\!1}
x^{\beta-1-2m}.
\end{equation}
}
\hspace*{-2.5mm}
Thus, we can derive the endpoint tail integral \eqref{aeq:TLambda(L)} as follows:  
{\small
\begin{equation}
T_\Lambda(L)
=
\sum_{m=0}^{\infty}\!
\frac{~2{2-2b\choose 2m+1}~}{\beta\!-\!2m}
L^{\beta-2m}.
\end{equation}
}
\hspace*{-3mm}
Although this expression is a non-analytic function of $L\!=\!\Lambda/k_s$, 
it does not represent a nonlocal clock term.\  
After multiplying by the overall factor $k_s^\beta$, each term becomes, 
{\small
\begin{equation}
k_s^\beta L^{\beta-2m}
=
k_s^\beta
\!\(\!\!\frac{\,\Lambda\,}{\,k_s\,}\!\!\)^{\!\!\beta-2m}\! 
=
\Lambda^{\beta-2m}k_s^{2m}.
\end{equation}
}
\hspace*{-2.5mm}
Thus, the endpoint-tail contribution to the cutoff-regulated integral is
{\small
\begin{equation}
\mathcal{J}_{\rm{anal}}(\Lambda,k_s)=\frac{k_s^\beta}{\,8\pi^2(1\!-\!b)\,}T_\Lambda(L)
=
\frac{1}{\,4\pi^2(1\!-\!b)\,}
\!\sum_{m=0}^{\infty}\!
\frac{{2-2b\choose 2m+1}}{~\beta\!-\!2m~}
\Lambda^{\beta-2m}k_s^{2m}.
\end{equation}
}
\hspace*{-2.5mm}
This is explicitly analytic in $k_s^2$.\  
We can further sum up the above analytic series as follows: 
{\small
\begin{equation}
\mathcal{J}_{\rm{anal}}(\Lambda,k_s)=
\frac{\Lambda^{\beta}}{~2\pi^2\beta~}
{}_3F_2\!\hsm 
\left[\!\hsm 
\begin{array}{c}
-\frac{1}{\,2\,}\!+\!b,\; b,\; -\frac{3}{2}\!+\!a\!+\!b\\
\frac{3}{\,2\,},\; -\frac{1}{2}\!+\!a\!+\!b
\end{array}
\hsm\middle|\hsm \bigg(\hsm\!\frac{\,k_s\,}{\,\Lambda\,}\!\!\bigg)^{\!\!2}
\right] \! .
\label{analytic}
\end{equation}
}
\hspace*{-2.5mm}
Hence, these endpoint-tail terms are contact contributions associated with the soft-hard separation $\Lambda$
and do not belong to the nonlocal clock signal.

\vs 

Hence, the cutoff-regulated integral is organized as follows:
{\small
\begin{equation}
\mathcal{J}_{<\Lambda}(k_s)
=
\mathcal{J}_{\rm{anal}}(\Lambda,k_s)
\hsm +\hsm 
\mathcal{J}_{\rm{CC}}(k_s)\,,
\end{equation}
}
\hspace*{-2.5mm}
where $\mathcal{J}_{\rm{anal}}(\Lambda,k_s)$ is the $k_s$-analytic endpoint-tail contribution as shown in Eq.\,\eqref{analytic}.\  
The contribution to the cosmological collider signal is given by the cutoff-independent non-analytic term, 
{\small
\begin{equation}
\mathcal{J}_{\rm{CC}}(k_s)
=
C_{\rm{CC}}^{}\,k_s^\beta \,.
\end{equation}
}
\hspace*{-2.5mm}
The coefficient $C_{\rm{CC}}^{}$ is fixed by the finite combination $S_{\rm{finite}}\!-\!A_2$ as follows: 
{\small
\begin{align}
C_{\rm{CC}}
=
\frac{1}{\,8\pi^2(1\!-\!b)\,}\!
\bigg[\hsm 
&
B(2\!-\!2a,2a\!+\!2b\!-\!4)
\!-\!
B(3\!-\!2b,2a\!+\!2b\!-\!4)
\!-\!
B(2\!-\!2a,3\!-\!2b)
\hsm\bigg].
\end{align}
}
\hspace*{-2.5mm}
Using Gamma-function identities, we further simplify this expression to the following form: 
{\small
\begin{equation}
C_{\rm{CC}}
=
\frac{~
\Gamma\!\left(a\!+\!b\!-\!\frac{\,3\,}{2}\right)
\Gamma\!\left(\frac{\,3\,}{2}\!-\!a\right)
\Gamma\!\left(\frac{\,3\,}{2}\!-\!b\right)
~}{(4\pi)^{3/2}\Gamma(a)\Gamma(b)\Gamma(3\!-\!a\!-\!b)
}.
\end{equation}
}
\hspace*{-2.5mm}
This agrees with the coefficient obtained from the full momentum integral
{\small
\begin{equation}
\mathcal{J}(k_s)
=
\int
\!\!\frac{\td^3q}{(2\pi)^3}
|\mathbf q|^{-2a}
|\mathbf q\!+\!\mathbf k_s|^{-2b}
=
C_{\rm{CC}}\,k_s^{3-2a-2b},
\end{equation}
}
\hspace*{-2.5mm}
which we derived in Eq.\eqref{loop_integral_form}.\ 
In the above, the integral is evaluated in its convergence domain and then analytically 
continued to the physical values of $a$ and $b\hp$. 

\vs 

Thus, the cutoff construction is useful for exposing the soft-region origin of the factorized clock signal.\ 
But, the physical non-analytic coefficient 
is cutoff-independent and is proportional to $k_s^{3-2a-2b}$, 
whereas all the $L$-dependent endpoint-tail terms become analytic powers of $k_s$ after combining with the overall prefactor.\  
For generic complex exponents, this remaining non-analytic term is the cosmological collider clock signal.

\vs 

For the benchmark value $\tilde\nu\!=\!30$ and the branch $a\!=\!b\!=\!\ii\hp2\tilde\nu$, 
the above analysis also illustrates the numerical importance of separating the endpoint tail 
from the genuine nonlocal term.\  
With these, we evaluate a benchmark value of the physical nonlocal coefficient,    
\begin{equation}
|C_{\rm{CC}}|\simeq 2\!\times\! 10^{-6}.
\end{equation}
In contrast, if the endpoint-tail contribution $\mathcal{J}_{\rm{anal}}(\Lambda,k_s)$ were incorrectly included 
into a clock contribution, the ratio of the analytic term to the nonlocal CC signal would take 
the following values for different choices of the cutoff scale:
{\small
\beqs
\begin{align}
&\left|\frac{\,\mathcal{J}_{\rm{anal}}(\Lambda,k_s)\,}{\mathcal{J}_{\rm{CC}}(k_s)}\right|_{\Lambda=k_s}
\!= 2.2\,,
\qquad\qquad\,\,\,\,
\left|\frac{\,\mathcal{J}_{\rm{anal}}(\Lambda,k_s)\,}{\mathcal{J}_{\rm{CC}}(k_s)}\right|_{\Lambda=2k_s}
\!=17\,,
\\
&\left|\frac{\,\mathcal{J}_{\rm{anal}}(\Lambda,k_s)\,}{\mathcal{J}_{\rm{CC}}(k_s)}\right|_{\Lambda=5k_s}
\!\!\!=4.4\!\times\!10^2\,,
\qquad\,\,
\left|\frac{\,\mathcal{J}_{\rm{anal}}(\Lambda,k_s)\,}{\mathcal{J}_{\rm{CC}}(k_s)}\right|_{\Lambda=10k_s}
\!\!\!=4.6\!\times\!10^3\,.
\end{align}
\eeqs
}
\hspace*{-2.5mm}
These estimates are obtained from the exact endpoint-tail expression \eqref{analytic}.\  
It shows that even though the value of the auxiliary cutoff $\Lambda\!\!\sim\hsm\! k_s$ 
is varied within one order of magnitude, the endpoint-tail contribution grows rapidly 
as the $\Lambda$ is increased.\ 
This growth is not a physical enhancement of the cosmological collider signal.\  In fact, 
it just reflects the cutoff-dependent analytic part generated by the artificial soft-hard separation.\  
Hence the cutoff-dependent endpoint tail contribution should not be used to define the clock coefficient.

\section{\hspace*{-1.5mm}Brief Review of Cosmological Collider Physics}
\label{A Brief Review of Cosmological Collider Physics}
\label{app:D}

In this Appendix, we give a brief review of the cosmological collider physics
and define the relevant observables and notations used in the main text.\ 

During inflation, the inflaton field can be decomposed into a homogeneous background $\phi_0(t)$ and quantum fluctuations 
$\delta\phi(t,\mathbf{x})\hp$:
\begin{equation}
\phi(t,\mathbf{x})=\phi_0(t) + \delta\phi(t,\mathbf{x})\hp .
\end{equation}
The background field $\phi_0(t)$ slowly rolls during inflation. On the other hand, the fluctuation 
$\delta\phi$ can be treated as an approximately massless scalar field due to the flatness of the inflaton potential.\  
As a result, in single-field inflation the quantum evolution of $\delta\phi(t,\mathbf{x})$ is nearly Gaussian.

\begin{figure}[H]
\centering
\includegraphics[width=0.50\linewidth]{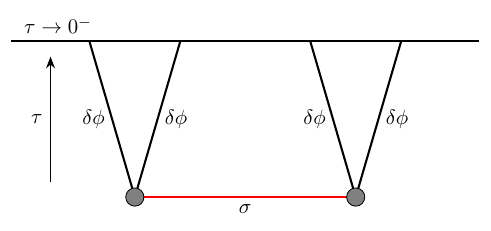}
\vspace*{-3mm}
\caption{\small An illustrative diagram of a four-point correlator of the inflaton $\delta\phi$ with the exchange of a massive scalar $\sigma$, namely $\langle \delta\phi(\mathbf{k}_1)\delta\phi(\mathbf{k}_2)\delta\phi(\mathbf{k}_3)\delta\phi(\mathbf{k}_4)\rangle$. The horizontal line represents the future boundary of de Sitter spacetime, $\tau\rightarrow 0^-$, and the two shaded blobs denote the interaction vertices.}
\label{4ptscalar_illustration}
\label{fig:13}    
\end{figure}

However, the inflaton needs to interact with other fields 
because reheating requires the inflaton decays through the inflation's couplings with 
the Standard Model (SM) fields.\  Such interactions can generate non-Gaussianity in the inflaton fluctuations, characterized by their $n$-point ($n\!\hsm\geqq\!3\hp$) correlators evaluated at the future boundary of inflationary spacetime,
\begin{equation}
\langle\delta\phi(\mathbf{k}_1)\delta\phi(\mathbf{k}_2)\cdots\delta\phi(\mathbf{k}_n)\rangle
\equiv
\lim_{\tau\rightarrow0^-}
\langle\delta\phi(\tau,\mathbf{k}_1)\delta\phi(\tau,\mathbf{k}_2)\cdots\delta\phi(\tau,\mathbf{k}_n)\rangle\,,
\end{equation}
namely in the late-time limit $\tau\rightarrow 0^-$. A schematic example is shown in Fig.\,\ref{4ptscalar_illustration}.

\vs 

In the spatially flat gauge, the primordial curvature perturbation $\zeta$ is related to the inflaton fluctuation by
\begin{equation}
    \zeta \simeq -\frac{\,H^{}\,}{\,\dot{\phi_0}\,} \delta \phi(\mathbf{x})\,.
\end{equation}
Consequently, primordial non-Gaussianity, as characterized by the $n$-point correlators of the curvature perturbation $\zeta$, 
can be directly probed by large-scale structure observations and therefore encodes information about interactions during inflation.\ 
The corresponding $n$-point correlators of the inflaton 
can be computed in the well-known in-in formalism\,\cite{Weinberg2005, Weinberg2006}, 
or equivalently in the Schwinger-Keldysh (SK) path-integral formalism.\  
See \cite{Chen:2010xka,Wang:2013zva,Chen170310166} for reviews.

Depending on the interactions and on the nature of the particles involved, the resulting non-Gaussianity can exhibit a variety of shapes. Among the different types of primordial non-Gaussianity, cosmological collider signals have attracted particular attention in recent years\,\cite{Chen:2009we,Chen:2009zp,Chen:2012ge,Baumann:2011nk,Pi:2012gf,Noumi:2012vr,Gong:2013sma,Arkani-Hamed2015,Chen:2015lza,Baumann160703735,Chen:2016uwp,Kumar:2017ecc,Kumar:2019ebj,Alexander:2019vtb}.\ A cosmological collider (CC) signal, also known as a clock signal, is a non-analytic momentum-dependent contribution to the $n$-point correlators 
$\langle\delta\phi^n\rangle$ in a special kinematic regime. Schematically, this is
\begin{equation}
\label{ccsignal_form}
\begin{aligned}
\langle \delta\phi^n \rangle_\text{soft~limit}
&\thicksim c_1^{} f_\rm{BG}^{}(\mathbf{k})\!+\!c_2^{} f_\rm{CC}^{}(\mathbf{k})\mathbf{P}(\mathbf{k})^{\ii\hp\omega}e^{\ii\hp\varphi} 
\!+\! \rm{H.c.}
\end{aligned}
\end{equation}
In the above, $c_1^{}$ and $c_2^{}$ are constant coefficients, $\omega$ is a constant determined by the mass of the particle, 
and $\varphi$ denotes the relevant phase angle.\ The first term, $f_\text{BG}(\mathbf{k})$, 
denotes the background non-Gaussian contribution, which is analytic in the momenta $\mathbf{k}$ and typically 
takes the form of a low-order polynomial.\ In contrast, the second term, $f_\text{CC}(\mathbf{k})\mathbf{P}(\mathbf{k})^{\ii\hp \omega}$, represents the CC signal.\ 
In Eq.\eqref{ccsignal_form}, $f_\text{CC}(\mathbf{k})$ is also analytic in the momenta, which may contain the information of the particle spin, 
whereas $\mathbf{P}(\mathbf{k})$ denotes a function of the momenta proportional to a soft momentum, which in the 3pt case is a ratio of two characteristic scales 
$k_\rm{long}/k_\rm{short}$.\ The CC signal originates from the non-analytic factor $\mathbf{P}(\mathbf{k})^{\ii\hp \omega}$.

For a three-point correlator
$\langle\delta\phi(\mathbf{k}_1)\delta\phi(\mathbf{k}_2)\delta\phi(\mathbf{k}_3)\rangle$, the \textit{soft limit} refers to the squeezed configuration $k_1 \ll k_2 \simeq k_3$, as shown in Fig.\ref{3ptmomentum}. The long-wavelength mode $k_\text{long}$ corresponds to $k_3$, while the short-wavelength modes $k_\text{short}$ correspond to $k_{1}$ and $k_{2}$. 

For a four-point correlator with the momenta satisfying $\langle\delta\phi(\mathbf{k}_1)\delta\phi(\mathbf{k}_2)\delta\phi(\mathbf{k}_3)\delta\phi(\mathbf{k}_4)\rangle$, with $\mathbf{k}_s\equiv\mathbf{k}_1+\mathbf{k}_2$, the \textit{soft limit} refers to the kinematic configuration $k_s\!\ll\! k_{1}+k_{2}, k_{3}+k_{4}\hp$, as shown in Fig.\ref{4ptmomentum}. The long-wavelength (soft) mode $k_\text{long}$ corresponds to $k_s$, while the short-wavelength (hard) modes $k_\text{short}$ correspond to $k_{1},\,k_{2},\,k_{3}$ and $k_{4}$.

The squeezed limit could cleanly separate the non-analytic CC signals 
from the analytic contributions by the contact interactions and local operators, 
which typically have a smooth power-law dependence on momenta.\

\subsection{\hspace*{-1.5mm}Origin of the Cosmological Collider Signatures}
\label{app:D.1}
\label{The origin of the Cosmological Collider Signal}

In general, cosmological collider signals arise from interactions between the inflaton and massive particles with masses $m \gtrsim H$. As a result, the masses and spins of such particles can be encoded in the resulting CC signals\,\cite{Chen:2009zp,Arkani-Hamed2015}. In this respect, an analogy may be drawn with particle accelerators in flat spacetime, where the presence of intermediate particles is revealed through characteristic structures of scattering amplitudes.\ 
For illustration, we give a diagram showing a 
four-point correlator of the inflaton mediated by a massive scalar boson $\sigma$.

\vs 

To study cosmological collider signals, it is essential to understand the behavior of massive fields propagating in the bulk spacetime of de Sitter, where the conformal time lies in the range of $-\infty<\tau<0$. For a free massive field, the nontrivial de Sitter geometry generally leads to mode functions expressed in terms of special functions, such as Hankel or Whittaker functions. For concreteness, we consider the scalar field $\sigma$ with mass $m$. Its mode function satisfies the equation of motion, 
\begin{equation}
\label{eq:EOM:sigma}
\ddot{\sigma}_k \!+\! 3H \dot{\sigma}_k \!+ \!\frac{k^2}{a^2} \sigma_k \!+\! m^2 \sigma_k = 0\,,
\end{equation}
where $\sigma_k$ denotes the mode function, $H$ is the Hubble parameter during inflation, and $a(\tau)=-1/(H\tau)$ is the scale factor.\ The solution of Eq.\eqref{eq:EOM:sigma} is given by 
\begin{equation}
\sigma_k(\tau) = \frac{\sqrt{\pi\,}H}{2}e^{-\pi\tnu/2}(-\tau)^{3/2}\text{H}_{\ii \hp\tnu}^{(1)}(-k\tau)\,,
\end{equation}
where $\tnu=\sqrt{(m/H)^2\!-\!9/4\,}$, 
and $\text{H}_{n}^{(1)}(z)$ 
denotes the Hankel function of the first kind. In the Schwinger-Keldysh (SK) formalism, the bulk propagators of this field take 
the following form\,\cite{Chen170310166}:
\beqs
\begin{align}
D_{\oplus\oplus }\(k ; \tau_{1}, \tau_{2}\)&=\sigma_k(\tau_1)\sigma^*_k(\tau_2)\,\theta\(\tau_{1}\!-\!\tau_{2}\)\!+\!\sigma^*_k(\tau_1)\sigma_k(\tau_2)\,\theta\(\tau_{2}\!-\!\tau_{1}\), 
\\
D_{\ominus\oplus }\(k ; \tau_{1}, \tau_{2}\)&=\sigma^*_k(\tau_1)\sigma_k(\tau_2), 
\\
D_{\oplus \ominus}\(k ; \tau_{1}, \tau_{2}\)&=\sigma_k(\tau_1)\sigma^*_k(\tau_2), 
\\
D_{\ominus\ominus }\(k ; \tau_{1}, \tau_{2}\)&=\sigma^*_k(\tau_1)\sigma_k(\tau_2)\,\theta\(\tau_{1}\!-\!\tau_{2}\)\!+\!\sigma_k(\tau_1)\sigma^*_k(\tau_2)\,\theta\(\tau_{2}\!-\!\tau_{1}\),
\end{align}
\eeqs
where $\sigma_k(\tau_1)\sigma^*_k(\tau_2)$ and
its complex conjugate are given by
\beqs
\begin{align}
    \sigma_k(\tau_1)\sigma^*_k(\tau_2)&=\frac{\pi H^2}{4} e^{-\pi\tnu} \(\tau_{1} \tau_{2}\)^{3 / 2} \mathrm{H}_{\ii \hp\tnu}^{(1)}(-k \tau_{1}) \mathrm{H}_{-\ii\hp\tnu^{*}}^{(2)}(-k \tau_{2}),
\\
    \sigma^*_k(\tau_1)\sigma_k(\tau_2)&=\frac{\pi H^2}{4} e^{-\pi\tnu} (\tau_{1} \tau_{2})^{3 / 2} \mathrm{H}_{-\ii\hp\tnu^{*}}^{(1)}(-k \tau_{1})\mathrm{H}_{\ii \hp\tnu}^{(2)}(-k \tau_{2}).
\end{align}
\eeqs

For the present study of the cosmological collider signals, 
it is useful to make it clear how the relevant information is encoded in the propagator of a massive field.\ 
The key feature is the non-analytic behavior of the propagator, which contains an imaginary power-law dependence on time and can be extracted in the late-time limit $|k\tau_1|, |k\tau_2| \to 0^-$.\ 
This is also referred to as the soft limit of the propagator.\ 
In this limit, the propagator can be decomposed into two contributions relevant 
for the cosmological collider signals, known as the nonlocal and local parts\,\cite{Tong:2021wai}, 
for instance,
\begin{equation}
\lim_{|k\tau_1|,|k\tau_2|\to 0^-}\!\!D_{\oplus \ominus}(k,\tau_1,\tau_2)
=
    \left[D_{\oplus \ominus}(k,\tau_1,\tau_2)\right]_\text{(NLoc)}
    \!+\!
    \left[D_{\oplus \ominus}(k,\tau_1,\tau_2)\right]_\text{(Loc)}\,,
\end{equation}
where the $\left[D_{\oplus \ominus}(k,\tau_1,\tau_2)\right]_\text{(NLoc)}$ and $\left[D_{\oplus \ominus}(k,\tau_1,\tau_2)\right]_\text{(Loc)}$ are given by 
{\small
\beqs
\begin{align}
\label{local nonlocal scalar propagator}
    \left[D_{\oplus \ominus}(k,\tau_1,\tau_2)\right]_\text{(NLoc)}
    &=
    \frac{\,H^2(\tau_1\tau_2)^{3/2}\,}{4\pi}
    \Bigg[
    \big[\Gamma(-\ii\hp\tnu)\big]^2
    \(\!\frac{\,k^2\tau_1\tau_2\,}{4}\!\)^{\!\!\ii\hp\tnu}
    \!\!+
    \big[\Gamma(\ii\hp\tnu)\big]^2
    \(\!\frac{\,k^2\tau_1\tau_2\,}{4}\!\)^{\!\!-\ii\hp\tnu}
    \Bigg],
    \\
    \left[D_{\oplus \ominus}(k,\tau_1,\tau_2)\right]_\text{(Loc)}
    &=
    \frac{\,H^2(\tau_1\tau_2)^{3/2}\,}{4\pi}
    \Gamma(-\ii\hp\tnu)\Gamma(\ii\hp\tnu)
    \Bigg[
    e^{\pi\tnu}\(\!\frac{\,\tau_1\,}{\tau_2}\!\)^{\!\!\ii\hp\tnu}
    \!\!+
    e^{-\pi\tnu}\(\!\frac{\,\tau_1\,}{\tau_2}\!\)^{\!\!-\ii\hp\tnu}
     \Bigg].
\end{align}
\eeqs
}
\hspace*{-2.5mm}
The nonlocal and local parts have distinct physical interpretations.\ First, the nonlocal part is non-analytic in $k$. In particular, the term proportional to $k^{\ii\hp 2\tnu}$ gives rise, after Fourier transformation, to a long-range correlation of the form 
$|\mathbf{x}_1\hsm -\mathbf{x}_2|^{\ii\hp 2\tnu}$. By contrast, the local part is independent of $k$, and its Fourier transform yields a contact term proportional to 
$\delta^3(\mathbf{x}_1\! -\hsm\mathbf{x}_2)$, as expected for a local contribution.\ 
Second, the time dependence of the nonlocal part reflects the accumulated dynamical phase of two propagating particles, whereas the time dependence of the local part corresponds to the dynamical phase of a single particle propagating from $\tau_2$ to $\tau_1$ (or from $\tau_1$ to $\tau_2$, with an additional suppression factor $e^{-2\pi\tnu}$).

For the inflaton fluctuation, treated as a massless scalar field in de\,Sitter spacetime, 
the mode function takes the form:
\begin{equation}
\delta\phi_{k}^{}(\tau)=\frac{H}{\,\sqrt{2k^3\,}\,}(1\!+\hsm \ii\hp k\tau)e^{-\ii\hp k\tau} \,.
\end{equation}
The corresponding bulk-to-bulk propagator is
given by
\begin{equation}
G_{\oplus \ominus}^{}(k;\tau_1,\tau_2)
=
\delta\phi_{k}(\tau_1)\delta\phi_{k}^*(\tau_2)
=
\frac{\,H^2\,}{\,2k^3\,}
\left[1\!+\hsm \ii\hp k(\tau_1\!-\!\tau_2)\!+\!k^2\tau_1\tau_2\right]
e^{-\ii\hp k(\tau_1^{}\hsm -\tau_2^{})} \,.
\end{equation}

Since we evaluate the correlator $\langle\delta\phi(\mathbf{k}_1)\delta\phi(\mathbf{k}_2)\cdots\delta\phi(\mathbf{k}_n)\rangle$ in the late-time limit $\tau\to 0^-$, we also need the bulk-to-boundary propagators of the inflaton $\delta\phi$, for which one time leg, say $\tau_2$, is taken to the final time slice, namely, $\tau_2=\tau_{\!f}^{}\to 0^-$. They are defined as
\begin{equation}
G_{a}^{}(\mathbf{k},\tau_1) \equiv \lim_{\tau_{\!f}\rightarrow0} G_{a\oplus}^{}(\mathbf{k};\tau_1,\tau_{\!f}) \hp.
\end{equation}
Then, the plus- and minus-type bulk-to-boundary propagators take the following form:
{\small
\beqs
\label{inflaton_bulk-to-boundary}
\begin{align}
G_{\oplus}^{}(\mathbf{k},\tau) &=
\lim_{\tau_{\!f}\rightarrow0} \frac{H^2}{\,2k^3\,}
\left[1\!-\!\ii\hp k(\tau\!-\!\tau_{\!f}^{})\!+\! k^2\tau\tau_{\!f}^{}\right]
e^{\ii\hp k(\tau-\tau_{\!f}^{})}
=
\frac{H^2}{\,2k^3\,}\left[1\!-\!\ii\hp k\tau\right]e^{\ii\hp k\tau}\hp,
\\
G_{\ominus}^{}(\mathbf{k},\tau) &=\lim_{\tau_{\!f}\rightarrow0} 
\frac{H^2}{\,2k^3\,}
\left[1\!+\!\ii\hp k(\tau\!-\!\tau_{\!f}^{})\!+\! k^2\tau\tau_{\!f}^{}\right]
e^{-\ii\hp k(\tau-\tau_{\!f}^{})}
=
\frac{H^2}{\,2k^3\,}\left[1\!+\!\ii\hp k\tau\right]e^{-\ii\hp k\tau}\hp.
\end{align}
\eeqs
}
\hspace{-2.5mm}
Hence, in the late-time limit, the massive field oscillates as 
$\tau^{\ii \hp\tnu}\!\sim\! e^{-\ii\hp\tnu H t}$, since in de Sitter spacetime the conformal time and cosmic time are related by 
$\tau\!=\!-H^{-1}e^{-Ht}$, whereas the massless inflaton mode oscillates as $e^{\ii\hp k\tau}$. During inflation, interactions between the inflaton and the heavy field can give rise to a resonance production or 
decay of the heavy particle\,\cite{Chen:2015lza,Tong:2021wai}.\ 
For computing the correlator induced by such a heavy field, one needs to deal with the following integral:
\begin{equation}
\int\! \d\tau\,\tau^{\alpha}
e^{\ii\hp k\tau}\tau^{\pm \ii\hp\tnu}
    \sim
    \frac{\,\sqrt{2\pi\tnu\,}\,}{k}e^{\pm \ii\hp\frac{\pi}{4}}\tau_*^{\alpha}
    e^{\ii\hp k\tau_*}\tau_*^{\pm \ii\hp\tnu}\,,
\end{equation}
where $\alpha$ is a constant, and the dominant contribution comes from the stationary point $\tau_*$ satisfying $|k\tau_*|\!\simeq\! \tnu$. Physically, $\tau_*$ corresponds to the resonance time, at which a resonant process takes place, such as the decay of a massive particle into massless inflaton modes, or, conversely the production of a heavy particle from inflaton modes through the interaction.\ Through this resonance, the oscillatory behavior of the heavy-particle mode function, characterized by $\tau^{\ii \hp\tnu}$, is transferred to the oscillatory behavior of the cosmological collider signal, which then appears as a non-analytic momentum dependence of the form $k^{\ii \hp\tnu}$, as indicated in Eq.\eqref{ccsignal_form}.

As an on-shell effect, the cosmological collider signal may be viewed as the inflationary analog of the mass-pole structure associated with Mandelstam variables in the flat-spacetime scattering amplitudes.\ One of its key properties, namely the oscillation frequency $\omega$ in Eq.\eqref{ccsignal_form}, is insensitive to the detailed form of the couplings between the inflaton and the heavy field, and depends primarily on the heavy-particle mass, $\omega=\tnu\!\sim\! m/H$. For this reason, the cosmological collider signals provide a nearly model-independent probe of very massive particles in the early universe.

In contrast, the overall size of the signal, namely the coefficient 
$c_2^{} f_{\rm{CC}}^{}(\mathbf{k})$ 
in Eq.\eqref{ccsignal_form}, generally does depend on the details of the interaction and thus can carry complementary information about the coupling structure.\ We will return to this point in the following sections.

\subsection{\hspace*{-1.5mm}Local and Nonlocal Cosmological Collider Signatures}
\label{app:D.2}
\label{The Local and the nonlocal Signal}

As discussed in the previous section, the late-time behavior of the propagator for a massive field can be separated into two parts, namely the local and nonlocal contributions, as shown in Eq.\eqref{local nonlocal scalar propagator}. Correspondingly, cosmological collider signals can also be classified into two types: local signals and nonlocal signals.

\begin{figure}[t]
\centering
\includegraphics[width=0.4\linewidth]{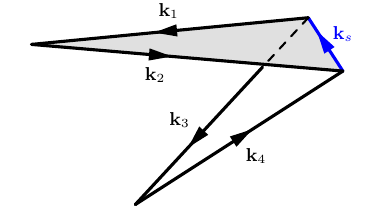}
\vspace*{-3.5mm}
\caption{\small  Illustration of the momentum configuration of the four-point correlator 
in the squeezed limit.}
\label{4ptmomentum}
\label{fig:14}
\end{figure}

The most commonly studied example is the 
four-point correlator,
$\langle\delta\phi(\mathbf{k}_1)\delta\phi(\mathbf{k}_2)\delta\phi(\mathbf{k}_3)$ $\delta\phi(\mathbf{k}_4)\rangle$, 
with the momentum conservation
$\mathbf{k}_1\!+\mathbf{k}_2\!+\mathbf{k}_3\!+\mathbf{k}_4\!=0\hp$.\
We further define the following:
\begin{equation}
\mathbf{k}_s\equiv\mathbf{k}_1\!+\!\mathbf{k}_2,\qquad r_1\equiv k_s/k_{12},\qquad r_2\equiv k_s/k_{34}\,.
\end{equation}
For the four-point correlator, 
the squeezed limit is given by 
$k_s\!\ll\! k_{12}, k_{34}\hp$, 
or equivalently, $r_1^{},r_2^{}\!\ll\!1\hp$.\ 
This squeezed configuration is illustrated in Fig.\,\ref{4ptmomentum}, where the blue momentum vector $\mathbf{k}_s$ is highly compressed.

\vs 

The local and nonlocal signals exhibit different momentum dependences in the 
four-point correlator.\ For the nonlocal signal, the function $\mathbf{P}(\mathbf{k})$ 
in Eq.\eqref{ccsignal_form} takes 
the form of
\begin{equation}
    \mathbf{P}(\mathbf{k})=\frac{\,k_s^2\,}{k_{12}k_{34}}=r_1r_2\,.
\end{equation}
In contrast, for the local signal we have
\begin{equation}
    \mathbf{P}(\mathbf{k})=\frac{\,k_{34}\,}{k_{12}}=\frac{\,r_1\,}{r_2}\,.
\end{equation}
In general, the squeezed limit behavior of the four-point correlator can be written as follows:
\begin{equation}
\langle\delta\phi^4\rangle_\text{squeezed}
\thicksim
A_{\mathrm{NLoc}}\(r_{1}, r_{2}\)\(r_{1} r_{2}\)^{ \pm \ii \hp \omega}
+
A_{\mathrm{Loc}}(r_{1}, r_{2})
\!\(\!\frac{\,r_{1}\,}{r_{2}\!}
\)^{\!\pm\ii\hp\omega}
\!\!+
A_{\mathrm{BG}}(r_{1}, r_{2})\,,
\end{equation}
where $A_{\mathrm{NLoc}}$, $A_{\mathrm{Loc}}$, and $A_{\rm{BG}}$ are analytic functions of $r_1$ and $r_2\hp$.

\vs 
The local and nonlocal signals have distinctive physical origins.\
This distinction can be understood most transparently 
from the resonance picture discussed in Section\,\ref{app:D.1}.\ 
In this picture, the nonlocal signal originates from the gravitational production of 
a pair of massive particles, followed by their resonant decay into inflaton modes 
at separated interaction vertices.\
This contribution is associated with the nonlocal part of the massive-field propagator,
\begin{equation}
\left[D_{\oplus \ominus}(\mathbf{k},\tau_1,\tau_2)\right]_{\rm{NLoc}}
\sim
\left(\!\frac{\,k_s^2\tau_1\tau_2\,}{4}\!\right)^{\!\!\ii\hp\tilde{\nu}} .
\end{equation}
After imposing the resonance conditions at the two interaction vertices, 
this time dependence is mapped into the momentum-space clock signal,
\begin{equation}
\left(\!\frac{k_s^2}{\,k_{12}k_{34}\,}\!\right)^{\!\!\ii\hp\tilde{\nu}} .
\end{equation}
By contrast, the local signal corresponds to the resonant production and 
subsequent decay of a single massive particle at interaction vertices.\  
It is associated with the local part of the massive-field propagator,
\begin{equation}
\left[D_{\oplus \ominus}(\mathbf{k},\tau_1,\tau_2)\right]_{\rm{Loc}}
\sim
\left(\!\frac{\,\tau_1\,}{\,\tau_2\,}\!\right)^{\!\!\ii\hp\tilde{\nu}} .
\end{equation}
The resonance conditions then map this time dependence into the local clock signal,
\begin{equation}
\left(\!\frac{\,k_{12}\,}{\,k_{34}\,}\!\right)^{\!\!\ii\hp\tilde{\nu}} .
\end{equation}
Hence, the local and nonlocal signals arise from different bulk histories, 
and their oscillatory phases track different intervals of massive-particle propagation 
in the inflationary bulk.\ 
Most previous studies have focused on extracting the nonlocal signals in the 
four-point correlator.

Compared with the four-point case, the structure of the three-point signal is 
more subtle.\ 
A useful way to understand this is to connect 
the three-point correlator to the folded limit of the four-point correlator 
by taking $k_4^{}\!\ito 0\hp$.\ 
In this limit, the momentum ratio 
$r_2^{}\!=\hsm k_s/k_{34}$ approaches unity,
\begin{equation}
    \lim_{k_4\to 0}r_2
    =
    \lim_{k_4\to 0}\frac{\,k_s\,}{k_{34}}
    =
    \lim_{k_4\to 0}\frac{\,|\mathbf{k}_3\!+\!\mathbf{k}_4|\,}{k_3\!+\!k_4}
    =1\,.
\end{equation}
In consequence, both the local and nonlocal contributions become functions of the same momentum ratio $r_1=k_3/k_{12}$.\ 
In particular, after taking the folded limit, both contributions exhibit the same characteristic 
non-analytic dependence, $r_1^{\pm i\omega}$.\ 
Hence, unlike in the four-point correlator, the local and nonlocal signals in the three-point correlator 
can no longer be distinguished from their squeezed limit momentum-dependence alone.

\vs 

\begin{figure}[t]
\centering
\includegraphics[width=0.45\linewidth]{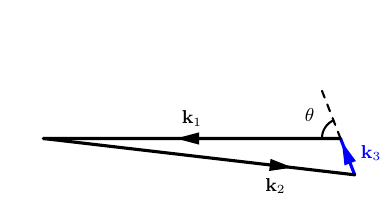}
\vspace*{-3.5mm}
\caption{\small  The momentum configuration of the three-point correlator in the squeezed limit.}
\label{3ptmomentum}
\label{fig:15}
\end{figure}

Accordingly, the squeezed limit behavior of the three-point correlator can be written 
as follows:
\begin{equation}
\langle\delta\phi^3\rangle_\text{squeezed}
\thicksim
\big[B_{\mathrm{NLoc}}^{}(r_1^{})\!+\!B_{\mathrm{Loc}}^{}(r_1^{})\big]
r_{1}^{ \pm\ii\hp\omega} \hsm +\hsm
B_{\rm{BG}}^{}(r_1^{})\hp ,
\end{equation}
where $B_{\rm{NLoc}}^{}$, $B_{\mathrm{Loc}}^{}$, 
and $B_{\mathrm{BG}}$ are analytic functions of $r_1^{}$.\ 
In the above,  $B_{\mathrm{NLoc}}^{}$ and $B_{\mathrm{Loc}}^{}$
denote the overall amplitude of the cosmological collider signal, 
which in general receives contributions from both the local and nonlocal parts.\ 
The squeezed limit of the three-point correlator corresponds to
\begin{equation}
r_1=\frac{k_3}{\,k_{12}\,}\to 0\,,
\end{equation}
which is illustrated in Fig.\,\ref{3ptmomentum}.

\vs 

Nevertheless, the local and nonlocal contributions remain physically and calculationally distinct.\ 
They may be defined separately by extracting the local or nonlocal part of the massive propagator, 
or equivalently by taking the folded limit of the corresponding local and nonlocal pieces of the four-point correlator.\  
What is lost in the three-point limit is not the distinction between their physical origins, 
but only the kinematic distinction in their final momentum-dependence.\

\newpage


\bibliographystyle{JHEP}
\bibliography{hinflation}

\end{document}